\documentclass[11pt,epic,eepic,psfig,epsfig]{article}
\usepackage{epic,eepic,epsfig,amssymb}
\usepackage{latexsym,amssymb}

\newcommand{\old}[1]{{}}

\newcommand{\Proc}[1]{#1\+}

\newcommand{\Endstep}[1]{\<\-}
\newcommand{\Returns}{{\small\bf returns}}
\newcommand{\Inputs}{{\small\bf input}}
\newcommand{\Procbegin}{{\small\bf begin}}
\newcommand{\If}{{\small\bf if}\ \=\+}
\newcommand{\Then}{{\small\bf then}\ \=\+}
\newcommand{\Else}{\<{\small\bf else}\ \>}
\newcommand{\Elseif}{\<\<{\small\bf elseif}\-\-\ \=\+}
\newcommand{\Endif}{\<\<{\small\bf end\ if\ }\-\-}

\newcommand{\Endproc}[1]{{\small\bf end} #1\-}

\newcommand{\For}{{\small\bf for}\ \=\+}
\newcommand{\Forevery}{{\small\bf for\ every}\ \=\+}

\newcommand{\While}{{\small\bf while}\ \=\+}

\newcommand{\Endfor}{\<{\small\bf end\ for}\ \-}

\newcommand{\Endwhile}{\<{\small\bf end\ while}\ \- }

\newenvironment{program}{
    \begin{minipage}{\textwidth}
    \begin{tabbing}
    \ \ \ \ \=\kill
  }{
    \end{tabbing}
    \end{minipage}
  }

\newcommand{\beq}{\begin{equation}}
\newcommand{\eeq}{\end{equation}}
\newcommand{\be}{\begin{enumerate}}
\newcommand{\ee}{\end{enumerate}}
\newcommand{\bi}{\begin{itemize}}
\newcommand{\ei}{\end{itemize}}
\newcommand{\bc}{\begin{center}}
\newcommand{\ec}{\end{center}}

\newtheorem{theorem}{Theorem}[section]
\newtheorem{lem}[theorem]{Lemma}
\newtheorem{defin}[theorem]{Definition}
\newtheorem{cor}[theorem]{Corollary}
\newtheorem{obs}[theorem]{Observation}
\newtheorem{remark}[theorem]{Remark}
\newtheorem{prop}[theorem]{Property}
\newtheorem{propos}[theorem]{Proposition}

\def\endpf{{\ \hfill\hbox{\vrule width1.0ex height1.0ex}\parfillskip 0pt}}
\newenvironment{prf}{\noindent{\bf Proof:}}{\endpf}

\begin{document}
\title{Minimum diameter and cycle-diameter orientations on planar
graphs\thanks{This is part of the Ph.D. dissertation 
``Approximation algorithms for three optimization problems on graphs''
by Nili Guttmann-Beck, 2005.  The special case of series-parallel
graphs is solved in \cite{GH11}.}}

 \author{Nili Guttmann-Beck\thanks{ Department of Computer Science, The Academic College of Tel-Aviv Yaffo, Yaffo, Israel.
 Email: becknili@mta.ac.il ;}
\and Refael Hassin\thanks{
School of Mathematical Sciences, Tel-Aviv University,
Tel-Aviv 69978, Israel.  Email: hassin@post.tau.ac.il.
} }

\date{October 2005}
 \maketitle

\begin{abstract}
Let $G$ be an edge weighted undirected graph. For every pair of
nodes consider the shortest cycle containing these nodes in $G$. The
cycle diameter of $G$ is the maximum length of a cycle in this set.
Let $H$ be a directed graph obtained by directing the edges of $G$.
The cycle diameter of $H$ is similarly defined except for that
cycles are replaced by directed closed walks. Is there always an
orientation $H$ of $G$ whose cycle diameter is bounded by a constant
times the cycle diameter of $G$?  We prove this property for
 planar
graphs.  These results have implications on the problem of
approximating an orientation with minimum diameter.
\end{abstract}

\section{Hereditary order on cycles}

 Let $G=(V,E)$ be a 2-edge connected  undirected graph. Choose one of its nodes,
mark it as $z$. For every node $v\in V\backslash\{z\}$ find the
shortest undirected cycle connecting $v$ and $z$, mark this cycle
as $C(v)$, we say that $v$ is served by $C(v)$. Let $\mathcal{C} =
\{C(v) | v \in V \}$ and let $G_\mathcal{C}$ be the graph induced
by the edges in $\mathcal{C}$. Let $V_1,\ldots,V_l$ be the node
sets of the connected components of $G_\mathcal{C}$ induced by $V
\backslash \{z \}$. Let $G_i \ i=1,\ldots,l$ be the subgraphs of
$G_\mathcal{C}$ induced by $V_i \cup \{ z \}$. Each of theses
subgraphs will be oriented independently of the others. Clearly,
the bound holds for the whole graph if it holds for every
component. Hence w.l.o.g we can assume that $l=1$.

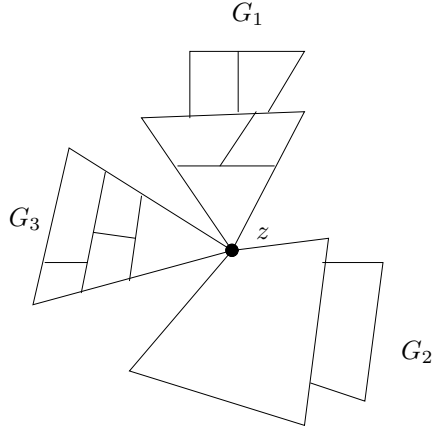
\begin{figure}
\begin{center}
\setlength{\unitlength}{0.00069991in}
\begingroup\makeatletter\ifx\SetFigFont\undefined%
\gdef\SetFigFont#1#2#3#4#5{%
  \reset@font\fontsize{#1}{#2pt}%
  \fontfamily{#3}\fontseries{#4}\fontshape{#5}%
  \selectfont}%
\fi\endgroup%
{\renewcommand{\dashlinestretch}{30}
\begin{picture}(3185,3192)(0,-10)
\put(1680,1317){\blacken\ellipse{90}{90}}
\put(1680,1317){\ellipse{90}{90}}
\path(1680,1317)(1005,2307)(2220,2352)(1680,1317)
\path(1275,1947)(1995,1947)
\path(1590,1947)(1860,2352)
\path(1365,2307)(1365,2802)(2220,2802)(1950,2352)
\path(1725,2352)(1725,2802)
\path(2355,1227)(2805,1227)(2670,192)(2265,327)
\path(1680,1317)(465,2082)(195,912)(1680,1317)
\path(735,1902)(555,1002)
\path(1005,1722)(915,1092)
\path(645,1452)(960,1407)
\path(285,1227)(600,1227)
\path(1680,1317)(2400,1407)(2220,12)
	(2220,12)(915,417)(1680,1317)
\put(1680,3027){\makebox(0,0)[lb]{\smash{{\SetFigFont{10}{12.0}{\rmdefault}{\mddefault}{\updefault}$G_1$}}}}
\put(2940,507){\makebox(0,0)[lb]{\smash{{\SetFigFont{10}{12.0}{\rmdefault}{\mddefault}{\updefault}$G_2$}}}}
\put(15,1497){\makebox(0,0)[lb]{\smash{{\SetFigFont{10}{12.0}{\rmdefault}{\mddefault}{\updefault}$G_3$}}}}
\put(1860,1407){\makebox(0,0)[lb]{\smash{{\SetFigFont{10}{12.0}{\rmdefault}{\mddefault}{\updefault}$z$}}}}
\end{picture}
}
\end{center}
\caption{$G_1$,$G_2$ and $G_3$.} \label{gdeffig}
\end{figure}

\noindent In Figure \ref{gdeffig}  $G_1,G_2$ and $G_3$ are
illustrated.

\noindent For every edge $e \in E$ let $l(e) \geq 0$ be the length
of the edge.
\be
\item A path $P$ is an ordered  set of nodes $(v_1, \ldots, v_n)$ and distinct
edges $(v_1,v_2),(v_2,v_3), \ldots, (v_{n-1},v_n)$. When
$v_1=v_n$, $P$ is a cycle.
\item For every $G'$ a subgraph of $G$, let $E(G')$ $(V(G'))$ be the subgraph's edge set (node set).
\item For every path $P$, $l(P)= \sum_{e \in E(P)} l(e)$.
\item For every two subgraphs $P$ and $Q$,  $P \backslash Q$ is
the subgraph induces by $E(P) \backslash E(Q)$.
\item For every two nodes $a,b$ on a given path $P$, let $[a,b]$ be the subpath between
them (the identity of $P$ will be clear from the context).
 \ee

\noindent Also assume w.l.o.g that for every $C_1,C_2 \in
\mathcal{C}$, $l(C_1) \neq l(C_2)$.

\noindent Let {\em $C_0$} be the shortest cycle in $\mathcal{C}$.

\begin{lem}
\label{goodc0} Let $C \in \mathcal{C}$ . \bi
\item $C_0 \backslash C$ is a path (or a cycle) with at most two nodes in $C_0 \cap C$.
\item $C \backslash C_0$ is a path (or a cycle) with at most two nodes in $C_0 \cap C$.
\ei
\end{lem}

\begin{prf}
Similar to the proof of Lemma \ref{goodC}.
%
%
\end{prf}
\begin{defin}
\label{sond} {\em Given two nodes $u \neq v\in V(C_0) $ let $P$ be
the $v-u$ path in $C_0$ which includes $z$. Define $S_{u,v}:= \{ C
\in \mathcal{C} | C \cap C_0 = P \}$. If $S_{u,v} \neq \phi$ then
the shortest cycle in $S_{u,v}$ is the {\em son} of $C_0$ {\em
defined by $S_{u,v}$}.
Given a node $v \in V(C_0)$, define $S_{v,v}$ to be the set $ \{ C
\in \mathcal{C} , C \neq C_0| C \backslash C_0$ is a cycle
touching $C_0$ in the node $v \}$. If $S_{v,v}\neq \phi $ then the
shortest cycle in $S_{v,v}$ is the {\em son} of $C_0$ {\em defined
by $S_{v,v}$.}}
\end{defin}

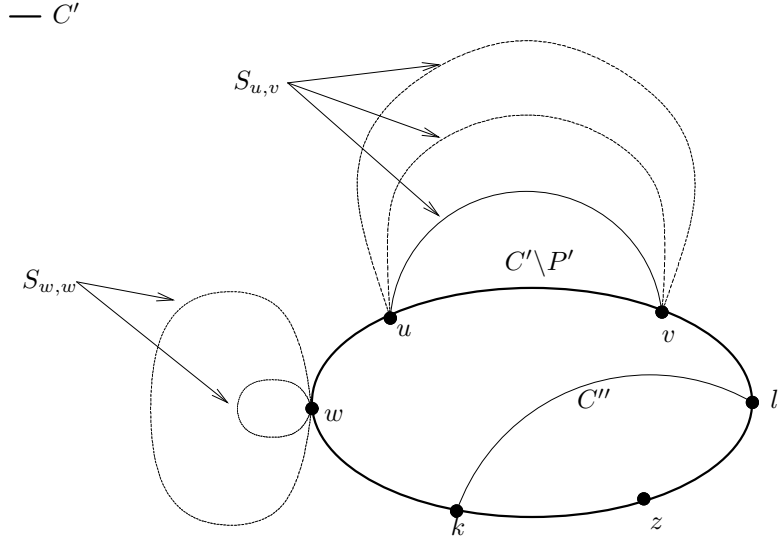
\begin{figure}
\begin{center}
\setlength{\unitlength}{0.00069991in}
\begingroup\makeatletter\ifx\SetFigFont\undefined%
\gdef\SetFigFont#1#2#3#4#5{%
  \reset@font\fontsize{#1}{#2pt}%
  \fontfamily{#3}\fontseries{#4}\fontshape{#5}%
  \selectfont}%
\fi\endgroup%
{\renewcommand{\dashlinestretch}{30}
\begin{picture}(5838,4007)(0,-10)
\put(4877.000,-318.750){\arc{3057.803}{3.4843}{5.2364}}
\put(3956.604,1562.834){\arc{2034.333}{3.2126}{6.1677}}
\put(3437,195){\blacken\ellipse{90}{90}}
\put(3437,195){\ellipse{90}{90}}
\put(5642,1005){\blacken\ellipse{90}{90}}
\put(5642,1005){\ellipse{90}{90}}
\put(4832,285){\blacken\ellipse{90}{90}}
\put(4832,285){\ellipse{90}{90}}
\put(2357,960){\blacken\ellipse{90}{90}}
\put(2357,960){\ellipse{90}{90}}
\put(2942,1635){\blacken\ellipse{90}{90}}
\put(2942,1635){\ellipse{90}{90}}
\thicklines
\put(3999,1005){\ellipse{3284}{1710}}
\thinlines
\put(4967,1680){\blacken\ellipse{90}{90}}
\put(4967,1680){\ellipse{90}{90}}
\thicklines
\path(107,3885)(332,3885)
\thinlines
\path(2177,3390)(3302,3525)
\path(3186.429,3480.916)(3302.000,3525.000)(3179.280,3540.489)
\path(2177,3390)(3302,2985)
\path(3178.932,2997.420)(3302.000,2985.000)(3199.255,3053.873)
\path(2177,3390)(3302,2400)
\path(3192.096,2456.754)(3302.000,2400.000)(3231.733,2501.797)
\path(602,1905)(1322,1770)
\path(1198.527,1762.628)(1322.000,1770.000)(1209.584,1821.601)
\path(602,1905)(1727,1005)
\path(1614.555,1056.537)(1727.000,1005.000)(1652.037,1103.389)
\dashline{1200.000}(2942,1635)(2941,1637)(2940,1642)
	(2937,1651)(2932,1665)(2926,1684)
	(2917,1708)(2908,1737)(2897,1770)
	(2885,1807)(2872,1846)(2858,1886)
	(2845,1928)(2832,1968)(2819,2008)
	(2807,2047)(2796,2084)(2786,2119)
	(2776,2153)(2767,2185)(2759,2215)
	(2751,2244)(2744,2272)(2738,2298)
	(2732,2324)(2727,2350)(2722,2375)
	(2717,2400)(2713,2423)(2709,2447)
	(2706,2471)(2703,2495)(2700,2519)
	(2698,2544)(2697,2569)(2695,2595)
	(2695,2621)(2695,2647)(2696,2674)
	(2698,2701)(2700,2728)(2704,2756)
	(2708,2784)(2713,2811)(2719,2839)
	(2726,2866)(2734,2894)(2743,2921)
	(2753,2947)(2764,2974)(2775,3000)
	(2788,3025)(2802,3051)(2817,3075)
	(2832,3100)(2849,3124)(2867,3148)
	(2886,3172)(2904,3194)(2922,3215)
	(2942,3237)(2964,3258)(2986,3280)
	(3009,3302)(3034,3323)(3060,3345)
	(3087,3367)(3116,3389)(3145,3411)
	(3176,3432)(3207,3453)(3240,3474)
	(3274,3495)(3308,3514)(3343,3534)
	(3379,3552)(3415,3570)(3451,3587)
	(3488,3602)(3525,3617)(3562,3631)
	(3599,3644)(3635,3655)(3672,3666)
	(3708,3675)(3744,3683)(3780,3690)
	(3816,3695)(3851,3699)(3886,3703)
	(3920,3704)(3955,3705)(3989,3704)
	(4024,3703)(4059,3699)(4094,3695)
	(4130,3690)(4165,3683)(4201,3675)
	(4238,3666)(4274,3655)(4311,3644)
	(4348,3631)(4385,3617)(4421,3602)
	(4458,3587)(4495,3570)(4531,3552)
	(4566,3534)(4601,3514)(4636,3495)
	(4669,3474)(4702,3453)(4734,3432)
	(4764,3411)(4794,3389)(4822,3367)
	(4849,3345)(4875,3323)(4900,3302)
	(4923,3280)(4946,3258)(4967,3237)
	(4987,3215)(5006,3194)(5023,3172)
	(5044,3147)(5063,3121)(5080,3095)
	(5097,3068)(5112,3042)(5126,3014)
	(5140,2987)(5152,2959)(5163,2930)
	(5173,2902)(5181,2873)(5189,2843)
	(5196,2814)(5201,2785)(5206,2755)
	(5209,2726)(5212,2697)(5213,2668)
	(5214,2640)(5214,2612)(5213,2585)
	(5212,2558)(5209,2532)(5207,2507)
	(5204,2481)(5200,2456)(5196,2432)
	(5192,2407)(5187,2381)(5182,2355)
	(5176,2328)(5169,2301)(5162,2273)
	(5155,2244)(5146,2214)(5137,2182)
	(5126,2148)(5115,2113)(5103,2075)
	(5090,2036)(5077,1995)(5063,1953)
	(5049,1912)(5034,1871)(5021,1831)
	(5008,1795)(4997,1763)(4987,1736)
	(4979,1714)(4974,1698)(4970,1688)
	(4968,1683)(4967,1680)
\dashline{1200.000}(2942,1635)(2942,1637)(2941,1642)
	(2941,1651)(2940,1665)(2938,1684)
	(2937,1709)(2934,1740)(2932,1776)
	(2930,1817)(2928,1861)(2925,1909)
	(2923,1958)(2922,2009)(2921,2060)
	(2920,2110)(2920,2159)(2921,2206)
	(2922,2251)(2924,2294)(2927,2335)
	(2931,2374)(2935,2411)(2941,2445)
	(2947,2478)(2954,2509)(2963,2539)
	(2972,2567)(2982,2594)(2994,2620)
	(3007,2645)(3021,2670)(3035,2693)
	(3050,2715)(3066,2738)(3084,2759)
	(3102,2781)(3122,2803)(3143,2824)
	(3165,2845)(3188,2865)(3212,2885)
	(3238,2905)(3264,2925)(3292,2944)
	(3320,2962)(3350,2980)(3380,2997)
	(3411,3013)(3442,3029)(3474,3043)
	(3506,3057)(3539,3070)(3571,3082)
	(3604,3093)(3637,3103)(3670,3112)
	(3702,3120)(3734,3127)(3767,3133)
	(3799,3139)(3830,3143)(3862,3146)
	(3893,3148)(3924,3150)(3955,3150)
	(3986,3150)(4017,3148)(4048,3146)
	(4079,3143)(4111,3139)(4143,3133)
	(4175,3127)(4208,3120)(4240,3112)
	(4273,3103)(4306,3093)(4338,3082)
	(4371,3071)(4403,3058)(4436,3044)
	(4468,3029)(4499,3014)(4530,2998)
	(4560,2981)(4589,2964)(4617,2945)
	(4645,2927)(4671,2908)(4697,2888)
	(4721,2868)(4744,2848)(4766,2828)
	(4787,2807)(4807,2786)(4825,2765)
	(4843,2743)(4859,2722)(4874,2700)
	(4888,2677)(4903,2652)(4917,2625)
	(4929,2598)(4940,2570)(4949,2541)
	(4958,2510)(4965,2477)(4972,2442)
	(4977,2405)(4981,2366)(4984,2325)
	(4987,2281)(4988,2235)(4989,2187)
	(4989,2136)(4988,2085)(4987,2033)
	(4985,1981)(4983,1931)(4981,1883)
	(4978,1838)(4976,1799)(4973,1764)
	(4971,1736)(4970,1714)(4969,1698)
	(4968,1688)(4967,1682)(4967,1680)
\dashline{1200.000}(2110,1710)(2092,1725)(2074,1739)
	(2055,1753)(2035,1765)(2014,1776)
	(1992,1786)(1969,1795)(1944,1804)
	(1919,1811)(1894,1817)(1868,1822)
	(1841,1826)(1814,1829)(1787,1831)
	(1760,1832)(1733,1831)(1707,1830)
	(1681,1828)(1656,1825)(1631,1822)
	(1607,1817)(1585,1812)(1563,1806)
	(1542,1800)(1521,1793)(1502,1785)
	(1482,1776)(1462,1766)(1443,1756)
	(1425,1744)(1407,1732)(1389,1718)
	(1372,1703)(1356,1687)(1340,1671)
	(1324,1653)(1310,1634)(1296,1614)
	(1282,1594)(1270,1572)(1258,1550)
	(1247,1528)(1237,1505)(1228,1482)
	(1219,1458)(1212,1434)(1205,1410)
	(1198,1385)(1192,1360)(1187,1335)
	(1182,1311)(1178,1286)(1174,1260)
	(1171,1234)(1167,1206)(1164,1178)
	(1162,1149)(1160,1119)(1158,1088)
	(1156,1057)(1155,1025)(1155,992)
	(1154,960)(1155,928)(1155,895)
	(1156,863)(1158,832)(1160,801)
	(1162,771)(1164,742)(1167,714)
	(1171,686)(1174,660)(1178,634)
	(1182,609)(1187,585)(1192,560)
	(1198,535)(1205,510)(1212,486)
	(1219,462)(1228,438)(1237,415)
	(1247,392)(1258,370)(1270,348)
	(1282,326)(1296,306)(1310,286)
	(1324,267)(1340,249)(1356,232)
	(1372,217)(1389,202)(1407,188)
	(1425,176)(1443,164)(1462,154)
	(1482,144)(1502,135)(1521,127)
	(1542,120)(1563,114)(1585,108)
	(1607,103)(1631,98)(1656,95)
	(1681,92)(1707,90)(1733,89)
	(1760,88)(1787,89)(1814,91)
	(1841,94)(1868,98)(1894,103)
	(1919,109)(1944,116)(1969,125)
	(1992,134)(2014,144)(2035,155)
	(2055,167)(2074,181)(2092,195)
	(2110,210)(2127,228)(2144,247)
	(2160,268)(2175,290)(2189,314)
	(2203,339)(2216,365)(2228,393)
	(2240,421)(2251,450)(2261,480)
	(2271,510)(2280,540)(2288,570)
	(2295,599)(2302,627)(2308,655)
	(2313,681)(2318,706)(2322,730)
	(2326,752)(2329,773)(2332,792)
	(2335,810)(2338,834)(2341,856)
	(2344,876)(2346,895)(2348,912)
	(2350,928)(2350,944)(2351,960)
	(2350,976)(2350,992)(2348,1008)
	(2346,1025)(2344,1044)(2341,1064)
	(2338,1086)(2335,1110)(2332,1128)
	(2329,1147)(2326,1168)(2322,1190)
	(2318,1214)(2313,1239)(2308,1265)
	(2302,1293)(2295,1321)(2288,1350)
	(2280,1380)(2271,1410)(2261,1440)
	(2251,1470)(2240,1499)(2228,1527)
	(2216,1555)(2203,1581)(2189,1606)
	(2175,1630)(2160,1652)(2144,1673)
	(2127,1692)(2110,1710)
\dashline{1200.000}(2215,1147)(2200,1154)(2184,1159)
	(2166,1164)(2148,1167)(2128,1171)
	(2108,1173)(2086,1174)(2065,1175)
	(2043,1174)(2021,1173)(2001,1171)
	(1981,1167)(1963,1164)(1945,1159)
	(1929,1154)(1915,1147)(1901,1141)
	(1888,1133)(1876,1124)(1865,1115)
	(1854,1104)(1845,1093)(1836,1080)
	(1828,1067)(1822,1054)(1816,1040)
	(1811,1027)(1808,1013)(1805,999)
	(1803,986)(1802,973)(1802,960)
	(1802,947)(1803,934)(1805,921)
	(1808,907)(1811,893)(1816,880)
	(1822,866)(1828,853)(1836,840)
	(1845,827)(1854,816)(1865,805)
	(1876,796)(1888,787)(1901,779)
	(1915,772)(1929,766)(1945,761)
	(1963,756)(1981,753)(2001,749)
	(2021,747)(2043,746)(2065,745)
	(2086,746)(2108,747)(2128,749)
	(2148,753)(2166,756)(2184,761)
	(2200,766)(2215,772)(2230,780)
	(2244,789)(2258,799)(2271,810)
	(2282,822)(2293,835)(2303,848)
	(2311,860)(2319,873)(2325,885)
	(2331,896)(2335,906)(2339,915)
	(2342,922)(2346,932)(2349,940)
	(2351,947)(2352,954)(2353,960)
	(2352,966)(2351,973)(2349,980)
	(2346,988)(2342,997)(2339,1005)
	(2335,1014)(2331,1024)(2325,1035)
	(2319,1047)(2311,1060)(2303,1073)
	(2293,1085)(2282,1098)(2271,1110)
	(2258,1121)(2244,1131)(2230,1140)(2215,1147)
\put(2987,1500){\makebox(0,0)[lb]{\smash{{\SetFigFont{10}{12.0}{\rmdefault}{\mddefault}{\itdefault}$u$}}}}
\put(4967,1455){\makebox(0,0)[lb]{\smash{{\SetFigFont{10}{12.0}{\rmdefault}{\mddefault}{\itdefault}$v$}}}}
\put(2447,870){\makebox(0,0)[lb]{\smash{{\SetFigFont{10}{12.0}{\rmdefault}{\mddefault}{\itdefault}$w$}}}}
\put(3797,1995){\makebox(0,0)[lb]{\smash{{\SetFigFont{10}{12.0}{\rmdefault}{\mddefault}{\updefault}$C'\backslash P'$}}}}
\put(3392,15){\makebox(0,0)[lb]{\smash{{\SetFigFont{10}{12.0}{\rmdefault}{\mddefault}{\updefault}$k$}}}}
\put(4877,60){\makebox(0,0)[lb]{\smash{{\SetFigFont{10}{12.0}{\rmdefault}{\mddefault}{\updefault}$z$}}}}
\put(4337,960){\makebox(0,0)[lb]{\smash{{\SetFigFont{10}{12.0}{\rmdefault}{\mddefault}{\updefault}$C''$}}}}
\put(422,3840){\makebox(0,0)[lb]{\smash{{\SetFigFont{10}{12.0}{\rmdefault}{\mddefault}{\updefault}$C'$}}}}
\put(5777,960){\makebox(0,0)[lb]{\smash{{\SetFigFont{10}{12.0}{\rmdefault}{\mddefault}{\updefault}$l$}}}}
\put(1772,3345){\makebox(0,0)[lb]{\smash{{\SetFigFont{10}{12.0}{\rmdefault}{\mddefault}{\updefault}$S_{u,v}$}}}}
\put(197,1860){\makebox(0,0)[lb]{\smash{{\SetFigFont{10}{12.0}{\rmdefault}{\mddefault}{\updefault}$S_{w,w}$}}}}
\end{picture}
}
\end{center}
\caption{$S_{u,v}$ and $S_{w,w}$.} \label{sonsfig}
\end{figure}

\begin{obs}
\label{firstpart} According to Lemma \ref{goodc0}, $\{ S_{u,v}:
u,v \in C_0 \}$ is a partition of $\mathcal{C} \backslash C_0$.
\end{obs}

\begin{defin} \label{sondm} {\em Let  $C'$ be a son of
$C''$, defined by the set $S_{k,l}$. Mark $P= C' \backslash C''$
(this is a path as we will prove in  Lemma \ref{goodC} ). Given
nodes $u \neq v \in V(P)$  let $P'$ be the $u-v$ path in $C'$
which includes $z$ (clearly $(C' \backslash P') \subseteq P$) .
Define $S'_{u,v}:= \{ C \in S_{k,l} \backslash \{ C'\}| C \cap C'
= P' \} $. If $S'_{u,v} \neq \phi $ then the shortest cycle in
$S'_{u,v}$ is the {\em son} of $C'$ {\em defined by $S'_{u,v}$}.
Similarly, given a node $v \in V(P)$, define $S'_{v,v}:= \{ C \in
S_{k,l}, C \neq C' | C \supset C', C \backslash C'$ is a cycle
touching $C'$ in the node $v \}$. If $S'_{v,v} \neq \phi $ then
the shortest cycle in $S'_{v,v}$ is the {\em son} of $C'$ {\em
defined by $S'_{v,v}$}} (See Figure \ref{sonsfig}).
\end{defin}

\begin{defin}
{\em If $C_1$ is a son of $C_2$ then $C_2$ is a {\em father} of
$C_1$.}
\end{defin}

\begin{obs}
\label{fss}A father  is shorter than its sons.
\end{obs}

\begin{lem}
\label{goodC} If $\hat{C}$ is the son of $C'$ defined by
$S_{k,l}$, and $C^* \in S_{k,l}$ then \bi
\item $\hat{C} \backslash C^*$ is a path (or a cycle).
\item $C^* \backslash \hat{C}$ is a path (or a cycle).
\ei
\end{lem}

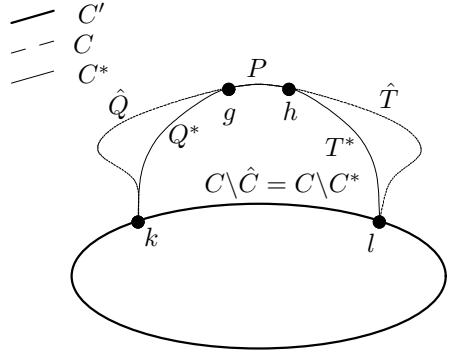
\begin{figure}
\begin{center}
\setlength{\unitlength}{0.00069991in}
\begingroup\makeatletter\ifx\SetFigFont\undefined%
\gdef\SetFigFont#1#2#3#4#5{%
  \reset@font\fontsize{#1}{#2pt}%
  \fontfamily{#3}\fontseries{#4}\fontshape{#5}%
  \selectfont}%
\fi\endgroup%
{\renewcommand{\dashlinestretch}{30}
\begin{picture}(3350,2644)(0,-10)
\thicklines
\put(1910,584){\ellipse{2790}{1080}}
\thinlines
\put(1010,989){\blacken\ellipse{90}{90}}
\put(1010,989){\ellipse{90}{90}}
\put(2810,989){\blacken\ellipse{90}{90}}
\put(2810,989){\ellipse{90}{90}}
\put(2135,1979){\blacken\ellipse{90}{90}}
\put(2135,1979){\ellipse{90}{90}}
\put(1685,1979){\blacken\ellipse{90}{90}}
\put(1685,1979){\ellipse{90}{90}}
\path(1010,989)(1010,993)(1010,1000)
	(1010,1014)(1011,1033)(1011,1058)
	(1012,1086)(1013,1115)(1014,1145)
	(1015,1173)(1016,1200)(1018,1225)
	(1020,1248)(1022,1269)(1024,1288)
	(1026,1307)(1029,1324)(1033,1341)
	(1036,1359)(1040,1376)(1045,1393)
	(1050,1411)(1056,1430)(1063,1448)
	(1070,1467)(1079,1486)(1088,1505)
	(1098,1525)(1108,1543)(1119,1562)
	(1131,1580)(1143,1598)(1156,1615)
	(1169,1632)(1183,1648)(1198,1664)
	(1211,1678)(1226,1693)(1241,1708)
	(1258,1723)(1275,1738)(1293,1753)
	(1312,1769)(1331,1784)(1351,1800)
	(1371,1815)(1390,1829)(1410,1844)
	(1430,1857)(1449,1870)(1467,1882)
	(1484,1893)(1501,1904)(1517,1913)
	(1532,1922)(1546,1930)(1565,1941)
	(1584,1950)(1602,1958)(1619,1966)
	(1635,1972)(1651,1978)(1667,1983)
	(1681,1987)(1695,1991)(1708,1994)
	(1720,1996)(1731,1998)(1742,2000)
	(1752,2001)(1764,2003)(1777,2005)
	(1790,2006)(1803,2008)(1817,2010)
	(1831,2011)(1845,2013)(1859,2014)
	(1872,2015)(1885,2016)(1898,2016)
	(1910,2016)(1922,2016)(1935,2016)
	(1948,2015)(1961,2014)(1975,2013)
	(1989,2011)(2003,2010)(2017,2008)
	(2030,2006)(2043,2005)(2056,2003)
	(2068,2001)(2078,2000)(2089,1998)
	(2100,1996)(2112,1994)(2125,1991)
	(2139,1987)(2153,1983)(2169,1978)
	(2185,1972)(2201,1966)(2218,1958)
	(2236,1950)(2255,1941)(2274,1930)
	(2288,1922)(2303,1913)(2319,1904)
	(2336,1893)(2353,1882)(2371,1870)
	(2390,1857)(2410,1844)(2430,1829)
	(2449,1815)(2469,1800)(2489,1784)
	(2508,1769)(2527,1753)(2545,1738)
	(2562,1723)(2579,1708)(2594,1693)
	(2609,1678)(2623,1664)(2637,1648)
	(2651,1632)(2664,1615)(2677,1598)
	(2689,1580)(2701,1562)(2712,1543)
	(2722,1525)(2732,1505)(2741,1486)
	(2750,1467)(2757,1448)(2764,1430)
	(2770,1411)(2775,1393)(2780,1376)
	(2784,1359)(2788,1341)(2791,1324)
	(2794,1307)(2796,1288)(2798,1269)
	(2800,1248)(2802,1225)(2804,1200)
	(2805,1173)(2806,1145)(2807,1115)
	(2808,1086)(2809,1058)(2809,1033)
	(2810,1014)(2810,1000)(2810,993)(2810,989)
\dashline{180.000}(65,2249)(380,2339)
\thicklines
\path(65,2474)(380,2564)
\thinlines
\path(65,2024)(380,2114)
\dashline{105.000}(1010,989)(1010,993)(1011,1000)
	(1012,1014)(1013,1032)(1014,1056)
	(1015,1083)(1015,1111)(1015,1139)
	(1015,1166)(1013,1191)(1011,1214)
	(1007,1235)(1003,1253)(997,1270)
	(990,1285)(982,1299)(973,1311)
	(960,1325)(946,1339)(930,1352)
	(913,1364)(894,1377)(875,1390)
	(854,1402)(834,1414)(814,1427)
	(795,1438)(778,1450)(761,1461)
	(747,1473)(735,1484)(725,1495)
	(718,1506)(713,1516)(710,1526)
	(709,1536)(709,1547)(711,1558)
	(714,1570)(720,1582)(727,1595)
	(735,1608)(746,1621)(758,1634)
	(771,1648)(786,1661)(802,1674)
	(819,1687)(837,1699)(857,1712)
	(877,1723)(898,1735)(920,1746)
	(939,1756)(960,1765)(981,1775)
	(1003,1784)(1026,1794)(1051,1803)
	(1076,1813)(1102,1823)(1129,1832)
	(1156,1842)(1183,1851)(1211,1861)
	(1239,1870)(1266,1879)(1293,1887)
	(1319,1895)(1344,1903)(1369,1910)
	(1393,1917)(1415,1923)(1436,1929)
	(1456,1935)(1476,1940)(1494,1945)
	(1519,1952)(1543,1958)(1565,1964)
	(1586,1969)(1606,1974)(1625,1978)
	(1643,1982)(1659,1985)(1675,1988)
	(1689,1991)(1701,1993)(1713,1995)
	(1724,1997)(1734,1998)(1743,2000)
	(1752,2001)(1764,2003)(1777,2005)
	(1790,2006)(1803,2008)(1817,2010)
	(1831,2011)(1845,2013)(1859,2014)
	(1872,2015)(1885,2016)(1898,2016)
	(1910,2016)(1922,2016)(1935,2016)
	(1948,2015)(1961,2014)(1975,2013)
	(1989,2011)(2003,2010)(2017,2008)
	(2030,2006)(2043,2005)(2056,2003)
	(2068,2001)(2077,2000)(2086,1998)
	(2096,1997)(2107,1995)(2119,1993)
	(2131,1991)(2145,1988)(2161,1985)
	(2177,1982)(2195,1978)(2214,1974)
	(2234,1969)(2255,1964)(2277,1958)
	(2301,1952)(2326,1945)(2344,1940)
	(2364,1935)(2384,1929)(2405,1923)
	(2427,1917)(2451,1910)(2476,1903)
	(2501,1895)(2527,1887)(2554,1879)
	(2581,1870)(2609,1861)(2637,1851)
	(2664,1842)(2691,1832)(2718,1823)
	(2744,1813)(2769,1803)(2794,1794)
	(2817,1784)(2839,1775)(2860,1765)
	(2881,1756)(2900,1746)(2922,1735)
	(2943,1723)(2963,1712)(2983,1699)
	(3001,1687)(3019,1674)(3035,1661)
	(3050,1648)(3064,1634)(3077,1621)
	(3088,1608)(3097,1595)(3105,1582)
	(3111,1570)(3116,1558)(3119,1547)
	(3121,1536)(3121,1526)(3120,1516)
	(3118,1506)(3113,1495)(3106,1484)
	(3097,1473)(3086,1461)(3074,1450)
	(3060,1438)(3045,1427)(3029,1414)
	(3012,1402)(2995,1390)(2979,1377)
	(2963,1364)(2947,1352)(2933,1339)
	(2920,1325)(2908,1311)(2898,1299)
	(2889,1285)(2880,1270)(2873,1253)
	(2865,1235)(2858,1214)(2852,1191)
	(2845,1166)(2839,1139)(2833,1111)
	(2827,1083)(2822,1056)(2817,1032)
	(2814,1014)(2812,1000)(2811,993)(2810,989)
\put(1820,2069){\makebox(0,0)[lb]{\smash{{\SetFigFont{10}{12.0}{\rmdefault}{\mddefault}{\updefault}$P$}}}}
\put(785,1799){\makebox(0,0)[lb]{\smash{{\SetFigFont{10}{12.0}{\rmdefault}{\mddefault}{\updefault}$\hat{Q}$}}}}
\put(1235,1574){\makebox(0,0)[lb]{\smash{{\SetFigFont{10}{12.0}{\rmdefault}{\mddefault}{\updefault}$Q^*$}}}}
\put(2405,1484){\makebox(0,0)[lb]{\smash{{\SetFigFont{10}{12.0}{\rmdefault}{\mddefault}{\updefault}$T^*$}}}}
\put(2810,1844){\makebox(0,0)[lb]{\smash{{\SetFigFont{10}{12.0}{\rmdefault}{\mddefault}{\updefault}$\hat{T}$}}}}
\put(2720,764){\makebox(0,0)[lb]{\smash{{\SetFigFont{10}{12.0}{\rmdefault}{\mddefault}{\updefault}$l$}}}}
\put(560,2474){\makebox(0,0)[lb]{\smash{{\SetFigFont{10}{12.0}{\rmdefault}{\mddefault}{\updefault}$C'$}}}}
\put(560,2024){\makebox(0,0)[lb]{\smash{{\SetFigFont{10}{12.0}{\rmdefault}{\mddefault}{\updefault}$C^*$}}}}
\put(524,2249){\makebox(0,0)[lb]{\smash{{\SetFigFont{10}{12.0}{\rmdefault}{\mddefault}{\updefault}$C$}}}}
\put(1055,809){\makebox(0,0)[lb]{\smash{{\SetFigFont{10}{12.0}{\rmdefault}{\mddefault}{\updefault}$k$}}}}
\put(1505,1214){\makebox(0,0)[lb]{\smash{{\SetFigFont{10}{12.0}{\rmdefault}{\mddefault}{\updefault}$C\backslash \hat{C}=C \backslash C^*$}}}}
\put(2090,1754){\makebox(0,0)[lb]{\smash{{\SetFigFont{10}{12.0}{\rmdefault}{\mddefault}{\updefault}$h$}}}}
\put(1640,1754){\makebox(0,0)[lb]{\smash{{\SetFigFont{10}{12.0}{\rmdefault}{\mddefault}{\updefault}$g$}}}}
\end{picture}
}
\end{center}
\caption{Figure for Lemma \ref{goodC}.} \label{goodCfig}
\end{figure}

\begin{prf}
Mark $\hat{P} = \hat{C} \backslash C'$ and $P^* = C^* \backslash C'$ .
From Observation (\ref{fss}), $l(C' \backslash \hat{C}) < l(\hat{P}) <
l(P^*)$. Therefore $U(\hat{C}) \subseteq V(\hat{P})$ and $U(C^*)
\subseteq V(P^*)$.

\noindent When $\hat{P} \cap P^*$ contains no edges then the claim
is obvious. Suppose otherwise, then $\hat{P} \cap P^*$ contains a
path, mark this path as $P$ ($P \subset (\hat{C} \cap C^*)
\backslash C'$). Mark the end points of $P$ by $g$ and $h$ ( see
Figure \ref{goodCfig}).
 Let $\hat{Q}$ (respectively
$Q^*$) be the $k-g$ path in $\hat{C}$ ($C^*$), and let $\hat{T}$
(respectively $T^*$) be the $h-l$ path in $\hat{C}$ ($C^*$) . Let
$v \in U(\hat{C})$,  w.l.o.g suppose that $v \in \hat{Q}$. Since
$\hat{C} \in \mathcal{C}$ $l(\hat{T})< l(T^*)$, indicating that
$U(C^*) \subseteq T^*$. But then since $C^* \in \mathcal{C}$,
$l(Q^*) < l(\hat{Q})$. In this case the cycle  $\hat{C} \backslash \hat{Q}
\cup Q^*$ is shorter than $\hat{C}$ and in $S$, contradiction the
fact that $\hat{C}$ was defined as a son by $S_{k,l}$.\end{prf}

\begin{obs}
\label{secondpart} According to Lemma \ref{goodC},  using  the
notation of Definition \ref{sondm}, $ \{ S'_{u,v}: u,v \in V(P) \}
$ is a partition of $S_{k,l} \backslash C'$.
\end{obs}

\begin{cor}
\label{noo} Every cycle $C \in \mathcal{C} \backslash \{ C_0 \}$
has a unique father.
\end{cor}

\begin{defin}
{\em If $C_1$ is a father of $C_2$ we mark  $C_1 \prec C_2$. If
$C_1 \prec C_2$ and $C_2 \prec C_3$ then $C_1 \prec C_3$. $\prec$
defines a hereditary order on $\mathcal{C}$ whose root is $C_0$.
If $C_1 \prec C_2$ we  say that $C_1$ is an {\em ancestor} of
$C_2$ , and $C_2$ is a {\em descendant} of $C_1$.}
\end{defin}

\begin{defin}
\label{psdef} {\em For every cycle $C\in\mathcal{C}\setminus\{C_0\}$
define (see Figure  \ref{sonsfig} (right)):
\be
\item $F(C)$,  the father of $C$.
\item $P_f(C) = F(C) \setminus C $ (the {\sl father} path).
\item $P_s(C) = C \setminus F(C)$ (the {\sl son} path).
\item $P_c(C) = C \cap F(C)$ (the {\sl common} path).
\ee }
\end{defin}

\begin{defin}
{\em If $C_1$ is a father of $C_2$ we mark  $C_1 \prec C_2$. If $C_1
\prec C_2$ and $C_2 \prec C_3$ then $C_1 \prec C_3$. $\prec$ defines
a hereditary order on $\mathcal{C}$ whose root is $C_0$. If $C_1
\prec C_2$ we  say that $C_1$ is an {\em ancestor} of $C_2$ , and
$C_2$ is a {\em descendant} of $C_1$.}
\end{defin}


\begin{prop}
\label{vins} For every cycle $C \in \mathcal{C}$, $U(C)$ induces a
path  $\subseteq P_s(C)$ $($since  nodes in $P_c(C)$ are in a
shorter cycle for example $F(C))$.
\end{prop}

\begin{prop}
\label{zinc} For every cycle $C \in \mathcal{C} \backslash \{C_0\}$,
$z \in P_c(C)$ (since $z \in C \cap F(C)$).
\end{prop}

\begin{defin}
{\em Suppose that $C_1,C_2 \in \mathcal{C}$ $C_1 \not\prec C_2$ and
$C_2 \not\prec C_1$. $C_a$ is {\em the lowest common ancestor} of
$C_1$ and $C_2$ if:
 \bi
  \item $C_a \prec C_1$.
  \item $C_a \prec C_2$.
  \item There is no cycle $C \in \mathcal{C}$ such that $C \prec C_1$,$C \prec C_2$, and
  $C_a \prec C$.
  \ei}
 \end{defin}

\begin{figure}
\begin{center}
\setlength{\unitlength}{0.00069991in}
\begingroup\makeatletter\ifx\SetFigFont\undefined%
\gdef\SetFigFont#1#2#3#4#5{%
  \reset@font\fontsize{#1}{#2pt}%
  \fontfamily{#3}\fontseries{#4}\fontshape{#5}%
  \selectfont}%
\fi\endgroup%
{\renewcommand{\dashlinestretch}{30}
\begin{picture}(6536,4419)(0,-10)
\put(2528,1320){\ellipse{5040}{2070}}
\put(863,2085){\blacken\ellipse{90}{90}}
\put(863,2085){\ellipse{90}{90}}
\put(2168,2355){\blacken\ellipse{90}{90}}
\put(2168,2355){\ellipse{90}{90}}
\put(1583,3210){\blacken\ellipse{90}{90}}
\put(1583,3210){\ellipse{90}{90}}
\put(1448,3210){\blacken\ellipse{90}{90}}
\put(1448,3210){\ellipse{90}{90}}
\put(953,3210){\blacken\ellipse{90}{90}}
\put(953,3210){\ellipse{90}{90}}
\put(2888,2355){\blacken\ellipse{90}{90}}
\put(2888,2355){\ellipse{90}{90}}
\put(4193,2085){\blacken\ellipse{90}{90}}
\put(4193,2085){\ellipse{90}{90}}
\put(1178,4065){\blacken\ellipse{90}{90}}
\put(1178,4065){\ellipse{90}{90}}
\put(1808,4065){\blacken\ellipse{90}{90}}
\put(1808,4065){\ellipse{90}{90}}
\put(3518,3210){\blacken\ellipse{90}{90}}
\put(3518,3210){\ellipse{90}{90}}
\put(2528,285){\blacken\ellipse{90}{90}}
\put(2528,285){\ellipse{90}{90}}
\put(2078,3210){\blacken\ellipse{90}{90}}
\put(2078,3210){\ellipse{90}{90}}
\path(863,2085)(863,3210)(2168,3210)(2168,2355)
\path(2888,2355)(2888,3210)(4193,3210)(4193,2085)
\path(1583,3210)(1583,4065)(2078,4065)(2078,3210)
\path(953,3210)(953,4065)(1448,4065)(1448,3210)
\put(2528,15){\makebox(0,0)[lb]{\smash{{\SetFigFont{10}{12.0}{\rmdefault}{\mddefault}{\updefault}$z$}}}}
\put(2123,2130){\makebox(0,0)[lb]{\smash{{\SetFigFont{10}{12.0}{\rmdefault}{\mddefault}{\updefault}$b$}}}}
\put(4148,1860){\makebox(0,0)[lb]{\smash{{\SetFigFont{10}{12.0}{\rmdefault}{\mddefault}{\updefault}$d$}}}}
\put(818,1860){\makebox(0,0)[lb]{\smash{{\SetFigFont{10}{12.0}{\rmdefault}{\mddefault}{\updefault}$a$}}}}
\put(2843,2130){\makebox(0,0)[lb]{\smash{{\SetFigFont{10}{12.0}{\rmdefault}{\mddefault}{\updefault}$c$}}}}
\put(908,2985){\makebox(0,0)[lb]{\smash{{\SetFigFont{10}{12.0}{\rmdefault}{\mddefault}{\updefault}$e$}}}}
\put(1403,2985){\makebox(0,0)[lb]{\smash{{\SetFigFont{10}{12.0}{\rmdefault}{\mddefault}{\updefault}$f$}}}}
\put(1538,2985){\makebox(0,0)[lb]{\smash{{\SetFigFont{10}{12.0}{\rmdefault}{\mddefault}{\updefault}$g$}}}}
\put(1988,2985){\makebox(0,0)[lb]{\smash{{\SetFigFont{10}{12.0}{\rmdefault}{\mddefault}{\updefault}$h$}}}}
\put(3473,2985){\makebox(0,0)[lb]{\smash{{\SetFigFont{10}{12.0}{\rmdefault}{\mddefault}{\updefault}$i$}}}}
\put(1133,4200){\makebox(0,0)[lb]{\smash{{\SetFigFont{10}{12.0}{\rmdefault}{\mddefault}{\updefault}$j$}}}}
\put(1763,4200){\makebox(0,0)[lb]{\smash{{\SetFigFont{10}{12.0}{\rmdefault}{\mddefault}{\updefault}$k$}}}}
\put(5183,4245){\makebox(0,0)[lb]{\smash{{\SetFigFont{10}{12.0}{\rmdefault}{\mddefault}{\updefault}The cycles}}}}
\put(5183,4020){\makebox(0,0)[lb]{\smash{{\SetFigFont{10}{12.0}{\rmdefault}{\mddefault}{\updefault}$C_0$ $zabcdz$}}}}
\put(5183,3795){\makebox(0,0)[lb]{\smash{{\SetFigFont{10}{12.0}{\rmdefault}{\mddefault}{\updefault}$C_1$ $zaefghbcdz$}}}}
\put(5183,3570){\makebox(0,0)[lb]{\smash{{\SetFigFont{10}{12.0}{\rmdefault}{\mddefault}{\updefault}$C_2$ $zaejfghbcdz$}}}}
\put(5183,3345){\makebox(0,0)[lb]{\smash{{\SetFigFont{10}{12.0}{\rmdefault}{\mddefault}{\updefault}$C_3$ $zaefgkhbcdz$}}}}
\put(5183,3120){\makebox(0,0)[lb]{\smash{{\SetFigFont{10}{12.0}{\rmdefault}{\mddefault}{\updefault}$C_4$ $zabcidz$}}}}
\put(2483,735){\makebox(0,0)[lb]{\smash{{\SetFigFont{10}{12.0}{\rmdefault}{\mddefault}{\updefault}$C_0$}}}}
\put(3338,2535){\makebox(0,0)[lb]{\smash{{\SetFigFont{10}{12.0}{\rmdefault}{\mddefault}{\updefault}$C_4$}}}}
\put(1313,2535){\makebox(0,0)[lb]{\smash{{\SetFigFont{10}{12.0}{\rmdefault}{\mddefault}{\updefault}$C_1$}}}}
\put(1043,3570){\makebox(0,0)[lb]{\smash{{\SetFigFont{10}{12.0}{\rmdefault}{\mddefault}{\updefault}$C_2$}}}}
\put(1673,3570){\makebox(0,0)[lb]{\smash{{\SetFigFont{10}{12.0}{\rmdefault}{\mddefault}{\updefault}$C_3$}}}}
\end{picture}
}
\end{center}
\caption{Examples of  lowest common ancestor.} \label{lcafig}
\end{figure}
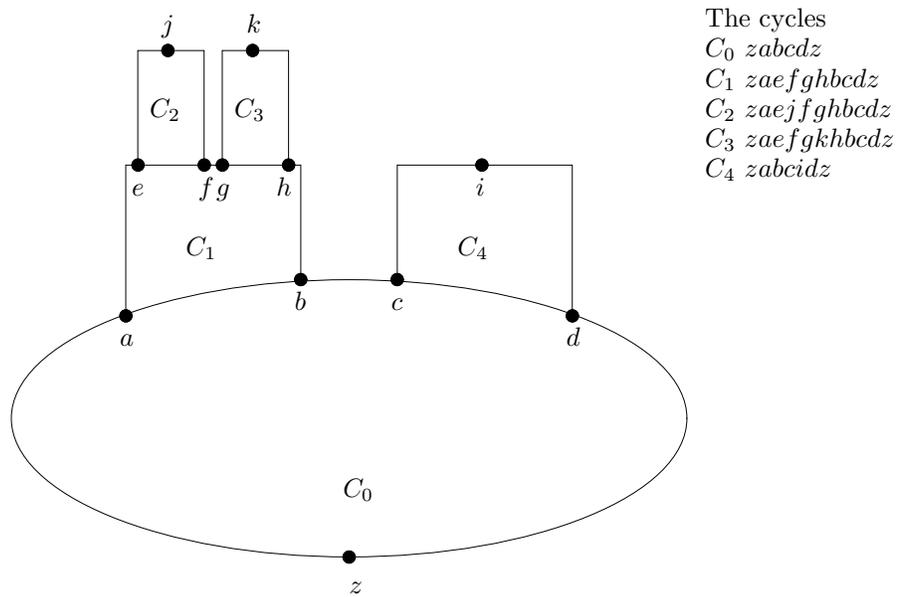

For example in Figure \ref{lcafig} $C_0$ is the lowest common
ancestor of $C_1$ and $C_4$, $C_0$ is also the lowest common
ancestor of $C_2$ and $C_4$, and $C_1$ is the lowest common ancestor
of $C_2$ and $C_3$.

\begin{defin}
{\em Suppose that $C_1,C_2 \in \mathcal{C}$ $C_1 \not\prec C_2$ and
$C_2 \not\prec C_1$. Let $C_a$ be the lowest common ancestor of
$C_1$ and $C_2$. Set $P_1 = C_a \backslash C_1$, $P_2 = C_a
\backslash C_2$. $C_1$ and $C_2$ are {\em crossing} (with respect to
$C_a$) if $E(P_1 \cap P_2)$, $E(P_1 \backslash P_2)$, $E(P_2
\backslash P_1)$ are all nonempty. }
 \end{defin}
\begin{figure}
\begin{center}
\setlength{\unitlength}{0.00069991in}
\begingroup\makeatletter\ifx\SetFigFont\undefined%
\gdef\SetFigFont#1#2#3#4#5{%
  \reset@font\fontsize{#1}{#2pt}%
  \fontfamily{#3}\fontseries{#4}\fontshape{#5}%
  \selectfont}%
\fi\endgroup%
{\renewcommand{\dashlinestretch}{30}
\begin{picture}(7133,4554)(0,-10)
\put(3883.079,2042.509){\arc{3146.161}{3.1400}{6.0539}}
\put(5129.580,2199.483){\arc{3323.443}{3.2626}{6.4613}}
\put(4425,240){\blacken\ellipse{90}{90}}
\put(4425,240){\ellipse{90}{90}}
\put(2310,2040){\blacken\ellipse{90}{90}}
\put(2310,2040){\ellipse{90}{90}}
\put(5415,2400){\blacken\ellipse{90}{90}}
\put(5415,2400){\ellipse{90}{90}}
\put(4425,1365){\ellipse{5400}{2250}}
\put(3480,2400){\blacken\ellipse{90}{90}}
\put(3480,2400){\ellipse{90}{90}}
\put(6765,1905){\blacken\ellipse{90}{90}}
\put(6765,1905){\ellipse{90}{90}}
\put(4200,3570){\blacken\ellipse{90}{90}}
\put(4200,3570){\ellipse{90}{90}}
\put(4965,3165){\blacken\ellipse{90}{90}}
\put(4965,3165){\ellipse{90}{90}}
\dottedline{45}(5145,3660)(4740,3435)
\path(4830.330,3519.502)(4740.000,3435.000)(4859.468,3467.052)
\dottedline{45}(5595,3120)(5325,2805)
\path(5380.317,2915.635)(5325.000,2805.000)(5425.873,2876.587)
\dottedline{45}(5145,3660)(5505,3345)
\path(5394.936,3401.443)(5505.000,3345.000)(5434.446,3446.598)
\put(4425,15){\makebox(0,0)[lb]{\smash{{\SetFigFont{10}{12.0}{\rmdefault}{\mddefault}{\updefault}$z$}}}}
\put(2265,1815){\makebox(0,0)[lb]{\smash{{\SetFigFont{10}{12.0}{\rmdefault}{\mddefault}{\updefault}$k$}}}}
\put(3435,2175){\makebox(0,0)[lb]{\smash{{\SetFigFont{10}{12.0}{\rmdefault}{\mddefault}{\updefault}$g$}}}}
\put(6720,1680){\makebox(0,0)[lb]{\smash{{\SetFigFont{10}{12.0}{\rmdefault}{\mddefault}{\updefault}$h$}}}}
\put(4155,3705){\makebox(0,0)[lb]{\smash{{\SetFigFont{10}{12.0}{\rmdefault}{\mddefault}{\updefault}$x$}}}}
\put(4560,2940){\makebox(0,0)[lb]{\smash{{\SetFigFont{10}{12.0}{\rmdefault}{\mddefault}{\updefault}$vc(Q')$}}}}
\put(2265,2850){\makebox(0,0)[lb]{\smash{{\SetFigFont{10}{12.0}{\rmdefault}{\mddefault}{\updefault}$P'$}}}}
\put(5235,3615){\makebox(0,0)[lb]{\smash{{\SetFigFont{10}{12.0}{\rmdefault}{\mddefault}{\updefault}$Q'$}}}}
\put(3390,2940){\makebox(0,0)[lb]{\smash{{\SetFigFont{10}{12.0}{\rmdefault}{\mddefault}{\updefault}$Q''$}}}}
\put(15,4380){\makebox(0,0)[lb]{\smash{{\SetFigFont{10}{12.0}{\rmdefault}{\mddefault}{\updefault}$C$ $zkglhz$}}}}
\put(15,4155){\makebox(0,0)[lb]{\smash{{\SetFigFont{10}{12.0}{\rmdefault}{\mddefault}{\updefault}$C'$ $zkxlhz$}}}}
\put(15,3930){\makebox(0,0)[lb]{\smash{{\SetFigFont{10}{12.0}{\rmdefault}{\mddefault}{\updefault}$C''$ $zkgxhz$}}}}
\put(15,3705){\makebox(0,0)[lb]{\smash{{\SetFigFont{10}{12.0}{\rmdefault}{\mddefault}{\updefault}$\hat{C}$ $zkxglhz$}}}}
\put(15,3480){\makebox(0,0)[lb]{\smash{{\SetFigFont{10}{12.0}{\rmdefault}{\mddefault}{\updefault}The shortcut $xg$}}}}
\put(5505,3165){\makebox(0,0)[lb]{\smash{{\SetFigFont{10}{12.0}{\rmdefault}{\mddefault}{\updefault}$R(Q')$}}}}
\put(6540,3210){\makebox(0,0)[lb]{\smash{{\SetFigFont{10}{12.0}{\rmdefault}{\mddefault}{\updefault}$P''$}}}}
\put(4695,2175){\makebox(0,0)[lb]{\smash{{\SetFigFont{10}{12.0}{\rmdefault}{\mddefault}{\updefault}$l=end(Q')$}}}}
\end{picture}
}
\end{center}
\caption{Figure for definitions \ref{scdef}, \ref{defcnode} and
\ref{Rdef}.} \label{crfig}
\end{figure}
\noindent Figure \ref{crfig} illustrates the following three
definitions:
\begin{defin}
\label{scdef} {\em The algorithm CANCEL-CROSSING described in
section \ref{secalgo} replaces cycles. Suppose that the algorithm
replaces the cycle $C'=C(v)$, which crossed another cycle $C''$ with
respect to a common  ancestor $C$. Let $\hat{C}$ be the cycle
serving $v$ after the change. The path $\hat{C} \backslash (C' \cup
C)$ is a {\em shortcut}.}
\end{defin}
\begin{defin}
\label{defcnode} {\em Let $C \in \mathcal{C}$ be a cycle. Suppose
that  $C$ contains four nodes in $P_s(C)$ ordered $k,g,l,h$, such
that $S_{k,l}$ and $S_{g,h}$ are non empty. In this case, every $C'
\in S_{k,l}$ and $C'' \in S_{g,h}$ are crossing with respect to $C$.
Since the graph is planar this means that $(C' \cap C'') \backslash
C$ contains at least one node. Choose such a node $x$ (if there is
more than one select one arbitrarily). We call such a node a {\em
cross-node} of $C'$ and $C''$  (with respect to $C$).}
\end{defin}
\begin{defin}
\label{Rdef} {\em Using the notation of Definition \ref{defcnode},
for a path $P$ between $x$ and $[g,l]$, let  $end(P) = P \cap
[g,l]$. Note that $end(P)$ is a node in $P$. Let $R(P) =
[vc(P),end(P)]$ be the maximal subpath of $P$ contained in a
shortcut defined before the processing of $C$. $(vc(P)=end(P)$ is
possible $)$.}
\end{defin}

\begin{defin}
For a cycle $C \in \mathcal{C}$ and nodes $v_1,v_2 \in C$ let
$P^{v_1,v_2}(C)$ be  the $v_1-v_2$ subpath of $C$ which includes
$z$.
\end{defin}

\begin{defin} Let $(P_1,\ldots,P_n) $ be the concatenation of paths
$P_1,\ldots,P_n$, in this order.
\end{defin}

\begin{defin}
\label{nvhdef} {\em Using the notation of Definition \ref{defcnode},
if $U(C') \subset [k,x]$, $U(C'') \subset [x,h]$, and $\frac{2}{3}
l(P_s(C)) \geq \max \{ l(k,l),l(g,h)\}$, then $C'$ and $C''$ are
{\em not-very-heavy-outer-crossing}\footnote{ According to Lemma
\ref{empV} bellow either $((U(C') \subset [k,x]) \wedge (U(C'')
\subset [x,h]))$ or $((U(C') \subset [x,l]) \wedge (U(C'') \subset
[g,x]))$.} .}
\end{defin}

\newpage
\section{Cancellation of Crossings}
We suggest now a way to change the cycles to avoid some cases of
crossing.
\subsection{Outline of the algorithm}
\label{secoutine} The algorithm builds the hereditary structure in
order of generations. During the process it cancels all
not-very-heavy-outer-crossings. In each stage the algorithm
considers a generation, and cancels all
not-very-heavy-outer-crossing with respect to cycles in this
generation. Afterwards, the next generation is defined and the
algorithm  proceeds to handle crossings with respect to the new
generation. We  note that once a cycle has been defined as a son of
another (already processed) cycle, it will not be changed.

\noindent We denote by $SC(C)$ the set of edges contained in
shortcuts used by a cycle $C \in \mathcal{C}$. Before the first
application of CANCEL-CROSSING , $SC(C) = \phi$ for every $C \in
\mathcal{C}$.

\subsection{The algorithm} \label{secalgo}

\begin{figure}[phtb]
\framebox[\textwidth]{
\begin{em}
\begin{program}
\Proc{CANCEL-CROSSING} \\
\Inputs\  \\
1.  An undirected graph $G=(V,E)$. \\
2.  A set of cycles $\mathcal{C}$.\\
\Returns\\
A new set of cycles $\mathcal{C}$ with some crossings removed.\\
\Procbegin \\
 $D:= \{C_0 \} $. \\
 \While $(D \neq \phi)$\\
 $NG := \phi$ $[NG$ is the next generation.$]$\\
 \Forevery cycle $C \in D$ $[$Processing of $C.]$ \\
Find the partition $\{ S_{u,v}: u,v \in P_s(C) \}$ described in\\
Observations  \ref{firstpart} and \ref{secondpart}.\\
\While $(\exists$ cycles $C'$ and $C''$ that are not-very-heavy-outer-crossing\\
         with respect to $C$) \\
 UNCROSS-CYCLES$(C',C'')$.\\
\Endwhile\\
\Forevery $S_{u,v} \neq \phi$\\
$C_{u,v} :=$ the son of $C$ defined by $S_{u,v}$. \\
$NG := NG \cup C_{u,v}$. \\
\Endfor\\
$[C$ is considered Processed $]$. \\
\Endfor\\
$D := NG$\\
\Endwhile\\
\Endproc{CANCEL-CROSSING}
\end{program}
\end{em}
} \caption{Procedure CANCEL-CROSSING} \label{cancelcrosfun}
\end{figure}

\begin{figure}[phtb]
\framebox[\textwidth]{
\begin{em}
\begin{program}
\Proc{UNCROSS-CYCLES} \\
\Inputs\  \\
Not-very-heavy-outer-crossing cycles $C' \in S_{k,l}$ and $C'' \in S_{g,h}$ (see Definition \ref{nvhdef}).\\
\Procbegin \\
$x:=$  the cross-node. \\
$P':=$ the  $x-k$ path in $C'$.\\
$P'':=$ the $x-h$ path in $C''$.\\
$Q':=$ the  $x-l$ path in $C'$.\\
$Q'':=$ the $x-g$ path in $C''$.\\
$[$ See Figure \ref{algofig} .$]$\\
$U^* := \{ Q' \} .$ \\
$B'(1):= \{ c | c $ creates n.v.h.c (not very heavy crossing) with $C'$ and $c$ is the right cycle $\}$.\\
$X'(1) := \{ x(c) | x(c)$ is the cross-node between $C'$ and $c$ , $c \in B'(1) \} $. \\
$P'(1):= \{ p(c) =  [x,x(c)] \cup [x(c),y] | x(c) \in X'(1), [x(c),y]$ is the left path between $x(c)$ and $[g,l] \}$ . \\
For every $p(c) \in P'(1)$ $R(p(c)) = (R(Q') \cap [x,x(c)]) \cup (R(c) \cap [x(c),y]) . $\\
$U^* = U^* \cup P'(1) . $\\
\While $(B'(i) \neq \phi)$\\
\If $( i \% 2 ) $ \\
\Then \\
$B'(i+1):= \{ c | c $ creates n.v.h.v with $c_i(c) \in B'(i)$ and $c$ is the left cycle $\}$.\\
$X'(i+1) := \{ x(c) | x(c)$ is the cross-node between $c_i(c)$ and $c$ ,$c \in B'(i+1) \} $. \\
$P'(i+1):= \{ p(c) |$ the sub-path of $p(c_i(c))$ till $x(c) \cup [x(c),y] | $\\
 $[x(c),y]$ is the right path between $x(c)$ and $[g,l] \}$ . \\
For every $p(c) \in P'(i+1)$ $R(p(c)) = ((R(c_i(c)) \cap p(c(i))) \cup (R(c) \cap [x(c),y]) . $\\
$U^* = U^* \cup P'(i+1) . $\\
\Else \\
$B'(i+1):= \{ c | c $ creates n.v.h.c with $c_i(c) \in B'(i)$ and $c$ is the right cycle $\}$.\\
$X'(i+1) := \{ x(c) | x(c)$ is the cross-node between $c_i(c)$ and $c$,  $c \in B'(i+1) \} $. \\
$P'(i+1):= \{ p(c) |$ the sub-path of $p(c_i(c))$ till $x(c)  \cup [x(c),y] | $\\
$ [x(c),y]$ is the left path between $x(c)$ and $[g,l] \}$ . \\
For every $p(c) \in P'(i+1)$ $R(p(c)) = ((R(c_i(c)) \cap p(c(i))) \cup (R(c) \cap [x(c),y]) . $\\
$U^* = U^* \cup P'(i+1) . $\\
\Endif \\
\Endwhile \\
$P^* := \arg\min_{P \in U^*}  \{ l(P)  + l( P \backslash  R(P) ) \}$ .\\
$U^{**}$ and $P^{**}$ are defined in a similar way for $C''$.\\
\If $( l(P^*)  + l( P^* \backslash  R(P^*) ) <  l(P^{**})  + l( P^{**} \backslash R(P^{**}) )) $ \\
\Then \\
$SP :=$ the shortest path between $x$ and $end(P^*)$. \\
\Forevery $v \in U(C'')$\\
$C(v) :=(P'',SP,P^{end(P^*),h}(C))$.
$SC(C(v)) := (SC(C'') \backslash Q'') \cup SP $.\\
\Endfor\\
\Forevery $v \in U(C')$ \\
$C(v) :=(P',SP,P^{k,end(P^*)}(C))$ .
$SC(C(v)) := (SC(C') \backslash (Q' \backslash SP)) \cup (SP \backslash Q') $. \\
\Endfor \\
$[$When $P^* = Q'$ the cycle for nodes in  $U_{P'}$ doesn't change .$]$\\
\Else  \\
$SP :=$ the shortest path between $x$ and $end(P^{**})$. \\
\Forevery $v \in U(C')$ \\
$C(v) :=(P',SP,P^{end(P^{**}),h}(C))$.
$SC(C(v)) := (SC(C') \backslash Q') \cup SP$.\\
\Endfor \\
\Forevery $v \in U(C'')$ \\
$C(v) :=(P'',SP,P^{k,end(P^{**})}(C))$.
$SC(C(v)) := (SC(C'') \backslash (Q'' \backslash SP)) \cup (SP \backslash Q'') $.\\
\Endfor \\
\Endif\\
\Endproc{UNCROSS-CYCLES}
\end{program}
\end{em}
} \caption{Procedure UNCROSS-CYCLES} \label{uncrosscyclesfun}
\end{figure}

\begin{figure}
\begin{center}
\setlength{\unitlength}{0.00087489in}
\begingroup\makeatletter\ifx\SetFigFont\undefined%
\gdef\SetFigFont#1#2#3#4#5{%
  \reset@font\fontsize{#1}{#2pt}%
  \fontfamily{#3}\fontseries{#4}\fontshape{#5}%
  \selectfont}%
\fi\endgroup%
{\renewcommand{\dashlinestretch}{30}
\begin{picture}(5337,1899)(0,-10)
\put(2768.828,-463.235){\arc{4340.443}{3.4980}{4.5836}}
\put(2211.173,-463.235){\arc{4340.441}{4.8412}{5.9268}}
\put(2490,1689){\blacken\ellipse{90}{90}}
\put(2490,1689){\ellipse{90}{90}}
\put(735,294){\blacken\ellipse{90}{90}}
\put(735,294){\ellipse{90}{90}}
\put(1545,294){\blacken\ellipse{90}{90}}
\put(1545,294){\ellipse{90}{90}}
\put(2490,294){\blacken\ellipse{90}{90}}
\put(2490,294){\ellipse{90}{90}}
\put(3480,294){\blacken\ellipse{90}{90}}
\put(3480,294){\ellipse{90}{90}}
\put(4245,294){\blacken\ellipse{90}{90}}
\put(4245,294){\ellipse{90}{90}}
\put(2805,1239){\blacken\ellipse{90}{90}}
\put(2805,1239){\ellipse{90}{90}}
\put(2670,1419){\blacken\ellipse{90}{90}}
\put(2670,1419){\ellipse{90}{90}}
\put(2310,1419){\blacken\ellipse{90}{90}}
\put(2310,1419){\ellipse{90}{90}}
\put(2738,654){\blacken\ellipse{90}{90}}
\put(2738,654){\ellipse{90}{90}}
\path(2490,1689)(1545,294)
\path(2670,1059)(2760,1194)
\path(2760.000,1139.917)(2760.000,1194.000)(2710.077,1173.199)
\path(2670,924)(2715,744)
\path(2671.344,794.932)(2715.000,744.000)(2729.552,809.485)
\path(2738,654)(2985,1014)
\path(2805,1239)(2490,1689)
\path(2490,294)(3480,294)
\dashline{90.000}(3480,294)(5325,294)
\dashline{90.000}(2805,1239)(3480,294)
\dashline{90.000}(2490,294)(285,294)
\dashline{90.000}(2490,294)(2738,654)
\dashline{90.000}(150,1824)(510,1824)
\put(15,294){\makebox(0,0)[lb]{\smash{{\SetFigFont{12}{14.4}{\rmdefault}{\mddefault}{\updefault}$C$}}}}
\put(2445,1779){\makebox(0,0)[lb]{\smash{{\SetFigFont{12}{14.4}{\rmdefault}{\mddefault}{\updefault}$x$}}}}
\put(915,1104){\makebox(0,0)[lb]{\smash{{\SetFigFont{12}{14.4}{\rmdefault}{\mddefault}{\updefault}$P'$}}}}
\put(3795,1104){\makebox(0,0)[lb]{\smash{{\SetFigFont{12}{14.4}{\rmdefault}{\mddefault}{\updefault}$P''$}}}}
\put(2445,924){\makebox(0,0)[lb]{\smash{{\SetFigFont{12}{14.4}{\rmdefault}{\mddefault}{\updefault}$P^*$}}}}
\put(600,1779){\makebox(0,0)[lb]{\smash{{\SetFigFont{8}{9.6}{\rmdefault}{\mddefault}{\updefault}Previous shortcut}}}}
\put(3120,924){\makebox(0,0)[lb]{\smash{{\SetFigFont{12}{14.4}{\rmdefault}{\mddefault}{\updefault}$Q'$}}}}
\put(1725,924){\makebox(0,0)[lb]{\smash{{\SetFigFont{12}{14.4}{\rmdefault}{\mddefault}{\updefault}$Q''$}}}}
\put(690,69){\makebox(0,0)[lb]{\smash{{\SetFigFont{12}{14.4}{\rmdefault}{\mddefault}{\updefault}$k$}}}}
\put(1500,69){\makebox(0,0)[lb]{\smash{{\SetFigFont{12}{14.4}{\rmdefault}{\mddefault}{\updefault}$g$}}}}
\put(2310,69){\makebox(0,0)[lb]{\smash{{\SetFigFont{12}{14.4}{\rmdefault}{\mddefault}{\updefault}$end(P^*)$}}}}
\put(3480,69){\makebox(0,0)[lb]{\smash{{\SetFigFont{12}{14.4}{\rmdefault}{\mddefault}{\updefault}$l$}}}}
\put(4200,69){\makebox(0,0)[lb]{\smash{{\SetFigFont{12}{14.4}{\rmdefault}{\mddefault}{\updefault}$h$}}}}
\put(2895,1194){\makebox(0,0)[lb]{\smash{{\SetFigFont{12}{14.4}{\rmdefault}{\mddefault}{\updefault}$vc(Q')$}}}}
\put(2040,1374){\makebox(0,0)[lb]{\smash{{\SetFigFont{12}{14.4}{\rmdefault}{\mddefault}{\updefault}$y''$}}}}
\put(2805,1374){\makebox(0,0)[lb]{\smash{{\SetFigFont{12}{14.4}{\rmdefault}{\mddefault}{\updefault}$y'$}}}}
\put(2130,609){\makebox(0,0)[lb]{\smash{{\SetFigFont{12}{14.4}{\rmdefault}{\mddefault}{\updefault}$vc(P^*)$}}}}
\end{picture}
}
\end{center}
\caption{Figure for UNCROSS-CYCLES.} \label{algofig}
\end{figure}

\subsection{Analysis}

\noindent Throughout this section we use the notation of Algorithm
UNCROSS-CYCLES.



\begin{prop}
\label{Cst}  Consider the beginning of the processing step of $C \in
\mathcal{C}$. For every  two nodes $u,v \in P_s(C)$, the $u-v$
subpath of $P_s(C)$ is shorter than any path between these two nodes
used by a descendant of $C$.
\end{prop}


\begin{obs}
\label{shortsc} A shortcut is always the shortest path between its
ends.
\end{obs}


\begin{defin}
{\em \label{cyclesdefin} For every node $v$ define $C_v^*$ to be the
original cycle serving $v$. For a set of nodes $U$ such that
$C_v^*=C_u^* \ \forall u,v \in U$, define $C_U^*$ as their common
original cycle. }
\end{defin}

\begin{obs}
\label{robs}  $[x,vc(Q')] \subset C^*_{U(C')}$  and symmetrically
$[x,vc(Q'')] \subset C^*_{U(C'')}$ . (see Figure \ref{algofig}).
\end{obs}

\begin{obs}
\label{goodu} From the way the algorithm works at any  stage of the
algorithm, all the nodes in $U(C^*)$ of an original cycle $C^*$ are
served by the same cycle.
\end{obs}

\begin{lem}
\label{nonewcros} The set of crossing after the processing of a
cycle $C$ by CANCEL-CROSSING does not contain any crossing that did
not exist before this step.

 \noindent Moreover, if $C'$ is a
descendant of $C$ which outer-crosses more than one not-very-heavy
cycle (with respect to $C$), and $C'$ is the left cycle in all these
crossings (and similarly if it is in the right side in all these
crossings), then the total effect of applying UNCROSS-CYCLES to all
these crossings is the same as if it is only called once with the
crossing point which is closest to $U(C')$ on the right side of
$U(C')$.
\end{lem}

\begin{figure}
\begin{center}
\setlength{\unitlength}{0.00087489in}
\begingroup\makeatletter\ifx\SetFigFont\undefined%
\gdef\SetFigFont#1#2#3#4#5{%
  \reset@font\fontsize{#1}{#2pt}%
  \fontfamily{#3}\fontseries{#4}\fontshape{#5}%
  \selectfont}%
\fi\endgroup%
{\renewcommand{\dashlinestretch}{30}
\begin{picture}(4797,1435)(0,-10)
\put(2175,1045){\blacken\ellipse{90}{90}}
\put(2175,1045){\ellipse{90}{90}}
\put(285,280){\blacken\ellipse{90}{90}}
\put(285,280){\ellipse{90}{90}}
\put(2580,640){\blacken\ellipse{90}{90}}
\put(2580,640){\ellipse{90}{90}}
\put(1725,280){\blacken\ellipse{64}{64}}
\put(1725,280){\ellipse{64}{64}}
\put(2265,280){\blacken\ellipse{90}{90}}
\put(2265,280){\ellipse{90}{90}}
\put(2760,280){\blacken\ellipse{90}{90}}
\put(2760,280){\ellipse{90}{90}}
\put(3435,280){\blacken\ellipse{90}{90}}
\put(3435,280){\ellipse{90}{90}}
\put(4200,280){\blacken\ellipse{90}{90}}
\put(4200,280){\ellipse{90}{90}}
\path(105,280)(4785,280)
\path(1725,280)(1725,283)(1726,289)
	(1728,300)(1730,317)(1733,338)
	(1737,365)(1742,395)(1748,428)
	(1754,461)(1760,494)(1767,527)
	(1774,558)(1781,587)(1789,614)
	(1797,640)(1806,663)(1815,686)
	(1825,707)(1836,728)(1847,748)
	(1860,767)(1872,786)(1886,804)
	(1900,822)(1916,840)(1932,859)
	(1950,877)(1969,896)(1989,915)
	(2011,934)(2033,953)(2056,972)
	(2080,991)(2105,1009)(2130,1026)
	(2156,1044)(2182,1060)(2208,1076)
	(2235,1091)(2261,1105)(2287,1119)
	(2313,1131)(2340,1143)(2366,1154)
	(2393,1165)(2417,1174)(2442,1183)
	(2468,1192)(2495,1200)(2522,1207)
	(2550,1215)(2579,1221)(2608,1228)
	(2638,1233)(2669,1239)(2700,1244)
	(2731,1248)(2763,1251)(2795,1254)
	(2827,1256)(2859,1258)(2890,1259)
	(2921,1259)(2951,1259)(2981,1258)
	(3011,1257)(3040,1254)(3068,1252)
	(3096,1248)(3123,1244)(3150,1240)
	(3177,1235)(3204,1229)(3231,1223)
	(3258,1216)(3286,1208)(3314,1199)
	(3342,1190)(3371,1180)(3399,1169)
	(3428,1157)(3457,1144)(3485,1131)
	(3514,1117)(3541,1103)(3569,1088)
	(3596,1072)(3622,1057)(3647,1040)
	(3671,1024)(3695,1007)(3717,990)
	(3739,973)(3760,956)(3780,938)
	(3799,921)(3818,902)(3835,884)
	(3853,865)(3870,845)(3887,824)
	(3904,803)(3921,779)(3938,754)
	(3956,728)(3974,699)(3992,668)
	(4011,636)(4031,601)(4051,565)
	(4071,528)(4092,491)(4111,454)
	(4130,418)(4147,385)(4162,356)
	(4174,331)(4184,311)(4192,297)
	(4196,288)(4199,282)(4200,280)
\path(2265,280)(2265,281)(2266,284)
	(2267,292)(2270,305)(2274,324)
	(2279,346)(2284,370)(2290,394)
	(2296,417)(2302,438)(2307,457)
	(2313,474)(2319,489)(2326,503)
	(2333,516)(2340,527)(2348,539)
	(2357,550)(2367,561)(2378,572)
	(2390,582)(2404,592)(2419,602)
	(2435,611)(2452,620)(2470,628)
	(2489,636)(2509,642)(2530,648)
	(2551,654)(2572,658)(2595,662)
	(2614,665)(2634,668)(2655,671)
	(2677,673)(2700,674)(2724,676)
	(2750,677)(2775,678)(2802,679)
	(2829,679)(2855,679)(2882,678)
	(2908,677)(2934,676)(2958,674)
	(2982,673)(3005,671)(3027,668)
	(3048,665)(3068,662)(3092,658)
	(3115,654)(3138,648)(3160,642)
	(3182,636)(3203,628)(3223,620)
	(3243,611)(3261,602)(3278,592)
	(3294,582)(3308,572)(3321,561)
	(3333,550)(3343,539)(3353,527)
	(3361,516)(3369,503)(3376,489)
	(3383,474)(3390,457)(3396,438)
	(3403,417)(3409,394)(3415,370)
	(3421,346)(3426,324)(3430,305)
	(3433,292)(3434,284)(3435,281)(3435,280)
\dashline{195.000}(285,280)(285,283)(286,289)
	(288,300)(290,317)(293,338)
	(297,365)(302,395)(308,428)
	(314,461)(320,494)(327,527)
	(334,558)(341,587)(349,614)
	(357,640)(366,663)(375,686)
	(385,707)(396,728)(407,748)
	(420,767)(432,786)(446,804)
	(460,822)(476,840)(492,859)
	(510,877)(529,896)(549,915)
	(571,934)(593,953)(616,972)
	(640,991)(665,1009)(690,1026)
	(716,1044)(742,1060)(768,1076)
	(795,1091)(821,1105)(847,1119)
	(873,1131)(900,1143)(926,1154)
	(953,1165)(977,1174)(1002,1183)
	(1028,1192)(1055,1200)(1082,1207)
	(1110,1215)(1139,1221)(1168,1228)
	(1198,1233)(1229,1239)(1260,1244)
	(1291,1248)(1323,1251)(1355,1254)
	(1387,1256)(1419,1258)(1450,1259)
	(1481,1259)(1511,1259)(1541,1258)
	(1571,1257)(1600,1254)(1628,1252)
	(1656,1248)(1683,1244)(1710,1240)
	(1737,1235)(1764,1229)(1791,1223)
	(1818,1216)(1846,1208)(1874,1199)
	(1902,1190)(1931,1180)(1959,1169)
	(1988,1157)(2017,1144)(2045,1131)
	(2074,1117)(2101,1103)(2129,1088)
	(2156,1072)(2182,1057)(2207,1040)
	(2231,1024)(2255,1007)(2277,990)
	(2299,973)(2320,956)(2340,938)
	(2359,921)(2378,902)(2395,884)
	(2413,865)(2430,845)(2447,824)
	(2464,803)(2481,779)(2498,754)
	(2516,728)(2534,699)(2552,668)
	(2571,636)(2591,601)(2611,565)
	(2631,528)(2652,491)(2671,454)
	(2690,418)(2707,385)(2722,356)
	(2734,331)(2744,311)(2752,297)
	(2756,288)(2759,282)(2760,280)
\put(780,1270){\makebox(0,0)[lb]{\smash{{\SetFigFont{12}{14.4}{\rmdefault}{\mddefault}{\updefault}$C'$}}}}
\put(15,280){\makebox(0,0)[lb]{\smash{{\SetFigFont{12}{14.4}{\rmdefault}{\mddefault}{\updefault}$C$}}}}
\put(3390,1270){\makebox(0,0)[lb]{\smash{{\SetFigFont{12}{14.4}{\rmdefault}{\mddefault}{\updefault}$C''$}}}}
\put(3210,685){\makebox(0,0)[lb]{\smash{{\SetFigFont{12}{14.4}{\rmdefault}{\mddefault}{\updefault}$\tilde{C}$}}}}
\put(2130,1090){\makebox(0,0)[lb]{\smash{{\SetFigFont{12}{14.4}{\rmdefault}{\mddefault}{\updefault}$x$}}}}
\put(2535,685){\makebox(0,0)[lb]{\smash{{\SetFigFont{12}{14.4}{\rmdefault}{\mddefault}{\updefault}$\tilde{x}$}}}}
\put(240,55){\makebox(0,0)[lb]{\smash{{\SetFigFont{12}{14.4}{\rmdefault}{\mddefault}{\updefault}$k$}}}}
\put(1680,55){\makebox(0,0)[lb]{\smash{{\SetFigFont{12}{14.4}{\rmdefault}{\mddefault}{\updefault}$g$}}}}
\put(2760,55){\makebox(0,0)[lb]{\smash{{\SetFigFont{12}{14.4}{\rmdefault}{\mddefault}{\updefault}$l$}}}}
\put(2220,55){\makebox(0,0)[lb]{\smash{{\SetFigFont{12}{14.4}{\rmdefault}{\mddefault}{\updefault}$f$}}}}
\put(4155,55){\makebox(0,0)[lb]{\smash{{\SetFigFont{12}{14.4}{\rmdefault}{\mddefault}{\updefault}$h$}}}}
\end{picture}
}
\end{center}
\caption{Figure for Lemma \ref{nonewcros}.} \label{nonewfig}
\end{figure}
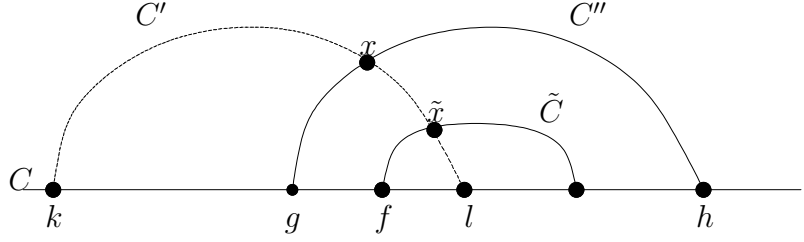

\begin{prf}
Suppose that UNCROSS-CYCLES  creates a new crossing. Let $C'$ and
$C''$ be two cycles which crossed with respect to $C$ (before $C$
was processed) and suppose that when UNCROSS-CYCLES is activated on
them,  $C''$ is changed. W.l.o.g suppose that $SP=P^*=Q'$ is the
shortcut (a similar proof applies when $SP=P^*\neq Q'$). Suppose
that the new $C''$ creates a crossing with a cycle $\tilde{C}$ (see
Figure \ref{nonewfig}).
Since $\tilde{C}$ and $C''$ didn't cross before the processing of
$C$ and they are crossing now:
  $(C \backslash \tilde{C}) \subseteq [g,h]$ and  $l \in (C \backslash \tilde{C})$. Therefore $\tilde{C}$ and $C'$
were crossing with respect to $C$ before the processing of $C$.
Moreover since  $C''$ is not-very-heavy and $\tilde{C} \subset C''$,
$\tilde{C}$ is not-very-heavy. Since $C'$ was outer-crossing with
$C''$, $U_{C'} \subseteq [x,k]$ so $C'$ and $\tilde{C}$ are
outer-crossing. Altogether we get that $C'$ and $\tilde{C}$ were
not-very-heavy-outer-crossing with respect to $C$ before the
processing of $C$.
 Let $\tilde{x}$ be a
cross-node of $\tilde{C}$ and $C'$ and let $f$ be the endpoint of
$(C \backslash \tilde{C})$ such that $f \in [g,l].$ When
UNCROSS-CYCLES is activated on $C'$ and $\tilde{C}$ it compares
among other also the paths $[\tilde{x},f]$ and $[\tilde{x},l]$, and
chooses the latter as the shortcut. Hence,  the cycle for the nodes
in $U_{\tilde{C}}$ will also be changed during the processing of
$C$, and the new cycle does not cross the new $C''$.

Also note that if UNCROSS-CYCLES was first activated on $C'$ and
$\breve{C}$ then $C'$ and $C''$ are still
not-very-heavy-outer-crossing, and then UNCROSS-CYCLES will be
activated on them, resulting in the same change of cycles that would
have happened if UNCROSS-CYCLES was first activated on $C'$ and
$C''$.
\end{prf}

\begin{theorem}
\label{bpst} Consider the beginning of the processing step of a
cycle $C$ by CANCEL-CROSSING. Let $E'$ be the set of edges contained
in cycles which have been processed before $C$. Let $C^*$ be an
original cycle and let  $\breve{C} \in \mathcal{C}$ be the cycle
which currently serves $U(C^*)$ (see Corollary \ref{goodu}). Let $A$
be the set of edges in the (processed) ancestors cycles of
$\breve{C}$.
 Consider $\breve{P}$ a subpath of $(\breve{C} \cap C^*)\backslash (E' \cap A)$ which contains $U(C^*)$.
 Define \bi
\item  $N(\breve{P}) := \breve{C} \backslash \breve{P}$.
 \item $O(\breve{P}) := C^* \backslash \breve{P}$.
 \item  $S(\breve{P}) := SC(\breve{C})$.
 \item $T(\breve{P}):= S(\breve{P}) \cap E'$.
 \item $A(\breve{P}):= S(\breve{P}) \backslash E' = S(\breve{P}) \backslash T(\breve{P})$
(The shortcuts that are in cycles which have not yet been
processed).
 \ei
Then:
\[ l(N(\breve{P})) + 2l(A(\breve{P})) \leq 9 l(O(\breve{P})) .\]
\end{theorem}

\begin{prf}
Denote by $C$ the cycle processed during a given execution of
CANCEL-CROSSING.  Let $\hat{N}(\breve{P})$, $\hat{S}(\breve{P})$,
and $\hat{T}(\breve{P})$ be the path and sets of shortcuts at the
beginning of this step. Let $N(\breve{P})$, $S(\breve{P})$, and
$T(\breve{P})$ be the path and sets of shortcuts at the end of it.
We will show that if the claim of the theorem holds at the beginning
of the given step, it still holds at the end of it.

\noindent W.l.o.g we consider $\breve{C}$ which is a descendant of
$C$ (other cycles are not changed while processing $C$.

\noindent Consider a call to UNCROSS-CYCLES to process
not-very-heavy-outer-crossing cycles $C'$ and $C''$ (see Figure
\ref{algofig}). By Lemma \ref{nonewcros} we can assume that
UNCROSS-CYCLES is not activated again on the right side of $C'$ and
that it is activated at most once on the left side of $C'$
(similarly for $C''$). The possibility of another activation of
UNCROSS-CYCLES on $C'$ (similarly on $C''$) affects the analysis at
only  one specific place and we will relate to it during the proof.
Since we assume that $\breve{P} \subset C^* \cap \breve{C}$ all the
applied changes are outside of $\breve{P}$. Since this theorem
considers the changes in the cycles, it is enough to prove it on any
$\breve{P}$ which is out of the changing part in each one of the
cycles. To simplify the notation we will consider $\breve{P} = P'$
and $\breve{P} = P''$.


\noindent The proof is given for the case that $P^*$ is the shortcut
(similar proof follows for the case $P^{**}$ is the shortcut) thus
$( l(P^*) + l( P^* \backslash ( R(P^*) \cup R(Q'))) < l(P^{**}) + l(
P^{**} \backslash (R(P^{**}) \cup R(Q'')))) $. Since we use $P^*$ as
the left hand side of the inequality the claim clearly holds since
the algorithm uses $SC$ which is not longer. A similar proof follows
when $P^{**}$ is the shortcut.

\noindent Denote: $v' = vc(Q') $ ($v'=l$ when $Q'$ doesn't intersect
previous shortcuts and $v'=x$ when $Q'$ is contained in previous
shortcuts), and $v'' = vc(Q'')$.

\noindent {\bf We first consider the case  $ P^* = Q'$} In this case
$P'$ doesn't change so we only need to prove the theorem for $P''$.
Since $Q'' \in U^{**}$: $l(Q') + l(Q' \backslash R(Q')) < l(Q'') +
l(Q'' \backslash R(Q'') )$, giving:
\begin{equation}
\label{v'v''eq} 2l(x,v') + l(v',l) < 2l(x,v'') + l(v'',g).
\end{equation}

Consider the possible locations of $v'$ and $v''$: \be

\item $v' \neq l$.  By definition of $v'$:  $Q' \cap \hat{S}(P')
\neq \phi$.

\be

\item \label{boundt} $v'' \neq g$ $(Q'' \cap \hat{S}(P'') \neq
\phi)$.

Let $w \in P_s(C)$ be the node for which $[h,w] \subset O(P'')$ and
$[h,w]$ is maximal ($h=w$ is possible).

 By definition
$[x,v''] \subset O(P'')$ giving that
\begin{equation}
\label{P8v'opt} l(O(P'')) = l(x,v'') +  l(O([x,v''],P'',[h,w])) +
l(x,v'') + l(h,w) .
\end{equation}

 According to the induction hypothesis
\[ l(\hat{N}([x,v''],P'',[h,w])) + 2l(\hat{A}([x,v''],P'',[h,w])) \leq 9 l(O([x,v''],P'',[h,w])), \]
Note that $\hat{A}([x,v''],P'',[h,w]) = \hat{A}(P'')$ , therefore
\begin{equation}
\label{P8v'ind}
 l(\hat{N}([x,v''],P'',[h,w])) + 2l(\hat{A}(P'')) \leq 9 l(O([x,v''],P'',[h,w])) .
\end{equation}
>From the way the algorithm works
\begin{equation}
\label{P8v'n} l(N(P'')) = l(\hat{N}([x,v''],P'',[h,w])) - l(v'',g) +
l(x,v') + l(v',l) + l(g,l) + l(h,w).
\end{equation}

Mark the endpoints of $P_s(C)$ as $a,b$ (the nodes on $P_s(C)$ are
ordered $a,k,g,l,h,b$).

In this case since we are processing $C$ $[a,g] \cup [w,b]$ is
removed from $A(P'')$ and since $P''$ is not using the path
$[v'',g]$ any more:
\[\hat{A}(P'') \backslash A(P'')= [a,g] \cup [w,b] \cup [v'',g].\]

On the other end $[x,v'] \cup [v',l]$ is the new shortcut so:

\[ A(P'') \backslash \hat{A}(P'') = [x,v'] \cup  [v',l]. \]

Altogether
\begin{equation}
\label{P8v'a}
 l(A((P''))) = l( \hat{A}(P'')) +l(x,v') +l(v',l) -
(  l(a,g) + l(w,b) + l(v'',g)) .
\end{equation}

 Since $l(g,h) \leq \frac{2}{3} l(P_s(C))$
\begin{equation}
\label{goodgl2} l(g,l) \leq l(g,h) \leq 2[l(a,g) +l(h,w)+ l(w,b)]
\end{equation}

Using (\ref{P8v'n}), (\ref{P8v'a}), (\ref{v'v''eq}),
(\ref{P8v'ind}),(\ref{goodgl2}), and (\ref{P8v'opt}) in this order:
\begin{eqnarray*}
l(N(P'')) + 2l(A(P'')) &=& [l(\hat{N}([x,v''],P'',[h,w])) - l(v'',g)
+
l(x,v') + l(v',l) + l(g,l) + l(h,w)] + \\
&& 2[l( \hat{A}(P''))  +l(x,v') +l(v',l) - (l(a,g) + l(w,b)+ l(v'',g))] \\
&=& l(\hat{N}([x,v''],P'',[h,w])) + 2l(\hat{A}(P'')) - 3l(v'',g) +
3l(x,v') + 3l(v',l) +\\
&& l(g,l) + l(h,w)-2l(a,g) - 2l(w,b) \\
&\leq& l(\hat{N}([x,v''],P'',[h,w])) + 2l(\hat{A}(P'')) + 6l(x,v'') + l(g,l) + l(h,w)-\\
&&2l(a,g) - 2l(w,b) \\
&\leq& 9l(O([x,v''],P'',[h,w])) + 6l(x,v'')+ l(g,l) + l(h,w)-2l(a,g) - 2l(w,b) \\
&\leq& 9l(O([x,v''],P'',[h,w])) + 6l(x,v'') + 3l(h,w) \\
&\leq& 9 l(O(P'')).
\end{eqnarray*}

Suppose that $C''$ participates in not-very-heavy-outer-crossings on
both its sides (see Figure \ref{twocrossfig}). Moreover, suppose
that both crossings satisfy the assumption of this case (using
shortcuts from previous steps). Note that in this case
\[ l(g,l) + l(q,h) \leq l(g,h) \leq \frac{2}{3} l(P_s(C)) .\]
Therefore, even if both applications of UNCROSS-CYCLES result in
changing $C''$ the inequality $l(N(P'')) + 2l(A(P'')) \leq
9l(O(P'')) $ is proven in a similar way.

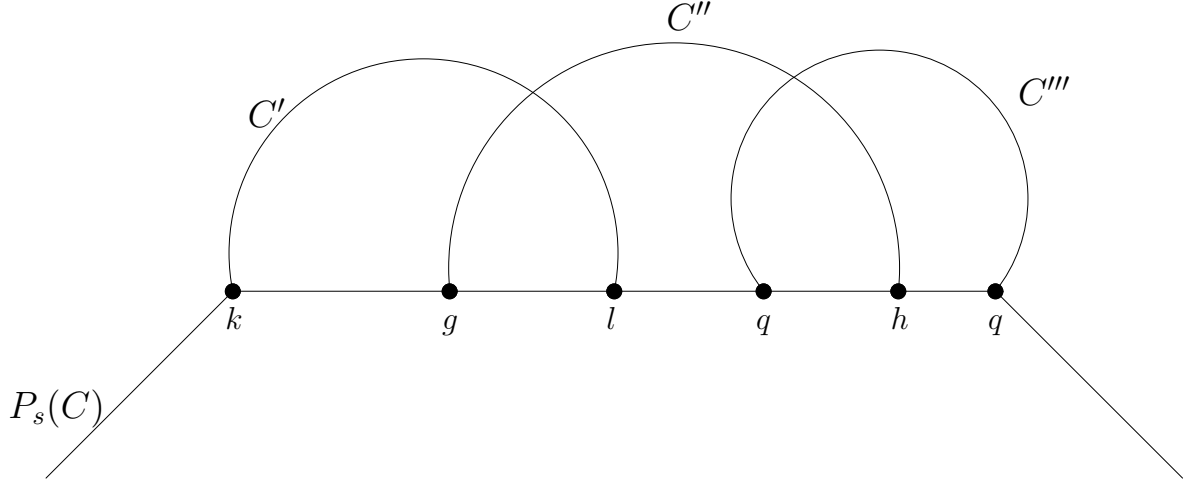
\begin{figure}
\begin{center}
\setlength{\unitlength}{0.00087489in}
\begingroup\makeatletter\ifx\SetFigFont\undefined%
\gdef\SetFigFont#1#2#3#4#5{%
  \reset@font\fontsize{#1}{#2pt}%
  \fontfamily{#3}\fontseries{#4}\fontshape{#5}%
  \selectfont}%
\fi\endgroup%
{\renewcommand{\dashlinestretch}{30}
\begin{picture}(7092,2982)(0,-10)
\put(2512.500,1367.081){\arc{2340.678}{2.9437}{6.4811}}
\put(4020.000,1276.773){\arc{2714.433}{3.0384}{6.3864}}
\put(5257.500,1696.167){\arc{1787.931}{2.4658}{6.9589}}
\put(5370,1137){\blacken\ellipse{90}{90}}
\put(5370,1137){\ellipse{90}{90}}
\put(1365,1137){\blacken\ellipse{90}{90}}
\put(1365,1137){\ellipse{90}{90}}
\put(3660,1137){\blacken\ellipse{90}{90}}
\put(3660,1137){\ellipse{90}{90}}
\put(2670,1137){\blacken\ellipse{90}{90}}
\put(2670,1137){\ellipse{90}{90}}
\put(4560,1137){\blacken\ellipse{90}{90}}
\put(4560,1137){\ellipse{90}{90}}
\put(5955,1137){\blacken\ellipse{90}{90}}
\put(5955,1137){\ellipse{90}{90}}
\path(240,12)(1365,1137)(2670,1137)
	(5955,1137)(7080,12)
\put(5325,912){\makebox(0,0)[lb]{\smash{{\SetFigFont{12}{14.4}{\rmdefault}{\mddefault}{\updefault}$h$}}}}
\put(1455,2127){\makebox(0,0)[lb]{\smash{{\SetFigFont{14}{16.8}{\rmdefault}{\mddefault}{\updefault}$C'$}}}}
\put(3975,2712){\makebox(0,0)[lb]{\smash{{\SetFigFont{14}{16.8}{\rmdefault}{\mddefault}{\updefault}$C''$}}}}
\put(6090,2262){\makebox(0,0)[lb]{\smash{{\SetFigFont{14}{16.8}{\rmdefault}{\mddefault}{\updefault}$C'''$}}}}
\put(3615,912){\makebox(0,0)[lb]{\smash{{\SetFigFont{12}{14.4}{\rmdefault}{\mddefault}{\updefault}$l$}}}}
\put(1320,912){\makebox(0,0)[lb]{\smash{{\SetFigFont{12}{14.4}{\rmdefault}{\mddefault}{\updefault}$k$}}}}
\put(2625,912){\makebox(0,0)[lb]{\smash{{\SetFigFont{12}{14.4}{\rmdefault}{\mddefault}{\updefault}$g$}}}}
\put(4515,912){\makebox(0,0)[lb]{\smash{{\SetFigFont{12}{14.4}{\rmdefault}{\mddefault}{\updefault}$q$}}}}
\put(5910,912){\makebox(0,0)[lb]{\smash{{\SetFigFont{12}{14.4}{\rmdefault}{\mddefault}{\updefault}$q$}}}}
\put(15,372){\makebox(0,0)[lb]{\smash{{\SetFigFont{14}{16.8}{\rmdefault}{\mddefault}{\updefault}$P_s(C)$}}}}
\end{picture}
}
\end{center}
\caption{Figure for the case $C''$ participates in two crossings on
both its sides.} \label{twocrossfig}
\end{figure}

\item \label {veqg} $v''=g$ $(Q'' \subset O(P''))$. Let $u\in
[k,g]$ such that $[u,g] \subset O(P'')$ and $[u,g]$ is maximal. We
get:
\begin{equation}
\label{P8v'gopt} l(O(P'')) = l(O(P'',Q'',[u,g])) + l(Q'') + l(g,u).
\end{equation}
 According to the induction hypothesis
\begin{equation}
\label{P8v'gind}
 l(\hat{N}(P'',Q'',[u,g])) + 2l(\hat{A}(P'',Q'',[u,g])) \leq 9 l(O(P'',Q'',[g,u])).
\end{equation}
>From the way the algorithm works:
\begin{equation}
\label{P8v'ga}  l(A(P'')) \leq l(\hat{A}(P'',Q'',[u,g])) +l(Q'),
\end{equation}
it is possible that $P''$ used shortcuts contained in $P_s(C)$,
during the processing of $C$ these paths are removed from $A(P'')$
and then the inequality above is strict. Also:
\begin{equation}
\label{P8v'gn} l(N(P'')) = l(\hat{N}(P'',Q'',[u,g]))+ l(Q') + l(g,l)
+ l(g,u).
\end{equation}
Since $[x,v'] \cup [v',l] = Q'$ and using (\ref{v'v''eq}):
\[2l(x,v') +l(v',l) \leq 2l(x,v'') + l(v'',g) \leq 2l(Q').\]  Hence  we get
\begin{equation}
\label{P8v'gasum}  l(Q') = l(x,v') + l(v',l) \leq 2l(Q'').
\end{equation}
>From Property \ref{Cst} and \ref{P8v'gasum}
\begin{equation}
\label{P8v'gc} l(g,l) \leq l(Q') + l(Q'') \leq 3l(Q'') .
\end{equation}

Using (\ref{P8v'gn}),(\ref{P8v'ga}), (\ref{P8v'gasum}),
(\ref{P8v'gc}), (\ref{P8v'gind}),  and (\ref{P8v'gopt}) in this
order:
\begin{eqnarray*}
l(N(P'')) + 2l(A(P'')) &\leq& [l(\hat{N}(P'',Q'',[u,g])) +
l(Q') + l(g,l)] + l(g,u) +  \\
&& 2[l(\hat{A}(P'',Q'',[u,g])) + l(Q')] \\
&=&  l(\hat{N}(P'',Q'',[u,g])) + 2l(\hat{A}(P'',Q'',[u,g] ))+ 3l(Q') + l(g,l) + l(g,u)\\
&\leq&  l(\hat{N}(P'',Q'',[u,g])) + 2l(\hat{A}(P'',Q'',[u,g] ))+ 6l(Q'') + l(g,l) + l(g,u)\\
&\leq&  l(\hat{N}(P'',Q'',[u,g])) + 2l(\hat{A}(P'',Q'',[u,g])) + 9l(Q'') + l(g,u)\\
&\leq&  9 l(O(P'',Q'',[g,u])) +  9l(Q'') + l(g,u)\\
&\leq&  9 l(O(P''))  . \\
\end{eqnarray*}
\ee

\item \label{v'eql} $v'=l$ $(Q' \subset O(P')).$

By (\ref{v'v''eq})
\begin{equation}
\label{good v'} 2l(Q') = 2l(x,v') \leq 2l(x,v'') + l(v'',g).
\end{equation}
By definition
\begin{equation}
\label{P8v''gopt} l(O(P'')) = l(O(P'',[x,v''])) + l(x,v'') .
\end{equation}
 According to the induction hypothesis
\begin{equation}
\label{P8v''gind}
 l(\hat{N}(P'',[x,v''])) + 2l(\hat{A}(P'',[x,v''])) \leq 9 l(O(P'',[x,v'']])).
\end{equation}
>From the way the algorithm works
\begin{equation}
\label{P8v''ga}  l(A(P'')) \leq l(\hat{A}(P'',[x,v''])) +l(Q')
-l(v'',g),
\end{equation} (again if $P_s(C)$ contains shortcuts used by $P''$ the inequality is strict) and
\begin{equation}
\label{P8v''gn} l(N(P'')) = l(\hat{N}(P'',[x,v']')) +l(Q') -
l(v'',g) + l(g,l).
\end{equation}
>From Property \ref{Cst} and Equation (\ref{good v'})
\begin{equation}
\label{P8v''gc} l(g,l) \leq l(x,v'') + l(v'',g) + l(Q') < 2l(x,v'')
+ 1.5l(v'',g) .
\end{equation}

Using (\ref{P8v''gn}),(\ref{P8v''ga}), (\ref{good v'}),
(\ref{P8v''gc}),(\ref{P8v''gind}), and (\ref{P8v''gopt}) in this
order:
\begin{eqnarray*}
l(N(P'')) + 2l(A(P'')) &\leq& [l(\hat{N}(P'',[x,v''])) +
l(Q') - l(v'',g) + l(g,l)] + 2[l(\hat{A}(P'',[x,v''])) +\\
&& l(Q') - l(v'',g)]\\
&=&  l(\hat{N}(P'',[x,v''])) + 2l(\hat{A}(P'',[x,v''])) + 3l(Q') - 3l(v'',g) + l(g,l) \\
&\leq& l(\hat{N}(P'',[x,v''])) + 2l(\hat{A}(P'',[x,v''])) + 3l(x,v'') - 1.5l(v'',g) + l(g,l) \\
&\leq& l(\hat{N}(P'',[x,v''])) + 2l(\hat{A}(P'',[x,v''])) + 5l(x,v'')  \\
&\leq& 9l(O(P'',[x,v'']))  + 5l(x,v'')  \\
&\leq& 9l(O(P'')) . \\
\end{eqnarray*}
\ee

 \noindent { \bf When $P^* \neq Q'$ .} The proof for $P'$ is similar
to the proof we have seen before (the cross-node moves to the other
end of $P^* \cap Q'$). For $P''$ there are two cases to be
considered: \bi \item $v'' \neq g$. Same proof as in Case
(\ref{boundt}) since in this case also we can bound $l(g,end(P^*))$
with respect to $l(T(P'') \backslash \hat{T}(P''))$. \item $v'' =
g$. Same proof as in Case (\ref{veqg}) since here also $l(P^*) <
2l(Q'')$ and $Q'' \subset O(P'')$. \ei

\end{prf}

\begin{cor}
\label{ncrocor} For every node $v$, let $C^*$ be the original cycle
which served $v$ and let $C$ be the cycle which serves $v$ after the
application of CANCEL-CROSSING. Then
\[ l(C) \leq 9l(C^*) .\]
\end{cor}

\begin{prf}
Let $\breve{P}$ be the path induced by $E(U(C^*))$.  From Theorem
\ref{bpst}
\[ l(N(\breve{P})) + 2l(S(\breve{P})) \leq 9 l(O(\breve{P}))+ 2l(T(\breve{P})).\]
Since $T(\breve{P}) \subset S(\breve{P})$ we get  $l(N(\breve{P}))
\leq 9 l(O(\breve{P}))$, giving
\[ l(C) = l(N(\breve{P})) + l(\breve{P}) \leq 9 l(O(\breve{P})) + l(\breve{P}) \leq 9 l(O(\breve{P})) + 9l(\breve{P}) = 9l(C^*)
.\]
\end{prf}

\subsection{In-out}

During the processing of $C$ let $C'$ and $C''$ be two crossing
cycles with respect to $C$ such that $P_s(C') = [k,x] \cup [x,l]$
and $P_s(C'' ) = [g,x] \cup [x,h]$, where $x$ is the cross-node and
$k,g,l,h$ are nodes in $P_s(C)$ arranged in that order.

\begin{defin}
{\em Define $OUT(C') = [k,x]$, $IN(C') = [x,l]$, $OUT(C'') = [x,h]$
and $IN(C'') = [g,x]$.}
\end{defin}

\begin{defin}
{\em Define $U_{IN(C')} = \{ v \in IN(C') | C(v) = C' \}$,
$U_{OUT(C')} = \{ v \in OUT(C') | C(v) = C' \}$, $U_{IN(C'')} = \{ v
\in IN(C'') | C(v) = C'' \}$, and $U_{OUT(C'')} = \{ v \in OUT(C'')
| C(v) = C'' \}$.}
\end{defin}


\begin{lem}
\label{sempv} Before UNCROSS-CYCLES is applied on $C'$ and $C''$, if
$U_{OUT(C')} \neq \phi$ then $U_{IN(C'')} = \phi$. (Similarly, if
$U_{OUT(C'')} \neq \phi$ then $U_{IN(C')} = \phi$. )
\end{lem}

\begin{prf}
Suppose that $U_{OUT(C')} \neq \phi$. If $IN(C')$ is part of the
original cycle serving $U_{OUT(C')}$ then obviously
\begin{equation}
\label {smpveq}  l(IN(C') ) + l(l,h) < l(OUT(C'')).
\end{equation}
If $IN(C')$ is not part of the original cycle then $IN(C')$ and
$[l,h]$ are part of a previous shortcut, and by Property
\ref{shortsc} Equation \ref{smpveq} holds.

\noindent In both cases Equation \ref{smpveq} holds, therefore the
nodes in $V(IN(C''))$ must  use $(IN(C'),[l,h])$ as part of their
cycle (and not $OUT(C'')$), giving that $U_{IN(C'')} = \phi$.
\end{prf}

\begin{lem}
\label{empV} Either $U_{OUT(C'))} = U(C')$ and $U_{OUT(C'')} =
U(C'')$ or  $U_{OUT(C'))} = U_{OUT(C''))} =\phi$.  In the first case
we say that $C'$ and $C''$ are {\em Outer-crossing}. In the second
case we say that $C'$ and $C''$ are {\em Inner-crossing}.
\end{lem}

\begin{prf}
If $U_{OUT(C')} \neq \phi $ then by Lemma \ref{sempv} $U_{IN(C'')} =
\phi$ and $U_{OUT(C'')} = U(C'')$. Since there are nodes which use
the cycle $C''$ we get that $U_{OUT(C'')} \neq \phi$, and  by Lemma
\ref{sempv} $U_{IN(C')} = \phi$, giving that $U_{OUT(C')} = U(C')$.

\noindent If $U_{IN(C')} \neq \phi$ then by Lemma \ref{sempv}
$U_{OUT(C'')} = \phi$ and $U_{IN(C'')} = U(C'')$. Since there are
nodes which use the cycle $C''$ we get that $U_{IN(C'')} \neq \phi$,
and by Lemma \ref{sempv} $U_{OUT(C')} = \phi$, giving that
$U_{IN(C')} = U(C')$.
\end{prf}

\newpage

\section{Orienting the graph}

After all the changes described in the previous section, we are
given $\mathcal{C}$, a set of unoriented cycles with hereditary
order.

\begin{figure}
\begin{center}
\setlength{\unitlength}{0.00069991in}
\begingroup\makeatletter\ifx\SetFigFont\undefined%
\gdef\SetFigFont#1#2#3#4#5{%
  \reset@font\fontsize{#1}{#2pt}%
  \fontfamily{#3}\fontseries{#4}\fontshape{#5}%
  \selectfont}%
\fi\endgroup%
{\renewcommand{\dashlinestretch}{30}
\begin{picture}(9370,6228)(0,-10)
\put(4643,1554){\blacken\ellipse{90}{90}}
\put(4643,1554){\ellipse{90}{90}}
\put(1358,3219){\blacken\ellipse{90}{90}}
\put(1358,3219){\ellipse{90}{90}}
\put(1358,3894){\blacken\ellipse{90}{90}}
\put(1358,3894){\ellipse{90}{90}}
\put(1358,4254){\blacken\ellipse{90}{90}}
\put(1358,4254){\ellipse{90}{90}}
\put(1358,4749){\blacken\ellipse{90}{90}}
\put(1358,4749){\ellipse{90}{90}}
\put(2618,4749){\blacken\ellipse{90}{90}}
\put(2618,4749){\ellipse{90}{90}}
\put(2618,4254){\blacken\ellipse{90}{90}}
\put(2618,4254){\ellipse{90}{90}}
\put(4013,5379){\blacken\ellipse{90}{90}}
\put(4013,5379){\ellipse{90}{90}}
\put(4013,4749){\blacken\ellipse{90}{90}}
\put(4013,4749){\ellipse{90}{90}}
\put(4013,4254){\blacken\ellipse{90}{90}}
\put(4013,4254){\ellipse{90}{90}}
\put(2168,3354){\blacken\ellipse{90}{90}}
\put(2168,3354){\ellipse{90}{90}}
\put(5318,4749){\blacken\ellipse{90}{90}}
\put(5318,4749){\ellipse{90}{90}}
\put(5318,4254){\blacken\ellipse{90}{90}}
\put(5318,4254){\ellipse{90}{90}}
\put(6668,4749){\blacken\ellipse{90}{90}}
\put(6668,4749){\ellipse{90}{90}}
\put(6668,4254){\blacken\ellipse{90}{90}}
\put(6668,4254){\ellipse{90}{90}}
\put(6668,3219){\blacken\ellipse{90}{90}}
\put(6668,3219){\ellipse{90}{90}}
\put(4013,2499){\ellipse{8010}{1890}}
\path(4013,4254)(6668,4254)
\path(2618,4749)(2618,4254)
\path(4013,4749)(6668,4749)
\path(5318,4749)(5318,4254)
\path(1358,4749)(4013,4749)
\path(1358,3894)(2168,3894)(2168,3354)
\blacken\path(3983.000,1591.500)(3833.000,1554.000)(3983.000,1516.500)(3983.000,1591.500)
\path(3833,1554)(4148,1554)
\blacken\path(1395.500,3744.000)(1358.000,3894.000)(1320.500,3744.000)(1395.500,3744.000)
\path(1358,3894)(1358,3219)
\blacken\path(1395.500,4104.000)(1358.000,4254.000)(1320.500,4104.000)(1395.500,4104.000)
\path(1358,4254)(1358,3894)
\blacken\path(1395.500,4599.000)(1358.000,4749.000)(1320.500,4599.000)(1395.500,4599.000)
\path(1358,4749)(1358,4254)
\blacken\path(1395.500,5229.000)(1358.000,5379.000)(1320.500,5229.000)(1395.500,5229.000)
\path(1358,5379)(1358,4749)
\path(1358,5379)(4013,5379)
\blacken\path(3863.000,5341.500)(4013.000,5379.000)(3863.000,5416.500)(3863.000,5341.500)
\blacken\path(3975.500,4899.000)(4013.000,4749.000)(4050.500,4899.000)(3975.500,4899.000)
\path(4013,4749)(4013,5379)
\blacken\path(3975.500,4404.000)(4013.000,4254.000)(4050.500,4404.000)(3975.500,4404.000)
\path(4013,4254)(4013,4749)
\blacken\path(3975.500,3549.000)(4013.000,3399.000)(4050.500,3549.000)(3975.500,3549.000)
\path(4013,3399)(4013,4254)
\blacken\path(3773.000,3406.500)(3923.000,3444.000)(3773.000,3481.500)(3773.000,3406.500)
\path(3923,3444)(3833,3444)
\blacken\path(6705.500,4104.000)(6668.000,4254.000)(6630.500,4104.000)(6705.500,4104.000)
\path(6668,4254)(6668,3219)
\blacken\path(6705.500,4599.000)(6668.000,4749.000)(6630.500,4599.000)(6705.500,4599.000)
\path(6668,4749)(6668,4254)
\blacken\path(6705.500,5229.000)(6668.000,5379.000)(6630.500,5229.000)(6705.500,5229.000)
\path(6668,5379)(6668,4749)
\blacken\path(4163.000,5416.500)(4013.000,5379.000)(4163.000,5341.500)(4163.000,5416.500)
\path(4013,5379)(6668,5379)
\path(1358,4254)(2618,4254)
\blacken\path(2468.000,4216.500)(2618.000,4254.000)(2468.000,4291.500)(2468.000,4216.500)
\path(2618,4254)(4013,4254)
\blacken\path(3863.000,4216.500)(4013.000,4254.000)(3863.000,4291.500)(3863.000,4216.500)
\put(3833,1689){\makebox(0,0)[lb]{\smash{{\SetFigFont{10}{12.0}{\rmdefault}{\mddefault}{\updefault}$C_0$}}}}
\put(1988,4074){\makebox(0,0)[lb]{\smash{{\SetFigFont{10}{12.0}{\rmdefault}{\mddefault}{\updefault}1}}}}
\put(3338,4074){\makebox(0,0)[lb]{\smash{{\SetFigFont{10}{12.0}{\rmdefault}{\mddefault}{\updefault}1}}}}
\put(4643,1284){\makebox(0,0)[lb]{\smash{{\SetFigFont{10}{12.0}{\rmdefault}{\mddefault}{\updefault}$z$}}}}
\put(4688,4074){\makebox(0,0)[lb]{\smash{{\SetFigFont{10}{12.0}{\rmdefault}{\mddefault}{\updefault}1}}}}
\put(6038,4074){\makebox(0,0)[lb]{\smash{{\SetFigFont{10}{12.0}{\rmdefault}{\mddefault}{\updefault}1}}}}
\put(5318,5469){\makebox(0,0)[lb]{\smash{{\SetFigFont{10}{12.0}{\rmdefault}{\mddefault}{\updefault}1.5}}}}
\put(3338,4839){\makebox(0,0)[lb]{\smash{{\SetFigFont{10}{12.0}{\rmdefault}{\mddefault}{\updefault}3}}}}
\put(4688,4839){\makebox(0,0)[lb]{\smash{{\SetFigFont{10}{12.0}{\rmdefault}{\mddefault}{\updefault}3}}}}
\put(6038,4839){\makebox(0,0)[lb]{\smash{{\SetFigFont{10}{12.0}{\rmdefault}{\mddefault}{\updefault}2}}}}
\put(2753,4434){\makebox(0,0)[lb]{\smash{{\SetFigFont{10}{12.0}{\rmdefault}{\mddefault}{\updefault}1}}}}
\put(5453,4434){\makebox(0,0)[lb]{\smash{{\SetFigFont{10}{12.0}{\rmdefault}{\mddefault}{\updefault}5}}}}
\put(1178,4749){\makebox(0,0)[lb]{\smash{{\SetFigFont{10}{12.0}{\rmdefault}{\mddefault}{\updefault}$d$}}}}
\put(1178,3219){\makebox(0,0)[lb]{\smash{{\SetFigFont{10}{12.0}{\rmdefault}{\mddefault}{\updefault}$a$}}}}
\put(1178,3849){\makebox(0,0)[lb]{\smash{{\SetFigFont{10}{12.0}{\rmdefault}{\mddefault}{\updefault}$b$}}}}
\put(1178,4254){\makebox(0,0)[lb]{\smash{{\SetFigFont{10}{12.0}{\rmdefault}{\mddefault}{\updefault}$c$}}}}
\put(1988,4839){\makebox(0,0)[lb]{\smash{{\SetFigFont{10}{12.0}{\rmdefault}{\mddefault}{\updefault}3}}}}
\put(3833,4074){\makebox(0,0)[lb]{\smash{{\SetFigFont{10}{12.0}{\rmdefault}{\mddefault}{\updefault}$i$}}}}
\put(3833,4614){\makebox(0,0)[lb]{\smash{{\SetFigFont{10}{12.0}{\rmdefault}{\mddefault}{\updefault}$j$}}}}
\put(3833,5469){\makebox(0,0)[lb]{\smash{{\SetFigFont{10}{12.0}{\rmdefault}{\mddefault}{\updefault}$k$}}}}
\put(2618,5469){\makebox(0,0)[lb]{\smash{{\SetFigFont{10}{12.0}{\rmdefault}{\mddefault}{\updefault}3}}}}
\put(6803,3219){\makebox(0,0)[lb]{\smash{{\SetFigFont{10}{12.0}{\rmdefault}{\mddefault}{\updefault}$n$}}}}
\put(1988,3399){\makebox(0,0)[lb]{\smash{{\SetFigFont{10}{12.0}{\rmdefault}{\mddefault}{\updefault}$e$}}}}
\put(2573,4074){\makebox(0,0)[lb]{\smash{{\SetFigFont{10}{12.0}{\rmdefault}{\mddefault}{\updefault}$f$}}}}
\put(2573,4839){\makebox(0,0)[lb]{\smash{{\SetFigFont{10}{12.0}{\rmdefault}{\mddefault}{\updefault}$g$}}}}
\put(3833,3534){\makebox(0,0)[lb]{\smash{{\SetFigFont{10}{12.0}{\rmdefault}{\mddefault}{\updefault}$h$}}}}
\put(5273,4074){\makebox(0,0)[lb]{\smash{{\SetFigFont{10}{12.0}{\rmdefault}{\mddefault}{\updefault}$l$}}}}
\put(5273,4794){\makebox(0,0)[lb]{\smash{{\SetFigFont{10}{12.0}{\rmdefault}{\mddefault}{\updefault}$m$}}}}
\put(6803,4209){\makebox(0,0)[lb]{\smash{{\SetFigFont{10}{12.0}{\rmdefault}{\mddefault}{\updefault}$o$}}}}
\put(6803,4749){\makebox(0,0)[lb]{\smash{{\SetFigFont{10}{12.0}{\rmdefault}{\mddefault}{\updefault}$r$}}}}
\put(8063,6054){\makebox(0,0)[lb]{\smash{{\SetFigFont{10}{12.0}{\rmdefault}{\mddefault}{\updefault}The cycles}}}}
\put(4463,69){\makebox(0,0)[lb]{\smash{{\SetFigFont{10}{12.0}{\rmdefault}{\mddefault}{\updefault}$S_{hn}=\{C_6,C_7,C_8,C_9\}$}}}}
\put(98,69){\makebox(0,0)[lb]{\smash{{\SetFigFont{10}{12.0}{\rmdefault}{\mddefault}{\updefault}$S_{ae}=\{C_1\}$}}}}
\put(1898,69){\makebox(0,0)[lb]{\smash{{\SetFigFont{10}{12.0}{\rmdefault}{\mddefault}{\updefault}$S_{ah}=\{C_2,C_3,C_4,C_5\}$}}}}
\put(8063,5829){\makebox(0,0)[lb]{\smash{{\SetFigFont{10}{12.0}{\rmdefault}{\mddefault}{\updefault}$C_0$ $zaehnz$}}}}
\put(8063,5604){\makebox(0,0)[lb]{\smash{{\SetFigFont{10}{12.0}{\rmdefault}{\mddefault}{\updefault}$C_1$ $zabehnz$}}}}
\put(8063,5379){\makebox(0,0)[lb]{\smash{{\SetFigFont{10}{12.0}{\rmdefault}{\mddefault}{\updefault}$C_2$ $zabcfihnz$}}}}
\put(8063,5154){\makebox(0,0)[lb]{\smash{{\SetFigFont{10}{12.0}{\rmdefault}{\mddefault}{\updefault}$C_3$ $zabcdgfihnz$}}}}
\put(8063,4929){\makebox(0,0)[lb]{\smash{{\SetFigFont{10}{12.0}{\rmdefault}{\mddefault}{\updefault}$C_4$ $zabcfgjihnz$}}}}
\put(8063,4704){\makebox(0,0)[lb]{\smash{{\SetFigFont{10}{12.0}{\rmdefault}{\mddefault}{\updefault}$C_5$ $zabcdkjihnz$}}}}
\put(8063,4479){\makebox(0,0)[lb]{\smash{{\SetFigFont{10}{12.0}{\rmdefault}{\mddefault}{\updefault}$C_6$ $zaehilonz$}}}}
\put(8063,4254){\makebox(0,0)[lb]{\smash{{\SetFigFont{10}{12.0}{\rmdefault}{\mddefault}{\updefault}$C_7$ $zaehilmronz$}}}}
\put(8063,4029){\makebox(0,0)[lb]{\smash{{\SetFigFont{10}{12.0}{\rmdefault}{\mddefault}{\updefault}$C_8$ $zaehijmronz$}}}}
\put(8063,3804){\makebox(0,0)[lb]{\smash{{\SetFigFont{10}{12.0}{\rmdefault}{\mddefault}{\updefault}$C_9$ $zaehijkronz$}}}}
\end{picture}
}
\end{center}
\caption{An example.} \label{deffig}
\end{figure}

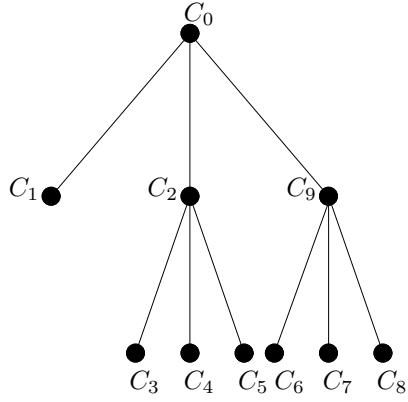
\begin{figure}
\begin{center}
\setlength{\unitlength}{0.00069991in}
\begingroup\makeatletter\ifx\SetFigFont\undefined%
\gdef\SetFigFont#1#2#3#4#5{%
  \reset@font\fontsize{#1}{#2pt}%
  \fontfamily{#3}\fontseries{#4}\fontshape{#5}%
  \selectfont}%
\fi\endgroup%
{\renewcommand{\dashlinestretch}{30}
\begin{picture}(2990,2925)(0,-10)
\put(1365,2670){\blacken\ellipse{134}{134}}
\put(1365,2670){\ellipse{134}{134}}
\put(1365,1455){\blacken\ellipse{134}{134}}
\put(1365,1455){\ellipse{134}{134}}
\put(330,1455){\blacken\ellipse{134}{134}}
\put(330,1455){\ellipse{134}{134}}
\put(2400,1455){\blacken\ellipse{134}{134}}
\put(2400,1455){\ellipse{134}{134}}
\put(960,285){\blacken\ellipse{134}{134}}
\put(960,285){\ellipse{134}{134}}
\put(1365,285){\blacken\ellipse{134}{134}}
\put(1365,285){\ellipse{134}{134}}
\put(1770,285){\blacken\ellipse{134}{134}}
\put(1770,285){\ellipse{134}{134}}
\put(1995,285){\blacken\ellipse{134}{134}}
\put(1995,285){\ellipse{134}{134}}
\put(2400,285){\blacken\ellipse{134}{134}}
\put(2400,285){\ellipse{134}{134}}
\put(2805,285){\blacken\ellipse{134}{134}}
\put(2805,285){\ellipse{134}{134}}
\path(1365,2670)(330,1455)
\path(1365,2670)(1365,1455)
\path(1365,1455)(1365,285)
\path(1365,2670)(2400,1455)
\path(2400,1410)(1995,285)
\path(2400,1455)(2805,285)
\path(1365,1455)(960,285)
\path(1365,1455)(1770,285)
\path(2400,1455)(2400,285)
\put(1320,2760){\makebox(0,0)[lb]{\smash{{\SetFigFont{10}{12.0}{\rmdefault}{\mddefault}{\updefault}$C_0$}}}}
\put(1050,1455){\makebox(0,0)[lb]{\smash{{\SetFigFont{10}{12.0}{\rmdefault}{\mddefault}{\updefault}$C_2$}}}}
\put(15,1455){\makebox(0,0)[lb]{\smash{{\SetFigFont{10}{12.0}{\rmdefault}{\mddefault}{\updefault}$C_1$}}}}
\put(2085,1455){\makebox(0,0)[lb]{\smash{{\SetFigFont{10}{12.0}{\rmdefault}{\mddefault}{\updefault}$C_9$}}}}
\put(1725,15){\makebox(0,0)[lb]{\smash{{\SetFigFont{10}{12.0}{\rmdefault}{\mddefault}{\updefault}$C_5$}}}}
\put(2355,15){\makebox(0,0)[lb]{\smash{{\SetFigFont{10}{12.0}{\rmdefault}{\mddefault}{\updefault}$C_7$}}}}
\put(915,15){\makebox(0,0)[lb]{\smash{{\SetFigFont{10}{12.0}{\rmdefault}{\mddefault}{\updefault}$C_3$}}}}
\put(1320,15){\makebox(0,0)[lb]{\smash{{\SetFigFont{10}{12.0}{\rmdefault}{\mddefault}{\updefault}$C_4$}}}}
\put(1995,15){\makebox(0,0)[lb]{\smash{{\SetFigFont{10}{12.0}{\rmdefault}{\mddefault}{\updefault}$C_6$}}}}
\put(2760,15){\makebox(0,0)[lb]{\smash{{\SetFigFont{10}{12.0}{\rmdefault}{\mddefault}{\updefault}$C_8$}}}}
\end{picture}
}
\end{center}
\caption{The hereditary tree for Figure \ref{deffig}.}
\label{deftree}
\end{figure}

\noindent In Figure \ref{deffig} an example of a partial oriented
graph, in intermediate stage of the algorithm, is given with its
corresponding $\mathcal{C} = \{ C_0,\ldots,C_9 \}$. An edge whose
length  is not indicated is of  zero length. In this graph: \bi
\item $C_1,C_2$ and $C_9$ are sons of $C_0$.
\item $C_3,C_4$ and $C_5$ are sons of $C_2$.
\item $C_6,C_7$ and $C_8$ are sons of  $C_9$.
\ei

\begin{defin}
{\em  For a cycle $C \in \mathcal{C}$ define its {\it generation
level} $g(C)$ as the distance in the hereditary tree from $C_0$. In
Figure \ref{deftree} we see that $g(C_0) = 0$, $g(C_2)=1$ and
$g(C_3) = 2$.}
\end{defin}

\begin{defin}
{\em If $C'$ is a contained brother of $C$ we mark it $C'\sqsubseteq
C$. If $C$ is not a contained brother of any cycle then it is {\em
uncontained}. In Figure \ref{deffig} $C_2$ and $C_9$ are
uncontained. $C_3,C_4$ are contained brothers of $C_5$.}
\end{defin}

\begin{defin}
{\em   For a cycle $C \in \mathcal{C} \backslash \{ C_0 \}$ define
{\it the level of containment} $lc(C)$ in the following manner: \bi
\item $lc(C) = 0$ if $C$ is uncontained.
\item $lc(C) = i$ if $C$ is a maximal contained brother of a cycle
$C'$ and $lc(C') = i-1$. \ei In Figure \ref{deffig} $lc(C_6) = 0$,
$lc(C_7) = 1$ and $lc(C_8) = 2$. }
\end{defin}

 The algorithm described below orients the  $P_s$ of cycles in order of their
 generation level. Starting by orienting $C_0$ then orienting all the
 cycles with generation level 1, and continuing with higher generation level.
In each generation the orientation is performed in order of  level
of containment.
 Again, starting by orienting the cycles with level of containment zero and continuing
by increasing  level of containment.

\begin{defin} {\em For every $v \in V$ we mark with $C_v \in \mathcal{C}$ the
shortest cycle in $\mathcal{C}$ containing $v$. For every cycle $C
\in \mathcal{C}$, let $U(C) = \{ v \in V : C_v = C \}$.
 By construction $U(C) \neq \phi$.}
\end{defin}

\begin{obs}
$U(C)$ induces a path.
\end{obs}

\begin{defin}
{\em The  algorithm will orient the edges in $U(C)$  to form a
directed path.
  We call the orientation induced on $C$  {\em the orientation of
 $C$}.}
\end{defin}
\begin {defin}
{\em A cycle $C$ is {\em oriented  backwards (forwards)} if $U(C)$
and $U(F(C))$ induce different (identical) directions on $P_c(C)$.
In Figure \ref{deffig} $C_2$ is oriented  forwards
$(U(C_2)=[a,b,c,f,i,h])$ and $C_9$ is oriented backwards
$(U(C_9)=[n,o,r,k,j,i,h])$. }
\end{defin}

\begin{defin}
{\em $C'$, a contained brother of $C$ is {\em  maximal contained}
when there is no $\hat{C}$ such that $ C' \sqsubseteq \hat{C}
\sqsubseteq C$. In Figure \ref{deffig} $C_1$ is a maximal contained
brother of $C_2$.}
\end{defin}

\begin{figure}
\begin{center}
\setlength{\unitlength}{0.00069991in}
\begingroup\makeatletter\ifx\SetFigFont\undefined%
\gdef\SetFigFont#1#2#3#4#5{%
  \reset@font\fontsize{#1}{#2pt}%
  \fontfamily{#3}\fontseries{#4}\fontshape{#5}%
  \selectfont}%
\fi\endgroup%
{\renewcommand{\dashlinestretch}{30}
\begin{picture}(9699,3780)(0,-10)
\put(5457,285){\blacken\ellipse{90}{90}}
\put(5457,285){\ellipse{90}{90}}
\put(912,285){\blacken\ellipse{90}{90}}
\put(912,285){\ellipse{90}{90}}
\put(2082,285){\blacken\ellipse{90}{90}}
\put(2082,285){\ellipse{90}{90}}
\put(2487,285){\blacken\ellipse{90}{90}}
\put(2487,285){\ellipse{90}{90}}
\put(3972,285){\blacken\ellipse{90}{90}}
\put(3972,285){\ellipse{90}{90}}
\put(4557,285){\blacken\ellipse{90}{90}}
\put(4557,285){\ellipse{90}{90}}
\put(6267,285){\blacken\ellipse{90}{90}}
\put(6267,285){\ellipse{90}{90}}
\put(912,780){\blacken\ellipse{90}{90}}
\put(912,780){\ellipse{90}{90}}
\put(2802,645){\blacken\ellipse{90}{90}}
\put(2802,645){\ellipse{90}{90}}
\put(912,1230){\blacken\ellipse{90}{90}}
\put(912,1230){\ellipse{90}{90}}
\put(2082,1230){\blacken\ellipse{90}{90}}
\put(2082,1230){\ellipse{90}{90}}
\put(2802,1230){\blacken\ellipse{90}{90}}
\put(2802,1230){\ellipse{90}{90}}
\put(3972,1230){\blacken\ellipse{90}{90}}
\put(3972,1230){\ellipse{90}{90}}
\put(4512,1230){\blacken\ellipse{90}{90}}
\put(4512,1230){\ellipse{90}{90}}
\put(3297,1230){\blacken\ellipse{90}{90}}
\put(3297,1230){\ellipse{90}{90}}
\put(5457,1230){\blacken\ellipse{90}{90}}
\put(5457,1230){\ellipse{90}{90}}
\put(6267,1230){\blacken\ellipse{90}{90}}
\put(6267,1230){\ellipse{90}{90}}
\put(6897,1230){\blacken\ellipse{90}{90}}
\put(6897,1230){\ellipse{90}{90}}
\put(7707,1230){\blacken\ellipse{90}{90}}
\put(7707,1230){\ellipse{90}{90}}
\put(8472,1230){\blacken\ellipse{90}{90}}
\put(8472,1230){\ellipse{90}{90}}
\put(282,3480){\blacken\ellipse{90}{90}}
\put(282,3480){\ellipse{90}{90}}
\put(282,285){\blacken\ellipse{90}{90}}
\put(282,285){\ellipse{90}{90}}
\put(9307,1190){\blacken\ellipse{90}{90}}
\put(9307,1190){\ellipse{90}{90}}
\put(7707,3480){\blacken\ellipse{90}{90}}
\put(7707,3480){\ellipse{90}{90}}
\put(9327,3480){\blacken\ellipse{90}{90}}
\put(9327,3480){\ellipse{90}{90}}
\put(4557,780){\blacken\ellipse{90}{90}}
\put(4557,780){\ellipse{90}{90}}
\put(2802,285){\blacken\ellipse{90}{90}}
\put(2802,285){\ellipse{90}{90}}
\put(6897,285){\blacken\ellipse{90}{90}}
\put(6897,285){\ellipse{90}{90}}
\put(8472,285){\blacken\ellipse{90}{90}}
\put(8472,285){\ellipse{90}{90}}
\put(9327,285){\blacken\ellipse{90}{90}}
\put(9327,285){\ellipse{90}{90}}
\put(7707,285){\blacken\ellipse{90}{90}}
\put(7707,285){\ellipse{90}{90}}
\path(2082,1230)(2802,1230)(2802,285)
\path(5457,285)(5457,1230)(6267,1230)(6267,285)
\path(6267,1230)(6897,1230)(6897,285)
\path(6897,1230)(7707,1230)(7707,285)
\thicklines
\path(4557,285)(4557,780)
\thinlines
\path(7707,3480)(9327,3480)(9327,285)
\path(7707,1230)(9327,1230)
\path(8472,1230)(8472,285)
\thicklines
\path(912,285)(912,780)
\thinlines
\path(282,285)(282,3480)(7707,3480)(7707,285)
\path(912,285)(912,1230)(2082,1230)(2082,285)
\path(282,285)(912,285)
\blacken\path(762.000,247.500)(912.000,285.000)(762.000,322.500)(762.000,247.500)
\path(12,285)(282,285)
\blacken\path(132.000,247.500)(282.000,285.000)(132.000,322.500)(132.000,247.500)
\path(912,285)(2082,285)
\blacken\path(1932.000,247.500)(2082.000,285.000)(1932.000,322.500)(1932.000,247.500)
\path(2082,285)(2487,285)
\blacken\path(2337.000,247.500)(2487.000,285.000)(2337.000,322.500)(2337.000,247.500)
\path(2487,285)(3297,1230)(3972,1230)(3972,285)
\path(2487,285)(2802,285)
\blacken\path(2652.000,247.500)(2802.000,285.000)(2652.000,322.500)(2652.000,247.500)
\path(2802,285)(3972,285)
\blacken\path(3822.000,247.500)(3972.000,285.000)(3822.000,322.500)(3822.000,247.500)
\path(3972,1230)(4557,1230)(4557,285)
\path(3972,285)(4557,285)
\blacken\path(4407.000,247.500)(4557.000,285.000)(4407.000,322.500)(4407.000,247.500)
\path(4557,285)(5457,285)
\blacken\path(5307.000,247.500)(5457.000,285.000)(5307.000,322.500)(5307.000,247.500)
\path(5457,285)(6267,285)
\blacken\path(6117.000,247.500)(6267.000,285.000)(6117.000,322.500)(6117.000,247.500)
\path(6267,285)(6897,285)
\blacken\path(6747.000,247.500)(6897.000,285.000)(6747.000,322.500)(6747.000,247.500)
\path(6942,285)(7707,285)
\blacken\path(7557.000,247.500)(7707.000,285.000)(7557.000,322.500)(7557.000,247.500)
\path(7707,285)(8472,285)
\blacken\path(8322.000,247.500)(8472.000,285.000)(8322.000,322.500)(8322.000,247.500)
\path(8472,285)(9327,285)
\blacken\path(9177.000,247.500)(9327.000,285.000)(9177.000,322.500)(9177.000,247.500)
\path(9327,285)(9687,285)
\blacken\path(9537.000,247.500)(9687.000,285.000)(9537.000,322.500)(9537.000,247.500)
\put(5727,1365){\makebox(0,0)[lb]{\smash{{\SetFigFont{10}{12.0}{\rmdefault}{\mddefault}{\updefault}$C_5$}}}}
\put(1362,1365){\makebox(0,0)[lb]{\smash{{\SetFigFont{10}{12.0}{\rmdefault}{\mddefault}{\updefault}$C_1$}}}}
\put(2442,1365){\makebox(0,0)[lb]{\smash{{\SetFigFont{10}{12.0}{\rmdefault}{\mddefault}{\updefault}$C_2$}}}}
\put(3522,1365){\makebox(0,0)[lb]{\smash{{\SetFigFont{10}{12.0}{\rmdefault}{\mddefault}{\updefault}$C_3$}}}}
\put(4130,1365){\makebox(0,0)[lb]{\smash{{\SetFigFont{10}{12.0}{\rmdefault}{\mddefault}{\updefault}$C_4$}}}}
\put(6447,1365){\makebox(0,0)[lb]{\smash{{\SetFigFont{10}{12.0}{\rmdefault}{\mddefault}{\updefault}$C_6$}}}}
\put(2037,15){\makebox(0,0)[lb]{\smash{{\SetFigFont{10}{12.0}{\rmdefault}{\mddefault}{\updefault}$v_2$}}}}
\put(2442,15){\makebox(0,0)[lb]{\smash{{\SetFigFont{10}{12.0}{\rmdefault}{\mddefault}{\updefault}$v_3$}}}}
\put(3927,15){\makebox(0,0)[lb]{\smash{{\SetFigFont{10}{12.0}{\rmdefault}{\mddefault}{\updefault}$v_4$}}}}
\put(4512,15){\makebox(0,0)[lb]{\smash{{\SetFigFont{10}{12.0}{\rmdefault}{\mddefault}{\updefault}$u$}}}}
\put(5412,15){\makebox(0,0)[lb]{\smash{{\SetFigFont{10}{12.0}{\rmdefault}{\mddefault}{\updefault}$v_5$}}}}
\put(6222,15){\makebox(0,0)[lb]{\smash{{\SetFigFont{10}{12.0}{\rmdefault}{\mddefault}{\updefault}$v_6$}}}}
\put(597,510){\makebox(0,0)[lb]{\smash{{\SetFigFont{10}{12.0}{\rmdefault}{\mddefault}{\updefault}$e_v$}}}}
\put(4692,510){\makebox(0,0)[lb]{\smash{{\SetFigFont{10}{12.0}{\rmdefault}{\mddefault}{\updefault}$e_u$}}}}
\put(732,15){\makebox(0,0)[lb]{\smash{{\SetFigFont{10}{12.0}{\rmdefault}{\mddefault}{\updefault}$v_1=v$}}}}
\put(7212,1365){\makebox(0,0)[lb]{\smash{{\SetFigFont{10}{12.0}{\rmdefault}{\mddefault}{\updefault}$C_7$}}}}
\put(7932,1365){\makebox(0,0)[lb]{\smash{{\SetFigFont{10}{12.0}{\rmdefault}{\mddefault}{\updefault}$C_8$}}}}
\put(8742,1365){\makebox(0,0)[lb]{\smash{{\SetFigFont{10}{12.0}{\rmdefault}{\mddefault}{\updefault}$C_9$}}}}
\put(6852,15){\makebox(0,0)[lb]{\smash{{\SetFigFont{10}{12.0}{\rmdefault}{\mddefault}{\updefault}$v_7$}}}}
\put(7662,15){\makebox(0,0)[lb]{\smash{{\SetFigFont{10}{12.0}{\rmdefault}{\mddefault}{\updefault}$v_8$}}}}
\put(8427,15){\makebox(0,0)[lb]{\smash{{\SetFigFont{10}{12.0}{\rmdefault}{\mddefault}{\updefault}$v_9$}}}}
\put(9282,15){\makebox(0,0)[lb]{\smash{{\SetFigFont{10}{12.0}{\rmdefault}{\mddefault}{\updefault}$v_{10}$}}}}
\put(3477,3615){\makebox(0,0)[lb]{\smash{{\SetFigFont{10}{12.0}{\rmdefault}{\mddefault}{\updefault}$C'$}}}}
\put(8427,3615){\makebox(0,0)[lb]{\smash{{\SetFigFont{10}{12.0}{\rmdefault}{\mddefault}{\updefault}$C''$}}}}
\end{picture}
}
\end{center}
\caption{Three blocks.} \label{intervalfig}
\end{figure}
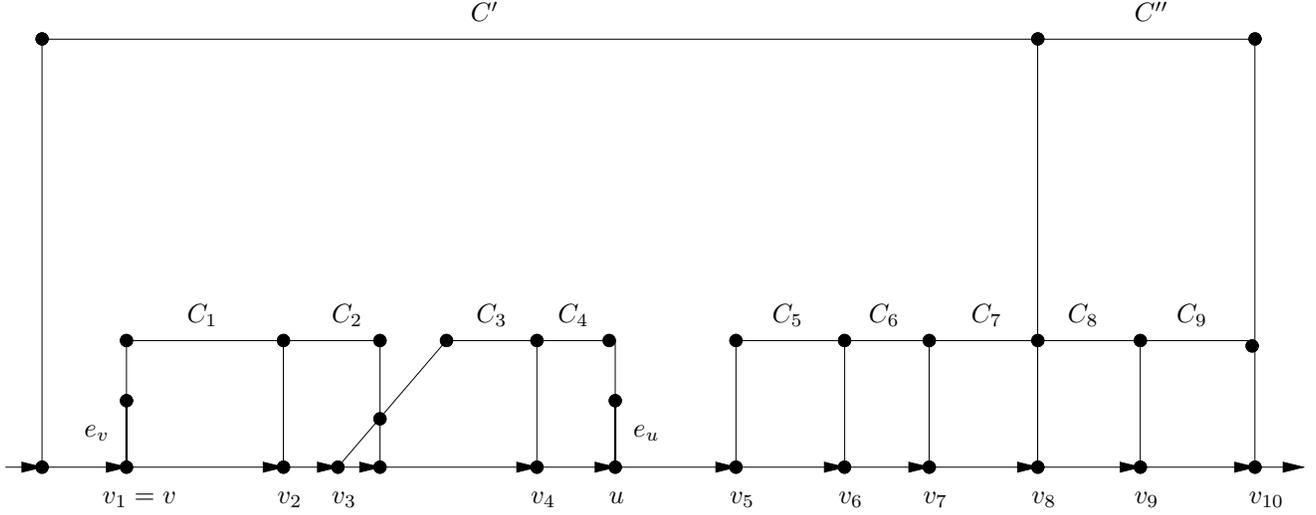

\begin{defin}
\label{maxintdef} {\em We say that brothers $C_l,\ldots,C_k$ $(l
\leq k)$ create a {\em block} $(C_l,\ldots,C_k)$ when: \bi
\item $lc(C_l)=lc(C_{l+1})=\cdots=lc(C_k)$.
\item If $lc(C_l) > 0$ then  there is $C'$ such that (for $l \leq i \leq
k$) $lc(C') = lc(C_i) -1$ and $C_i \sqsubseteq C'$ .
\item $C_i$ and $C_{i+1}$ are neighbor or crossing brothers  (for $l \leq i \leq
k-1$). \ei In Figure \ref{intervalfig} we see three blocks,
$(C_1,C_2,C_3,C_4)$, $(C_5,C_6,C_7)$ and $(C_8,C_9)$. The cycles
$C_5,\ldots,C_9$ are divided into two blocks since $C_5,C_6,C_7$ are
contained brothers of $C'$ and $C_8,C_9$ are contained brothers of
$C''$.}
\end{defin}

\begin{defin}
{\em Let $C_1,C_2,\ldots,C_m$ be sons of $C$. Mark with $v_i$ the
"left" end node of $P_f(C_i)$.  $C_1,C_2,\ldots,C_m$ are {\em
ordered according to the direction of $C$} if $v_1,v_2,\ldots,v_m$
are ordered in this way. Whenever we write $C_1,C_2,\ldots,C_m$ we
assume the cycles are ordered according to direction of their
father. In Figure \ref{intervalfig} the nodes $v_1,\ldots,v_9$ are
marked,  the cycles $C_1,\ldots,C_9$ are ordered according to the
direction of their father. }
\end{defin}

\begin{obs}
{\em For every cycle $C$, $P_f(C) \subseteq P_s(F(C))$.}
\end{obs}

\begin{defin}
{\em If $l(P_f(C))> \frac{1}{3}l(P_s(F(C))$  $C$ is {\em heavy}, if
$l(P_f(C)) > \frac{2}{3} l(P_s(F(C))$  $C$ is {\em very heavy}, if
$l(P_f(C)) \leq \frac{1}{3}l(P_s(F(C))$ $C$ is {\em light}}.
\end{defin}

\begin{defin}
\label{csetodef} {\em A cycle {\em sets the orientation of an edge}
when it gives the first orientation of the edge or when it changes a
previous orientation of this edge.}
\end{defin}

\begin{obs}
\label{croosobs}
 {\em If $C_1$ and $C_2$ are crossing brothers with $lc(C_1) =
 lc(C_2)$ then either:
 \bi
 \item $C_1$ and $C_2$ are inner crossing. The cross-node's cycle $C'$
 is a containing brother of $C_1$ and $C_2$ ($C_1 \sqsubseteq C'$, $C_2 \sqsubseteq
C'$), $C_1,C_2$ are the only maximal uncontained brothers of $C'$
and $l(C') < min \{ l(C_1),l(C_2) \}$. In this case we say that $C'$
is {\em the closest containing brother} of $C_1$ and $C_2$. (In
Figure \ref{inncrossfig} $C_1,C_2$ and $C'$ are illustrated). Or
\item $C_1$ and $C_2$ are outer crossing and  at least
one of them is very heavy (otherwise $C_1$ and $C_2$ were
not-very-heavy-outer-crossing and were changed in previous section).
In this work we handle the case when at most  one such crossing can
exist in each level of containment \footnote{A special procedure can
be written for the  case when two such crossing exist.}.
 \ei}
\end{obs}

\begin{defin}
\label{scbdef} {\em Suppose that:
 \bi
  \item $C \sqsubseteq B$,
  \item  $P_s(C) \cap P_s(B) \neq \phi$,
  \item $C$ is the only contained brother of $B$ in its level of containment,
  \item  $B$ is oriented forwards.
  \ei Then $C$ is a {\em special contained brother } of $B$.
  In Figure \ref{deffig} $C_1$ is a special contained brother of $C_2$.}
\end{defin}


\subsection{General description} The algorithm orients the cycles of
$G$ according to generation order. In each application of  the
procedure DIRECT-BROTHERS, it orients $P_s(C)$ for all the cycles
$C$ which belong to  a certain generation.  The algorithm may change
a previously defined direction but only in the procedure DIRECT-ONE.

In the following  proofs we define for every  $C \in \mathcal{C}
\backslash \{ C_0 \}$ several directed paths. The concatenation of
these paths creates a directed cycle connecting $z$ and $U(C)$. We
 prove that the length of these paths is bounded with respect to
the length of $C$.

In the algorithm there are cycles whose undirected cycle (the one
that connects them to $z$) is changed. Again the length of the new
cycle is bounded with respect to the original cycle's length.

Ideally we would like to orient the $P_s$ of cycles alternately (see
Figure \ref{dirmanyfig}(a) and (b)). However, there are limitations
implied by containing brothers and previous generations which were
oriented before. Hence, some neighbor brothers will have to be
oriented in the same direction (forwards or backwards). Consider
Figure \ref{bigexfig}. In this figure $D_i$ is the oriented cycle
connecting $U(C_i)$ and $z$ for $i = 1,\ldots,11$.
$D_8,\ldots,D_{11}$ in this figure are all oriented in the same
direction. We use the notation {\em directed cycle} for a directed
walk that starts and ends in $z$. In this notation a directed cycle
may use an edge twice.

In Figure \ref{bigexfig} $D_1,D_2$ and $D_3,D_4,D_5$ are oriented
alternately. In this figure we can see how the orientation changed
the undirected cycles.

We sometimes use the same name $C$ for the undirected cycle and the
directed cycle which connects $U(C)$ and $z$. From the context it
will always be clear which cycle is referred to.

\begin{figure}
\begin{center}
\setlength{\unitlength}{0.00069991in}
\begingroup\makeatletter\ifx\SetFigFont\undefined%
\gdef\SetFigFont#1#2#3#4#5{%
  \reset@font\fontsize{#1}{#2pt}%
  \fontfamily{#3}\fontseries{#4}\fontshape{#5}%
  \selectfont}%
\fi\endgroup%
{\renewcommand{\dashlinestretch}{30}
\begin{picture}(9212,6822)(0,-10)
\put(3788,2049){\blacken\ellipse{90}{90}}
\put(3788,2049){\ellipse{90}{90}}
\put(1088,4164){\blacken\ellipse{90}{90}}
\put(1088,4164){\ellipse{90}{90}}
\put(4598,2184){\blacken\ellipse{90}{90}}
\put(4598,2184){\ellipse{90}{90}}
\put(3023,2004){\blacken\ellipse{90}{90}}
\put(3023,2004){\ellipse{90}{90}}
\put(2348,2049){\blacken\ellipse{90}{90}}
\put(2348,2049){\ellipse{90}{90}}
\put(773,2409){\blacken\ellipse{90}{90}}
\put(773,2409){\ellipse{90}{90}}
\put(5273,1284){\blacken\ellipse{90}{90}}
\put(5273,1284){\ellipse{90}{90}}
\put(4598,1284){\blacken\ellipse{90}{90}}
\put(4598,1284){\ellipse{90}{90}}
\put(3788,1284){\blacken\ellipse{90}{90}}
\put(3788,1284){\ellipse{90}{90}}
\put(3023,1284){\blacken\ellipse{90}{90}}
\put(3023,1284){\ellipse{90}{90}}
\put(2348,1284){\blacken\ellipse{90}{90}}
\put(2348,1284){\ellipse{90}{90}}
\put(2348,654){\blacken\ellipse{90}{90}}
\put(2348,654){\ellipse{90}{90}}
\put(323,3759){\blacken\ellipse{90}{90}}
\put(323,3759){\ellipse{90}{90}}
\put(5273,2409){\blacken\ellipse{90}{90}}
\put(5273,2409){\ellipse{90}{90}}
\put(4148,5514){\blacken\ellipse{90}{90}}
\put(4148,5514){\ellipse{90}{90}}
\put(4148,6549){\blacken\ellipse{90}{90}}
\put(4148,6549){\ellipse{90}{90}}
\put(3023,3219){\ellipse{6030}{2430}}
\put(2573,4434){\blacken\ellipse{90}{90}}
\put(2573,4434){\ellipse{90}{90}}
\put(3563,6549){\blacken\ellipse{90}{90}}
\put(3563,6549){\ellipse{90}{90}}
\put(3563,5514){\blacken\ellipse{90}{90}}
\put(3563,5514){\ellipse{90}{90}}
\put(4733,5514){\blacken\ellipse{90}{90}}
\put(4733,5514){\ellipse{90}{90}}
\put(5183,4029){\blacken\ellipse{90}{90}}
\put(5183,4029){\ellipse{90}{90}}
\put(2978,5514){\blacken\ellipse{90}{90}}
\put(2978,5514){\ellipse{90}{90}}
\put(2573,5514){\blacken\ellipse{90}{90}}
\put(2573,5514){\ellipse{90}{90}}
\path(1988,4344)(2033,4344)
\blacken\path(1883.000,4306.500)(2033.000,4344.000)(1883.000,4381.500)(1883.000,4306.500)
\path(3068,4434)(3113,4434)
\blacken\path(2963.000,4396.500)(3113.000,4434.000)(2963.000,4471.500)(2963.000,4396.500)
\path(4103,2094)(4058,2094)
\blacken\path(4208.000,2131.500)(4058.000,2094.000)(4208.000,2056.500)(4208.000,2131.500)
\path(2033,2072)(1988,2072)
\blacken\path(2138.000,2109.500)(1988.000,2072.000)(2138.000,2034.500)(2138.000,2109.500)
\path(5273,1284)(5273,654)
\blacken\path(5235.500,804.000)(5273.000,654.000)(5310.500,804.000)(5235.500,804.000)
\path(5273,654)(2348,654)
\blacken\path(2498.000,691.500)(2348.000,654.000)(2498.000,616.500)(2498.000,691.500)
\path(2348,654)(773,654)
\blacken\path(923.000,691.500)(773.000,654.000)(923.000,616.500)(923.000,691.500)
\path(2348,654)(2348,1284)
\blacken\path(2385.500,1134.000)(2348.000,1284.000)(2310.500,1134.000)(2385.500,1134.000)
\path(2348,1284)(2348,2049)
\blacken\path(2385.500,1899.000)(2348.000,2049.000)(2310.500,1899.000)(2385.500,1899.000)
\path(5273,1284)(4598,1284)
\blacken\path(4748.000,1321.500)(4598.000,1284.000)(4748.000,1246.500)(4748.000,1321.500)
\path(4598,1284)(3788,1284)
\blacken\path(3938.000,1321.500)(3788.000,1284.000)(3938.000,1246.500)(3938.000,1321.500)
\path(3788,1284)(3023,1284)
\blacken\path(3173.000,1321.500)(3023.000,1284.000)(3173.000,1246.500)(3173.000,1321.500)
\path(3023,1284)(2348,1284)
\blacken\path(2498.000,1321.500)(2348.000,1284.000)(2498.000,1246.500)(2498.000,1321.500)
\path(3383,2004)(3338,2004)
\blacken\path(3488.000,2041.500)(3338.000,2004.000)(3488.000,1966.500)(3488.000,2041.500)
\path(2618,2004)(2573,2004)
\blacken\path(2723.000,2041.500)(2573.000,2004.000)(2723.000,1966.500)(2723.000,2041.500)
\path(1088,4164)(1088,5514)
\blacken\path(1125.500,5364.000)(1088.000,5514.000)(1050.500,5364.000)(1125.500,5364.000)
\path(1088,5514)(2573,5514)
\blacken\path(2423.000,5476.500)(2573.000,5514.000)(2423.000,5551.500)(2423.000,5476.500)
\path(773,654)(773,2409)
\blacken\path(810.500,2259.000)(773.000,2409.000)(735.500,2259.000)(810.500,2259.000)
\path(2573,5514)(2573,4434)
\blacken\path(2535.500,4584.000)(2573.000,4434.000)(2610.500,4584.000)(2535.500,4584.000)
\blacken\path(2723.000,5551.500)(2573.000,5514.000)(2723.000,5476.500)(2723.000,5551.500)
\path(2573,5514)(2978,5514)
\blacken\path(3173.000,5551.500)(3023.000,5514.000)(3173.000,5476.500)(3173.000,5551.500)
\path(3023,5514)(3563,5514)
\blacken\path(3713.000,5551.500)(3563.000,5514.000)(3713.000,5476.500)(3713.000,5551.500)
\path(3563,5514)(4148,5514)
\blacken\path(4298.000,5551.500)(4148.000,5514.000)(4298.000,5476.500)(4298.000,5551.500)
\path(4148,5514)(4733,5514)
\blacken\path(4883.000,5551.500)(4733.000,5514.000)(4883.000,5476.500)(4883.000,5551.500)
\path(4733,5514)(5183,5514)
\blacken\path(5220.500,5364.000)(5183.000,5514.000)(5145.500,5364.000)(5220.500,5364.000)
\path(5183,5514)(5183,4029)
\blacken\path(3128.000,6586.500)(2978.000,6549.000)(3128.000,6511.500)(3128.000,6586.500)
\path(2978,6549)(3563,6549)
\path(3563,6549)(4148,6549)
\blacken\path(3998.000,6511.500)(4148.000,6549.000)(3998.000,6586.500)(3998.000,6511.500)
\blacken\path(4298.000,6586.500)(4148.000,6549.000)(4298.000,6511.500)(4298.000,6586.500)
\path(4148,6549)(4733,6549)
\blacken\path(4770.500,6399.000)(4733.000,6549.000)(4695.500,6399.000)(4770.500,6399.000)
\path(4733,6549)(4733,5514)
\blacken\path(2940.500,5664.000)(2978.000,5514.000)(3015.500,5664.000)(2940.500,5664.000)
\path(2978,5514)(2978,6549)
\blacken\path(3600.500,6399.000)(3563.000,6549.000)(3525.500,6399.000)(3600.500,6399.000)
\path(3563,6549)(3563,5514)
\path(4148,6549)(4148,5514)
\blacken\path(4110.500,5664.000)(4148.000,5514.000)(4185.500,5664.000)(4110.500,5664.000)
\path(5273,2409)(5273,1284)
\blacken\path(5235.500,1434.000)(5273.000,1284.000)(5310.500,1434.000)(5235.500,1434.000)
\blacken\path(4635.500,2034.000)(4598.000,2184.000)(4560.500,2034.000)(4635.500,2034.000)
\path(4598,2184)(4598,1284)
\blacken\path(3825.500,1899.000)(3788.000,2049.000)(3750.500,1899.000)(3825.500,1899.000)
\path(3788,2049)(3788,1284)
\blacken\path(3060.500,1854.000)(3023.000,2004.000)(2985.500,1854.000)(3060.500,1854.000)
\path(3023,2004)(3023,1284)
\put(1043,3939){\makebox(0,0)[lb]{\smash{{\SetFigFont{10}{12.0}{\rmdefault}{\mddefault}{\updefault}$a$}}}}
\put(5228,2544){\makebox(0,0)[lb]{\smash{{\SetFigFont{10}{12.0}{\rmdefault}{\mddefault}{\updefault}$d$}}}}
\put(4553,2319){\makebox(0,0)[lb]{\smash{{\SetFigFont{10}{12.0}{\rmdefault}{\mddefault}{\updefault}$e$}}}}
\put(3743,2184){\makebox(0,0)[lb]{\smash{{\SetFigFont{10}{12.0}{\rmdefault}{\mddefault}{\updefault}$f$}}}}
\put(2978,2139){\makebox(0,0)[lb]{\smash{{\SetFigFont{10}{12.0}{\rmdefault}{\mddefault}{\updefault}$g$}}}}
\put(2303,2184){\makebox(0,0)[lb]{\smash{{\SetFigFont{10}{12.0}{\rmdefault}{\mddefault}{\updefault}$h$}}}}
\put(773,2544){\makebox(0,0)[lb]{\smash{{\SetFigFont{10}{12.0}{\rmdefault}{\mddefault}{\updefault}$i$}}}}
\put(4553,1059){\makebox(0,0)[lb]{\smash{{\SetFigFont{10}{12.0}{\rmdefault}{\mddefault}{\updefault}$r$}}}}
\put(3743,1059){\makebox(0,0)[lb]{\smash{{\SetFigFont{10}{12.0}{\rmdefault}{\mddefault}{\updefault}$s$}}}}
\put(2978,1059){\makebox(0,0)[lb]{\smash{{\SetFigFont{10}{12.0}{\rmdefault}{\mddefault}{\updefault}$t$}}}}
\put(2123,1239){\makebox(0,0)[lb]{\smash{{\SetFigFont{10}{12.0}{\rmdefault}{\mddefault}{\updefault}$u$}}}}
\put(188,3804){\makebox(0,0)[lb]{\smash{{\SetFigFont{10}{12.0}{\rmdefault}{\mddefault}{\updefault}$z$}}}}
\put(2528,4209){\makebox(0,0)[lb]{\smash{{\SetFigFont{10}{12.0}{\rmdefault}{\mddefault}{\updefault}$b$}}}}
\put(5138,3804){\makebox(0,0)[lb]{\smash{{\SetFigFont{10}{12.0}{\rmdefault}{\mddefault}{\updefault}$c$}}}}
\put(2573,5649){\makebox(0,0)[lb]{\smash{{\SetFigFont{10}{12.0}{\rmdefault}{\mddefault}{\updefault}$j$}}}}
\put(2978,5289){\makebox(0,0)[lb]{\smash{{\SetFigFont{10}{12.0}{\rmdefault}{\mddefault}{\updefault}$k$}}}}
\put(4688,5289){\makebox(0,0)[lb]{\smash{{\SetFigFont{10}{12.0}{\rmdefault}{\mddefault}{\updefault}$n$}}}}
\put(3563,5289){\makebox(0,0)[lb]{\smash{{\SetFigFont{10}{12.0}{\rmdefault}{\mddefault}{\updefault}$l$}}}}
\put(4103,5289){\makebox(0,0)[lb]{\smash{{\SetFigFont{10}{12.0}{\rmdefault}{\mddefault}{\updefault}$m$}}}}
\put(3518,6684){\makebox(0,0)[lb]{\smash{{\SetFigFont{10}{12.0}{\rmdefault}{\mddefault}{\updefault}$o$}}}}
\put(4103,6684){\makebox(0,0)[lb]{\smash{{\SetFigFont{10}{12.0}{\rmdefault}{\mddefault}{\updefault}$p$}}}}
\put(5408,1239){\makebox(0,0)[lb]{\smash{{\SetFigFont{10}{12.0}{\rmdefault}{\mddefault}{\updefault}$q$}}}}
\put(2303,429){\makebox(0,0)[lb]{\smash{{\SetFigFont{10}{12.0}{\rmdefault}{\mddefault}{\updefault}$v$}}}}
\put(7028,2544){\makebox(0,0)[lb]{\smash{{\SetFigFont{10}{12.0}{\rmdefault}{\mddefault}{\updefault}$D_0$ $zabcdefghiz$}}}}
\put(7028,2319){\makebox(0,0)[lb]{\smash{{\SetFigFont{10}{12.0}{\rmdefault}{\mddefault}{\updefault}$D_1$ $zajbcdefghiz$}}}}
\put(7028,2094){\makebox(0,0)[lb]{\smash{{\SetFigFont{10}{12.0}{\rmdefault}{\mddefault}{\updefault}$D_2$ $zabcnmlkjbcdefghiz$}}}}
\put(7028,1869){\makebox(0,0)[lb]{\smash{{\SetFigFont{10}{12.0}{\rmdefault}{\mddefault}{\updefault}$D_3$ $zabcnmlokjbcdefghiz$}}}}
\put(7028,1194){\makebox(0,0)[lb]{\smash{{\SetFigFont{10}{12.0}{\rmdefault}{\mddefault}{\updefault}$D_6$ $zabcdqviz$}}}}
\put(7028,969){\makebox(0,0)[lb]{\smash{{\SetFigFont{10}{12.0}{\rmdefault}{\mddefault}{\updefault}$D_7$ $zabcdqvuhiz$}}}}
\put(7028,744){\makebox(0,0)[lb]{\smash{{\SetFigFont{10}{12.0}{\rmdefault}{\mddefault}{\updefault}$D_8$ $zabcdqrefghiz$}}}}
\put(7028,519){\makebox(0,0)[lb]{\smash{{\SetFigFont{10}{12.0}{\rmdefault}{\mddefault}{\updefault}$D_9$ $zabcdqrsfghiz$}}}}
\put(7028,69){\makebox(0,0)[lb]{\smash{{\SetFigFont{10}{12.0}{\rmdefault}{\mddefault}{\updefault}$D_{11}$ $zabcdqrstuhiz$}}}}
\put(7028,1644){\makebox(0,0)[lb]{\smash{{\SetFigFont{10}{12.0}{\rmdefault}{\mddefault}{\updefault}$D_4$ $zabcnmlopmlkjbcdefghiz$}}}}
\put(7028,294){\makebox(0,0)[lb]{\smash{{\SetFigFont{10}{12.0}{\rmdefault}{\mddefault}{\updefault}$D_{10}$ $zabcdqrstghiz$}}}}
\put(7028,6324){\makebox(0,0)[lb]{\smash{{\SetFigFont{10}{12.0}{\rmdefault}{\mddefault}{\updefault}Undirected cycles}}}}
\put(7028,6099){\makebox(0,0)[lb]{\smash{{\SetFigFont{10}{12.0}{\rmdefault}{\mddefault}{\updefault}$C_0$ $zabcdefghiz$}}}}
\put(7028,5874){\makebox(0,0)[lb]{\smash{{\SetFigFont{10}{12.0}{\rmdefault}{\mddefault}{\updefault}$C_1$ $zajbcdefghiz$}}}}
\put(7028,5649){\makebox(0,0)[lb]{\smash{{\SetFigFont{10}{12.0}{\rmdefault}{\mddefault}{\updefault}$C_2$ $zabjklmncdefghiz$}}}}
\put(7028,5424){\makebox(0,0)[lb]{\smash{{\SetFigFont{10}{12.0}{\rmdefault}{\mddefault}{\updefault}$C_3$ $zabjkolmncdefghiz$}}}}
\put(7028,4974){\makebox(0,0)[lb]{\smash{{\SetFigFont{10}{12.0}{\rmdefault}{\mddefault}{\updefault}$C_5$ $zabjklmpncdefghiz$}}}}
\put(7028,4749){\makebox(0,0)[lb]{\smash{{\SetFigFont{10}{12.0}{\rmdefault}{\mddefault}{\updefault}$C_6$ $zabcdqviz$}}}}
\put(7028,4524){\makebox(0,0)[lb]{\smash{{\SetFigFont{10}{12.0}{\rmdefault}{\mddefault}{\updefault}$C_7$ $zabcdqvuhiz$}}}}
\put(7028,4299){\makebox(0,0)[lb]{\smash{{\SetFigFont{10}{12.0}{\rmdefault}{\mddefault}{\updefault}$C_8$ $zabcdqrefghiz$}}}}
\put(7028,4074){\makebox(0,0)[lb]{\smash{{\SetFigFont{10}{12.0}{\rmdefault}{\mddefault}{\updefault}$C_9$ $zabcdersfghiz$}}}}
\put(7028,3624){\makebox(0,0)[lb]{\smash{{\SetFigFont{10}{12.0}{\rmdefault}{\mddefault}{\updefault}$C_{11}$ $zabcdefgtuhiz$}}}}
\put(7028,5199){\makebox(0,0)[lb]{\smash{{\SetFigFont{10}{12.0}{\rmdefault}{\mddefault}{\updefault}$C_4$ $zabjklopmncdefghiz$}}}}
\put(7028,3849){\makebox(0,0)[lb]{\smash{{\SetFigFont{10}{12.0}{\rmdefault}{\mddefault}{\updefault}$C_{10}$ $zabcdefstghiz$}}}}
\put(7028,2769){\makebox(0,0)[lb]{\smash{{\SetFigFont{10}{12.0}{\rmdefault}{\mddefault}{\updefault}Directed cycles}}}}
\put(7028,1419){\makebox(0,0)[lb]{\smash{{\SetFigFont{10}{12.0}{\rmdefault}{\mddefault}{\updefault}$D_5$ $zabcnpmlkjbcdefghiz$}}}}
\put(143,3264){\makebox(0,0)[lb]{\smash{{\SetFigFont{10}{12.0}{\rmdefault}{\mddefault}{\updefault}$C_0$}}}}
\put(1673,4794){\makebox(0,0)[lb]{\smash{{\SetFigFont{10}{12.0}{\rmdefault}{\mddefault}{\updefault}$C_1$}}}}
\put(3743,4794){\makebox(0,0)[lb]{\smash{{\SetFigFont{10}{12.0}{\rmdefault}{\mddefault}{\updefault}$C_2$}}}}
\put(3248,5964){\makebox(0,0)[lb]{\smash{{\SetFigFont{10}{12.0}{\rmdefault}{\mddefault}{\updefault}$C_3$}}}}
\put(3833,5964){\makebox(0,0)[lb]{\smash{{\SetFigFont{10}{12.0}{\rmdefault}{\mddefault}{\updefault}$C_4$}}}}
\put(4418,5964){\makebox(0,0)[lb]{\smash{{\SetFigFont{10}{12.0}{\rmdefault}{\mddefault}{\updefault}$C_5$}}}}
\put(1493,384){\makebox(0,0)[lb]{\smash{{\SetFigFont{10}{12.0}{\rmdefault}{\mddefault}{\updefault}$C_6$}}}}
\put(3158,384){\makebox(0,0)[lb]{\smash{{\SetFigFont{10}{12.0}{\rmdefault}{\mddefault}{\updefault}$C_7$}}}}
\put(4913,1689){\makebox(0,0)[lb]{\smash{{\SetFigFont{10}{12.0}{\rmdefault}{\mddefault}{\updefault}$C_8$}}}}
\put(4148,1689){\makebox(0,0)[lb]{\smash{{\SetFigFont{10}{12.0}{\rmdefault}{\mddefault}{\updefault}$C_9$}}}}
\put(3338,1689){\makebox(0,0)[lb]{\smash{{\SetFigFont{10}{12.0}{\rmdefault}{\mddefault}{\updefault}$C_{10}$}}}}
\put(2573,1689){\makebox(0,0)[lb]{\smash{{\SetFigFont{10}{12.0}{\rmdefault}{\mddefault}{\updefault}$C_{11}$}}}}
\end{picture}
}
\end{center}
\caption{An example orientation.} \label{bigexfig}
\end{figure}

The algorithm orients the paths to satisfy the following properties:
\begin{prop} \label{longfwd}
A heavy cycle is oriented forwards.
\end{prop}
\begin{prop} \label{n2back}In a block with more than two brothers, the
first two are not both oriented backwards. The same holds for the
last two brothers.
\end{prop}
\begin{prop}
Inner-crossing cycles are oriented in the same direction (forwards
or backwards) like their closest containing brother (see Definition
\ref{croosobs}).
\end{prop}
\begin{prop}
Outer-crossing cycles are oriented forwards.
\end{prop}
\begin{prop}
\label{scbprop} If $C$ is a special contained brother, it is
oriented forwards. Its containing brother $B$ with $lc(C) = lc(B)+1$
is also oriented forwards (see Definition \ref{scbdef}).
\end{prop}

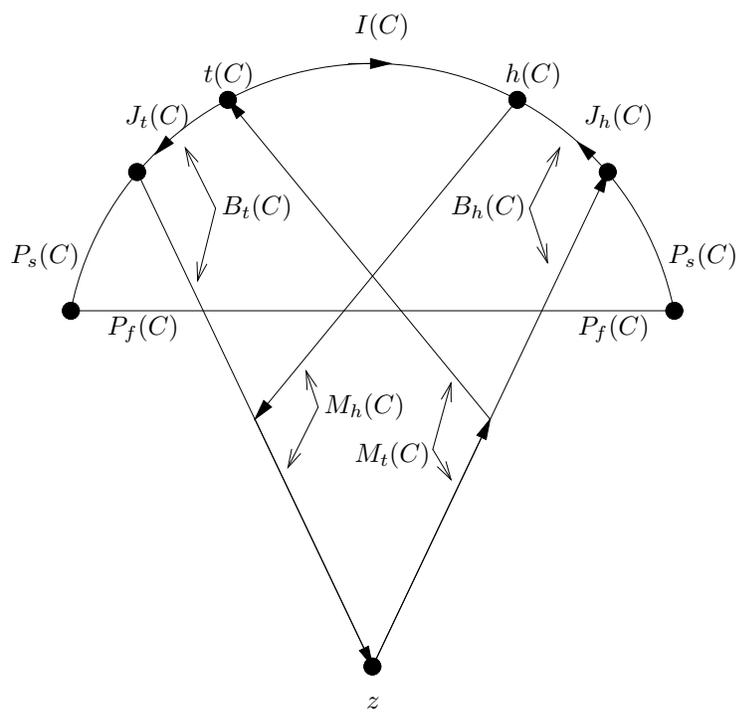
\begin{figure}
\begin{center}
\setlength{\unitlength}{0.00069991in}
\begingroup\makeatletter\ifx\SetFigFont\undefined%
\gdef\SetFigFont#1#2#3#4#5{%
  \reset@font\fontsize{#1}{#2pt}%
  \fontfamily{#3}\fontseries{#4}\fontshape{#5}%
  \selectfont}%
\fi\endgroup%
{\renewcommand{\dashlinestretch}{30}
\begin{picture}(5411,5229)(0,-10)
\put(2715.000,2535.549){\arc{4588.902}{3.3388}{6.0860}}
\put(2715,330){\blacken\ellipse{126}{126}}
\put(2715,330){\ellipse{126}{126}}
\put(4470,4020){\blacken\ellipse{126}{126}}
\put(4470,4020){\ellipse{126}{126}}
\put(960,4020){\blacken\ellipse{126}{126}}
\put(960,4020){\ellipse{126}{126}}
\put(1635,4560){\blacken\ellipse{126}{126}}
\put(1635,4560){\ellipse{126}{126}}
\put(3795,4560){\blacken\ellipse{126}{126}}
\put(3795,4560){\ellipse{126}{126}}
\put(465,2985){\blacken\ellipse{126}{126}}
\put(465,2985){\ellipse{126}{126}}
\put(4965,2985){\blacken\ellipse{126}{126}}
\put(4965,2985){\ellipse{126}{126}}
\path(1140,4200)(1095,4155)
\blacken\path(1174.550,4287.583)(1095.000,4155.000)(1227.583,4234.550)(1174.550,4287.583)
\path(1185,4245)(1095,4155)
\blacken\path(4377.583,4165.450)(4245.000,4245.000)(4324.550,4112.417)(4377.583,4165.450)
\path(4245,4245)(4335,4155)
\path(1545,3750)(1320,4200)
\path(1413.915,4096.023)(1320.000,4200.000)(1346.833,4062.482)
\path(3885,3750)(4110,4200)
\path(4083.167,4062.482)(4110.000,4200.000)(4016.085,4096.023)
\path(1545,3750)(1410,3210)
\path(1406.362,3350.064)(1410.000,3210.000)(1479.123,3331.874)
\path(3885,3750)(4020,3345)
\path(3941.734,3461.214)(4020.000,3345.000)(4012.885,3484.931)
\path(3592,2175)(1635,4560)
\blacken\path(1759.140,4467.828)(1635.000,4560.000)(1701.160,4420.253)(1759.140,4467.828)
\path(2715,330)(4470,4020)
\blacken\path(4439.439,3868.434)(4470.000,4020.000)(4371.709,3900.647)(4439.439,3868.434)
\path(2715,330)(3592,2175)
\blacken\path(3561.472,2023.427)(3592.000,2175.000)(3493.736,2055.625)(3561.472,2023.427)
\blacken\path(2616.709,449.353)(2715.000,330.000)(2684.439,481.566)(2616.709,449.353)
\path(2715,330)(960,4020)
\path(2715,330)(1838,2175)
\path(3795,4560)(1838,2175)
\blacken\path(1904.160,2314.747)(1838.000,2175.000)(1962.140,2267.172)(1904.160,2314.747)
\path(2310,2265)(2220,2535)
\path(2298.266,2418.786)(2220.000,2535.000)(2227.115,2395.069)
\path(2310,2265)(2085,1815)
\path(2111.833,1952.518)(2085.000,1815.000)(2178.915,1918.977)
\path(3165,1950)(3300,2445)
\path(3300.658,2304.890)(3300.000,2445.000)(3228.301,2324.624)
\path(3165,1950)(3300,1725)
\path(3198.387,1821.468)(3300.000,1725.000)(3262.699,1860.055)
\path(2535,4830)(2850,4830)
\blacken\path(2700.000,4792.500)(2850.000,4830.000)(2700.000,4867.500)(2700.000,4792.500)
\path(465,2985)(4965,2985)
\put(2580,5055){\makebox(0,0)[lb]{\smash{{\SetFigFont{10}{12.0}{\rmdefault}{\mddefault}{\updefault}$I(C)$}}}}
\put(1590,3705){\makebox(0,0)[lb]{\smash{{\SetFigFont{10}{12.0}{\rmdefault}{\mddefault}{\updefault}$B_t(C)$}}}}
\put(2670,15){\makebox(0,0)[lb]{\smash{{\SetFigFont{10}{12.0}{\rmdefault}{\mddefault}{\updefault}$z$}}}}
\put(2355,2220){\makebox(0,0)[lb]{\smash{{\SetFigFont{10}{12.0}{\rmdefault}{\mddefault}{\updefault}$M_h(C)$}}}}
\put(735,2805){\makebox(0,0)[lb]{\smash{{\SetFigFont{10}{12.0}{\rmdefault}{\mddefault}{\updefault}$P_f(C)$}}}}
\put(4245,2805){\makebox(0,0)[lb]{\smash{{\SetFigFont{10}{12.0}{\rmdefault}{\mddefault}{\updefault}$P_f(C)$}}}}
\put(1455,4695){\makebox(0,0)[lb]{\smash{{\SetFigFont{10}{12.0}{\rmdefault}{\mddefault}{\updefault}$t(C)$}}}}
\put(3705,4695){\makebox(0,0)[lb]{\smash{{\SetFigFont{10}{12.0}{\rmdefault}{\mddefault}{\updefault}$h(C)$}}}}
\put(3300,3705){\makebox(0,0)[lb]{\smash{{\SetFigFont{10}{12.0}{\rmdefault}{\mddefault}{\updefault}$B_h(C)$}}}}
\put(870,4380){\makebox(0,0)[lb]{\smash{{\SetFigFont{10}{12.0}{\rmdefault}{\mddefault}{\updefault}$J_t(C)$}}}}
\put(4290,4380){\makebox(0,0)[lb]{\smash{{\SetFigFont{10}{12.0}{\rmdefault}{\mddefault}{\updefault}$J_h(C)$}}}}
\put(4920,3345){\makebox(0,0)[lb]{\smash{{\SetFigFont{10}{12.0}{\rmdefault}{\mddefault}{\updefault}$P_s(C)$}}}}
\put(15,3345){\makebox(0,0)[lb]{\smash{{\SetFigFont{10}{12.0}{\rmdefault}{\mddefault}{\updefault}$P_s(C)$}}}}
\put(2580,1860){\makebox(0,0)[lb]{\smash{{\SetFigFont{10}{12.0}{\rmdefault}{\mddefault}{\updefault}$M_t(C)$}}}}
\end{picture}
}
\end{center}
\caption{Paths associated with $C$.} \label{btbhfig}
\end{figure}

\begin{defin}
{\em Consider an oriented cycle $C$ such that $U(C)$ is already
oriented. Define the following paths (see  Figure \ref{btbhfig}):
 \be
 \item Let $I(C)$ be the maximal directed path in $P_s(C)$ which contains
$U(C)$.
\item Let $t(C)$, $h(C)$ be the tail and head of $I(C)$,
respectively.
 \item Let $M_t(C)$  be a directed path from $z$ to
$t(C)$ ($M_t(C)$ is constructed during the proof).
 \item Let $M_h(C)$ be a directed path from $h(C)$ to $z$ ($M_h(C)$ is constructed
during the proof).
 \item Let $M(C) = M_t(C) \cup M_h(C)$.
 \item Let $J_t(C)$ be the  maximal directed subpath of $P_s(C) \backslash
I(C)$ with tail $t(C)$. $J_t(C) = \phi$ if $t(C) \in F(C)$.
 \item Let $J_h(C)$ be the maximal directed subpath of $P_s(C) \backslash I(C)$
with head $h(C)$. $J_h(C) = \phi$ if $h(C) \in F(C)$.
 \ee}
\end{defin}

\begin{defin}
{\em When $I(C) \neq P_s(C)$ we define $B_t(C)$ and $B_h(C)$ the
{\it bypasses of $C$} in the following manner (these paths will be
constructed during the proof):
 \be
\item When $h(C) \not\in F(C)$, $B_h(C)$ is a path from $z$ to
$h(C)$  containing $J_h(C)$.
 \item When $t(C) \not\in F(C)$,
$B_t(C)$ is a path from $t(C)$ to $z$  containing $J_t(C)$.
 \ee }
\end{defin}

\begin{defin}
{\em  {\em $C_1$ is preferred to $C_2$ } if any of the following
holds:
 \be
  \item $C_1$ and $C_2$ both oriented forwards or both oriented backwards, and
  $l(C_1)  < l(C_2)$.
 \item $C_1$ is oriented forwards, $C_2$ is oriented backwards and
 $l(C_1)  < l(C_2) + 2l(P_c(C_2) \backslash C_1))$.
 \item $C_1$ is oriented backwards, $C_2$ is oriented forwards and
 $l(C_1) +2l(P_c(C_1) \backslash C_2)) < l(C_2)$.
 \ee}
 \end{defin}

\begin{defin}
{\em Two cycles $C_1$ and $C_2$ are {\em competing on a path $P$}
when $P \subseteq (P_s(C_1) \cap P_s(C_2))$ and $I(C_1)$, $I(C_2)$
induce different orientations on $P$. $C_1$ is the {\em winner} of
the competition if the final orientation of $P$ is the one induced
by $I(C_1)$.}
\end{defin}

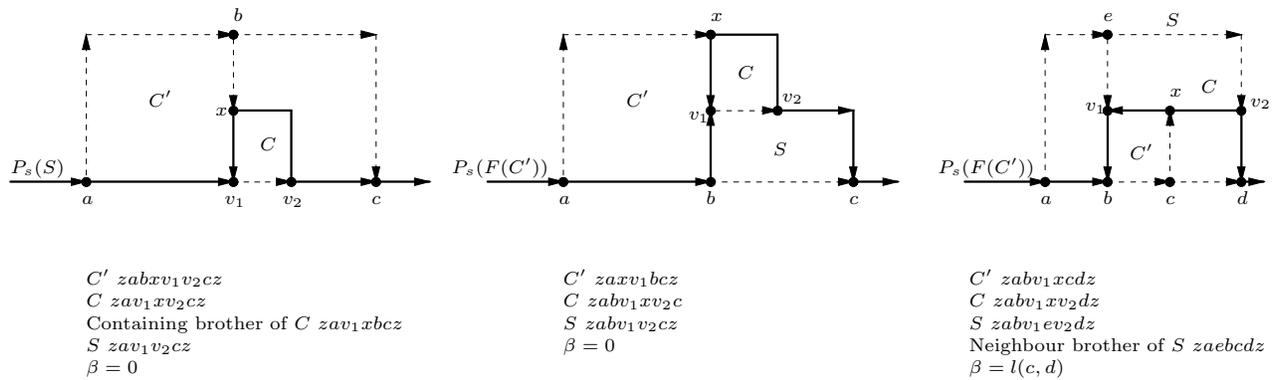
\begin{figure}
\begin{center}
\setlength{\unitlength}{0.00052493in}
\begingroup\makeatletter\ifx\SetFigFont\undefined%
\gdef\SetFigFont#1#2#3#4#5{%
  \reset@font\fontsize{#1}{#2pt}%
  \fontfamily{#3}\fontseries{#4}\fontshape{#5}%
  \selectfont}%
\fi\endgroup%
{\renewcommand{\dashlinestretch}{30}
\begin{picture}(13071,3789)(0,-10)
\put(978,2004){\blacken\ellipse{90}{90}}
\put(978,2004){\ellipse{90}{90}}
\put(3903,2004){\blacken\ellipse{90}{90}}
\put(3903,2004){\ellipse{90}{90}}
\put(2463,2004){\blacken\ellipse{90}{90}}
\put(2463,2004){\ellipse{90}{90}}
\put(2463,3489){\blacken\ellipse{90}{90}}
\put(2463,3489){\ellipse{90}{90}}
\put(2463,2724){\blacken\ellipse{90}{90}}
\put(2463,2724){\ellipse{90}{90}}
\put(3048,2004){\blacken\ellipse{90}{90}}
\put(3048,2004){\ellipse{90}{90}}
\put(7278,2004){\blacken\ellipse{90}{90}}
\put(7278,2004){\ellipse{90}{90}}
\put(8718,2004){\blacken\ellipse{90}{90}}
\put(8718,2004){\ellipse{90}{90}}
\put(7278,3489){\blacken\ellipse{90}{90}}
\put(7278,3489){\ellipse{90}{90}}
\put(7953,2724){\blacken\ellipse{90}{90}}
\put(7953,2724){\ellipse{90}{90}}
\put(7278,2724){\blacken\ellipse{90}{90}}
\put(7278,2724){\ellipse{90}{90}}
\put(5793,2004){\blacken\ellipse{90}{90}}
\put(5793,2004){\ellipse{90}{90}}
\put(11283,2724){\blacken\ellipse{90}{90}}
\put(11283,2724){\ellipse{90}{90}}
\put(11283,3489){\blacken\ellipse{90}{90}}
\put(11283,3489){\ellipse{90}{90}}
\put(11913,2724){\blacken\ellipse{90}{90}}
\put(11913,2724){\ellipse{90}{90}}
\put(12633,2724){\blacken\ellipse{90}{90}}
\put(12633,2724){\ellipse{90}{90}}
\put(10653,2004){\blacken\ellipse{90}{90}}
\put(10653,2004){\ellipse{90}{90}}
\put(11283,2004){\blacken\ellipse{90}{90}}
\put(11283,2004){\ellipse{90}{90}}
\put(11913,2004){\blacken\ellipse{90}{90}}
\put(11913,2004){\ellipse{90}{90}}
\put(12633,2004){\blacken\ellipse{90}{90}}
\put(12633,2004){\ellipse{90}{90}}
\thicklines
\path(213,2004)(978,2004)
\blacken\thinlines
\path(828.000,1966.500)(978.000,2004.000)(828.000,2041.500)(828.000,1966.500)
\thicklines
\path(978,2004)(2463,2004)
\blacken\thinlines
\path(2313.000,1966.500)(2463.000,2004.000)(2313.000,2041.500)(2313.000,1966.500)
\thicklines
\path(3903,2004)(4443,2004)
\blacken\thinlines
\path(4293.000,1966.500)(4443.000,2004.000)(4293.000,2041.500)(4293.000,1966.500)
\thicklines
\path(3048,2004)(3903,2004)
\blacken\thinlines
\path(3753.000,1966.500)(3903.000,2004.000)(3753.000,2041.500)(3753.000,1966.500)
\thicklines
\path(2463,2724)(2463,2004)
\blacken\thinlines
\path(2425.500,2154.000)(2463.000,2004.000)(2500.500,2154.000)(2425.500,2154.000)
\dashline{60.000}(2463,2004)(3048,2004)
\blacken\path(2898.000,1966.500)(3048.000,2004.000)(2898.000,2041.500)(2898.000,1966.500)
\thicklines
\path(2463,2724)(3048,2724)
\path(3048,2724)(3048,2004)
\thinlines
\dashline{60.000}(978,1959)(978,3489)
\blacken\path(1015.500,3339.000)(978.000,3489.000)(940.500,3339.000)(1015.500,3339.000)
\dashline{60.000}(978,3489)(2463,3489)
\blacken\path(2313.000,3451.500)(2463.000,3489.000)(2313.000,3526.500)(2313.000,3451.500)
\dashline{60.000}(2463,3444)(2463,2724)
\blacken\path(2425.500,2874.000)(2463.000,2724.000)(2500.500,2874.000)(2425.500,2874.000)
\dashline{60.000}(2463,3489)(3903,3489)
\blacken\path(3753.000,3451.500)(3903.000,3489.000)(3753.000,3526.500)(3753.000,3451.500)
\dashline{60.000}(3903,3489)(3903,2004)
\blacken\path(3865.500,2154.000)(3903.000,2004.000)(3940.500,2154.000)(3865.500,2154.000)
\dashline{60.000}(5793,1959)(5793,3489)
\blacken\path(5830.500,3339.000)(5793.000,3489.000)(5755.500,3339.000)(5830.500,3339.000)
\dashline{60.000}(5793,3489)(7278,3489)
\blacken\path(7128.000,3451.500)(7278.000,3489.000)(7128.000,3526.500)(7128.000,3451.500)
\dashline{60.000}(7323,2724)(7953,2724)
\blacken\path(7803.000,2686.500)(7953.000,2724.000)(7803.000,2761.500)(7803.000,2686.500)
\dashline{60.000}(7278,2004)(8718,2004)
\blacken\path(8568.000,1966.500)(8718.000,2004.000)(8568.000,2041.500)(8568.000,1966.500)
\thicklines
\path(5028,2004)(5793,2004)
\blacken\thinlines
\path(5643.000,1966.500)(5793.000,2004.000)(5643.000,2041.500)(5643.000,1966.500)
\thicklines
\path(5793,2004)(7278,2004)
\blacken\thinlines
\path(7128.000,1966.500)(7278.000,2004.000)(7128.000,2041.500)(7128.000,1966.500)
\blacken\path(7315.500,2574.000)(7278.000,2724.000)(7240.500,2574.000)(7315.500,2574.000)
\thicklines
\path(7278,2724)(7278,2004)
\path(7278,3444)(7278,2724)
\blacken\thinlines
\path(7240.500,2874.000)(7278.000,2724.000)(7315.500,2874.000)(7240.500,2874.000)
\thicklines
\path(7323,3489)(7953,3489)
\path(7953,3489)(7953,2724)
\path(7953,2724)(8718,2724)
\blacken\thinlines
\path(8568.000,2686.500)(8718.000,2724.000)(8568.000,2761.500)(8568.000,2686.500)
\thicklines
\path(8718,2724)(8718,2004)
\blacken\thinlines
\path(8680.500,2154.000)(8718.000,2004.000)(8755.500,2154.000)(8680.500,2154.000)
\thicklines
\path(8718,2004)(9168,2004)
\blacken\thinlines
\path(9018.000,1966.500)(9168.000,2004.000)(9018.000,2041.500)(9018.000,1966.500)
\blacken\path(11950.500,2574.000)(11913.000,2724.000)(11875.500,2574.000)(11950.500,2574.000)
\dashline{60.000}(11913,2724)(11913,2004)
\thicklines
\path(12633,2724)(12633,2004)
\blacken\thinlines
\path(12595.500,2154.000)(12633.000,2004.000)(12670.500,2154.000)(12595.500,2154.000)
\thicklines
\path(9843,2004)(10653,2004)
\blacken\thinlines
\path(10503.000,1966.500)(10653.000,2004.000)(10503.000,2041.500)(10503.000,1966.500)
\thicklines
\path(10653,2004)(11283,2004)
\blacken\thinlines
\path(11133.000,1966.500)(11283.000,2004.000)(11133.000,2041.500)(11133.000,1966.500)
\thicklines
\path(11283,2724)(11283,2004)
\blacken\thinlines
\path(11245.500,2154.000)(11283.000,2004.000)(11320.500,2154.000)(11245.500,2154.000)
\blacken\path(11433.000,2761.500)(11283.000,2724.000)(11433.000,2686.500)(11433.000,2761.500)
\thicklines
\path(11283,2724)(11913,2724)
\path(11913,2724)(12633,2724)
\path(12633,2004)(12858,2004)
\blacken\thinlines
\path(12708.000,1966.500)(12858.000,2004.000)(12708.000,2041.500)(12708.000,1966.500)
\dashline{60.000}(10653,2004)(10653,3489)
\blacken\path(10690.500,3339.000)(10653.000,3489.000)(10615.500,3339.000)(10690.500,3339.000)
\dashline{60.000}(10653,3489)(11283,3489)
\blacken\path(11133.000,3451.500)(11283.000,3489.000)(11133.000,3526.500)(11133.000,3451.500)
\dashline{60.000}(11283,3489)(12633,3489)
\blacken\path(12483.000,3451.500)(12633.000,3489.000)(12483.000,3526.500)(12483.000,3451.500)
\dashline{60.000}(12633,3489)(12633,2724)
\blacken\path(12595.500,2874.000)(12633.000,2724.000)(12670.500,2874.000)(12595.500,2874.000)
\dashline{60.000}(11283,3444)(11283,2724)
\blacken\path(11245.500,2874.000)(11283.000,2724.000)(11320.500,2874.000)(11245.500,2874.000)
\dashline{60.000}(11283,2004)(11913,2004)
\blacken\path(11763.000,1966.500)(11913.000,2004.000)(11763.000,2041.500)(11763.000,1966.500)
\dashline{60.000}(11913,2004)(12633,2004)
\blacken\path(12483.000,1966.500)(12633.000,2004.000)(12483.000,2041.500)(12483.000,1966.500)
\put(1608,2769){\makebox(0,0)[lb]{\smash{{\SetFigFont{7}{8.4}{\rmdefault}{\mddefault}{\updefault}$C'$}}}}
\put(3858,1779){\makebox(0,0)[lb]{\smash{{\SetFigFont{7}{8.4}{\rmdefault}{\mddefault}{\updefault}$c$}}}}
\put(2463,3624){\makebox(0,0)[lb]{\smash{{\SetFigFont{7}{8.4}{\rmdefault}{\mddefault}{\updefault}$b$}}}}
\put(2373,1779){\makebox(0,0)[lb]{\smash{{\SetFigFont{7}{8.4}{\rmdefault}{\mddefault}{\updefault}$v_1$}}}}
\put(2958,1779){\makebox(0,0)[lb]{\smash{{\SetFigFont{7}{8.4}{\rmdefault}{\mddefault}{\updefault}$v_2$}}}}
\put(4668,2094){\makebox(0,0)[lb]{\smash{{\SetFigFont{7}{8.4}{\rmdefault}{\mddefault}{\updefault}$P_s(F(C'))$}}}}
\put(978,744){\makebox(0,0)[lb]{\smash{{\SetFigFont{7}{8.4}{\rmdefault}{\mddefault}{\updefault}$C$ $zav_1xv_2cz$}}}}
\put(978,969){\makebox(0,0)[lb]{\smash{{\SetFigFont{7}{8.4}{\rmdefault}{\mddefault}{\updefault}$C'$ $zabxv_1v_2cz$ }}}}
\put(978,69){\makebox(0,0)[lb]{\smash{{\SetFigFont{7}{8.4}{\rmdefault}{\mddefault}{\updefault}$\beta=0$}}}}
\put(978,294){\makebox(0,0)[lb]{\smash{{\SetFigFont{7}{8.4}{\rmdefault}{\mddefault}{\updefault}$S$ $zav_1v_2cz$ }}}}
\put(978,519){\makebox(0,0)[lb]{\smash{{\SetFigFont{7}{8.4}{\rmdefault}{\mddefault}{\updefault}Containing brother of $C$ $zav_1xbcz$}}}}
\put(213,2094){\makebox(0,0)[lb]{\smash{{\SetFigFont{7}{8.4}{\rmdefault}{\mddefault}{\updefault}$P_s(S)$}}}}
\put(933,1779){\makebox(0,0)[lb]{\smash{{\SetFigFont{7}{8.4}{\rmdefault}{\mddefault}{\updefault}$a$}}}}
\put(2283,2679){\makebox(0,0)[lb]{\smash{{\SetFigFont{7}{8.4}{\rmdefault}{\mddefault}{\updefault}$x$}}}}
\put(2733,2319){\makebox(0,0)[lb]{\smash{{\SetFigFont{7}{8.4}{\rmdefault}{\mddefault}{\updefault}$C$}}}}
\put(6423,2769){\makebox(0,0)[lb]{\smash{{\SetFigFont{7}{8.4}{\rmdefault}{\mddefault}{\updefault}$C'$}}}}
\put(5748,1779){\makebox(0,0)[lb]{\smash{{\SetFigFont{7}{8.4}{\rmdefault}{\mddefault}{\updefault}$a$}}}}
\put(8673,1779){\makebox(0,0)[lb]{\smash{{\SetFigFont{7}{8.4}{\rmdefault}{\mddefault}{\updefault}$c$}}}}
\put(7278,3624){\makebox(0,0)[lb]{\smash{{\SetFigFont{7}{8.4}{\rmdefault}{\mddefault}{\updefault}$x$}}}}
\put(7908,2274){\makebox(0,0)[lb]{\smash{{\SetFigFont{7}{8.4}{\rmdefault}{\mddefault}{\updefault}$S$}}}}
\put(7548,3039){\makebox(0,0)[lb]{\smash{{\SetFigFont{7}{8.4}{\rmdefault}{\mddefault}{\updefault}$C$}}}}
\put(7998,2814){\makebox(0,0)[lb]{\smash{{\SetFigFont{7}{8.4}{\rmdefault}{\mddefault}{\updefault}$v_2$}}}}
\put(7053,2634){\makebox(0,0)[lb]{\smash{{\SetFigFont{7}{8.4}{\rmdefault}{\mddefault}{\updefault}$v_1$}}}}
\put(7233,1779){\makebox(0,0)[lb]{\smash{{\SetFigFont{7}{8.4}{\rmdefault}{\mddefault}{\updefault}$b$}}}}
\put(5793,294){\makebox(0,0)[lb]{\smash{{\SetFigFont{7}{8.4}{\rmdefault}{\mddefault}{\updefault}$\beta=0$}}}}
\put(5793,969){\makebox(0,0)[lb]{\smash{{\SetFigFont{7}{8.4}{\rmdefault}{\mddefault}{\updefault}$C'$ $zaxv_1bcz$}}}}
\put(5793,744){\makebox(0,0)[lb]{\smash{{\SetFigFont{7}{8.4}{\rmdefault}{\mddefault}{\updefault}$C$ $zabv_1xv_2c$}}}}
\put(5793,519){\makebox(0,0)[lb]{\smash{{\SetFigFont{7}{8.4}{\rmdefault}{\mddefault}{\updefault}$S$ $zabv_1v_2cz$}}}}
\put(11058,2724){\makebox(0,0)[lb]{\smash{{\SetFigFont{7}{8.4}{\rmdefault}{\mddefault}{\updefault}$v_1$}}}}
\put(12723,2769){\makebox(0,0)[lb]{\smash{{\SetFigFont{7}{8.4}{\rmdefault}{\mddefault}{\updefault}$v_2$}}}}
\put(11913,2859){\makebox(0,0)[lb]{\smash{{\SetFigFont{7}{8.4}{\rmdefault}{\mddefault}{\updefault}$x$}}}}
\put(11238,3624){\makebox(0,0)[lb]{\smash{{\SetFigFont{7}{8.4}{\rmdefault}{\mddefault}{\updefault}$e$}}}}
\put(10608,1779){\makebox(0,0)[lb]{\smash{{\SetFigFont{7}{8.4}{\rmdefault}{\mddefault}{\updefault}$a$}}}}
\put(11238,1779){\makebox(0,0)[lb]{\smash{{\SetFigFont{7}{8.4}{\rmdefault}{\mddefault}{\updefault}$b$}}}}
\put(11868,1779){\makebox(0,0)[lb]{\smash{{\SetFigFont{7}{8.4}{\rmdefault}{\mddefault}{\updefault}$c$}}}}
\put(12588,1779){\makebox(0,0)[lb]{\smash{{\SetFigFont{7}{8.4}{\rmdefault}{\mddefault}{\updefault}$d$}}}}
\put(11508,2229){\makebox(0,0)[lb]{\smash{{\SetFigFont{7}{8.4}{\rmdefault}{\mddefault}{\updefault}$C'$}}}}
\put(12228,2904){\makebox(0,0)[lb]{\smash{{\SetFigFont{7}{8.4}{\rmdefault}{\mddefault}{\updefault}$C$}}}}
\put(11868,3579){\makebox(0,0)[lb]{\smash{{\SetFigFont{7}{8.4}{\rmdefault}{\mddefault}{\updefault}$S$}}}}
\put(9573,2094){\makebox(0,0)[lb]{\smash{{\SetFigFont{7}{8.4}{\rmdefault}{\mddefault}{\updefault}$P_s(F(C'))$}}}}
\put(9888,969){\makebox(0,0)[lb]{\smash{{\SetFigFont{7}{8.4}{\rmdefault}{\mddefault}{\updefault}$C'$ $zabv_1xcdz$}}}}
\put(9888,744){\makebox(0,0)[lb]{\smash{{\SetFigFont{7}{8.4}{\rmdefault}{\mddefault}{\updefault}$C$ $zabv_1xv_2dz$}}}}
\put(9888,519){\makebox(0,0)[lb]{\smash{{\SetFigFont{7}{8.4}{\rmdefault}{\mddefault}{\updefault}$S$ $zabv_1ev_2dz$}}}}
\put(9888,294){\makebox(0,0)[lb]{\smash{{\SetFigFont{7}{8.4}{\rmdefault}{\mddefault}{\updefault}Neighbour brother of $S$ $zaebcdz$}}}}
\put(9888,69){\makebox(0,0)[lb]{\smash{{\SetFigFont{7}{8.4}{\rmdefault}{\mddefault}{\updefault}$\beta=l(c,d)$}}}}
\end{picture}
}
\end{center}
\caption{Examples for DIRECT-ONE.} \label{dironefig}
\end{figure}

To motivate the last two definitions consider the comparison between
$C$ and $C'$ performed  in DIRECT-ONE and Figure \ref{dironefig}. In
this procedure, the cycles $C$ and $C'$ compete over the path
$[x,v_1]$. The preferred cycle  sets the orientation of this path.

\begin{obs}
\label{winpath} When two cycles compete the preferred cycle is the
winner.
\end{obs}

\clearpage

\subsection{The algorithm}

\begin{figure}[phtb]
\framebox[\textwidth]{
\begin{em}
\begin{program}
\Proc{Main} \\
\Inputs\  \\
1.  An undirected planar graph $G=(V,E)$. \\
2.  A set of cycles $\mathcal{C}$ and its hereditary order, with $g_{max}$ generations.\\
$[\mathcal{C}$ doesn't contain not-very-heavy-outer-crossing cycles.$]$\\
\Returns\\
An orientation of $G$\\
\Procbegin \\
  Orient $C_0$ in arbitrary orientation . $(\ast)$\\
  \For $(i=1, \ldots, g_{max})$ \\
\Forevery cycle $C$ such that $g(C) = i$\\
$C_1',C_2', \ldots ,C_m':=$  the uncontained sons of
$C$.\\
DIRECT-BROTHERS$(C_1',C_2', \ldots ,C_m')$.\\
\Endfor\\
\Endfor\\
\Endproc{Main}\\
\\
 $\ast$ The orientation of edges in $E$ is a  global variable,
 seen by all procedures.
\end{program}
\end{em}
} \caption{Procedure $Main$} \label{Mainproc}
\end{figure}

\begin{figure}[phtb]
\framebox[\textwidth]{
\begin{em}
\begin{program}
\Proc{DIRECT-BROTHERS} \\
\Inputs\  \\
An ordered set of brothers $C_1,C_2, \ldots ,C_m$ from the same level of containment.\\
\Procbegin \\
\Forevery  block $(C_l,\ldots,C_k)$ $[$see Definition \ref{maxintdef}.$]$\\
 DIRECT$(C_l,\ldots,C_k)$.\\
\For $(i=l,\ldots,k )$\\
  $\hat{C_1},\ldots,\hat{C_p} :=$ the maximal contained
brothers of $C_i$.\\
DIRECT-BROTHERS$(\hat{C_1},\hat{C_2}, \ldots ,\hat{C_p})$.\\
\Endfor\\
\Endfor\\
\Endproc{DIRECT-BROTHERS}
\end{program}
\end{em}
} \caption{Procedure DIRECT-BROTHERS} \label{Bigd}
\end{figure}

\begin{figure}[phtb]
\framebox[\textwidth]{
\begin{em}
\begin{program}
\Proc{DIRECT} \\
\Inputs\  \\
A block $(C_1, \ldots ,C_n)$.\\
\Procbegin\\
\If $(n=1)$ \\
\Then $ \{v, u\}  = P_f(C_1) \cap P_s(C_1)$ ordered according to the
orientation  of $F(C_1)$.\\
\Else \\
$v:= (P_f(C_1) \cap P_s(C_1)) \backslash P_f(C_2)$. \\
$u:= (P_f(C_{n-1}) \cap P_s(C_n)) \backslash P_f(C_{n-1})$. \\
\Endif\\
$e_v:=$ edge in $P_s(C_1)$ incident to $v$.\\
$e_u:=$ edge in $P_s(C_n)$ incident to $u$.\\
$[v,u,e_v$ and $e_u$ are described in Figure \ref{intervalfig}
 for $(C_1,C_2,C_3,C_4).]$\\
\If $(e_v$ is undirected $)$\\
 \Then $l_1:=0$.\\
\Elseif $(e_v$ is oriented  from $v)$\\
\Then $l_1: =1$.\\
\Else $l_1: =-1$.\\
\Endif\\
\If $(e_u$ is undirected $)$ \\
\Then $l_2:=0$.\\
\Elseif $(e_u$ is oriented  into $u)$\\
\Then $l_2 :=1$.\\
\Else $l_2 :=-1$.\\
\Endif\\
\If $((n=2) \wedge (C_1,C_2$ are inner-crossing $))$ \\
\Then DIRECT-INNER-CROSSING$(C_1,C_2,l_1,l_2)$ \\
\Else DIRECT-K$(C_1,\ldots,C_n,l_1,l_2)$.\\
\Endif\\
\Endproc{DIRECT} \\
\\
\Proc{DIRECT-K} \\
\Inputs\  \\
1. A block $(C_1, \ldots ,C_n)$. $[$There is no inner-crossing in this block.$]$\\
2. Parameters $l_1$,$l_2$.\\
\Procbegin\\
\If $(n=1)$ \\
\Then DIRECT-ONE$(C_1,l_1,l_2)$.\\
\Elseif  $(n=2)$\\
\Then DIRECT-TWO$(C_1,C_2,l_1,l_2)$. \\
\Else $(n > 2)$ DIRECT-MANY$(C_1,\ldots,C_n,l_1,l_2)$.\\
\Endif\\
\Endproc{DIRECT-K}
\end{program}
\end{em}
} \caption{Procedures DIRECT and DIRECT-K} \label{Dirfig}
\end{figure}

\begin{figure}[phtb]
\framebox[\textwidth]{
\begin{em}
\begin{program}
\Proc{DIRECT-MANY} \\
\Inputs\  \\
1. A block $(C_1, \ldots ,C_n)$, $n>2$. $[$There is no inner-crossing in this block.$]$\\
2. Parameters $l_1$,$l_2$.\\
$[$According to Observation \ref{croosobs} there may be at most
one outer-crossing.$]$\\
\Procbegin\\
 $m_1 :=\arg\max\{ l(P_f(C_i)) : i=1,\ldots,n \}$.\\
$m_2 :=\arg\max\{ l(P_f(C_i)) :i =1,\ldots,n, i \neq m_1\}$.\\
 \If $(C_{m_1}$ is light$)$\\
 \Then $[$According to Observation \ref{croosobs} also no
 outer-crossing in this case.$]$ \\
 \If $((l_1 \neq -1) \wedge (((n$ is even$) \wedge (l_2 \neq 1)) \vee ((n$ is odd$) \wedge (l_2 \neq -1))))$ \\
  \Then DIRECT-FORWARDS$(C_1, \ldots C_n)$. $[$See Figure \ref{dirmanyfig}(a) and (b).$]$\\
  \Elseif  $((l_1 \neq 1) \wedge ( ( (n$ is even$) \wedge (l_2 \neq -1)) \vee((n$ is odd$) \wedge (l_2 \neq 1))))$ \\
  \Then DIRECT-BACKWARDS$(C_1, \ldots C_n)$.$[$See Figure \ref{dirmanyfig}(c) and (d).$]$ \\
  \Elseif $((l_1 =1)  \wedge  (((n$ is even $) \wedge(l_2=1)) \vee ((n$ is odd $) \wedge (l_2 = -1))))$. \\
  \Then $[$see Figure \ref{dirmanyfig}(e) and (f) $]$ \\
  DIRECT-TWO$(C_1,C_2,1,1)$. \\
  DIRECT-BACKWARDS$(C_3,\ldots,C_n)$. \\
  \Elseif $((n$ is odd$) \wedge (l_1 =-1)  \wedge (l_2=1))$. \\
  \Then $[$See Figure \ref{dirmanyfig}(g).$]$ \\
  DIRECT-TWO$(C_{n-1},C_n,1,1)$. \\
  DIRECT-BACKWARDS$(C_1,\ldots,C_{n-2})$. \\
  \Elseif $((n$ is even$) \wedge (l_1 =-1)  \wedge (l_2=-1))$. \\
  \Then $[$See Figure \ref{dirmanyfig}(h).$]$ $(\ast)$ \\
  DIRECT-ONE$(C_1,-1,-1)$. \\
  DIRECT-TWO$(C_2,C_3,1,1)$. \\
  DIRECT-BACKWARDS$(C_4,\ldots,C_n)$. \\
  \Endif\\
  \Else  $\ (C_{m_1}$ is heavy$)$\\
  \If $(( C_{m_2} $ is light $) \wedge (C_{m_1}$ is not crossing$))$\\
    \Then\\
    \If $(m_1=1)$\\
    \Then
    DIRECT-ONE$(C_1,l_1,+1)$. \\
    DIRECT-K$(C_2,\ldots,C_n,-1,l_2)$.\\
    \Elseif $(m_1=n)$\\
    \Then
    DIRECT-ONE$(C_n,+1,l_2)$ .\\
    DIRECT-K$(C_1,\dots,C_{n-1},l_1,-1)$. \\
    \Else $(1 < m_1 < n)$ \\
    DIRECT-ONE$(C_{m_1},+1,+1)$. \\
    DIRECT-K$(C_1,\ldots,C_{m_1-1},l_1,-1)$. \\
    DIRECT-K$(C_{m_1+1},\ldots,C_n,-1,l_2)$. \\
    \Endif\\
    \Endif\\
    $\ast$ Special treatment to satisfy Property \ref{n2back}.
\end{program}
\end{em}
} \caption{Procedure DIRECT-MANY, first part} \label{Dirmanyfig_1}
\end{figure}

\begin{figure}[phtb]
\framebox[\textwidth]{
\begin{em}
\begin{program}
\Proc{DIRECT-MANY (cont.)} \\
    \Else $\ ((C_{m_2}$ is heavy $) \vee (C_{m_1}$ is crossing$))$\\
    $[$When $C_{m_1}$ is crossing, it is outer-crossing and very
    heavy, see Observation \ref{croosobs}$]$\\
    $[$In this case $C_{m_2}$ is either the crossing brother or light.$]$ \\
    \If $(C_{m_1}$ is crossing $)$ \\
    \Then $\{ j,k \} :=$  indices of $C_{m_1}$ and its crossing brother, $j=k-1$.  \\
    \Else $\{ j,k \}:= \{ m_1,m_2 \},\ j<k$.\\
    \Endif\\
    \If $(j=1)$\\
    \Then  DIRECT-ONE$(C_1,l_1,+1)$. \\
    \Else  DIRECT-ONE$(C_j,+1,+1)$. \\
    \Endif\\
    \If $(k=n)$\\
    \Then $\delta = l_2$. $[l_2$ indicates the orientation  of the last edge.$]$\\
    \Else $\delta= 1$. $[C_k$ is oriented  forwards and the last edge agree.$]$\\
    \Endif\\
    \If $((j=k-1) \wedge (C_{m_1}$ is not crossing $))$\\
    \Then DIRECT-ONE$(C_k,-1,\delta)$. $[-1$ to suit previous orientation by $C_{j}.]$\\
    \Else DIRECT-ONE$(C_k,+1,\delta)$. $[C_k$ is  oriented forwards and the first edge agrees$.]$\\
    \Endif\\
    \If $(j> 1)$\\
    \Then DIRECT-K$(C_1,\ldots,C_{j-1},l_1,-1)$.\\
    \Endif\\
    \If $(j<k-1)$\\
    \Then DIRECT-K$(C_{j+1},\ldots,C_{k-1},-1,-1)$.\\
    \Endif\\
    \If $(k<n)$
    \Then DIRECT-K$(C_{k+1},\ldots,C_n,-1,l_2)$.\\
    \Endif \\
   \Endproc{DIRECT-MANY}
\end{program}
\end{em}
} \caption{Procedure DIRECT-MANY, second part} \label{Dirmanyfig_2}
\end{figure}

\begin{figure}
\begin{center}
\setlength{\unitlength}{0.00069991in}
\begingroup\makeatletter\ifx\SetFigFont\undefined%
\gdef\SetFigFont#1#2#3#4#5{%
  \reset@font\fontsize{#1}{#2pt}%
  \fontfamily{#3}\fontseries{#4}\fontshape{#5}%
  \selectfont}%
\fi\endgroup%
{\renewcommand{\dashlinestretch}{30}
\begin{picture}(9003,10200)(0,-10)
\path(3837,10137)(3837,9462)
\blacken\path(3799.500,9612.000)(3837.000,9462.000)(3874.500,9612.000)(3799.500,9612.000)
\blacken\path(1099.500,9612.000)(1137.000,9462.000)(1174.500,9612.000)(1099.500,9612.000)
\path(1137,9462)(1137,10137)
\path(1812,9462)(1812,10137)
\blacken\path(1849.500,9987.000)(1812.000,10137.000)(1774.500,9987.000)(1849.500,9987.000)
\blacken\path(2449.500,9612.000)(2487.000,9462.000)(2524.500,9612.000)(2449.500,9612.000)
\path(2487,9462)(2487,10137)
\path(3162,9462)(3162,10137)
\blacken\path(3199.500,9987.000)(3162.000,10137.000)(3124.500,9987.000)(3199.500,9987.000)
\path(462,9462)(462,10137)
\blacken\path(499.500,9987.000)(462.000,10137.000)(424.500,9987.000)(499.500,9987.000)
\path(462,10137)(1137,10137)
\blacken\path(987.000,10099.500)(1137.000,10137.000)(987.000,10174.500)(987.000,10099.500)
\blacken\path(1287.000,10174.500)(1137.000,10137.000)(1287.000,10099.500)(1287.000,10174.500)
\path(1137,10137)(1812,10137)
\path(1812,10137)(2487,10137)
\blacken\path(2337.000,10099.500)(2487.000,10137.000)(2337.000,10174.500)(2337.000,10099.500)
\blacken\path(2637.000,10174.500)(2487.000,10137.000)(2637.000,10099.500)(2637.000,10174.500)
\path(2487,10137)(3162,10137)
\blacken\path(4549.500,9987.000)(4512.000,10137.000)(4474.500,9987.000)(4549.500,9987.000)
\path(4512,10137)(4512,9462)
\path(3162,10137)(3837,10137)
\blacken\path(3687.000,10099.500)(3837.000,10137.000)(3687.000,10174.500)(3687.000,10099.500)
\path(4512,10137)(3837,10137)
\blacken\path(3987.000,10174.500)(3837.000,10137.000)(3987.000,10099.500)(3987.000,10174.500)
\path(12,9462)(4962,9462)
\path(462,8787)(1137,8787)
\blacken\path(987.000,8749.500)(1137.000,8787.000)(987.000,8824.500)(987.000,8749.500)
\blacken\path(1287.000,8824.500)(1137.000,8787.000)(1287.000,8749.500)(1287.000,8824.500)
\path(1137,8787)(1812,8787)
\path(3837,8787)(3837,8112)
\blacken\path(3799.500,8262.000)(3837.000,8112.000)(3874.500,8262.000)(3799.500,8262.000)
\blacken\path(1099.500,8262.000)(1137.000,8112.000)(1174.500,8262.000)(1099.500,8262.000)
\path(1137,8112)(1137,8787)
\path(1812,8112)(1812,8787)
\blacken\path(1849.500,8637.000)(1812.000,8787.000)(1774.500,8637.000)(1849.500,8637.000)
\blacken\path(2449.500,8262.000)(2487.000,8112.000)(2524.500,8262.000)(2449.500,8262.000)
\path(2487,8112)(2487,8787)
\path(3162,8112)(3162,8787)
\blacken\path(3199.500,8637.000)(3162.000,8787.000)(3124.500,8637.000)(3199.500,8637.000)
\path(462,8112)(462,8787)
\blacken\path(499.500,8637.000)(462.000,8787.000)(424.500,8637.000)(499.500,8637.000)
\path(3162,8787)(3837,8787)
\blacken\path(3687.000,8749.500)(3837.000,8787.000)(3687.000,8824.500)(3687.000,8749.500)
\path(12,8112)(4962,8112)
\path(12,6762)(4962,6762)
\path(1812,8787)(2487,8787)
\blacken\path(2337.000,8749.500)(2487.000,8787.000)(2337.000,8824.500)(2337.000,8749.500)
\blacken\path(2637.000,8824.500)(2487.000,8787.000)(2637.000,8749.500)(2637.000,8824.500)
\path(2487,8787)(3162,8787)
\blacken\path(424.500,6912.000)(462.000,6762.000)(499.500,6912.000)(424.500,6912.000)
\path(462,6762)(462,7437)
\blacken\path(612.000,7474.500)(462.000,7437.000)(612.000,7399.500)(612.000,7474.500)
\path(462,7437)(1137,7437)
\path(1137,6762)(1137,7437)
\blacken\path(1174.500,7287.000)(1137.000,7437.000)(1099.500,7287.000)(1174.500,7287.000)
\path(1137,7437)(1812,7437)
\blacken\path(1662.000,7399.500)(1812.000,7437.000)(1662.000,7474.500)(1662.000,7399.500)
\blacken\path(1774.500,6912.000)(1812.000,6762.000)(1849.500,6912.000)(1774.500,6912.000)
\path(1812,6762)(1812,7437)
\blacken\path(1962.000,7474.500)(1812.000,7437.000)(1962.000,7399.500)(1962.000,7474.500)
\path(1812,7437)(2487,7437)
\path(2487,6762)(2487,7437)
\blacken\path(2524.500,7287.000)(2487.000,7437.000)(2449.500,7287.000)(2524.500,7287.000)
\path(2487,7437)(3162,7437)
\blacken\path(3012.000,7399.500)(3162.000,7437.000)(3012.000,7474.500)(3012.000,7399.500)
\blacken\path(3124.500,6912.000)(3162.000,6762.000)(3199.500,6912.000)(3124.500,6912.000)
\path(3162,6762)(3162,7437)
\blacken\path(3312.000,7474.500)(3162.000,7437.000)(3312.000,7399.500)(3312.000,7474.500)
\path(3162,7437)(3837,7437)
\blacken\path(3874.500,7287.000)(3837.000,7437.000)(3799.500,7287.000)(3874.500,7287.000)
\path(3837,7437)(3837,6762)
\blacken\path(4362.000,7399.500)(4512.000,7437.000)(4362.000,7474.500)(4362.000,7399.500)
\path(4512,7437)(3837,7437)
\path(4512,7437)(4512,6762)
\blacken\path(4474.500,6912.000)(4512.000,6762.000)(4549.500,6912.000)(4474.500,6912.000)
\blacken\path(424.500,5562.000)(462.000,5412.000)(499.500,5562.000)(424.500,5562.000)
\path(462,5412)(462,6087)
\blacken\path(612.000,6124.500)(462.000,6087.000)(612.000,6049.500)(612.000,6124.500)
\path(462,6087)(1137,6087)
\path(1137,5412)(1137,6087)
\blacken\path(1174.500,5937.000)(1137.000,6087.000)(1099.500,5937.000)(1174.500,5937.000)
\path(1137,6087)(1812,6087)
\blacken\path(1662.000,6049.500)(1812.000,6087.000)(1662.000,6124.500)(1662.000,6049.500)
\blacken\path(1774.500,5562.000)(1812.000,5412.000)(1849.500,5562.000)(1774.500,5562.000)
\path(1812,5412)(1812,6087)
\blacken\path(1962.000,6124.500)(1812.000,6087.000)(1962.000,6049.500)(1962.000,6124.500)
\path(1812,6087)(2487,6087)
\path(2487,5412)(2487,6087)
\blacken\path(2524.500,5937.000)(2487.000,6087.000)(2449.500,5937.000)(2524.500,5937.000)
\path(2487,6087)(3162,6087)
\blacken\path(3012.000,6049.500)(3162.000,6087.000)(3012.000,6124.500)(3012.000,6049.500)
\blacken\path(3124.500,5562.000)(3162.000,5412.000)(3199.500,5562.000)(3124.500,5562.000)
\path(3162,5412)(3162,6087)
\blacken\path(3312.000,6124.500)(3162.000,6087.000)(3312.000,6049.500)(3312.000,6124.500)
\path(3162,6087)(3837,6087)
\blacken\path(3874.500,5937.000)(3837.000,6087.000)(3799.500,5937.000)(3874.500,5937.000)
\path(3837,6087)(3837,5412)
\path(12,5412)(4962,5412)
\path(12,4062)(4962,4062)
\path(462,4062)(462,4737)
\blacken\path(497.250,4596.000)(462.000,4737.000)(426.750,4596.000)(497.250,4596.000)
\path(462,4737)(1137,4737)
\blacken\path(996.000,4701.750)(1137.000,4737.000)(996.000,4772.250)(996.000,4701.750)
\path(1137,4737)(1812,4737)
\blacken\path(1671.000,4701.750)(1812.000,4737.000)(1671.000,4772.250)(1671.000,4701.750)
\blacken\path(1953.000,4772.250)(1812.000,4737.000)(1953.000,4701.750)(1953.000,4772.250)
\path(1812,4737)(2487,4737)
\path(2487,4737)(3162,4737)
\blacken\path(3021.000,4701.750)(3162.000,4737.000)(3021.000,4772.250)(3021.000,4701.750)
\blacken\path(3303.000,4772.250)(3162.000,4737.000)(3303.000,4701.750)(3303.000,4772.250)
\path(3162,4737)(3837,4737)
\path(3837,4737)(4512,4737)
\blacken\path(4371.000,4701.750)(4512.000,4737.000)(4371.000,4772.250)(4371.000,4701.750)
\path(1137,4062)(1137,4737)
\blacken\path(1776.750,4203.000)(1812.000,4062.000)(1847.250,4203.000)(1776.750,4203.000)
\path(1812,4062)(1812,4737)
\path(2487,4062)(2487,4737)
\blacken\path(2522.250,4596.000)(2487.000,4737.000)(2451.750,4596.000)(2522.250,4596.000)
\blacken\path(3126.750,4203.000)(3162.000,4062.000)(3197.250,4203.000)(3126.750,4203.000)
\path(3162,4062)(3162,4737)
\blacken\path(3872.250,4596.000)(3837.000,4737.000)(3801.750,4596.000)(3872.250,4596.000)
\path(3837,4737)(3837,4062)
\path(4512,4737)(4512,4062)
\blacken\path(4476.750,4203.000)(4512.000,4062.000)(4547.250,4203.000)(4476.750,4203.000)
\blacken\path(1957.310,3381.330)(1812.000,3345.000)(1957.310,3308.670)(1957.310,3381.330)
\path(1812,3345)(2487,3345)
\path(462,2712)(462,3345)
\blacken\path(498.330,3199.690)(462.000,3345.000)(425.670,3199.690)(498.330,3199.690)
\path(1137,2712)(1137,3345)
\path(462,3345)(1137,3345)
\blacken\path(991.690,3308.670)(1137.000,3345.000)(991.690,3381.330)(991.690,3308.670)
\path(1137,3345)(1812,3345)
\blacken\path(1666.690,3308.670)(1812.000,3345.000)(1666.690,3381.330)(1666.690,3308.670)
\blacken\path(1775.670,2857.310)(1812.000,2712.000)(1848.330,2857.310)(1775.670,2857.310)
\path(1812,2712)(1812,3345)
\path(2487,2712)(2487,3345)
\blacken\path(2523.330,3199.690)(2487.000,3345.000)(2450.670,3199.690)(2523.330,3199.690)
\blacken\path(3125.670,2857.310)(3162.000,2712.000)(3198.330,2857.310)(3125.670,2857.310)
\path(3162,2712)(3162,3345)
\blacken\path(3873.330,3199.690)(3837.000,3345.000)(3800.670,3199.690)(3873.330,3199.690)
\path(3837,3345)(3837,2712)
\path(2487,3345)(3162,3345)
\blacken\path(3016.690,3308.670)(3162.000,3345.000)(3016.690,3381.330)(3016.690,3308.670)
\blacken\path(3307.310,3381.330)(3162.000,3345.000)(3307.310,3308.670)(3307.310,3381.330)
\path(3162,3345)(3837,3345)
\path(12,2712)(4962,2712)
\path(3162,1362)(3162,2037)
\path(3837,2037)(3837,1362)
\blacken\path(3799.500,1512.000)(3837.000,1362.000)(3874.500,1512.000)(3799.500,1512.000)
\path(3162,2037)(3837,2037)
\blacken\path(3687.000,1999.500)(3837.000,2037.000)(3687.000,2074.500)(3687.000,1999.500)
\path(2487,2037)(3162,2037)
\blacken\path(3012.000,1999.500)(3162.000,2037.000)(3012.000,2074.500)(3012.000,1999.500)
\path(2487,1362)(2487,2037)
\blacken\path(2524.500,1887.000)(2487.000,2037.000)(2449.500,1887.000)(2524.500,1887.000)
\blacken\path(1962.000,2074.500)(1812.000,2037.000)(1962.000,1999.500)(1962.000,2074.500)
\path(1812,2037)(2487,2037)
\blacken\path(1774.500,1512.000)(1812.000,1362.000)(1849.500,1512.000)(1774.500,1512.000)
\path(1812,1362)(1812,2037)
\path(1137,2037)(1812,2037)
\blacken\path(1662.000,1999.500)(1812.000,2037.000)(1662.000,2074.500)(1662.000,1999.500)
\path(1137,1362)(1137,2037)
\blacken\path(1174.500,1887.000)(1137.000,2037.000)(1099.500,1887.000)(1174.500,1887.000)
\blacken\path(612.000,2074.500)(462.000,2037.000)(612.000,1999.500)(612.000,2074.500)
\path(462,2037)(1137,2037)
\blacken\path(424.500,1512.000)(462.000,1362.000)(499.500,1512.000)(424.500,1512.000)
\path(462,1362)(462,2037)
\path(1812,12)(1812,687)
\blacken\path(424.500,162.000)(462.000,12.000)(499.500,162.000)(424.500,162.000)
\path(462,12)(462,687)
\blacken\path(612.000,724.500)(462.000,687.000)(612.000,649.500)(612.000,724.500)
\path(462,687)(1137,687)
\path(1137,12)(1137,687)
\blacken\path(1174.500,537.000)(1137.000,687.000)(1099.500,537.000)(1174.500,537.000)
\path(1137,687)(1812,687)
\blacken\path(1662.000,649.500)(1812.000,687.000)(1662.000,724.500)(1662.000,649.500)
\path(1812,687)(2487,687)
\blacken\path(2337.000,649.500)(2487.000,687.000)(2337.000,724.500)(2337.000,649.500)
\blacken\path(2449.500,162.000)(2487.000,12.000)(2524.500,162.000)(2449.500,162.000)
\path(2487,12)(2487,687)
\blacken\path(2637.000,724.500)(2487.000,687.000)(2637.000,649.500)(2637.000,724.500)
\path(2487,687)(3162,687)
\path(3162,12)(3162,687)
\blacken\path(3199.500,537.000)(3162.000,687.000)(3124.500,537.000)(3199.500,537.000)
\path(3162,687)(3837,687)
\blacken\path(3687.000,649.500)(3837.000,687.000)(3687.000,724.500)(3687.000,649.500)
\path(3837,687)(3837,12)
\blacken\path(3799.500,162.000)(3837.000,12.000)(3874.500,162.000)(3799.500,162.000)
\blacken\path(3987.000,724.500)(3837.000,687.000)(3987.000,649.500)(3987.000,724.500)
\path(3837,687)(4512,687)
\blacken\path(4549.500,537.000)(4512.000,687.000)(4474.500,537.000)(4549.500,537.000)
\path(4512,687)(4512,12)
\path(12,1362)(4962,1362)
\path(12,12)(4962,12)
\put(8742,9687){\makebox(0,0)[lb]{\smash{{\SetFigFont{10}{12.0}{\rmdefault}{\mddefault}{\updefault}$(a)$}}}}
\put(5862,9687){\makebox(0,0)[lb]{\smash{{\SetFigFont{10}{12.0}{\rmdefault}{\mddefault}{\updefault}$l_1\neq -1$, $l_2\neq 1$, $n$ is even}}}}
\put(5862,8337){\makebox(0,0)[lb]{\smash{{\SetFigFont{10}{12.0}{\rmdefault}{\mddefault}{\updefault}$l_1\neq -1$, $l_2\neq -1$, $n$ is odd}}}}
\put(8742,8337){\makebox(0,0)[lb]{\smash{{\SetFigFont{10}{12.0}{\rmdefault}{\mddefault}{\updefault}$(b)$}}}}
\put(8742,6987){\makebox(0,0)[lb]{\smash{{\SetFigFont{10}{12.0}{\rmdefault}{\mddefault}{\updefault}$(c)$}}}}
\put(5862,6987){\makebox(0,0)[lb]{\smash{{\SetFigFont{10}{12.0}{\rmdefault}{\mddefault}{\updefault}$l_1\neq 1$, $l_2\neq -1$, $n$ is even}}}}
\put(5862,5637){\makebox(0,0)[lb]{\smash{{\SetFigFont{10}{12.0}{\rmdefault}{\mddefault}{\updefault}$l_1\neq 1$, $l_2\neq 1$, $n$ is odd}}}}
\put(8742,5637){\makebox(0,0)[lb]{\smash{{\SetFigFont{10}{12.0}{\rmdefault}{\mddefault}{\updefault}$(d)$}}}}
\put(5862,4287){\makebox(0,0)[lb]{\smash{{\SetFigFont{10}{12.0}{\rmdefault}{\mddefault}{\updefault}$l_1=1$, $l_2=1$, $n$ is even}}}}
\put(8742,4287){\makebox(0,0)[lb]{\smash{{\SetFigFont{10}{12.0}{\rmdefault}{\mddefault}{\updefault}$(e)$}}}}
\put(5862,2937){\makebox(0,0)[lb]{\smash{{\SetFigFont{10}{12.0}{\rmdefault}{\mddefault}{\updefault}$l_1=1$, $l_2=-1$, $n$ is odd}}}}
\put(8742,2937){\makebox(0,0)[lb]{\smash{{\SetFigFont{10}{12.0}{\rmdefault}{\mddefault}{\updefault}$(f)$}}}}
\put(5862,1587){\makebox(0,0)[lb]{\smash{{\SetFigFont{10}{12.0}{\rmdefault}{\mddefault}{\updefault}$l_1=-1$, $l_2=1$, $n$ is odd}}}}
\put(8742,1587){\makebox(0,0)[lb]{\smash{{\SetFigFont{10}{12.0}{\rmdefault}{\mddefault}{\updefault}$(g)$}}}}
\put(5862,237){\makebox(0,0)[lb]{\smash{{\SetFigFont{10}{12.0}{\rmdefault}{\mddefault}{\updefault}$l_1=-1$, $l_2=-1$, $n$ is even}}}}
\put(8742,237){\makebox(0,0)[lb]{\smash{{\SetFigFont{10}{12.0}{\rmdefault}{\mddefault}{\updefault}$(h)$}}}}
\end{picture}
}
\end{center}
\caption{Examples for DIRECT-MANY orientations.} \label{dirmanyfig}
\end{figure}
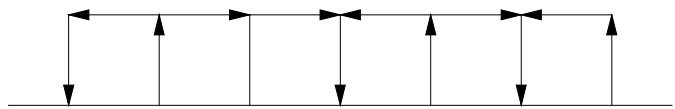

\begin{figure}[phtb]
\framebox[\textwidth]{
\begin{em}
\begin{program}
\Proc{DIRECT-FORWARDS} \\
\Inputs\  \\
Ordered neighbor brothers $(C_1, \ldots, C_n)$ with left points $v_1,\ldots,v_n$, respectively.\\
\Procbegin\\
$v_{n+1} := (P_f(C_n) \cap P_s(C_n)) \backslash P_s(C_{n-1})$.\\
\For $(i=1,\ldots, n)$\\
\If $(i$ is odd$)$ \\
\Then Orient $P_s(C_i)$  $v_{i} \rightarrow v_{i+1}$.\\
\Else Orient $P_s(C_i)$ $v_{i} \leftarrow v_{i+1}$.\\
\Endif\\
\Endfor\\
\Endproc{DIRECT-FORWARDS}\\
\\
\Proc{DIRECT-BACKWARDS} \\
\Inputs\  \\
Ordered neighbor brothers $(C_1, \ldots, C_n)$ with left points $v_1,\ldots,v_n$, respectively.\\
\Procbegin\\
$v_{n+1} := (P_f(C_n) \cap P_s(C_n)) \backslash P_s(C_{n-1})$.\\
\For $(i=1,\ldots,n)$\\
\If $(i$ is odd$)$ \\
\Then Orient $P_s(C_i)$  $v_{i} \leftarrow v_{i+1}$.\\
\Else Orient $P_s(C_i)$ $v_{i} \rightarrow v_{i+1}$.\\
\Endif\\
\Endfor\\
\Endproc{DIRECT-BACKWARDS}
\end{program}
\end{em}
} \caption{Procedures DIRECT-FORWARDS and DIRECT-BACKWARDS}
\label{dirodd}
\end{figure}

\begin{figure}[phtb]
\framebox[\textwidth]{
\begin{em}
\begin{program}
\Proc{DIRECT-INNER-CROSSING} \\
\Inputs\  \\
Inner-crossing brothers  $(C_1,C_2)$.\\
Parameters $l_1$ and $l_2$.\\
\Procbegin\\
Let $C'$ be the closest-containing brother of $C_1$ and $C_2$ $[$see Definition \ref{croosobs}$]$. \\
\If $(C'$ is oriented forwards$)$ \\
\Then  $[$see Figure \ref{inncrossfig}(a).$]$\\
DIRECT-ONE$(C_1,l_1,+1)$. \\
DIRECT-ONE$(C_2,+1,l_2)$.\\
\Else  $[$see Figure \ref{inncrossfig}(b).$]$\\
DIRECT-ONE$(C_1,l_1,-1)$. \\
DIRECT-ONE$(C_2,-1,l_2)$.\\
\Endif  \\
\Endproc{DIRECT-INNER-CROSSING}\\
\end{program}
\end{em}
} \caption{Procedure DIRECT-INNER-CROSSING} \label{dircros}
\end{figure}

\begin{figure}[phtb]
\framebox[\textwidth]{
\begin{em}
\begin{program}
\Proc{DIRECT-TWO} \\
\Inputs\  \\
Neighboring brothers $(C_1,C_2)$ with left ends $v_1,v_2$, respectively.\\
Parameters $l_1$ and $l_2$.\\
\Procbegin\\
 $u:= $ the end node of $P_s(C_1) \cap P_s(C_2)$ such that $u \not\in P_f(C_1)$.\\
 $v_3:= (P_f(C_2) \cap P_s(C_2)) \backslash P_s(C_1)$.\\
 \If $(C_1$ and $C_2$ are light$)$ \\
\Then\\
\If  $((l_1 \neq -1) \wedge (l_2 \neq 1)) $\\
\Then  Orient $P_s(C_1)$  $v_1 \rightarrow v_2$ .
Orient $P_s(C_2)$ $v_2 \leftarrow v_3$ .\\
\Elseif  $(((l_1 = 0) \wedge (l_2 = 1))  \vee  ((l_1 = -1) \wedge
(l_2 \neq -1)))$ \\
\Then  Orient $P_s(C_1)$ $v_1 \leftarrow v_2$ .
Orient $P_s(C_2)$  $v_2 \rightarrow v_3$ .\\
\Elseif  $((l_1 = 1) \wedge (l_2 = 1))$ \\
\Then \\
Orient $v_1 \rightarrow u \rightarrow v_3$. \\
\If  $(l(C_1) < l(C_2))$ \\
\Then Orient $u \rightarrow v_2$. \\
\Else Orient $u \leftarrow v_2$. \\
\Endif\\
\Else  $((l_1 = -1) \wedge (l_2 = -1)) $  \\
Orient $v_1 \leftarrow u \leftarrow v_3$. \\
\If  $(l(C_1) < l(C_2))$ \\
\Then Orient $u \leftarrow v_2$. \\
\Else Orient $u \rightarrow v_2$. \\
\Endif\\
\Endif\\
\Else \\
DIRECT-ONE $(C_1,l_1,+1)$.\\
DIRECT-ONE $(C_2,-1,l_2)$.\\
$[$Preventing a situation where both cycles are oriented backwards, \\
which is permitted only when both cycles are light.$]$\\
\Endif\\
\Endproc{DIRECT-TWO}
\end{program}
\end{em}
} \caption{Procedure DIRECT-TWO } \label{dirtwo}
\end{figure}

\begin{figure}[phtb]
\framebox[\textwidth]{
\begin{em}
\begin{program}
\Proc{DIRECT-ONE} \\
\Inputs\  \\
A cycle $C$.  Parameters $l_1$ and $l_2$.\\
\Procbegin\\
$v_1,v_2 :=$ the nodes in $P_s(C) \cap P_f(C)$ ordered according
to  orientation  of $F(C)$.\\
\If $((l_1 \neq -1) \wedge (l_2 \neq -1))$ \\
\Then Orient $C$ $v_1 \rightarrow v_2$.\\
\Elseif  $((C$ is light$) \wedge (C$ is not special-contained brother$) \wedge (l_1 \neq 1) \wedge (l_2 \neq 1))$\\
\Then Orient $C$ $v_1 \leftarrow v_2$. \\
\Else $[C$ must be oriented forwards to satisfy Property \ref{longfwd}, or Property \ref{scbprop}$]$ \\
\If $(l_1 = -1)$ \\
\Then  $C':=$  the cycle which sets the orientation  $l_1 = -1$ $[$see Definition \ref{csetodef}$]$.\\
$x:=$ the end node of the path $P_s(C') \cap P_s(C)$ such that $ x \not\in F(C)$.\\
$[$The path is oriented  $x \rightarrow v_1.]$ \\
$[$The possible positions of $C'$ are shown in Figure \ref{dironefig},\\
where the bold line indicates $C$.$]$\\
Orient $x \rightarrow v_2$.\\
\If $(C'$ is oriented  forwards$)$\\
\Then $\beta = 0$.\\
\Else $\beta = l(P_c(C') \backslash  C)$.\\
\Endif\\
\If $(l(C) < l(C') + 2\beta)$\\
\Then Orient $v_1 \rightarrow x$ $[$changing the current orientation$]$.\\
\If $(l(C') < l(C))$\\
\Then \\
\Forevery $D$: $(F(D) = C') \wedge (P_f(D) \subset
(P_s(C') \cap P_s(C)))$ \\
$F(D) := C$.\\
\Endfor\\
\Endif\\
\Endif\\
\Endif\\
\If $(l_2 = -1 )$ \\
\Then  $C'':=$  the cycle which sets the orientation $l_2 = -1$.\\
$y:=$ the end node of the path $P_s(C'') \cap P_s(C)$ such that $y \not\in F(C)$.\\
$[$The path is directed $v_2 \rightarrow y.]$  \\
Orient $v_1 \rightarrow y$.\\
\If $(C''$ is oriented  forwards$)$\\
\Then $\beta = 0$.\\
\Else $\beta = l(P_c(C'') \backslash  C)$.\\
\Endif\\
\If $(l(C) < l(C'')+ 2\beta)$\\
\Then Orient $y \rightarrow v_2$ $[$changing the current orientation$]$.\\
\If $(l(C'') < l(C))$\\
\Then \\
\Forevery $D$: $(F(D) = C'') \wedge (P_f(D) \subset
(P_s(C'') \cap P_s(C)))$ \\
$F(D) := C$.\\
\Endfor\\
\Endif\\
\Endif\\
\Endif\\
\Endif\\
\Endproc{DIRECT-ONE}
\end{program}
\end{em}
} \caption{Procedure DIRECT-ONE} \label{dirone}
\end{figure}

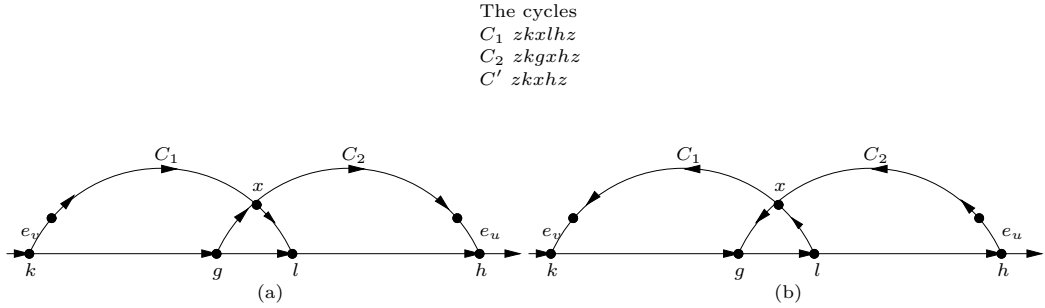
\begin{figure}
\begin{center}
\setlength{\unitlength}{0.00052493in}
\begingroup\makeatletter\ifx\SetFigFont\undefined%
\gdef\SetFigFont#1#2#3#4#5{%
  \reset@font\fontsize{#1}{#2pt}%
  \fontfamily{#3}\fontseries{#4}\fontshape{#5}%
  \selectfont}%
\fi\endgroup%
{\renewcommand{\dashlinestretch}{30}
\begin{picture}(10464,3078)(0,-10)
\put(6829.500,-69.553){\arc{2904.239}{3.5589}{5.8659}}
\put(8719.500,-69.553){\arc{2904.239}{3.5589}{5.8659}}
\put(1564.500,-69.553){\arc{2904.239}{3.5589}{5.8659}}
\put(3454.500,-69.553){\arc{2904.239}{3.5589}{5.8659}}
\put(5727,879){\blacken\ellipse{90}{90}}
\put(5727,879){\ellipse{90}{90}}
\put(9822,879){\blacken\ellipse{90}{90}}
\put(9822,879){\ellipse{90}{90}}
\put(462,879){\blacken\ellipse{90}{90}}
\put(462,879){\ellipse{90}{90}}
\put(4557,879){\blacken\ellipse{90}{90}}
\put(4557,879){\ellipse{90}{90}}
\put(237,519){\blacken\ellipse{90}{90}}
\put(237,519){\ellipse{90}{90}}
\put(2127,519){\blacken\ellipse{90}{90}}
\put(2127,519){\ellipse{90}{90}}
\put(2892,519){\blacken\ellipse{90}{90}}
\put(2892,519){\ellipse{90}{90}}
\put(4782,519){\blacken\ellipse{90}{90}}
\put(4782,519){\ellipse{90}{90}}
\put(5502,519){\blacken\ellipse{90}{90}}
\put(5502,519){\ellipse{90}{90}}
\put(7392,519){\blacken\ellipse{90}{90}}
\put(7392,519){\ellipse{90}{90}}
\put(8157,519){\blacken\ellipse{90}{90}}
\put(8157,519){\ellipse{90}{90}}
\put(10047,519){\blacken\ellipse{90}{90}}
\put(10047,519){\ellipse{90}{90}}
\put(2532,1014){\blacken\ellipse{90}{90}}
\put(2532,1014){\ellipse{90}{90}}
\put(7797,1014){\blacken\ellipse{90}{90}}
\put(7797,1014){\ellipse{90}{90}}
\path(5277,519)(5502,519)
\blacken\path(5352.000,481.500)(5502.000,519.000)(5352.000,556.500)(5352.000,481.500)
\path(5502,519)(7392,519)
\blacken\path(7242.000,481.500)(7392.000,519.000)(7242.000,556.500)(7242.000,481.500)
\path(7392,519)(8157,519)
\blacken\path(8007.000,481.500)(8157.000,519.000)(8007.000,556.500)(8007.000,481.500)
\path(8157,519)(10047,519)
\blacken\path(9897.000,481.500)(10047.000,519.000)(9897.000,556.500)(9897.000,481.500)
\path(597,1014)(687,1104)
\blacken\path(607.450,971.417)(687.000,1104.000)(554.417,1024.450)(607.450,971.417)
\path(1587,1374)(1722,1374)
\blacken\path(1572.000,1336.500)(1722.000,1374.000)(1572.000,1411.500)(1572.000,1336.500)
\path(3477,1374)(3612,1374)
\blacken\path(3462.000,1336.500)(3612.000,1374.000)(3462.000,1411.500)(3462.000,1336.500)
\path(2352,879)(2442,969)
\blacken\path(2362.450,836.417)(2442.000,969.000)(2309.417,889.450)(2362.450,836.417)
\path(237,519)(2127,519)
\blacken\path(1977.000,481.500)(2127.000,519.000)(1977.000,556.500)(1977.000,481.500)
\path(2127,519)(2892,519)
\blacken\path(2742.000,481.500)(2892.000,519.000)(2742.000,556.500)(2742.000,481.500)
\path(2892,519)(4782,519)
\blacken\path(4632.000,481.500)(4782.000,519.000)(4632.000,556.500)(4632.000,481.500)
\path(4782,519)(5187,519)
\blacken\path(5037.000,481.500)(5187.000,519.000)(5037.000,556.500)(5037.000,481.500)
\blacken\path(7002.000,1411.500)(6852.000,1374.000)(7002.000,1336.500)(7002.000,1411.500)
\path(6852,1374)(6987,1374)
\blacken\path(7651.550,966.583)(7572.000,834.000)(7704.583,913.550)(7651.550,966.583)
\path(7572,834)(7662,924)
\blacken\path(8892.000,1411.500)(8742.000,1374.000)(8892.000,1336.500)(8892.000,1411.500)
\path(8742,1374)(8877,1374)
\path(9732,969)(9642,1059)
\blacken\path(9774.583,979.450)(9642.000,1059.000)(9721.550,926.417)(9774.583,979.450)
\blacken\path(4334.417,1048.550)(4467.000,969.000)(4387.450,1101.583)(4334.417,1048.550)
\path(4467,969)(4377,1059)
\blacken\path(5941.550,1146.583)(5862.000,1014.000)(5994.583,1093.550)(5941.550,1146.583)
\path(5862,1014)(5952,1104)
\path(12,519)(237,519)
\blacken\path(87.000,481.500)(237.000,519.000)(87.000,556.500)(87.000,481.500)
\path(10047,519)(10452,519)
\blacken\path(10302.000,481.500)(10452.000,519.000)(10302.000,556.500)(10302.000,481.500)
\blacken\path(2605.934,897.640)(2712.000,834.000)(2648.360,940.066)(2605.934,897.640)
\path(2712,834)(2622,924)
\path(8022,789)(7932,879)
\blacken\path(8038.066,815.360)(7932.000,879.000)(7995.640,772.934)(8038.066,815.360)
\put(6762,1464){\makebox(0,0)[lb]{\smash{{\SetFigFont{7}{8.4}{\rmdefault}{\mddefault}{\updefault}$C_1$}}}}
\put(8652,1464){\makebox(0,0)[lb]{\smash{{\SetFigFont{7}{8.4}{\rmdefault}{\mddefault}{\updefault}$C_2$}}}}
\put(5412,699){\makebox(0,0)[lb]{\smash{{\SetFigFont{7}{8.4}{\rmdefault}{\mddefault}{\updefault}$e_v$}}}}
\put(10047,699){\makebox(0,0)[lb]{\smash{{\SetFigFont{7}{8.4}{\rmdefault}{\mddefault}{\updefault}$e_u$}}}}
\put(1497,1464){\makebox(0,0)[lb]{\smash{{\SetFigFont{7}{8.4}{\rmdefault}{\mddefault}{\updefault}$C_1$}}}}
\put(3387,1464){\makebox(0,0)[lb]{\smash{{\SetFigFont{7}{8.4}{\rmdefault}{\mddefault}{\updefault}$C_2$}}}}
\put(147,699){\makebox(0,0)[lb]{\smash{{\SetFigFont{7}{8.4}{\rmdefault}{\mddefault}{\updefault}$e_v$}}}}
\put(4782,699){\makebox(0,0)[lb]{\smash{{\SetFigFont{7}{8.4}{\rmdefault}{\mddefault}{\updefault}$e_u$}}}}
\put(2532,69){\makebox(0,0)[lb]{\smash{{\SetFigFont{7}{8.4}{\rmdefault}{\mddefault}{\updefault}(a)}}}}
\put(7752,69){\makebox(0,0)[lb]{\smash{{\SetFigFont{7}{8.4}{\rmdefault}{\mddefault}{\updefault}(b)}}}}
\put(192,294){\makebox(0,0)[lb]{\smash{{\SetFigFont{7}{8.4}{\rmdefault}{\mddefault}{\updefault}$k$}}}}
\put(4737,294){\makebox(0,0)[lb]{\smash{{\SetFigFont{7}{8.4}{\rmdefault}{\mddefault}{\updefault}$h$}}}}
\put(2082,294){\makebox(0,0)[lb]{\smash{{\SetFigFont{7}{8.4}{\rmdefault}{\mddefault}{\updefault}$g$}}}}
\put(2892,294){\makebox(0,0)[lb]{\smash{{\SetFigFont{7}{8.4}{\rmdefault}{\mddefault}{\updefault}$l$}}}}
\put(5457,294){\makebox(0,0)[lb]{\smash{{\SetFigFont{7}{8.4}{\rmdefault}{\mddefault}{\updefault}$k$}}}}
\put(7347,294){\makebox(0,0)[lb]{\smash{{\SetFigFont{7}{8.4}{\rmdefault}{\mddefault}{\updefault}$g$}}}}
\put(8157,294){\makebox(0,0)[lb]{\smash{{\SetFigFont{7}{8.4}{\rmdefault}{\mddefault}{\updefault}$l$}}}}
\put(10002,294){\makebox(0,0)[lb]{\smash{{\SetFigFont{7}{8.4}{\rmdefault}{\mddefault}{\updefault}$h$}}}}
\put(2487,1149){\makebox(0,0)[lb]{\smash{{\SetFigFont{7}{8.4}{\rmdefault}{\mddefault}{\updefault}$x$}}}}
\put(7752,1149){\makebox(0,0)[lb]{\smash{{\SetFigFont{7}{8.4}{\rmdefault}{\mddefault}{\updefault}$x$}}}}
\put(4782,2904){\makebox(0,0)[lb]{\smash{{\SetFigFont{7}{8.4}{\rmdefault}{\mddefault}{\updefault}The cycles}}}}
\put(4782,2679){\makebox(0,0)[lb]{\smash{{\SetFigFont{7}{8.4}{\rmdefault}{\mddefault}{\updefault}$C_1$ $zkxlhz$}}}}
\put(4782,2454){\makebox(0,0)[lb]{\smash{{\SetFigFont{7}{8.4}{\rmdefault}{\mddefault}{\updefault}$C_2$ $zkgxhz$}}}}
\put(4782,2229){\makebox(0,0)[lb]{\smash{{\SetFigFont{7}{8.4}{\rmdefault}{\mddefault}{\updefault}$C'$ $zkxhz$}}}}
\end{picture}
}
\end{center}
\caption{The orientation of DIRECT-INNER-CROSSING.}
\label{inncrossfig}
\end{figure}

\clearpage
\subsection{The main result}

The proof of next Proposition is given the following sections.
\begin{propos}
\label{indpro} It is possible to define directed paths
$M(S)$,$B_t(S)$ and $B_h(S)$ for every cycle $S$, such that:

\begin{description}
\item [A1] $l(M(S)) \leq 15 l(P_c(S)) + 2l(P_s(S)) + l(I(S)) $.
\item [A2] \begin{vbox}
{\bi
 \item If  $h(S) \not\in F(S)$ then $l(B_h(S)) +l(M_h(S)) \leq 15 l(P_c(S)) +
4l(P_s(S))$.
\item If $t(S) \not\in F(S)$ then $l(M_t(S)) + l(B_t(S))  \leq 15 l(P_c(S)) +
4l(P_s(S))$. \ei Moreover: \bi
\item  If $t(S),h(S) \not\in F(S)$ then  $l(B_h(S)) + l(B_t(S))  \leq 15 l(P_c(S)) +
6l(P_s(S)) - l(I(S))$. \ei}
\end{vbox}
 \item [A3] For every son $C$ of $S$, if $P_f(C)$ contains at
least one edge, then $P_f(C)$ and $ I(S)$ have a
 common edge. When $P_f(C)$ contains only one node, this node is in
$I(S)$.
 \item [A4] If $C$ is a son of $S$,
$P_f(C) \subset J_t(S) \cup I(S) \cup J_h(S)$.
 \item [A5] $M(S)$ is contained in the union of
$P_s(S)$ and all cycles which are preferred to $S$.
\item [A6] $M(S)$ is contained in
$P_s(S),P_s(F(S)),M(F(S)),B_t(F(S)),B_h(F(S))$, and a set $N(S)$ of
at most three other cycles.
\end{description}
\end{propos}
We will prove the persistence of these hypotheses by induction on
the index $i \in \{ 1,\ldots,g_{max} \}$ in Main, corresponding to a
cycle $S$ such that $g(S) \leq i$.

\noindent Initially   $I(C_0) = C_0$. $M(C_0) = \phi$. $B_t(C_0)=
\phi$. $B_h(C_0)= \phi$.

\begin{theorem}
In the directed graph returned by MAIN, every undirected cycle $S
\in \mathcal{C}$ has a directed cycle $\overrightarrow{S}$, such
that
\[ l(\overrightarrow{S}) \leq 15l(S). \]

\end{theorem}
\begin{prf}
By Proposition \ref{indpro} for every cycle  $S$ in the output of
MAIN there are directed paths $M(S),B_t(S),B_h(S)$ and $I(S)$ such
that
\begin{equation} \label{Meq}
l(M(S)) \leq 15 l(P_c(S)) + 2l(P_s(S)) + l(I(S)) .
\end{equation}
$I(S) \subseteq P_s(S)$ giving that  $l(I(S)) \leq l(P_s(S))$ and
therefore by (\ref{Meq}):
\[ l(M(S)) \leq 15 l(P_c(S)) + 3l(P_s(S)) . \]
The concatenation of $M(S)$ and $I(S)$ creates the directed cycle
$\overrightarrow{S}$ which connects $U(S)$  and $z$. Since $S =
P_c(S) \cup P_s(S)$
\[ l(\overrightarrow{S}) = l(M(S)) + l(I(S)) \leq 15 l(P_c(S)) +
4l(P_s(S)) \leq 15l(S)  \]

\end{prf}

\begin{theorem} For every node $v$, let $C^*(v)$ be the original
cycle which serves $v$ in $G$  and let $C(v)$ be the undirected
cycle which serves $v$ after the application of CANCEL-CROSSING and
MAIN. Then:
\[ l(C(v)) \leq 27l(C^*(v)) .\]
\end{theorem}

\begin{prf}
Let $\hat{C}(v)$ be the cycle which serves $v$ after CANCEL-CROSSING
is applied on the graph. According to Corollary \ref{ncrocor}
\[ l(\hat{C}(v)) \leq 9 l(C^*(v)). \]
During the application of MAIN $\hat{C}(v)$ may be farther changed.
According to Remark \ref{undrem} (below)
\[ l(C(v)) \leq 3 l(\hat{C}(v)). \]
Altogether
\[ l(C(v)) \leq 27 l(C^*(v)). \]

\end{prf}

\begin{cor} \label{cor405}For every node $v$, let $C^*(v)$ be the original
cycle which serves $v$ in $G$, and let $\overrightarrow{C}(v)$ be
the directed cycle which connects $v$ and $z$ in the graph returned
by MAIN. Then:
\[ l(\overrightarrow{C}(v)) \leq 15 \cdot 27 l(C(v)) = 405 l(C(v)). \]
\end{cor}

\begin{theorem} Let $D$ be the directed diameter of the graph returned by MAIN,
 \[ D \leq 1620 D_{H_{opt}}. \]
\end{theorem}

\begin{prf}
For every node $v$, let $C^*(v)$ be the original cycle which serves
$v$ in $G$, and let $\overrightarrow{C}(v)$ be the directed cycle
which connects $v$ and $z$ in the graph returned by MAIN. For every
two nodes $v_1,v_2 \in V$ the length of the directed path from $v_1$
to $v_2$ is bounded by the sum of the length of directed path from
$v_1$ to $z$ in $\overrightarrow{C}(v_1)$ and the length of the path
from $z$ to $v_2$ in $\overrightarrow{C}(v_2)$, thus:
\[ D \leq 2 max_{v \in V} \{ l(\overrightarrow{C}(v)) \} \]
>From Corollary \ref{cor405}
\[ D \leq 2 \cdot 405 max_{v \in V} \{ l(C(v)) \} \]
For every undirected cycle $C \in \mathcal{C}$ $l(C) \leq 2
D_{H_{opt}}$, giving: \[ D \leq  1620 D_{H_{opt}}. \]
\end{prf}

\subsection{The path $B(S)$}

\begin{lem}
\label{lbscor} In many cases we construct $B_h(S)$ and $B_t(S)$ in
the following manner. We first create a directed path $B(S)$ from
$M_t(S)$ to $M_h(S)$ such that $l(B(S)) \leq 2l(P_s(S)) - l(I(S))$.
When $h(S) \not\in F(S)$ we  define $B_h(S) \subseteq M_t(S) \cup
B(S)$ as follows: start in $z$, continue with $M_t(S)$ until it
meets $B(S)$, and then use $B(S)$ to reach $h(S)$ via $J_h(S)$.
Similarly when $t(S) \not\in F(S)$ define $B_t(S) \subseteq M_h(S)
\cup B(S)$. If $B(S)$  exists, $l(B(S)) \leq 2l(P_s(S)) - l(I(S))$,
and {\rm A1} holds, then {\rm A2} holds for the paths
$B_t(S),B_h(S)$ constructed above.
\end{lem}
\begin{prf} From A1
 \[l(M_h(S)) + l(M_t(S)) = l(M(S)) \leq 15 l(P_c(S)) + 2l(P_s(S)) + l(I(S)). \]
When $h(S) \not\in F(S)$ since $B_h(S) \subseteq M_t(S) \cup B(S)$:
\[ l(B_h(S)) + l(M_h(S)) \leq l(B(C)) + l(M_t(S)) + l(M_h(S)) .\]
>From the lemma's assumption we get that:
\[ l(B_h(S)) + l(M_h(S)) \leq 15 l(P_c(S)) + 4l(P_s(S)).\]
Similarly:
\[l(B_t(S)) + l(M_t(S))  \leq 15 l(P_c(S)) + 4l(P_s(S)).\]
By assumption, when $h(S) ,t(S) \not\in F(S)$: $B_h(S) \subseteq
B(S) \cup M_t(S)$ and $B_t(S) \subseteq B(S) \cup M_h(S)$, giving:
\[ l(B_h(S)) + l(B_t(S)) \leq l(M_h(S)) + l(M_t(S)) + 2lB(S) .\]
The required bound follows now from the lemma's assumptions.
\end{prf}

\subsection{Outline of the proof} In Section \ref{secieqps} we
prove that a cycle $C$ with $I(C)=P_s(C)$ satisfies the induction
hypotheses. This is sufficient to prove that when DIRECT-FORWARDS
and DIRECT-BACKWARDS are applied, the induction hypotheses are
satisfied.

In Section \ref{sectwo} we prove that when two neighbor brothers are
oriented in the same direction they satisfy the induction
hypotheses. This is sufficient to prove that when DIRECT-TWO and
DIRECT-MANY are applied to light cycles, the induction hypotheses
are satisfied.

When $C$ is a heavy cycle or  a special-contained brother DIRECT-ONE
is activated to orient it. In Section \ref{secdiron} we discuss
cycles oriented by DIRECT-ONE. If a cycle is neither heavy nor
special-contained, previous proofs apply. Otherwise DIRECT-ONE
defines the cycles $C'$ or $C''$. We continue by considering the
relation between $C$ and $C',C''$. Section \ref{secsamelevel}
handles the case where $C$ and $C'$ are brothers from the same level
on containment (and uses proofs given in Section \ref{secthree} and
Section \ref{secfour}). Section \ref{secsmalllevl} handles the case
where $C'$ is a brother from a previous level of containment.
Finally, Section \ref{secuncle} handles the case where $C''$ is from
a previous generation than $C$. This section is divided into two
parts. Section \ref{secfor} assumes that $F(C)$ is oriented
forwards, and Section \ref{secback} assumes that $F(C)$ is oriented
backwards.

In Section \ref{seccross} we prove that two inner-crossing brothers
satisfy the induction hypotheses. This is sufficient to prove that
when DIRECT-INNER-CROSSING is applied,  the induction hypotheses are
satisfied.

\subsection{$I(C)=P_s(C)$} \label{secieqps} We
first consider the simple cases when $I(C)=P_s(C)$.

Let $S=F(C)$. Mark $\{ v_1,v_2 \} = P_s(C) \cap P_f(C) $. Suppose
that $v_1,v_2$ are ordered according to direction of $S$.

In this proof and all similar proofs we mark $x=t(S),y=h(S)$, and
the end nodes $v,w$ of $P_s(S)$,  ordered $v,x,y,w$.

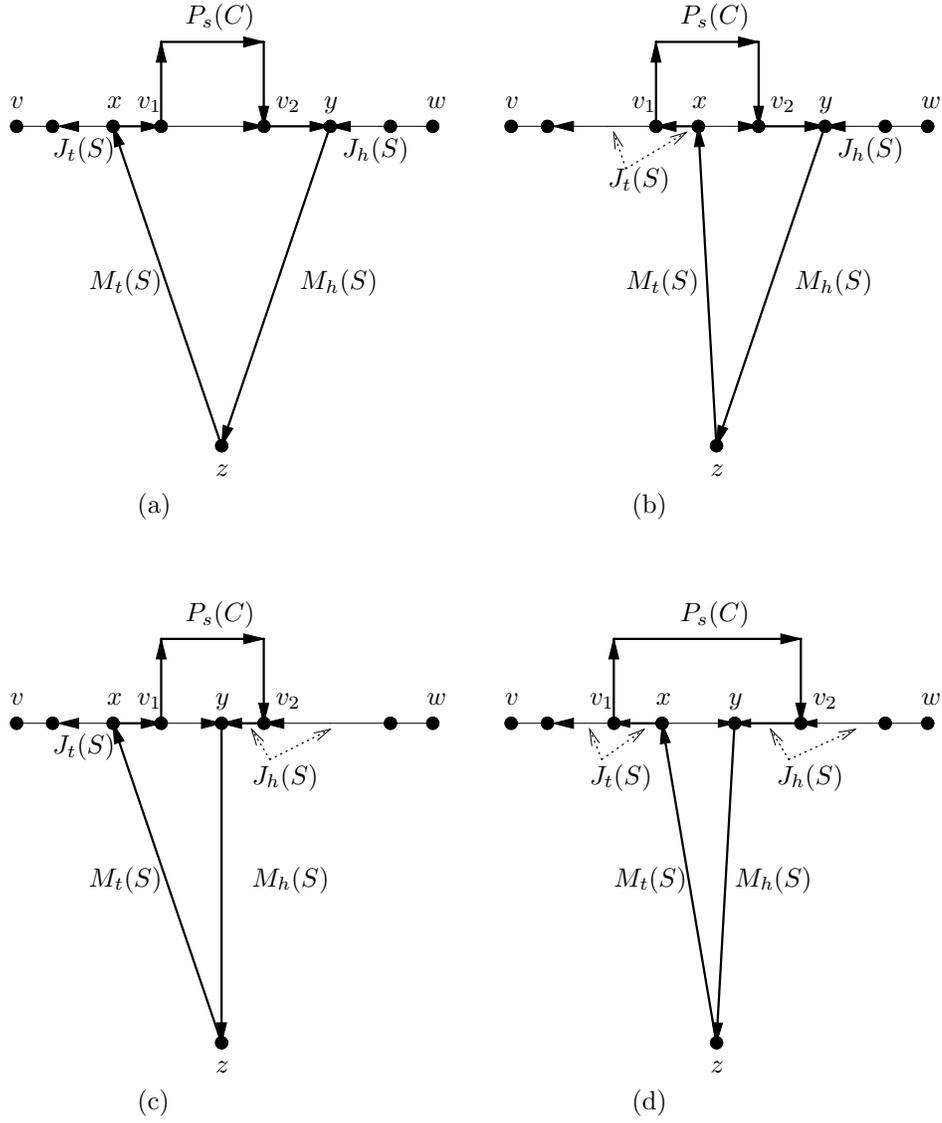
\begin{figure}
\begin{center}
\setlength{\unitlength}{0.00069991in}
\begingroup\makeatletter\ifx\SetFigFont\undefined%
\gdef\SetFigFont#1#2#3#4#5{%
  \reset@font\fontsize{#1}{#2pt}%
  \fontfamily{#3}\fontseries{#4}\fontshape{#5}%
  \selectfont}%
\fi\endgroup%
{\renewcommand{\dashlinestretch}{30}
\begin{picture}(6963,8343)(0,-10)
\put(1905,7404){\blacken\ellipse{90}{90}}
\put(1905,7404){\ellipse{90}{90}}
\put(2850,7404){\blacken\ellipse{90}{90}}
\put(2850,7404){\ellipse{90}{90}}
\put(780,7404){\blacken\ellipse{90}{90}}
\put(780,7404){\ellipse{90}{90}}
\put(2400,7404){\blacken\ellipse{90}{90}}
\put(2400,7404){\ellipse{90}{90}}
\put(1140,7404){\blacken\ellipse{90}{90}}
\put(1140,7404){\ellipse{90}{90}}
\put(60,7404){\blacken\ellipse{90}{90}}
\put(60,7404){\ellipse{90}{90}}
\put(3165,7404){\blacken\ellipse{90}{90}}
\put(3165,7404){\ellipse{90}{90}}
\put(330,7404){\blacken\ellipse{90}{90}}
\put(330,7404){\ellipse{90}{90}}
\put(5595,7404){\blacken\ellipse{90}{90}}
\put(5595,7404){\ellipse{90}{90}}
\put(6540,7404){\blacken\ellipse{90}{90}}
\put(6540,7404){\ellipse{90}{90}}
\put(6090,7404){\blacken\ellipse{90}{90}}
\put(6090,7404){\ellipse{90}{90}}
\put(4830,7404){\blacken\ellipse{90}{90}}
\put(4830,7404){\ellipse{90}{90}}
\put(3750,7404){\blacken\ellipse{90}{90}}
\put(3750,7404){\ellipse{90}{90}}
\put(6855,7404){\blacken\ellipse{90}{90}}
\put(6855,7404){\ellipse{90}{90}}
\put(4020,7404){\blacken\ellipse{90}{90}}
\put(4020,7404){\ellipse{90}{90}}
\put(5280,5019){\blacken\ellipse{90}{90}}
\put(5280,5019){\ellipse{90}{90}}
\put(5145,7404){\blacken\ellipse{90}{90}}
\put(5145,7404){\ellipse{90}{90}}
\put(2850,2949){\blacken\ellipse{90}{90}}
\put(2850,2949){\ellipse{90}{90}}
\put(780,2949){\blacken\ellipse{90}{90}}
\put(780,2949){\ellipse{90}{90}}
\put(1140,2949){\blacken\ellipse{90}{90}}
\put(1140,2949){\ellipse{90}{90}}
\put(60,2949){\blacken\ellipse{90}{90}}
\put(60,2949){\ellipse{90}{90}}
\put(3165,2949){\blacken\ellipse{90}{90}}
\put(3165,2949){\ellipse{90}{90}}
\put(330,2949){\blacken\ellipse{90}{90}}
\put(330,2949){\ellipse{90}{90}}
\put(6540,2949){\blacken\ellipse{90}{90}}
\put(6540,2949){\ellipse{90}{90}}
\put(3750,2949){\blacken\ellipse{90}{90}}
\put(3750,2949){\ellipse{90}{90}}
\put(6855,2949){\blacken\ellipse{90}{90}}
\put(6855,2949){\ellipse{90}{90}}
\put(4020,2949){\blacken\ellipse{90}{90}}
\put(4020,2949){\ellipse{90}{90}}
\put(1590,2949){\blacken\ellipse{90}{90}}
\put(1590,2949){\ellipse{90}{90}}
\put(1905,2949){\blacken\ellipse{90}{90}}
\put(1905,2949){\ellipse{90}{90}}
\put(1590,5019){\blacken\ellipse{90}{90}}
\put(1590,5019){\ellipse{90}{90}}
\put(1590,564){\blacken\ellipse{90}{90}}
\put(1590,564){\ellipse{90}{90}}
\put(5280,564){\blacken\ellipse{90}{90}}
\put(5280,564){\ellipse{90}{90}}
\put(4515,2949){\blacken\ellipse{90}{90}}
\put(4515,2949){\ellipse{90}{90}}
\put(5910,2949){\blacken\ellipse{90}{90}}
\put(5910,2949){\ellipse{90}{90}}
\put(5415,2949){\blacken\ellipse{90}{90}}
\put(5415,2949){\ellipse{90}{90}}
\put(4875,2949){\blacken\ellipse{90}{90}}
\put(4875,2949){\ellipse{90}{90}}
\path(330,7404)(60,7404)
\path(2850,7404)(3210,7404)
\blacken\path(525.000,7441.500)(375.000,7404.000)(525.000,7366.500)(525.000,7441.500)
\path(375,7404)(780,7404)
\blacken\path(2550.000,7441.500)(2400.000,7404.000)(2550.000,7366.500)(2550.000,7441.500)
\path(2400,7404)(2850,7404)
\path(4020,7404)(3750,7404)
\path(6540,7404)(6900,7404)
\blacken\path(4215.000,7441.500)(4065.000,7404.000)(4215.000,7366.500)(4215.000,7441.500)
\path(4065,7404)(4470,7404)
\blacken\path(6240.000,7441.500)(6090.000,7404.000)(6240.000,7366.500)(6240.000,7441.500)
\path(6090,7404)(6540,7404)
\path(4470,7404)(4830,7404)
\path(5145,7404)(5595,7404)
\blacken\path(5445.000,7366.500)(5595.000,7404.000)(5445.000,7441.500)(5445.000,7366.500)
\path(330,2949)(60,2949)
\path(2850,2949)(3210,2949)
\blacken\path(525.000,2986.500)(375.000,2949.000)(525.000,2911.500)(525.000,2986.500)
\path(375,2949)(780,2949)
\path(4020,2949)(3750,2949)
\path(6540,2949)(6900,2949)
\blacken\path(4215.000,2986.500)(4065.000,2949.000)(4215.000,2911.500)(4215.000,2986.500)
\path(4065,2949)(4470,2949)
\path(1140,2949)(1590,2949)
\blacken\path(1440.000,2911.500)(1590.000,2949.000)(1440.000,2986.500)(1440.000,2911.500)
\path(2400,2949)(2850,2949)
\blacken\path(2055.000,2986.500)(1905.000,2949.000)(2055.000,2911.500)(2055.000,2986.500)
\path(1905,2949)(2400,2949)
\blacken\path(6030.000,2979.000)(5910.000,2949.000)(6030.000,2919.000)(6030.000,2979.000)
\path(5910,2949)(6540,2949)
\path(4875,2949)(5415,2949)
\blacken\path(5295.000,2919.000)(5415.000,2949.000)(5295.000,2979.000)(5295.000,2919.000)
\path(1140,7404)(1905,7404)
\blacken\path(1755.000,7366.500)(1905.000,7404.000)(1755.000,7441.500)(1755.000,7366.500)
\thicklines
\path(780,7404)(1140,7404)
\blacken\thinlines
\path(990.000,7366.500)(1140.000,7404.000)(990.000,7441.500)(990.000,7366.500)
\thicklines
\path(2400,7404)(1590,5019)
\blacken\thinlines
\path(1602.729,5173.092)(1590.000,5019.000)(1673.745,5148.973)(1602.729,5173.092)
\thicklines
\path(5280,5019)(5145,7404)
\blacken\thinlines
\path(5190.917,7256.359)(5145.000,7404.000)(5116.037,7252.120)(5190.917,7256.359)
\blacken\path(4980.000,7441.500)(4830.000,7404.000)(4980.000,7366.500)(4980.000,7441.500)
\thicklines
\path(4830,7404)(5145,7404)
\path(4830,7404)(4830,8034)
\blacken\thinlines
\path(4867.500,7884.000)(4830.000,8034.000)(4792.500,7884.000)(4867.500,7884.000)
\thicklines
\path(4830,8034)(5595,8034)
\blacken\thinlines
\path(5445.000,7996.500)(5595.000,8034.000)(5445.000,8071.500)(5445.000,7996.500)
\thicklines
\path(5595,8034)(5595,7404)
\blacken\thinlines
\path(5557.500,7554.000)(5595.000,7404.000)(5632.500,7554.000)(5557.500,7554.000)
\thicklines
\path(5595,7404)(6090,7404)
\blacken\thinlines
\path(5940.000,7366.500)(6090.000,7404.000)(5940.000,7441.500)(5940.000,7366.500)
\thicklines
\path(6090,7404)(5280,5019)
\blacken\thinlines
\path(5292.729,5173.092)(5280.000,5019.000)(5363.745,5148.973)(5292.729,5173.092)
\thicklines
\path(1590,5019)(780,7404)
\blacken\thinlines
\path(863.745,7274.027)(780.000,7404.000)(792.729,7249.908)(863.745,7274.027)
\thicklines
\path(1140,7404)(1140,8034)
\blacken\thinlines
\path(1177.500,7884.000)(1140.000,8034.000)(1102.500,7884.000)(1177.500,7884.000)
\thicklines
\path(1140,8034)(1905,8034)
\blacken\thinlines
\path(1755.000,7996.500)(1905.000,8034.000)(1755.000,8071.500)(1755.000,7996.500)
\thicklines
\path(1905,8034)(1905,7404)
\blacken\thinlines
\path(1867.500,7554.000)(1905.000,7404.000)(1942.500,7554.000)(1867.500,7554.000)
\thicklines
\path(1905,7404)(2400,7404)
\blacken\thinlines
\path(2250.000,7366.500)(2400.000,7404.000)(2250.000,7441.500)(2250.000,7366.500)
\thicklines
\path(1590,564)(780,2949)
\blacken\thinlines
\path(863.745,2819.027)(780.000,2949.000)(792.729,2794.908)(863.745,2819.027)
\thicklines
\path(780,2949)(1140,2949)
\blacken\thinlines
\path(990.000,2911.500)(1140.000,2949.000)(990.000,2986.500)(990.000,2911.500)
\thicklines
\path(1140,2949)(1140,3579)
\blacken\thinlines
\path(1177.500,3429.000)(1140.000,3579.000)(1102.500,3429.000)(1177.500,3429.000)
\thicklines
\path(1140,3579)(1905,3579)
\blacken\thinlines
\path(1755.000,3541.500)(1905.000,3579.000)(1755.000,3616.500)(1755.000,3541.500)
\thicklines
\path(1905,3579)(1905,2949)
\blacken\thinlines
\path(1867.500,3099.000)(1905.000,2949.000)(1942.500,3099.000)(1867.500,3099.000)
\blacken\path(1740.000,2986.500)(1590.000,2949.000)(1740.000,2911.500)(1740.000,2986.500)
\thicklines
\path(1590,2949)(1905,2949)
\path(1590,2949)(1590,564)
\blacken\thinlines
\path(1552.500,714.000)(1590.000,564.000)(1627.500,714.000)(1552.500,714.000)
\thicklines
\path(5280,564)(4875,2949)
\blacken\thinlines
\path(4937.083,2807.395)(4875.000,2949.000)(4863.141,2794.839)(4937.083,2807.395)
\thicklines
\path(4875,2949)(4515,2949)
\blacken\thinlines
\path(4635.000,2979.000)(4515.000,2949.000)(4635.000,2919.000)(4635.000,2979.000)
\thicklines
\path(4515,2949)(4515,3579)
\blacken\thinlines
\path(4552.500,3429.000)(4515.000,3579.000)(4477.500,3429.000)(4552.500,3429.000)
\thicklines
\path(4515,3579)(5910,3579)
\blacken\thinlines
\path(5760.000,3541.500)(5910.000,3579.000)(5760.000,3616.500)(5760.000,3541.500)
\thicklines
\path(5910,3579)(5910,2949)
\blacken\thinlines
\path(5872.500,3099.000)(5910.000,2949.000)(5947.500,3099.000)(5872.500,3099.000)
\thicklines
\path(5910,2949)(5415,2949)
\blacken\thinlines
\path(5535.000,2979.000)(5415.000,2949.000)(5535.000,2919.000)(5535.000,2979.000)
\thicklines
\path(5415,2949)(5280,564)
\blacken\thinlines
\path(5251.037,715.880)(5280.000,564.000)(5325.917,711.641)(5251.037,715.880)
\dottedline{45}(1950,2679)(2400,2904)
\path(2306.085,2823.502)(2400.000,2904.000)(2279.252,2877.167)
\dottedline{45}(1950,2679)(1815,2904)
\path(1902.464,2816.536)(1815.000,2904.000)(1851.015,2785.666)
\dottedline{45}(4425,2679)(4740,2904)
\path(4659.789,2809.839)(4740.000,2904.000)(4624.915,2858.663)
\dottedline{45}(4425,2679)(4335,2904)
\path(4407.421,2803.725)(4335.000,2904.000)(4351.713,2781.441)
\dottedline{45}(5820,2679)(6315,2904)
\path(6218.170,2827.033)(6315.000,2904.000)(6193.342,2881.655)
\dottedline{45}(5820,2679)(5685,2904)
\path(5772.464,2816.536)(5685.000,2904.000)(5721.015,2785.666)
\path(4967.536,7271.536)(5055.000,7359.000)(4936.666,7322.985)
\dottedline{45}(5055,7359)(4605,7089)
\path(4581.408,7254.645)(4515.000,7359.000)(4524.487,7235.671)
\dottedline{45}(4515,7359)(4605,7089)
\put(2355,7539){\makebox(0,0)[lb]{\smash{{\SetFigFont{10}{12.0}{\rmdefault}{\mddefault}{\updefault}$y$}}}}
\put(600,6189){\makebox(0,0)[lb]{\smash{{\SetFigFont{10}{12.0}{\rmdefault}{\mddefault}{\updefault}$M_t(S)$}}}}
\put(2175,6189){\makebox(0,0)[lb]{\smash{{\SetFigFont{10}{12.0}{\rmdefault}{\mddefault}{\updefault}$M_h(S)$}}}}
\put(1320,8169){\makebox(0,0)[lb]{\smash{{\SetFigFont{10}{12.0}{\rmdefault}{\mddefault}{\updefault}$P_s(C)$}}}}
\put(15,7539){\makebox(0,0)[lb]{\smash{{\SetFigFont{10}{12.0}{\rmdefault}{\mddefault}{\updefault}$v$}}}}
\put(3120,7539){\makebox(0,0)[lb]{\smash{{\SetFigFont{10}{12.0}{\rmdefault}{\mddefault}{\updefault}$w$}}}}
\put(2490,7179){\makebox(0,0)[lb]{\smash{{\SetFigFont{10}{12.0}{\rmdefault}{\mddefault}{\updefault}$J_h(S)$}}}}
\put(330,7179){\makebox(0,0)[lb]{\smash{{\SetFigFont{10}{12.0}{\rmdefault}{\mddefault}{\updefault}$J_t(S)$}}}}
\put(6045,7539){\makebox(0,0)[lb]{\smash{{\SetFigFont{10}{12.0}{\rmdefault}{\mddefault}{\updefault}$y$}}}}
\put(5865,6189){\makebox(0,0)[lb]{\smash{{\SetFigFont{10}{12.0}{\rmdefault}{\mddefault}{\updefault}$M_h(S)$}}}}
\put(5010,8169){\makebox(0,0)[lb]{\smash{{\SetFigFont{10}{12.0}{\rmdefault}{\mddefault}{\updefault}$P_s(C)$}}}}
\put(3705,7539){\makebox(0,0)[lb]{\smash{{\SetFigFont{10}{12.0}{\rmdefault}{\mddefault}{\updefault}$v$}}}}
\put(6810,7539){\makebox(0,0)[lb]{\smash{{\SetFigFont{10}{12.0}{\rmdefault}{\mddefault}{\updefault}$w$}}}}
\put(5100,7539){\makebox(0,0)[lb]{\smash{{\SetFigFont{10}{12.0}{\rmdefault}{\mddefault}{\updefault}$x$}}}}
\put(4605,6189){\makebox(0,0)[lb]{\smash{{\SetFigFont{10}{12.0}{\rmdefault}{\mddefault}{\updefault}$M_t(S)$}}}}
\put(735,7539){\makebox(0,0)[lb]{\smash{{\SetFigFont{10}{12.0}{\rmdefault}{\mddefault}{\updefault}$x$}}}}
\put(960,7539){\makebox(0,0)[lb]{\smash{{\SetFigFont{10}{12.0}{\rmdefault}{\mddefault}{\updefault}$v_1$}}}}
\put(1995,7539){\makebox(0,0)[lb]{\smash{{\SetFigFont{10}{12.0}{\rmdefault}{\mddefault}{\updefault}$v_2$}}}}
\put(4650,7539){\makebox(0,0)[lb]{\smash{{\SetFigFont{10}{12.0}{\rmdefault}{\mddefault}{\updefault}$v_1$}}}}
\put(5685,7539){\makebox(0,0)[lb]{\smash{{\SetFigFont{10}{12.0}{\rmdefault}{\mddefault}{\updefault}$v_2$}}}}
\put(5235,4794){\makebox(0,0)[lb]{\smash{{\SetFigFont{10}{12.0}{\rmdefault}{\mddefault}{\updefault}$z$}}}}
\put(4650,4524){\makebox(0,0)[lb]{\smash{{\SetFigFont{10}{12.0}{\rmdefault}{\mddefault}{\updefault}(b)}}}}
\put(600,1734){\makebox(0,0)[lb]{\smash{{\SetFigFont{10}{12.0}{\rmdefault}{\mddefault}{\updefault}$M_t(S)$}}}}
\put(1320,3714){\makebox(0,0)[lb]{\smash{{\SetFigFont{10}{12.0}{\rmdefault}{\mddefault}{\updefault}$P_s(C)$}}}}
\put(3120,3084){\makebox(0,0)[lb]{\smash{{\SetFigFont{10}{12.0}{\rmdefault}{\mddefault}{\updefault}$w$}}}}
\put(5010,3714){\makebox(0,0)[lb]{\smash{{\SetFigFont{10}{12.0}{\rmdefault}{\mddefault}{\updefault}$P_s(C)$}}}}
\put(3705,3084){\makebox(0,0)[lb]{\smash{{\SetFigFont{10}{12.0}{\rmdefault}{\mddefault}{\updefault}$v$}}}}
\put(6810,3084){\makebox(0,0)[lb]{\smash{{\SetFigFont{10}{12.0}{\rmdefault}{\mddefault}{\updefault}$w$}}}}
\put(1545,3084){\makebox(0,0)[lb]{\smash{{\SetFigFont{10}{12.0}{\rmdefault}{\mddefault}{\updefault}$y$}}}}
\put(15,3084){\makebox(0,0)[lb]{\smash{{\SetFigFont{10}{12.0}{\rmdefault}{\mddefault}{\updefault}$v$}}}}
\put(735,3084){\makebox(0,0)[lb]{\smash{{\SetFigFont{10}{12.0}{\rmdefault}{\mddefault}{\updefault}$x$}}}}
\put(960,3084){\makebox(0,0)[lb]{\smash{{\SetFigFont{10}{12.0}{\rmdefault}{\mddefault}{\updefault}$v_1$}}}}
\put(1995,3084){\makebox(0,0)[lb]{\smash{{\SetFigFont{10}{12.0}{\rmdefault}{\mddefault}{\updefault}$v_2$}}}}
\put(4515,1734){\makebox(0,0)[lb]{\smash{{\SetFigFont{10}{12.0}{\rmdefault}{\mddefault}{\updefault}$M_t(S)$}}}}
\put(5415,1734){\makebox(0,0)[lb]{\smash{{\SetFigFont{10}{12.0}{\rmdefault}{\mddefault}{\updefault}$M_h(S)$}}}}
\put(1815,1734){\makebox(0,0)[lb]{\smash{{\SetFigFont{10}{12.0}{\rmdefault}{\mddefault}{\updefault}$M_h(S)$}}}}
\put(1545,4794){\makebox(0,0)[lb]{\smash{{\SetFigFont{10}{12.0}{\rmdefault}{\mddefault}{\updefault}$z$}}}}
\put(960,4524){\makebox(0,0)[lb]{\smash{{\SetFigFont{10}{12.0}{\rmdefault}{\mddefault}{\updefault}(a)}}}}
\put(5235,339){\makebox(0,0)[lb]{\smash{{\SetFigFont{10}{12.0}{\rmdefault}{\mddefault}{\updefault}$z$}}}}
\put(4650,69){\makebox(0,0)[lb]{\smash{{\SetFigFont{10}{12.0}{\rmdefault}{\mddefault}{\updefault}(d)}}}}
\put(1545,339){\makebox(0,0)[lb]{\smash{{\SetFigFont{10}{12.0}{\rmdefault}{\mddefault}{\updefault}$z$}}}}
\put(960,69){\makebox(0,0)[lb]{\smash{{\SetFigFont{10}{12.0}{\rmdefault}{\mddefault}{\updefault}(c)}}}}
\put(6180,7179){\makebox(0,0)[lb]{\smash{{\SetFigFont{10}{12.0}{\rmdefault}{\mddefault}{\updefault}$J_h(S)$}}}}
\put(4470,6954){\makebox(0,0)[lb]{\smash{{\SetFigFont{10}{12.0}{\rmdefault}{\mddefault}{\updefault}$J_t(S)$}}}}
\put(330,2724){\makebox(0,0)[lb]{\smash{{\SetFigFont{10}{12.0}{\rmdefault}{\mddefault}{\updefault}$J_t(S)$}}}}
\put(4335,3084){\makebox(0,0)[lb]{\smash{{\SetFigFont{10}{12.0}{\rmdefault}{\mddefault}{\updefault}$v_1$}}}}
\put(4830,3084){\makebox(0,0)[lb]{\smash{{\SetFigFont{10}{12.0}{\rmdefault}{\mddefault}{\updefault}$x$}}}}
\put(1815,2499){\makebox(0,0)[lb]{\smash{{\SetFigFont{10}{12.0}{\rmdefault}{\mddefault}{\updefault}$J_h(S)$}}}}
\put(6000,3084){\makebox(0,0)[lb]{\smash{{\SetFigFont{10}{12.0}{\rmdefault}{\mddefault}{\updefault}$v_2$}}}}
\put(5730,2499){\makebox(0,0)[lb]{\smash{{\SetFigFont{10}{12.0}{\rmdefault}{\mddefault}{\updefault}$J_h(S)$}}}}
\put(4335,2499){\makebox(0,0)[lb]{\smash{{\SetFigFont{10}{12.0}{\rmdefault}{\mddefault}{\updefault}$J_t(S)$}}}}
\put(5370,3084){\makebox(0,0)[lb]{\smash{{\SetFigFont{10}{12.0}{\rmdefault}{\mddefault}{\updefault}$y$}}}}
\end{picture}
}
\end{center}
\caption{Figure for Lemma \ref{lemf}.} \label{iceqpsfor}
\end{figure}

\begin{lem}
\label{lemf} Suppose that $S$ satisfies the induction hypotheses and
$P_s(C)$ is oriented $v_1 \rightarrow v_2$, then $C$ satisfies the
induction hypotheses. Moreover $|N(C)| = 0$.
\end{lem}
\begin{prf}
Since $P_s(C)$ is oriented  $v_1 \rightarrow v_2$, $I(C)=P_s(C)$. We
go over the induction hypotheses:
\begin{description}
\item [A1]
Since $S=F(C)$
\begin{equation}
\label{v1v2s} l(v_1,v_2) \leq l(P_s(C)).
\end{equation}

 By A3 there are four cases to be considered (described in
Figure \ref{iceqpsfor}):

\begin{description}
\item (a) $[v_1,v_2] \subseteq [x,y]$ ($P_f(C) \subset I(S)$).
This case is described in Figure \ref{iceqpsfor} (a).
 From A1
 \begin{eqnarray*}
 l(M(S)) &\leq& 15 l(P_c(S)) + 2l(P_s(S)) + l(I(S))\\
 &=& 15 l(P_c(S)) + 2l(v,x) + 3l(x,v_1) + 3l(v_1,v_2) +
3l(v_2,y) + 2l(y,w).\\
\end{eqnarray*}

Since $[x \rightarrow y]=I(S)$, $[x,y]$ is oriented $x \rightarrow
v_1 \rightarrow v_2 \rightarrow y$ (where $v_1 \rightarrow v_2$
refers to $P_f(C)).$  Let
\[ M_t(C) = M_t(S) \cup [x \rightarrow v_1], \ M_h(C) =M_h(S) \cup [v_2 \rightarrow y] ,\]
then:
\begin{eqnarray*}
l(M(C)) &=& l(M(S)) + l(x,v_1)+ l(v_2,y) \\
&\leq&  15 l(P_c(S)) + 2l(v,x) + 4l(x,v_1) + 3l(v_1,v_2) +
4l(v_2,y)+ 2l(y,w).\\
\end{eqnarray*}

 By definition,
\[ l(P_c(C)) = l(P_c(S)) + l(v,x) + l(x,v_1) + l(v_2,y) +
l(y,w).
\] Hence,
\[ l(M(C))  \leq  15 l(P_c(C)) + 3l(v_1,v_2)  .\]
In this case also $I(C) = P_s(C)$ and using Equation (\ref{v1v2s})
we obtain:
\[ l(M(C))  \leq  15 l(P_c(C)) + 2l(P_s(C)) + l(I(C)).  \]

\item (b) $x \in [v_1,v_2]$ and $v_2 \in [x,y]$, so the nodes are ordered
$v,v_1,x,v_2,y,w$. This case is described in Figure \ref{iceqpsfor}
(b). From A1
\begin{eqnarray*}
l(M(S)) &\leq& 15 l(P_c(S)) + 2l(P_s(S)) + l(I(S)) \\
&=& 15 l(P_c(S)) + 2l(v,v_1) + 2l(v_1,x) + 3l(x,v_2) + 3l(v_2,y) +
2l(y,w).
\end{eqnarray*}
 From A4 on $S$,  $v_1 \leftarrow x \rightarrow v_2
\rightarrow y$. Let \[ M_t(C) = M_t(S) \cup [x \rightarrow v_1], \
M_h(C) = M_h(S) \cup [v_2 \rightarrow y] ,\] then:
\begin{eqnarray*}
 l(M(C)) &=& l(M(S)) + l(x,v_1) + l(v_2,y) \\
&=&  15 l(P_c(S)) + 2l(v,v_1) + 3l(v_1,x) + 3l(x,v_2) +
4l(v_2,y) + 2l(y,w) . \\
\end{eqnarray*}
By definition,
\[ l(P_c(C)) = l(P_c(S)) + l(v,v_1)  + l(v_2,y) + l(y,w).\]
Hence,
\[ l(M(C))  \leq  15 l(P_c(C)) + 3l(v_1,x) + 3l(x,v_2)= 15 l(P_c(C)) + 3l(v_1,v_2)   \]
In this case also $I(C) = P_s(C)$ and using Equation (\ref{v1v2s})
we obtain:
\[ l(M(C))  \leq  15 l(P_c(C)) + 2l(P_s(C)) + l(I(C)).  \]

\item (c) $v_1 \in [x,y]$ and $ y \in [v_1,v_2]$. This case is described in Figure \ref{iceqpsfor} (c).
The proof is similar to the proof of previous case.

\item (d) $[x,y] \subset [v_1,v_2]$ $(I(S) \subset
P_f(C))$, so the nodes are ordered $v,v_1,x,y,v_2,w$. This case is
described in Figure \ref{iceqpsfor} (d). From A1
\begin{eqnarray*}
l(M(S)) &\leq& 15 l(P_c(S)) + 2l(P_s(S)) + l(I(S)) \\
&=&  15 l(P_c(S)) + 2l(v,v_1) + 2l(v_1,x) + 3l(x,y) + 2l(y,v_2) +
2l(v_2,w).
\end{eqnarray*}
>From  A4 $v_1 \leftarrow x \rightarrow y \leftarrow v_2$. Let
\[ M_t(C) = M_t(S) \cup [x \rightarrow v_1], \  M_h(C) = M_h(S) \cup
[v_2 \rightarrow y] ,\] then:
\begin{eqnarray*}
l(M(C_1)) &=& l(M(S)) + 2l(x,v_1) + l(v_2,y) \\
&\leq&   15 l(P_c(S)) + 2l(v,v_1) + 3l(v_1,x) + 3l(x,y) + 3l(y,v_2)
+ 2l(v_2,w).
\end{eqnarray*}

By definition
\[ l(P_c(C)) = l(P_c(S)) + l(v,v_1) + l(v_2,w) ).\] Hence,
\[ l(M(C))  \leq  15 l(P_c(C)) + 3l(v_1,x) +3l(x,y) + 3l(y,v_2)= 15 l(P_c(C)) + 3l(v_1,v_2)  . \]
By Equation (\ref{v1v2s}):
\[ l(M(C))  \leq  15 l(P_c(C)) + 2l(P_s(C)) + l(I(C)).  \]
 \end{description}
 \item [A2] In this case $I(C) = P_s(C)$.
\item [A3] Since $I(C) = P_s(C)$  and for every son $\hat{C}$ of $C$
$P_f(\hat{C}) \cap P_s(C) \neq \phi$ , $P_f(\hat{C}) \cap I(C) \neq
\phi$.
\item [A4] If $\hat{C}$ is a son of $C$, $P_f(\hat{C}) \subseteq P_s(C)=I(C)$.
\item [A5] $M(C)$ is contained in $P_s(C)$ and $P_s(S) \cup M(S)$ which are shorter then $C$.
\item [A6] $M(C) \subseteq P_s(C) \cup P_s(S) \cup M(S)$, so  $N(C)=\phi$.
\end{description}

\end{prf}

\begin{figure}
\begin{center}
\setlength{\unitlength}{0.00069991in}
\begingroup\makeatletter\ifx\SetFigFont\undefined%
\gdef\SetFigFont#1#2#3#4#5{%
  \reset@font\fontsize{#1}{#2pt}%
  \fontfamily{#3}\fontseries{#4}\fontshape{#5}%
  \selectfont}%
\fi\endgroup%
{\renewcommand{\dashlinestretch}{30}
\begin{picture}(7518,8343)(0,-10)
\put(1905,7404){\blacken\ellipse{90}{90}}
\put(1905,7404){\ellipse{90}{90}}
\put(2850,7404){\blacken\ellipse{90}{90}}
\put(2850,7404){\ellipse{90}{90}}
\put(780,7404){\blacken\ellipse{90}{90}}
\put(780,7404){\ellipse{90}{90}}
\put(2400,7404){\blacken\ellipse{90}{90}}
\put(2400,7404){\ellipse{90}{90}}
\put(1140,7404){\blacken\ellipse{90}{90}}
\put(1140,7404){\ellipse{90}{90}}
\put(60,7404){\blacken\ellipse{90}{90}}
\put(60,7404){\ellipse{90}{90}}
\put(3165,7404){\blacken\ellipse{90}{90}}
\put(3165,7404){\ellipse{90}{90}}
\put(330,7404){\blacken\ellipse{90}{90}}
\put(330,7404){\ellipse{90}{90}}
\put(5595,7404){\blacken\ellipse{90}{90}}
\put(5595,7404){\ellipse{90}{90}}
\put(6540,7404){\blacken\ellipse{90}{90}}
\put(6540,7404){\ellipse{90}{90}}
\put(6090,7404){\blacken\ellipse{90}{90}}
\put(6090,7404){\ellipse{90}{90}}
\put(3750,7404){\blacken\ellipse{90}{90}}
\put(3750,7404){\ellipse{90}{90}}
\put(6855,7404){\blacken\ellipse{90}{90}}
\put(6855,7404){\ellipse{90}{90}}
\put(4020,7404){\blacken\ellipse{90}{90}}
\put(4020,7404){\ellipse{90}{90}}
\put(2850,2949){\blacken\ellipse{90}{90}}
\put(2850,2949){\ellipse{90}{90}}
\put(780,2949){\blacken\ellipse{90}{90}}
\put(780,2949){\ellipse{90}{90}}
\put(1140,2949){\blacken\ellipse{90}{90}}
\put(1140,2949){\ellipse{90}{90}}
\put(60,2949){\blacken\ellipse{90}{90}}
\put(60,2949){\ellipse{90}{90}}
\put(3165,2949){\blacken\ellipse{90}{90}}
\put(3165,2949){\ellipse{90}{90}}
\put(330,2949){\blacken\ellipse{90}{90}}
\put(330,2949){\ellipse{90}{90}}
\put(6540,2949){\blacken\ellipse{90}{90}}
\put(6540,2949){\ellipse{90}{90}}
\put(3750,2949){\blacken\ellipse{90}{90}}
\put(3750,2949){\ellipse{90}{90}}
\put(6855,2949){\blacken\ellipse{90}{90}}
\put(6855,2949){\ellipse{90}{90}}
\put(1590,2949){\blacken\ellipse{90}{90}}
\put(1590,2949){\ellipse{90}{90}}
\put(2220,2949){\blacken\ellipse{90}{90}}
\put(2220,2949){\ellipse{90}{90}}
\put(1590,5019){\blacken\ellipse{90}{90}}
\put(1590,5019){\ellipse{90}{90}}
\put(4515,2949){\blacken\ellipse{90}{90}}
\put(4515,2949){\ellipse{90}{90}}
\put(5910,2949){\blacken\ellipse{90}{90}}
\put(5910,2949){\ellipse{90}{90}}
\put(4515,7404){\blacken\ellipse{90}{90}}
\put(4515,7404){\ellipse{90}{90}}
\put(5145,7404){\blacken\ellipse{90}{90}}
\put(5145,7404){\ellipse{90}{90}}
\put(5280,5019){\blacken\ellipse{90}{90}}
\put(5280,5019){\ellipse{90}{90}}
\put(1590,564){\blacken\ellipse{90}{90}}
\put(1590,564){\ellipse{90}{90}}
\put(5415,2949){\blacken\ellipse{90}{90}}
\put(5415,2949){\ellipse{90}{90}}
\put(4020,2949){\blacken\ellipse{90}{90}}
\put(4020,2949){\ellipse{90}{90}}
\put(3525,5964){\blacken\ellipse{90}{90}}
\put(3525,5964){\ellipse{90}{90}}
\put(3210,1734){\blacken\ellipse{90}{90}}
\put(3210,1734){\ellipse{90}{90}}
\put(7305,1464){\blacken\ellipse{90}{90}}
\put(7305,1464){\ellipse{90}{90}}
\put(5145,564){\blacken\ellipse{90}{90}}
\put(5145,564){\ellipse{90}{90}}
\put(3840,1329){\blacken\ellipse{90}{90}}
\put(3840,1329){\ellipse{90}{90}}
\put(4875,2949){\blacken\ellipse{90}{90}}
\put(4875,2949){\ellipse{90}{90}}
\path(330,7404)(60,7404)
\path(2850,7404)(3210,7404)
\blacken\path(525.000,7441.500)(375.000,7404.000)(525.000,7366.500)(525.000,7441.500)
\path(375,7404)(780,7404)
\blacken\path(2550.000,7441.500)(2400.000,7404.000)(2550.000,7366.500)(2550.000,7441.500)
\path(2400,7404)(2850,7404)
\path(5595,7404)(6090,7404)
\blacken\path(5940.000,7366.500)(6090.000,7404.000)(5940.000,7441.500)(5940.000,7366.500)
\path(4020,7404)(3750,7404)
\path(6540,7404)(6900,7404)
\blacken\path(6240.000,7441.500)(6090.000,7404.000)(6240.000,7366.500)(6240.000,7441.500)
\path(6090,7404)(6540,7404)
\path(780,2949)(1140,2949)
\blacken\path(990.000,2911.500)(1140.000,2949.000)(990.000,2986.500)(990.000,2911.500)
\path(330,2949)(60,2949)
\path(2850,2949)(3210,2949)
\blacken\path(525.000,2986.500)(375.000,2949.000)(525.000,2911.500)(525.000,2986.500)
\path(375,2949)(780,2949)
\path(4020,2949)(3750,2949)
\path(6540,2949)(6900,2949)
\path(4875,2949)(5415,2949)
\blacken\path(5295.000,2919.000)(5415.000,2949.000)(5295.000,2979.000)(5295.000,2919.000)
\drawline(5775,6909)(5775,6909)
\path(5910,2949)(5415,2949)
\blacken\path(5535.000,2979.000)(5415.000,2949.000)(5535.000,2919.000)(5535.000,2979.000)
\path(5145,7404)(4515,7404)
\blacken\path(4665.000,7441.500)(4515.000,7404.000)(4665.000,7366.500)(4665.000,7441.500)
\dottedline{45}(4245,6594)(3795,6504)
\path(3906.786,6556.951)(3795.000,6504.000)(3918.553,6498.117)
\dottedline{45}(4245,6594)(3930,5829)
\path(3947.950,5951.384)(3930.000,5829.000)(4003.430,5928.539)
\path(2220,2949)(1590,2949)
\blacken\path(1740.000,2986.500)(1590.000,2949.000)(1740.000,2911.500)(1740.000,2986.500)
\dottedline{45}(2355,2679)(2535,2949)
\path(2493.397,2832.513)(2535.000,2949.000)(2443.474,2865.795)
\dottedline{45}(2355,2679)(1905,2904)
\path(2025.748,2877.167)(1905.000,2904.000)(1998.915,2823.502)
\dottedline{45}(2040,1914)(3030,2094)
\path(2917.302,2043.018)(3030.000,2094.000)(2906.569,2102.050)
\dottedline{45}(2040,1914)(2355,2679)
\path(2337.050,2556.616)(2355.000,2679.000)(2281.570,2579.461)
\dottedline{45}(2040,1914)(2445,1284)
\path(2354.874,1368.719)(2445.000,1284.000)(2405.344,1401.164)
\dottedline{45}(6090,2544)(5685,2904)
\path(5794.620,2846.699)(5685.000,2904.000)(5754.758,2801.854)
\dottedline{45}(6090,2544)(6315,2904)
\path(6276.840,2786.340)(6315.000,2904.000)(6225.960,2818.140)
\dottedline{45}(5865,2094)(6090,2544)
\path(6063.167,2423.252)(6090.000,2544.000)(6009.502,2450.085)
\dottedline{45}(5865,2094)(7080,1824)
\path(6956.350,1820.746)(7080.000,1824.000)(6969.365,1879.317)
\dottedline{45}(5865,2094)(6180,1059)
\path(6116.360,1165.066)(6180.000,1059.000)(6173.761,1182.536)
\path(4875,2949)(4515,2949)
\blacken\path(4635.000,2979.000)(4515.000,2949.000)(4635.000,2919.000)(4635.000,2979.000)
\dottedline{45}(4650,2589)(4740,2904)
\path(4735.879,2780.375)(4740.000,2904.000)(4678.188,2796.859)
\dottedline{45}(4650,2589)(4380,2904)
\path(4480.873,2832.413)(4380.000,2904.000)(4435.317,2793.365)
\dottedline{45}(4290,1869)(4650,2589)
\path(4623.167,2468.252)(4650.000,2589.000)(4569.502,2495.085)
\dottedline{45}(4290,1869)(4020,2094)
\path(4131.392,2040.225)(4020.000,2094.000)(4092.981,1994.131)
\dottedline{45}(4290,1869)(4155,1194)
\path(4149.117,1317.553)(4155.000,1194.000)(4207.951,1305.786)
\dottedline{45}(4245,6594)(4515,7134)
\path(4488.167,7013.252)(4515.000,7134.000)(4434.502,7040.085)
\dottedline{45}(4515,7134)(4335,7359)
\path(4433.389,7284.037)(4335.000,7359.000)(4386.537,7246.555)
\dottedline{45}(4515,7134)(4965,7404)
\path(4877.536,7316.536)(4965.000,7404.000)(4846.666,7367.985)
\thicklines
\path(1590,5019)(780,7404)
\blacken\thinlines
\path(863.745,7274.027)(780.000,7404.000)(792.729,7249.908)(863.745,7274.027)
\thicklines
\path(780,7404)(1140,7404)
\blacken\thinlines
\path(990.000,7366.500)(1140.000,7404.000)(990.000,7441.500)(990.000,7366.500)
\thicklines
\path(1140,7404)(1905,7404)
\blacken\thinlines
\path(1755.000,7366.500)(1905.000,7404.000)(1755.000,7441.500)(1755.000,7366.500)
\blacken\path(1935.000,7914.000)(1905.000,8034.000)(1875.000,7914.000)(1935.000,7914.000)
\thicklines
\path(1905,8034)(1905,7404)
\blacken\thinlines
\path(1260.000,8064.000)(1140.000,8034.000)(1260.000,8004.000)(1260.000,8064.000)
\thicklines
\path(1140,8034)(1905,8034)
\blacken\thinlines
\path(1110.000,7524.000)(1140.000,7404.000)(1170.000,7524.000)(1110.000,7524.000)
\thicklines
\path(1140,7404)(1140,8034)
\path(1905,7404)(2400,7404)
\blacken\thinlines
\path(2250.000,7366.500)(2400.000,7404.000)(2250.000,7441.500)(2250.000,7366.500)
\thicklines
\path(2400,7404)(1590,5019)
\blacken\thinlines
\path(1602.729,5173.092)(1590.000,5019.000)(1673.745,5148.973)(1602.729,5173.092)
\thicklines
\path(5280,5019)(5145,7404)
\blacken\thinlines
\path(5190.917,7256.359)(5145.000,7404.000)(5116.037,7252.120)(5190.917,7256.359)
\thicklines
\path(5145,7404)(5595,7404)
\blacken\thinlines
\path(5445.000,7366.500)(5595.000,7404.000)(5445.000,7441.500)(5445.000,7366.500)
\blacken\path(5625.000,7914.000)(5595.000,8034.000)(5565.000,7914.000)(5625.000,7914.000)
\thicklines
\path(5595,8034)(5595,7404)
\blacken\thinlines
\path(4635.000,8064.000)(4515.000,8034.000)(4635.000,8004.000)(4635.000,8064.000)
\thicklines
\path(4515,8034)(5595,8034)
\blacken\thinlines
\path(4485.000,7524.000)(4515.000,7404.000)(4545.000,7524.000)(4485.000,7524.000)
\thicklines
\path(4515,7404)(4515,8034)
\blacken\thinlines
\path(4215.000,7441.500)(4065.000,7404.000)(4215.000,7366.500)(4215.000,7441.500)
\thicklines
\path(4065,7404)(4470,7404)
\path(4020,7359)(3525,5964)
\blacken\thinlines
\path(3539.820,6117.905)(3525.000,5964.000)(3610.503,6092.824)(3539.820,6117.905)
\thicklines
\path(3525,5964)(5280,5019)
\blacken\thinlines
\path(5130.151,5057.097)(5280.000,5019.000)(5165.708,5123.133)(5130.151,5057.097)
\blacken\path(3110.354,1615.776)(3210.000,1734.000)(3066.442,1676.577)(3110.354,1615.776)
\thicklines
\path(3210,1734)(1590,564)
\blacken\thinlines
\path(2970.207,2768.907)(2895.000,2904.000)(2897.785,2749.409)(2970.207,2768.907)
\thicklines
\path(2895,2904)(3210,1734)
\path(2850,2949)(2220,2949)
\blacken\thinlines
\path(2370.000,2986.500)(2220.000,2949.000)(2370.000,2911.500)(2370.000,2986.500)
\blacken\path(2250.000,3459.000)(2220.000,3579.000)(2190.000,3459.000)(2250.000,3459.000)
\thicklines
\path(2220,3579)(2220,2949)
\blacken\thinlines
\path(1260.000,3609.000)(1140.000,3579.000)(1260.000,3549.000)(1260.000,3609.000)
\thicklines
\path(1140,3579)(2220,3579)
\blacken\thinlines
\path(1110.000,3069.000)(1140.000,2949.000)(1170.000,3069.000)(1110.000,3069.000)
\thicklines
\path(1140,2949)(1140,3579)
\path(1140,2949)(1590,2949)
\blacken\thinlines
\path(1440.000,2911.500)(1590.000,2949.000)(1440.000,2986.500)(1440.000,2911.500)
\thicklines
\path(1590,2949)(1590,564)
\blacken\thinlines
\path(1552.500,714.000)(1590.000,564.000)(1627.500,714.000)(1552.500,714.000)
\thicklines
\path(5145,564)(7305,1464)
\blacken\thinlines
\path(7180.962,1371.692)(7305.000,1464.000)(7152.115,1440.923)(7180.962,1371.692)
\thicklines
\path(7305,1464)(6540,2949)
\blacken\thinlines
\path(6642.030,2832.827)(6540.000,2949.000)(6575.357,2798.480)(6642.030,2832.827)
\blacken\path(6030.000,2979.000)(5910.000,2949.000)(6030.000,2919.000)(6030.000,2979.000)
\thicklines
\path(5910,2949)(6540,2949)
\blacken\thinlines
\path(5947.500,3429.000)(5910.000,3579.000)(5872.500,3429.000)(5947.500,3429.000)
\thicklines
\path(5910,3579)(5910,2949)
\blacken\thinlines
\path(4665.000,3616.500)(4515.000,3579.000)(4665.000,3541.500)(4665.000,3616.500)
\thicklines
\path(4515,3579)(5910,3579)
\blacken\thinlines
\path(4485.000,3069.000)(4515.000,2949.000)(4545.000,3069.000)(4485.000,3069.000)
\thicklines
\path(4515,2949)(4515,3579)
\blacken\thinlines
\path(4215.000,2986.500)(4065.000,2949.000)(4215.000,2911.500)(4215.000,2986.500)
\thicklines
\path(4065,2949)(4470,2949)
\path(4020,2904)(3840,1329)
\blacken\thinlines
\path(3819.775,1482.288)(3840.000,1329.000)(3894.289,1473.772)(3819.775,1482.288)
\thicklines
\path(3840,1329)(5145,564)
\blacken\thinlines
\path(4996.631,607.507)(5145.000,564.000)(5034.560,672.209)(4996.631,607.507)
\put(2355,7539){\makebox(0,0)[lb]{\smash{{\SetFigFont{10}{12.0}{\rmdefault}{\mddefault}{\updefault}$y$}}}}
\put(600,6189){\makebox(0,0)[lb]{\smash{{\SetFigFont{10}{12.0}{\rmdefault}{\mddefault}{\updefault}$M_t(S)$}}}}
\put(2175,6189){\makebox(0,0)[lb]{\smash{{\SetFigFont{10}{12.0}{\rmdefault}{\mddefault}{\updefault}$M_h(S)$}}}}
\put(1320,8169){\makebox(0,0)[lb]{\smash{{\SetFigFont{10}{12.0}{\rmdefault}{\mddefault}{\updefault}$P_s(C)$}}}}
\put(15,7539){\makebox(0,0)[lb]{\smash{{\SetFigFont{10}{12.0}{\rmdefault}{\mddefault}{\updefault}$v$}}}}
\put(3120,7539){\makebox(0,0)[lb]{\smash{{\SetFigFont{10}{12.0}{\rmdefault}{\mddefault}{\updefault}$w$}}}}
\put(2490,7179){\makebox(0,0)[lb]{\smash{{\SetFigFont{10}{12.0}{\rmdefault}{\mddefault}{\updefault}$J_h(S)$}}}}
\put(330,7179){\makebox(0,0)[lb]{\smash{{\SetFigFont{10}{12.0}{\rmdefault}{\mddefault}{\updefault}$J_t(S)$}}}}
\put(6045,7539){\makebox(0,0)[lb]{\smash{{\SetFigFont{10}{12.0}{\rmdefault}{\mddefault}{\updefault}$y$}}}}
\put(3705,7539){\makebox(0,0)[lb]{\smash{{\SetFigFont{10}{12.0}{\rmdefault}{\mddefault}{\updefault}$v$}}}}
\put(6810,7539){\makebox(0,0)[lb]{\smash{{\SetFigFont{10}{12.0}{\rmdefault}{\mddefault}{\updefault}$w$}}}}
\put(735,7539){\makebox(0,0)[lb]{\smash{{\SetFigFont{10}{12.0}{\rmdefault}{\mddefault}{\updefault}$x$}}}}
\put(960,7539){\makebox(0,0)[lb]{\smash{{\SetFigFont{10}{12.0}{\rmdefault}{\mddefault}{\updefault}$v_1$}}}}
\put(1995,7539){\makebox(0,0)[lb]{\smash{{\SetFigFont{10}{12.0}{\rmdefault}{\mddefault}{\updefault}$v_2$}}}}
\put(5685,7539){\makebox(0,0)[lb]{\smash{{\SetFigFont{10}{12.0}{\rmdefault}{\mddefault}{\updefault}$v_2$}}}}
\put(5235,4794){\makebox(0,0)[lb]{\smash{{\SetFigFont{10}{12.0}{\rmdefault}{\mddefault}{\updefault}$z$}}}}
\put(4650,4524){\makebox(0,0)[lb]{\smash{{\SetFigFont{10}{12.0}{\rmdefault}{\mddefault}{\updefault}(b)}}}}
\put(3120,3084){\makebox(0,0)[lb]{\smash{{\SetFigFont{10}{12.0}{\rmdefault}{\mddefault}{\updefault}$w$}}}}
\put(5010,3714){\makebox(0,0)[lb]{\smash{{\SetFigFont{10}{12.0}{\rmdefault}{\mddefault}{\updefault}$P_s(C)$}}}}
\put(3705,3084){\makebox(0,0)[lb]{\smash{{\SetFigFont{10}{12.0}{\rmdefault}{\mddefault}{\updefault}$v$}}}}
\put(6810,3084){\makebox(0,0)[lb]{\smash{{\SetFigFont{10}{12.0}{\rmdefault}{\mddefault}{\updefault}$w$}}}}
\put(1545,3084){\makebox(0,0)[lb]{\smash{{\SetFigFont{10}{12.0}{\rmdefault}{\mddefault}{\updefault}$y$}}}}
\put(15,3084){\makebox(0,0)[lb]{\smash{{\SetFigFont{10}{12.0}{\rmdefault}{\mddefault}{\updefault}$v$}}}}
\put(735,3084){\makebox(0,0)[lb]{\smash{{\SetFigFont{10}{12.0}{\rmdefault}{\mddefault}{\updefault}$x$}}}}
\put(960,3084){\makebox(0,0)[lb]{\smash{{\SetFigFont{10}{12.0}{\rmdefault}{\mddefault}{\updefault}$v_1$}}}}
\put(1545,4794){\makebox(0,0)[lb]{\smash{{\SetFigFont{10}{12.0}{\rmdefault}{\mddefault}{\updefault}$z$}}}}
\put(960,4524){\makebox(0,0)[lb]{\smash{{\SetFigFont{10}{12.0}{\rmdefault}{\mddefault}{\updefault}(a)}}}}
\put(4650,69){\makebox(0,0)[lb]{\smash{{\SetFigFont{10}{12.0}{\rmdefault}{\mddefault}{\updefault}(d)}}}}
\put(1545,339){\makebox(0,0)[lb]{\smash{{\SetFigFont{10}{12.0}{\rmdefault}{\mddefault}{\updefault}$z$}}}}
\put(960,69){\makebox(0,0)[lb]{\smash{{\SetFigFont{10}{12.0}{\rmdefault}{\mddefault}{\updefault}(c)}}}}
\put(4335,3084){\makebox(0,0)[lb]{\smash{{\SetFigFont{10}{12.0}{\rmdefault}{\mddefault}{\updefault}$v_1$}}}}
\put(4830,3084){\makebox(0,0)[lb]{\smash{{\SetFigFont{10}{12.0}{\rmdefault}{\mddefault}{\updefault}$x$}}}}
\put(6000,3084){\makebox(0,0)[lb]{\smash{{\SetFigFont{10}{12.0}{\rmdefault}{\mddefault}{\updefault}$v_2$}}}}
\put(5370,3084){\makebox(0,0)[lb]{\smash{{\SetFigFont{10}{12.0}{\rmdefault}{\mddefault}{\updefault}$y$}}}}
\put(4335,7539){\makebox(0,0)[lb]{\smash{{\SetFigFont{10}{12.0}{\rmdefault}{\mddefault}{\updefault}$v_1$}}}}
\put(330,2724){\makebox(0,0)[lb]{\smash{{\SetFigFont{10}{12.0}{\rmdefault}{\mddefault}{\updefault}$J_t(S)$}}}}
\put(6180,7179){\makebox(0,0)[lb]{\smash{{\SetFigFont{10}{12.0}{\rmdefault}{\mddefault}{\updefault}$J_h(S)$}}}}
\put(5100,7494){\makebox(0,0)[lb]{\smash{{\SetFigFont{10}{12.0}{\rmdefault}{\mddefault}{\updefault}$x$}}}}
\put(2310,3084){\makebox(0,0)[lb]{\smash{{\SetFigFont{10}{12.0}{\rmdefault}{\mddefault}{\updefault}$v_2$}}}}
\put(1545,3714){\makebox(0,0)[lb]{\smash{{\SetFigFont{10}{12.0}{\rmdefault}{\mddefault}{\updefault}$P_s(C)$}}}}
\put(4875,8169){\makebox(0,0)[lb]{\smash{{\SetFigFont{10}{12.0}{\rmdefault}{\mddefault}{\updefault}$P_s(C)$}}}}
\put(5100,339){\makebox(0,0)[lb]{\smash{{\SetFigFont{10}{12.0}{\rmdefault}{\mddefault}{\updefault}$z$}}}}
\put(5280,6549){\makebox(0,0)[lb]{\smash{{\SetFigFont{10}{12.0}{\rmdefault}{\mddefault}{\updefault}$M_t(S)$}}}}
\put(4245,6369){\makebox(0,0)[lb]{\smash{{\SetFigFont{10}{12.0}{\rmdefault}{\mddefault}{\updefault}$B_t(S)$}}}}
\put(1635,1959){\makebox(0,0)[lb]{\smash{{\SetFigFont{10}{12.0}{\rmdefault}{\mddefault}{\updefault}$B_h(S)$}}}}
\put(2400,2544){\makebox(0,0)[lb]{\smash{{\SetFigFont{10}{12.0}{\rmdefault}{\mddefault}{\updefault}$J_h(S)$}}}}
\put(6135,2364){\makebox(0,0)[lb]{\smash{{\SetFigFont{10}{12.0}{\rmdefault}{\mddefault}{\updefault}$J_h(S)$}}}}
\put(5325,2004){\makebox(0,0)[lb]{\smash{{\SetFigFont{10}{12.0}{\rmdefault}{\mddefault}{\updefault}$B_h(S)$}}}}
\put(4695,2454){\makebox(0,0)[lb]{\smash{{\SetFigFont{10}{12.0}{\rmdefault}{\mddefault}{\updefault}$J_t(S)$}}}}
\put(4425,1734){\makebox(0,0)[lb]{\smash{{\SetFigFont{10}{12.0}{\rmdefault}{\mddefault}{\updefault}$B_t(S)$}}}}
\put(1005,2499){\makebox(0,0)[lb]{\smash{{\SetFigFont{10}{12.0}{\rmdefault}{\mddefault}{\updefault}$M_h(S)$}}}}
\put(4515,6909){\makebox(0,0)[lb]{\smash{{\SetFigFont{10}{12.0}{\rmdefault}{\mddefault}{\updefault}$J_t(S)$}}}}
\end{picture}
}
\end{center}
\caption{Figure for Lemma \ref{lemb}.} \label{iceqpsbac}
\end{figure}
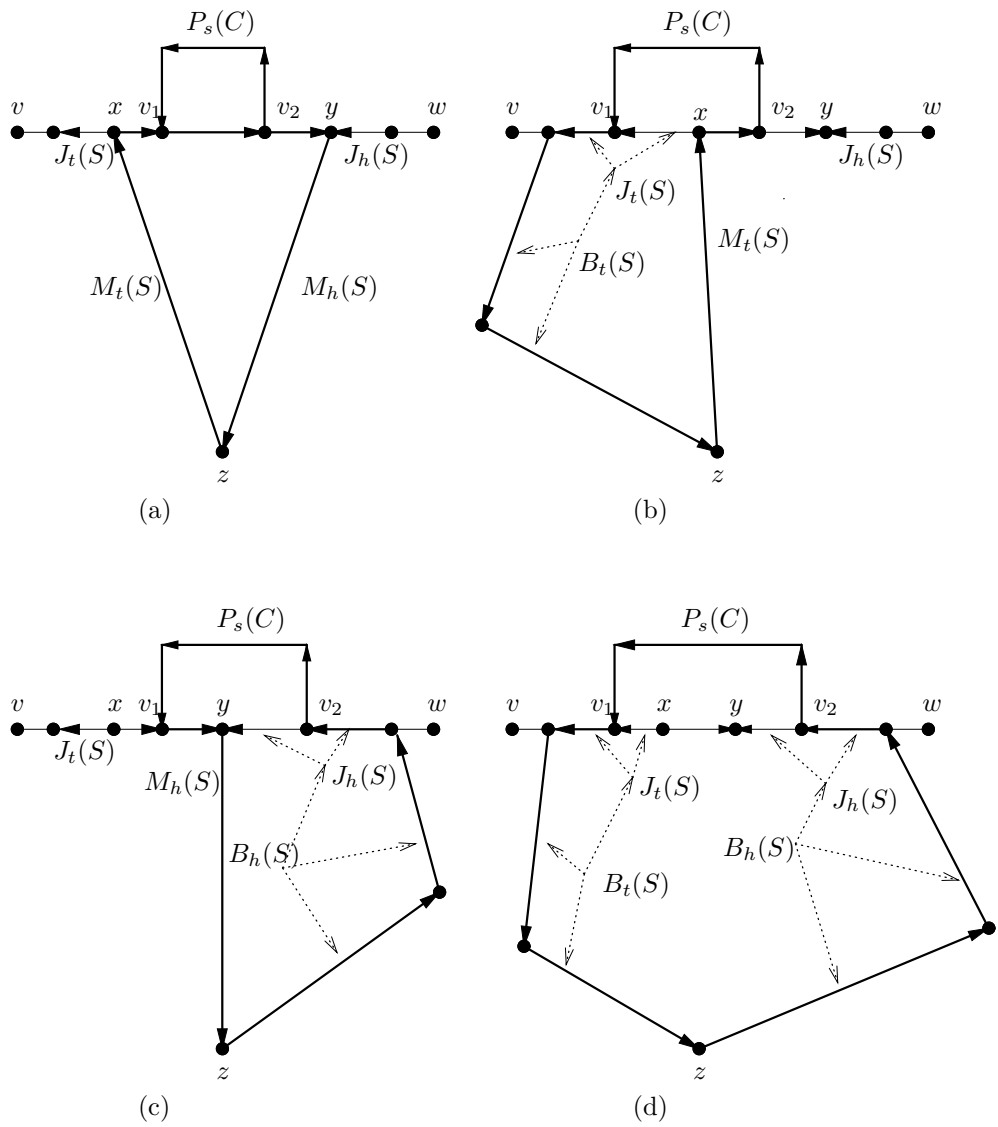

\begin{lem}
\label{lemb} Suppose that $S$ satisfies the induction hypotheses and
$P_s(C)$ is oriented $v_1 \leftarrow v_2$  then $C$ satisfies the
induction hypotheses. Moreover $|N(C)| = 0$.
\end{lem}
\begin{prf}
Since $P_s(C)$ is oriented in a consistent manner, $I(C)=P_s(C)$. We
go over the induction hypotheses:

\begin{description}
\item [A1]  Since $C$ is oriented backwards,
according to Property \ref{longfwd} $C$ is light, giving:
\begin{equation}
\label{goodfp} l(v_1,v_2) \leq 0.5 [l(v,v_1) + l(v_2,w)].
\end{equation}

 By A3 there are four cases to be considered (described in Figure \ref{iceqpsbac}):
\begin{description}
\item (a) $[v_1,v_2] \subset [x,y]$ ($P_f(C_2) \subset I(S)$). This case is described in Figure \ref{iceqpsbac}(a).
>From A1
\begin{eqnarray*}
 l(M(S)) &\leq& 15 l(P_c(S)) + 2l(P_s(S)) + l(I(S)) \\
&=&  15 l(P_c(S)) + 2l(v,x) + 3l(x,v_1) + 3l(v_1,v_2) + 3l(v_2,y) +
2l(y,w).
\end{eqnarray*}
Since $[x \rightarrow y]=I(S)$, $[x,y]$ is oriented $x \rightarrow
v_1 \rightarrow v_2 \rightarrow y$. Let
 \[ M_t(C) =M_t(S) \cup [x \rightarrow v_2], \ M_h(C) =[v_1,y] \cup M_h(S). \]
>From (\ref{goodfp}):
\begin{eqnarray*}
l(M(C)) &=& l(M(S)) + l(x,v_1)+ 2l(v_1,v_2) +  l(v_2,y)\\
& \leq&  15 l(P_c(S)) + 2l(v,x) + 4l(x,v_1) + 5l(v_1,v_2) +
4l(v_2,y) + 2l(y,w) \\
&\leq&  15 l(P_c(S)) + 4.5l(v,x) + 6.5l(x,v_1) + 6.5l(v_2,y) +
4.5l(y,w).
\\
\end{eqnarray*}
By definition
\[ l(P_c(C)) = l(P_c(S)) + l(v,x) + l(x,v_1) + l(v_2,y) + l(y,w),
\] giving
\[ l(M(C))  \leq  15 l(P_c(C)) \leq   l(P_c(C)) + 2l(P_s(C)) + l(I(C)) .\]

 \item (b) $x \in [v_1,v_2]$ and $ v_2 \in [x,y]$. This case is described in Figure \ref{iceqpsbac}(b).
 Since $x \in [v_1,v_2]$ obviously $x \not\in F(S)$, and therefore
by A2:
\[ l(M_t(S)) + l(B_t(S))  \leq 15 l(P_c(S)) + 4l(P_s(S)) .\]
>From  A4 we get that $v_1 \leftarrow x \rightarrow v_2 \rightarrow
y$ . Let
\[ M_t(C)=M_t(S) \cup [x \rightarrow v_2], \ M_h(C) =  B_t(S) \backslash [x \rightarrow v_1] .
\]
Then:
\begin{eqnarray*}
l(M(C)) &\leq& l(M_t(S))+l(x,v_2)  + l(B_t(S)) \\
&\leq& 15l(P_c(S)) + 4l(P_s(S)) + l(x,v_2) \\
 &\leq& 15l(P_c(S)) + 4l(v,v_1) + 4l(v_1,x) + 5l(x,v_2)
+ 4l(v_2,y) + 4l(y,w)  \\
& \leq&   15l(P_c(S)) + 6.5l(v,v_1) + 6.5l(v_2,y) + 6.5l(y,w)  ,\\
\end{eqnarray*}
where the last inequality follows from (\ref{goodfp}). By definition
\[ l(P_c(C)) = l(P_c(S)) + l(v,v_1)  + l(v_2,y) +
l(y,w),\] giving:
\[ l(M(C))  \leq  15 l(P_c(C))  \leq  15 l(P_c(C)) + 2l(P_s(C)) + l(I(C)).    \]

\item (c) $v_1 \in [x,y]$ and $ y \in [v_1,v_2]$. This case is described in Figure \ref{iceqpsbac}(c).
The proof is similar to the proof of previous case.

\item (d) $[x,y] \subset [v_1,v_2]$ $(I(S) \subset
P_f(C_1))$. This case is described in Figure \ref{iceqpsbac}(d).
 From  A2 (in this case $x,y \not\in F(S)$)
\[ l(B_h(S)) + l(B_t(S))  \leq 15 l(P_c(S)) + 6l(P_s(S)) .\]
>From A4 we get that $v_1 \leftarrow x \rightarrow y \leftarrow v_2$.
Let
\[ M_t(C) = B_h(S) \backslash [v_2,y],\]
 and
\[ M_h(C) =  B_t(S) \backslash [x,v_1].\] Then:

\begin{eqnarray*}
 l(M(C)) &\leq& l(B_h(S)) + l(B_t(S))\\
  &\leq& 15 l(P_c(S)) +6l(P_s(S))\\
  &\leq&   15l(P_c(S)) + 6l(v,v_1) + 6l(v_1,x) + 6l(x,y) +
6l(y,v_2) + 6l(v_2,w)\\
& \leq&   15l(P_c(S)) + 9l(v,v_1) +2l(v_1,x) + 2l(x,y) + 2l(y,v_2)  + 9l(v_2,w),\\
\end{eqnarray*}
where the last inequality follows from (\ref{goodfp}). By definition
\[ l(P_c(C)) = l(P_c(S)) + l(v,v_1)+ l(v_2,w),\] giving:
\[ l(M(C) \leq 15l(P_c(C)) + 2l(v_1,v_2) .\]
By Equation (\ref{v1v2s}):
 \[l(M(C)) \leq  15 l(P_c(C)) + 2l(P_s(C)) + l(I(C)).\]

 \end{description}

\item [A2] In this case $I(C) = P_s(C)$.
\item [A3] Since $I(C) = P_s(C)$  and for every son $\hat{C}$ of $C$
$P_f(\hat{C}) \cap P_s(C) \neq \phi$ , $P_f(\hat{C}) \cap I(C) \neq
\phi$.
\item [A4] If $\hat{C}$ is a son of $C$, $P_f(\hat{C}) \subseteq
P_s(C)=I(C)$.
\item [A5] $M(C)$ is contained in $P_s(C)$ and $P_s(S) \cup M(S)$ which are shorter then $C$.
\item [A6] $M(C) \subseteq P_s(C) \cup P_s(S) \cup M(S)$, so
$N(C)=\phi$.
\end{description}

\end{prf}

\begin{cor}
\label{dirbackfor}
 The cycles directed by DIRECT-FORWARDS and
DIRECT-BACKWARDS satisfy the induction hypotheses.
\end{cor}

\subsection{Two  brothers} \label{sectwo} In this section
we consider two neighbor brothers $C_1$ and $C_2$ ordered according
to direction of $S=F(C_1)=F(C_2)$, where the two brothers were
oriented in the same direction\footnote{ For example,  such
orientation can be the result of DIRECT-TWO with $l_1=l_2=1$ or
$l_1=l_2=-1.$} (so they were not oriented by DIRECT-FORWARDS or
DIRECT-BACKWARDS). The first lemma assumes that they were both
oriented forwards, and the second lemma assumes they were both
oriented backwards.

Mark \bi
\item $u:= $ the end node of the path $P_s(C_1) \cap P_s(C_2)$, such that $u \not\in S$.
\item $v_1:= S \cap (P_s(C_1) \backslash P_s(C_2))$.
\item $v_2:= S \cap (P_s(C_1) \cap P_s(C_2))$.
\item $v_3:= S \cap (P_s(C_2) \backslash P_s(C_1))$.
\ei For one of these brothers $I(C)=P_s(C)$, but for the other
brother $I(C) \subset P_s(C)$.

\begin{figure}
\begin{center}
\setlength{\unitlength}{0.00069991in}
\begingroup\makeatletter\ifx\SetFigFont\undefined%
\gdef\SetFigFont#1#2#3#4#5{%
  \reset@font\fontsize{#1}{#2pt}%
  \fontfamily{#3}\fontseries{#4}\fontshape{#5}%
  \selectfont}%
\fi\endgroup%
{\renewcommand{\dashlinestretch}{30}
\begin{picture}(5118,4061)(0,-10)
\put(60,2535){\blacken\ellipse{90}{90}}
\put(60,2535){\ellipse{90}{90}}
\put(1365,2535){\blacken\ellipse{90}{90}}
\put(1365,2535){\ellipse{90}{90}}
\put(5010,2535){\blacken\ellipse{90}{90}}
\put(5010,2535){\ellipse{90}{90}}
\put(2445,3795){\blacken\ellipse{90}{90}}
\put(2445,3795){\ellipse{90}{90}}
\put(2445,2535){\blacken\ellipse{90}{90}}
\put(2445,2535){\ellipse{90}{90}}
\put(600,2535){\blacken\ellipse{90}{90}}
\put(600,2535){\ellipse{90}{90}}
\put(2445,240){\blacken\ellipse{90}{90}}
\put(2445,240){\ellipse{90}{90}}
\put(3795,2535){\blacken\ellipse{90}{90}}
\put(3795,2535){\ellipse{90}{90}}
\put(2985,2535){\blacken\ellipse{90}{90}}
\put(2985,2535){\ellipse{90}{90}}
\put(4380,2535){\blacken\ellipse{90}{90}}
\put(4380,2535){\ellipse{90}{90}}
\path(60,2535)(600,2535)
\path(1365,2535)(2445,2535)
\blacken\path(2295.000,2497.500)(2445.000,2535.000)(2295.000,2572.500)(2295.000,2497.500)
\path(2445,3795)(2445,2535)
\blacken\path(2407.500,2685.000)(2445.000,2535.000)(2482.500,2685.000)(2407.500,2685.000)
\path(2445,2535)(2985,2535)
\blacken\path(2835.000,2497.500)(2985.000,2535.000)(2835.000,2572.500)(2835.000,2497.500)
\path(5010,2535)(4380,2535)
\path(4380,2535)(3795,2535)
\blacken\path(3945.000,2572.500)(3795.000,2535.000)(3945.000,2497.500)(3945.000,2572.500)
\dottedline{45}(3120,2895)(2490,3165)
\path(2612.115,3145.304)(2490.000,3165.000)(2588.480,3090.155)
\dottedline{45}(3120,2895)(2670,2580)
\path(2751.104,2673.392)(2670.000,2580.000)(2785.512,2624.239)
\dottedline{45}(3795,2175)(3390,2490)
\path(3503.140,2440.008)(3390.000,2490.000)(3466.304,2392.647)
\dottedline{45}(3795,2175)(4200,2490)
\path(4123.696,2392.647)(4200.000,2490.000)(4086.860,2440.008)
\thicklines
\path(2445,240)(600,2535)
\blacken\thinlines
\path(723.210,2441.590)(600.000,2535.000)(664.757,2394.598)(723.210,2441.590)
\thicklines
\path(600,2535)(1365,2535)
\blacken\thinlines
\path(1215.000,2497.500)(1365.000,2535.000)(1215.000,2572.500)(1215.000,2497.500)
\thicklines
\path(1365,2535)(1365,3795)
\blacken\thinlines
\path(1402.500,3645.000)(1365.000,3795.000)(1327.500,3645.000)(1402.500,3645.000)
\thicklines
\path(1365,3795)(2445,3795)
\blacken\thinlines
\path(2295.000,3757.500)(2445.000,3795.000)(2295.000,3832.500)(2295.000,3757.500)
\thicklines
\path(2445,3795)(3795,3795)
\blacken\thinlines
\path(3645.000,3757.500)(3795.000,3795.000)(3645.000,3832.500)(3645.000,3757.500)
\thicklines
\path(3795,3795)(3795,2535)
\blacken\thinlines
\path(3757.500,2685.000)(3795.000,2535.000)(3832.500,2685.000)(3757.500,2685.000)
\thicklines
\path(3840,2535)(2985,2535)
\blacken\thinlines
\path(3135.000,2572.500)(2985.000,2535.000)(3135.000,2497.500)(3135.000,2572.500)
\thicklines
\path(2985,2535)(2445,240)
\blacken\thinlines
\path(2442.853,394.602)(2445.000,240.000)(2515.859,377.424)(2442.853,394.602)
\put(555,2310){\makebox(0,0)[lb]{\smash{{\SetFigFont{10}{12.0}{\rmdefault}{\mddefault}{\updefault}$x$}}}}
\put(1320,2310){\makebox(0,0)[lb]{\smash{{\SetFigFont{10}{12.0}{\rmdefault}{\mddefault}{\updefault}$v_1$}}}}
\put(2400,2310){\makebox(0,0)[lb]{\smash{{\SetFigFont{10}{12.0}{\rmdefault}{\mddefault}{\updefault}$v_2$}}}}
\put(4965,2310){\makebox(0,0)[lb]{\smash{{\SetFigFont{10}{12.0}{\rmdefault}{\mddefault}{\updefault}$w$}}}}
\put(15,2310){\makebox(0,0)[lb]{\smash{{\SetFigFont{10}{12.0}{\rmdefault}{\mddefault}{\updefault}$v$}}}}
\put(2400,3885){\makebox(0,0)[lb]{\smash{{\SetFigFont{10}{12.0}{\rmdefault}{\mddefault}{\updefault}$u$}}}}
\put(1410,1635){\makebox(0,0)[lb]{\smash{{\SetFigFont{10}{12.0}{\rmdefault}{\mddefault}{\updefault}$M_t(S)$}}}}
\put(2895,1635){\makebox(0,0)[lb]{\smash{{\SetFigFont{10}{12.0}{\rmdefault}{\mddefault}{\updefault}$M_h(S)$}}}}
\put(2400,15){\makebox(0,0)[lb]{\smash{{\SetFigFont{10}{12.0}{\rmdefault}{\mddefault}{\updefault}$z$}}}}
\put(3750,2310){\makebox(0,0)[lb]{\smash{{\SetFigFont{10}{12.0}{\rmdefault}{\mddefault}{\updefault}$v_3$}}}}
\put(3030,2310){\makebox(0,0)[lb]{\smash{{\SetFigFont{10}{12.0}{\rmdefault}{\mddefault}{\updefault}$y$}}}}
\put(3300,3345){\makebox(0,0)[lb]{\smash{{\SetFigFont{10}{12.0}{\rmdefault}{\mddefault}{\updefault}$C_2$}}}}
\put(1635,3345){\makebox(0,0)[lb]{\smash{{\SetFigFont{10}{12.0}{\rmdefault}{\mddefault}{\updefault}$C_1$}}}}
\put(2535,3210){\makebox(0,0)[lb]{\smash{{\SetFigFont{10}{12.0}{\rmdefault}{\mddefault}{\updefault}$J_t(C_2)$}}}}
\put(3120,2805){\makebox(0,0)[lb]{\smash{{\SetFigFont{10}{12.0}{\rmdefault}{\mddefault}{\updefault}$B(C_2)$}}}}
\put(3660,1995){\makebox(0,0)[lb]{\smash{{\SetFigFont{10}{12.0}{\rmdefault}{\mddefault}{\updefault}$J_h(S)$}}}}
\end{picture}
}
\end{center}
\caption{Figure for Lemma \ref{lemtwof}.} \label{twobrofor}
\end{figure}
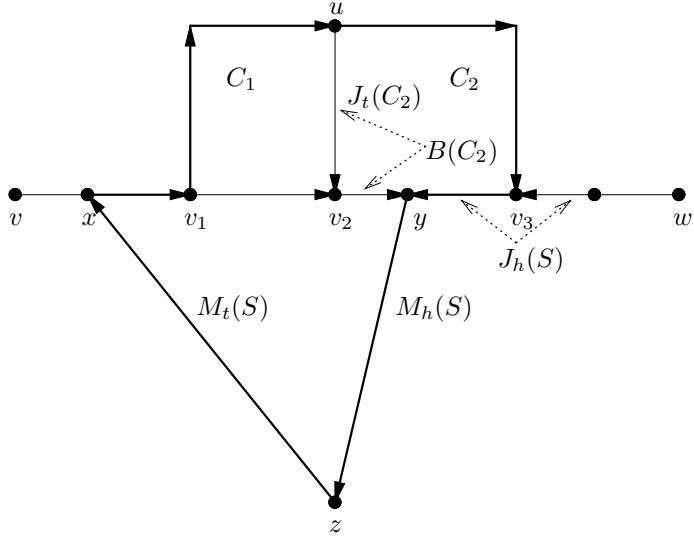

\begin{lem}
\label{lemtwof} Suppose that $S$ satisfies the induction hypotheses,
and $C_1$, $C_2$ were oriented (see Figure \ref{twobrofor}) $v_1
\rightarrow u \rightarrow v_2$, $ u \rightarrow v_3$ (i.e. both were
oriented forwards). Then $C_2$ satisfies the induction hypotheses.
Moreover $|N(C_2)| = 1$.
\end{lem}
\begin{prf}
By Observation  \ref{winpath}  the orientation  $u \rightarrow v_3$
indicates that $l(C_1) < l(C_2)$ (for example in the procedure
DIRECT-TWO). It follows that \footnote{ $[u,v_3]$ is the path in
$P_s(C_2)$, $[v_1,u]$ is the path in $P_s(C_1)$} $U(C_2) \subset
[u,v_3]$, and therefore:
\[ I(C_2) = [u \rightarrow v_3], \ t(C_2) = u, \ h(C_2)=v_3 .\]

Since  $l(C_1) < l(C_2)$
\begin{equation}
\label{c1sc2} l(v_1,u) + l(v_2,v_3) < l(v_1,v_2) + l(u,v_3).
\end{equation}

\begin{description}
\item [A1] By A3 on $S$ $x \in [v,v_2]$   and
$y \in [v_2,w]$. By A4 the direction is always $x \rightarrow v_1$
and $v_3 \rightarrow y$ so we can define:
\[ M_t(C_2) = M_t(S) \cup [x \rightarrow v_1] \cup [v_1 \rightarrow u], \ M_h(C_2) =[v_3 \rightarrow y] \cup M_h(S) .\]
 Using A1 on $S$
\begin{eqnarray*}
l(M(C_2)) &=& l(M(S)) + l(x,v_1) + l(v_1,u)+  l(v_3,y) \\
&\leq& 15l(P_c(S)) + 2l(P_s(S)) + l(I(S)) + l(x,v_1) + l(v_1
\rightarrow u) +  l(v_3,y)\\
&\leq& 15l(P_c(S)) + 4l(v,v_1) + 3l(v_1, v_2) + 3l(v_2,v_3) +
4l(v_3,w) +l(v_1 \rightarrow u ).
\end{eqnarray*}
Note that:
 \bi
  \item The lengths of $[v,v_1]$ and $[v_3,w]$
are taken four times to satisfy all four cases of locations for
$x,y$. In Figure \ref{twobrofor} the case $x \in [v,v_1], y \in
[v_2,v_3]$ is described. In this case the length of $[x,v_1]$ should
be counted four times: twice in $2l(P_s(S))$, once in $l(I(S))$ and
once in $l(M_t(C_2))$. This is taken into account by considering
$4l(v,v_1)$.
 \item The length of $[v_1,v_2]$ is taken only three times. When $x
 \not\in [v_1,v_2]$ then $[v_1,v_2]$ is counted three times in
 $2l(P_s(S)) + l(I(S))$. When $x \in [v_1,v_2]$ then $[v_1,x]$ is
 counted twice in $2l(P_s(S)) + l(I(S))$ and once more in
 $l(M_t(C_2))$. The path $[x,v_2]$ is counted three times in $2l(P_s(S)) +
 l(I(S))$, giving altogether three times $l(v_1,v_2)$.
\item In the same manner the length of $[v_2,v_3]$ is taken only
three times both when $y \in [v_2,v_3]$ and for $y \not\in
[v_2,v_3]$. \ei

>From (\ref{c1sc2})
\begin{eqnarray*}
l(M(C_2)) &\leq&  15 l(P_c(S)) + 4l(v,v_1) + 4l(v_1,v_2) +
 2l(v_2,v_3)+ 4l(v_3,w) + l(u,v_3) .
\end{eqnarray*}

 By definition
\[ l(P_c(C_2)) = l(P_c(S)) + l(v,v_1) + l(v_1,v_2)  + l(v_3,w) ,\] so that
\[ l(M(C_2)) \leq 15 l(P_c(C_2)) +2l(v_2,v_3) + l(u,v_3).\]
Since $C_2$ is a son of $S$, $l(v_2,v_3) \leq l(P_s(C_2))$. In this
case $I(C_2) = [u \rightarrow v_3]$ so
\[ l(M(C_2)) \leq 15 l(P_c(C_2)) +2l(P_s(C_2)) +l(I(C_2)).\]

\item [A2] When $y \in [v_3,w]$ we define
\[ B(C_2) = [u \rightarrow v_2 \rightarrow v_3] .\]
When $y \in [v_2,v_3]$ we define\footnote{Note that $y \in
M_h(C_2)$.}
\[ B(C_2) = [u \rightarrow v_2 \rightarrow y]. \]
In both cases $l(B(C_2)) \leq l(u,v_2) + l(v_2,v_3)$. From the
definition of $I(C_2)$ and since $C_2$ is a son of $S$,
\begin{eqnarray*}
l(B(C_2)) &\leq& l(u,v_2) + l(v_2,v_3) \\
&=& l(P_s(C_2)) - l(I(C_2)) + l(v_2,v_3) \\
&\leq& l(P_s(C_2)) - l(I(C_2)) +   l(P_s(C_2)) \\
&=& 2l(P_s(C_2)) - l(I(C_2)).
\end{eqnarray*}

\item [A3] Let $\hat{C}$ be a son of $C_2$.  Since $[u,v_2] \subseteq P_s(C_1)$ and $l(C_1) < l(C_2)$,
if $P_f(\hat{C}) \subseteq [u,v_2]$ then $\hat{C}$ should have been
a son of $C_1$ . Therefore $P_f(\hat{C}) \cap (u,v_3) \neq \phi$.
\item [A4] In this case $J_t(C_2) = [u \rightarrow v_2]$. Obviously for every
son $\hat{C}$ of $C_2$ $P_f(\hat{C}) \subset P_s(C_2) = J_t(C_2)
\cup I(C_2)$.
\item [A5] $M(C_2) \subseteq P_s(C_2) \cup C_1
\cup M(S)$. Since $l(C_1) < l(C_2)$ the condition is satisfied.
\item [A6] $M(C_2)$ is contained in $P_s(C_2)$, $C_1$ (a brother
of $C_2$), $S$, $B_t(S)$,$B_h(S)$ and $M(S)$, so $N(C_2) = \{C_1 \}
$.
\end{description}

\end{prf}

\begin{figure}
\begin{center}
\setlength{\unitlength}{0.00069991in}
\begingroup\makeatletter\ifx\SetFigFont\undefined%
\gdef\SetFigFont#1#2#3#4#5{%
  \reset@font\fontsize{#1}{#2pt}%
  \fontfamily{#3}\fontseries{#4}\fontshape{#5}%
  \selectfont}%
\fi\endgroup%
{\renewcommand{\dashlinestretch}{30}

}
\end{center}
\caption{Figure for Lemma \ref{lemtwob}.} \label{twobrobac}
\end{figure}

\begin{lem}
\label{lemtwob} Suppose that $S$ satisfies the induction hypotheses,
and $C_1$, $C_2$  were oriented (see Figure \ref{twobrobac}) $v_1
\leftarrow u \leftarrow v_2, u \leftarrow v_3$ (i.e. both were
oriented backwards). Then $C_2$ satisfies the induction hypotheses.
Moreover $|N(C_2)| = 1$.
\end{lem}
\begin{prf} By Observation  \ref{winpath}  the orientation  $u
\leftarrow v_3$ indicates that $l(C_1) < l(C_2)$ (for example in the
procedure DIRECT-TWO). It follows that $U(C_2) \subset [u,v_3]$ and
therefore
\[ I(C_2) = [v_3 \rightarrow u], \ t(C_2) = v_3, \ h(C_2)=u .\]

>From $l(C_1) < l(C_2)$ we get that (\ref{c1sc2}) still holds.

\begin{description}
\item [A1,A2] Since  $C_2$ was oriented
backwards, according to Property \ref{longfwd} $C_2$ is light:
\begin{equation}
\label{pfc2s} l(v_2,v_3) \leq 0.5 [l(v,v_2) + l(v_3,w)].
\end{equation}

By A3 there are four cases to be considered (described in Figure
\ref{twobrobac}):
 \begin{description}
\item (a)$[v_1,v_3] \subseteq [x,y]$. This case is described in Figure
\ref{twobrobac}(a). Since $[x \rightarrow y]=I(S)$, $[x,y]$ is
oriented $x \rightarrow v_1 \rightarrow v_2 \rightarrow v_3
\rightarrow y$. Let
\[ M_t(C_2) = M_t(S) \cup [x \rightarrow v_1 \rightarrow v_2 \rightarrow v_3],\
M_h(C_2) =[u \rightarrow v_1 \rightarrow v_2 \rightarrow v_3
\rightarrow y] \cup M_h(S) .\] By A1, Equations (\ref{c1sc2}) and
(\ref{pfc2s})
\begin{eqnarray*}
l(M(C_2)) &=& l(M(S)) + l(x,v_1) + 2 l(v_1,v_2) + 2l(v_2,v_3) + l(v_1,u) + l(v_3,y) \\
&\leq& 15 l(P_c(S)) + 2l(v,x) + 3l(x,v_1) + 3l(v_1,v_2) +
3l(v_2,v_3)+ 3l(v_3,y) + \\
&&  2l(y,w) +l(x,v_1) + 2 l(v_1,v_2) + 2l(v_2,v_3) + l(v_1,u) + l(v_3,y) \\
 &\leq& 15 l(P_c(S)) + 2l(v,x) + 4l(x,v_1) + 5l(v_1,v_2) +
5l(v_2,v_3)+ 4l(v_3,y) + \\
&& 2l(y,w) + l(v_1,u) \\
&\leq& 15 l(P_c(S)) + 2l(v,x) + 4l(x,v_1) + 6l(v_1,v_2) +
4l(v_2,v_3)+ 4l(v_3,y) + \\
&& 2l(y,w) + l(u,v_3) \\
&\leq& 15 l(P_c(S)) + 3l(v,x) + 5l(x,v_1) + 7l(v_1,v_2) +
2l(v_2,v_3)+ 5l(v_3,y) + \\
&& 3l(y,w) + l(u,v_3).
\end{eqnarray*}
 By definition
\[ l(P_c(C_2)) = l(P_c(S)) + l(v,x) + l(x,v_1) + l(v_1,v_2) +
l(v_3,y) + l(y,w) ,\] so that
\[ l(M(C_2)) \leq 15 l(P_c(C_2)) +2l(v_2,v_3) +l(u,v_3).\]
Since $C_2$ is a son of $S$, $l(v_2,v_3) \leq l(P_s(C_2))$. $I(C_2)
= [v_3 \rightarrow u]$ so
\[ l(M(C_2)) \leq 15 l(P_c(C_2)) +2l(P_s(C_2)) +l(I(C_2)).\]

In this case we define\footnote{Note that $v_2 \in M_t(C_2)$.}:
\[ B(C_2) = [v_2 \rightarrow u] .\]
>From the definition of $I(C_2)$ and since $C_2$ is a son of $S$
\[ l(B(C_2)) = l(v_2,u) = l(P_s(C_2)) - l(I(C_2)) \leq 2l(P_s(C_2)) - l(I(C_2)).\]

\item (b) $x \in [v_1,v_2]$ and $y\in [v_3,w]$.  This case is described in Figure
\ref{twobrobac}(b). From A4 the paths are oriented $ v_1 \leftarrow
x \rightarrow v_2 \rightarrow v_3 \rightarrow y$. From A2 (since $x
\not\in F(S)$ )
\[ l(M_t(S)) + l(B_t(S))  \leq 15 l(P_c(S)) +4l(P_s(S)). \] Let
\[ M_t(C_2) = M_t(S) \cup [x \rightarrow v_2 \rightarrow v_3],\
 M_h(C_2) =[u \rightarrow v_1] \cup (B_t(S)\backslash [x,v_1]) ,\]
then:
\begin{eqnarray*}
l(M(C_2)) &\leq& l(M_t(S)) + l(B_t(S))  +l(x,v_2) + l(v_2,v_3) + l(u,v_1)   \\
&\leq& 15 l(P_c(S)) + 4l(P_s(S))  + l(x,v_2) + l(v_2,v_3) + l(u,v_1)\\
&=& 15 l(P_c(S)) + 4l(v,v_1) + 4l(v_1,x) + 5l(x,v_2) + 5l(v_2,v_3)
+4l(v_3,y) +  \\
&& 4l(y,w) +l(u,v_1) \\
&\leq& 15 l(P_c(S)) + 4l(v,v_1) + 5l(v_1,x) + 6l(x,v_2) +
4l(v_2,v_3)+4l(v_3,y) +  \\
&& 4l(y,w) +l(u,v_3) \\
&\leq& 15 l(P_c(S)) + 5l(v,v_1) + 6l(v_1,x) + 7l(x,v_2) +
2l(v_2,v_3)+5l(v_3,y) +  \\
&& 5l(y,w) +l(u,v_3),
\end{eqnarray*}
where the last two inequalities follow from (\ref{c1sc2}) and
(\ref{pfc2s}). By definition
\[ l(P_c(C_2)) = l(P_c(S)) + l(v,v_1) + l(v_1,x) + l(x,v_2) +
l(v_3,y) + l(y,w) ,\] so that
\[ l(M(C_2)) \leq 15 l(P_c(C_2)) +2l(v_2,v_3) +l(u,v_3).\]
Since $C_2$ is a son of $S$, $l(v_2,v_3) \leq l(P_s(C_2)$. $I(C_2) =
[u \rightarrow v_3]$ so
\[ l(M(C_2)) \leq 15 l(P_c(C_2)) +2l(P_s(C_2)) +l(I(C_2)).\]

In this case again
\[ B(C_2) = [v_2 \rightarrow u] \]
so as in previous case,
\[ l(B(C_2)) \leq 2 l(P_s(C_2)) - l(I(C_2)) .\]

 \item (c) $x \in [v,v_1]$ and $y \in [v_2,v_3]$.  This case is described in Figure
\ref{twobrobac}(c). From A4  the paths are oriented: $ x \rightarrow
v_1 \rightarrow v_2 \rightarrow y \leftarrow v_3$ .

 From  A2 (since $y \not\in F(S)$ )
\[ l(M_h(S)) + l(B_h(S))  \leq 15 l(P_c(S)) + 4l(P_s(S)). \] Let:
\[ M_t(C_2) = (B_h(S) \backslash [v_3 \rightarrow y]),
 \  M_h(C_2) =[u \rightarrow v_1 \rightarrow v_2 \rightarrow y] \cup M_h(S) ,\]
giving:
\begin{eqnarray*}
l(M(C_2)) &\leq& l(M_h(S)) + l(B_h(S)) + l(u,v_1) + l(v_1,v_2) +l(v_2,y) \\
&\leq& 15 l(P_c(S)) + 4l(P_s(S)) + l(u,v_1) + l(v_1,v_2) + l(v_2,y)\\
&=& 15 l(P_c(S)) + 4l(v,x) + 4l(x,v_1) + 5l(v_1,v_2) + 5l(v_2,y) +
4l(y,v_3) + \\
&& 4l(v_3,w) +l(u,v_1) \\
&\leq& 15 l(P_c(S)) + 4l(v,x) + 4l(x,v_1) + 6l(v_1,v_2) + 4l(v_2,y)
+3l(y,v_3) + \\
&& 4l(v_3,w) +l(u,v_3) \\
&\leq& 15 l(P_c(S)) + 5l(v,x) + 5l(x,v_1) + 7l(v_1,v_2) + 2l(v_2,y)
+l(y,v_3) + \\
&& 5l(v_3,w) +l(u,v_3),
\end{eqnarray*}
where the last two inequalities follow from (\ref{c1sc2}) and
(\ref{pfc2s}). By definition
\[ l(P_c(C_2)) = l(P_c(S)) + l(v,x) + l(x,v_1) + l(v_1,v_2)  + l(v_3,w) ,\]
so that
\[ l(M(C_2)) \leq 15 l(P_c(C_2)) +2l(v_2,v_3) +l(u,v_3).\]
Since $S=F(C_2)$, $l(v_2,v_3) \leq l(P_s(C_2)$. $I(C_2) = [u
\rightarrow v_3]$ so
\[ l(M(C_2)) \leq 15 l(P_c(C_2)) +2l(P_s(C_2)) +l(I(C_2)).\]


We prove A2 directly  on $C_2$ (not using Lemma \ref{lbscor}). From
A1
\[ l(M(S)) \leq 15 l(P_c(S)) + 2l(P_s(S)) + l(I(S)). \]

In this case $t(C_2) = v_3 \in F(C_2)$, $h(C_2) = u \not\in F(C_2)$,
so only the first inequality of A2 should be proven. Let
\[ B_h(C_2) = M_t(S) \cup [x \rightarrow v_1 \rightarrow v_2 \rightarrow u], \]
giving
\begin{eqnarray*}
 l(B_h(C_2)) + l(M_h(C_2)) &=& l(M_t(S)) +l(M_h(S)) +
l(x,v_1) + 2l(v_1,v_2) + l(v_2,u) + \\
&& l(u,v_1)+ l(v_2,y)\\
&\leq& 15 l(P_c(S)) + 2l(P_s(S)) + l(I(S)) + l(x,v_1) + 2l(v_1,v_2) + \\
&& l(v_2,u) + l(u,v_1) + l(v_2,y)\\
&=& 15 l(P_c(S)) + 2l(v,x) + 4l(x,v_1) + 5l(v_1,v_2) + 4l(v_2,y)  + \\
&& 2l(y,v_3)+ 2l(v_3,w) + l(v_2,u) + l(u,v_1) \\
&\leq& 15 l(P_c(S)) + 2l(v,x) + 4l(x,v_1) + 6l(v_1,v_2) + 3l(v_2,y)+ \\
&& l(y,v_3) +2l(v_3,w) + l(v_2,u) + l(u,v_3)\\
&\leq& 15 l(P_c(S)) + 2.5l(v,x) + 4.5l(x,v_1) + 6.5l(v_1,v_2) + 2l(v_2,y)  \\
&& 2.5l(v_3,w) + l(v_2,u) + l(u,v_3) \\
&\leq& 15l(P_c(C_2)) + 2l(v_2,y) + l(P_s(C_2)),
\end{eqnarray*}
where the last inequalities follow from (\ref{c1sc2}), (\ref{pfc2s})
and the  definitions of $P_c(C_2)$ and $P_s(C_2)$. Since $S=F(C_2)$
$l(v_2,y) \leq l(v_2,v_3) \leq l(P_s(C_2))$, giving
\[  l(B_h(C_2)) + l(M_h(C_2)) \leq 15l(P_c(C_2)) +3l(P_s(C_2)). \]

\item (d) $x \in [v_1,v_2]$ and $ y \in [v_2,v_3] $.  This case is described in Figure
\ref{twobrobac}(d). From A4 the paths are oriented: $ v_1 \leftarrow
x \rightarrow v_2 \rightarrow y \leftarrow v_3$ . From A2 (since
$x,y \not\in F(S)$)

\[ l(B_h(S)) + l(B_t(S))  \leq 15 l(P_c(S)) + 6l(P_s(S)). \] Let:
\[ M_t(C_2) = B_h(S) \backslash [v_3 \rightarrow y]), \  M_h(C_2) = [u \rightarrow v_1] \cup (B_t(S) \backslash [x
\rightarrow v_1]) ,\] then
\begin{eqnarray*}
l(M(C_2)) &\leq& l(B_h(S)) + l(B_t(S)) + l(u,v_1) \\
&\leq& 15 l(P_c(S)) + 6l(P_s(S)) +  l(u,v_1) \\
&=& 15l(P_c(S)) + 6 l(v,v_1) + 6 l(v_1 \rightarrow x \rightarrow
v_2) + 6l(v_2 \rightarrow y \rightarrow v_3) + \\
&&6l(v_3,w) + l(u,v_1) \\
&\leq& 15l(P_c(S)) + 6 l(v,v_1) + 7 l(v_1,v_2) + 5l(v_2,v_3) +
6l(v_3,w) + l(u,v_3) \\
&\leq& 15l(P_c(S)) + 7.5 l(v,v_1) + 8.5 l(v_1,x) + 8.5l(x,v_2) +
2l(v_2,v_3) + \\
&&7.5l(v_3,w) + l(u,v_3) \\
&\leq& 15 l(P_c(C_2)) +2l(v_2,v_3) +l(u,v_3),
\end{eqnarray*}
where the third and fourth  inequalities follow from (\ref{c1sc2})
and (\ref{pfc2s}).

Since $C_2$ is a son of $S$ $l(v_2,v_3) \leq l(P_s(C_2)$, $I(C_2) =
[u \rightarrow v_3]$ so
\[ l(M(C_2)) \leq 15 l(P_c(C_2)) +2l(P_s(C_2)) +l(I(C_2)).\]


We prove A2 directly on $C_2$ (not using Lemma \ref{lbscor}). From
A2
\[ l(M_t(S)) + l(B_t(S)) \leq 15 l(P_c(S)) + 4l(P_s(S))  \]

In this case $t(C_2) = v_3 \in F(C_2)$, $h(C_2) = u \not\in F(C_2)$,
so only the first inequality of A2 should be proven. Let
\[ B_h(C_2) = M_t(S) \cup [x \rightarrow v_2 \rightarrow u], \]
giving
\begin{eqnarray*}
 l(B_h(C_2)) + l(M_h(C_2)) &\leq& l(M_t(S)) + l(B_t(S)) + l(x,v_2) + l(v_2,u) +
l(u,v_1) \\
&\leq&  15 l(P_c(S)) + 4l(P_s(S))+ l(x,v_2) + l(v_2,u) +
l(u,v_1) \\
&=&  15 l(P_c(S)) + 4l(v,v_1) + 4l(v_1,x) + 5l(x,v_2)+ 4l(v_2,v_3)+\\
&& 4l(v_3,w) + l(v_2,u) + l(u,v_1) \\
&\leq&  15 l(P_c(S)) + 4l(v,v_1) + 5l(v_1,x) + 6l(x,v_2)+ 3l(v_2,v_3)+\\
&& 4l(v_3,w) + l(v_2,u) + l(u,v_3) \\
&\leq&  15 l(P_c(S)) + 5l(v,v_1) + 6l(v_1,x) + 7l(x,v_2)+ l(v_2,v_3)+\\
&& 5l(v_3,w) + l(v_2,u) + l(u,v_3) \\
\end{eqnarray*}
where the last inequalities follow from (\ref{c1sc2}) and
(\ref{pfc2s}). From the definition of $P_c(C_2)$ and $P_s(C_2)$
\[ l(B_h(C_2)) + l(M_h(C_2)) \leq 15l(P_c(C_2)) + 2l(v_2,v_3) +
l(P_s(C_2)) \] Since $C_2$ is a son of $S$ $l(v_2,v_3) \leq
l(P_s(C_2))$ giving
\[  l(B_h(C_2)) + l(M_h(C_2)) \leq 15l(P_c(C_2)) +3l(P_s(C_2)). \]

 \end{description}

\item [A3] Let $\hat{C}$ be a son of $C_2$. If $P_f(\hat{C}) \subseteq
[u,v_2]$,  since $[u,v_2] \subseteq P_s(C_1)$ and $l(C_1) < l(C_2)$
$\hat{C}$ should have been a son of $C_1$ . Therefore $P_f(\hat{C})
\cap [u,v_3] \neq \phi$.
\item [A4] In this case $J_h(C_2) = [v_2 \rightarrow u]$ and obviously for
every son $\hat{C}$ of $C_2$ $P_f(\hat{C}) \subset P_s(C_2) =
J_h(C_2) \cup I(C_2)$.

\item [A5] $M(C_2) \subseteq P_s(C_2) \cup C_1 \cup M(S)$.
Since $l(C_1) < l(C_2)$ the condition is satisfied.
\item [A6] $M(C_2)$ is contained in $P_s(C_2)$, $C_1$ (a brother
of $C_2$),  $S$ and $M(S)$, so $N(C_2) = \{ C_1 \}$.
\end{description}

\end{prf}

\begin{cor}
\label{dirtwocor} If $C_1$ and $C_2$ are light, the cycles oriented
in DIRECT-TWO satisfy the induction hypotheses.
\end{cor}
\begin{prf}
In the first two cases of the procedure the cycles are directed to
satisfy $I(C_i)=P_s(C_i)$ and then according to Lemmas \ref{lemf}
and \ref{lemb} the induction hypotheses are satisfied.

Suppose that $l(C_1) < l(C_2)$. If $l_1=l_2=1$ then  $C_1$ and $C_2$
satisfy the induction hypotheses by Lemmas \ref{lemf} and
\ref{lemtwof}  respectively. If  $l_1=l_2=-1$ then the claim holds
by Lemmas  \ref{lemb} and \ref{lemtwob}, respectively. The case
$l(C_1) > l(C_2)$ is symmetrically proven.
\end{prf}

\newpage
\subsection{Three  brothers} \label{secthree} In this
section we consider three neighbor brothers $C_1,C_2,C_3$ where the
three brothers were all oriented forwards.\footnote{ For example,
such orientation can be the result of DIRECT-MANY when $n=3$, $C_2$
is heavy and $l_1=l_2=1$.} In the first lemma we handle the
orientation applied by the algorithm when $l(C_1) < l(C_2) <
l(C_3)$. In the second lemma we handle the orientation applied by
the algorithm when $l(C_1) < l(C_2)$ and $l(C_3) < l(C_2)$. Other
equivalent cases can be handled by similar lemmas.

 Mark  \bi
\item $u_1:= $ the end node of the path $P_s(C_1) \cap P_s(C_2)$, such that $u_1 \not\in S$.
\item $u_2:= $ the end node of the path $P_s(C_2) \cap P_s(C_3)$, such that $u_2 \not\in S$.
\item $v_1:= S \cap (P_s(C_1) \backslash P_s(C_2))$.
\item $v_2:= S \cap (P_s(C_1) \cap P_s(C_2))$.
\item $v_3:= S \cap (P_s(C_2) \cap P_s(C_3))$.
\item $v_3:= S \cap (P_s(C_3) \backslash P_s(C_2))$.
\ei

\begin{figure}
\begin{center}
\setlength{\unitlength}{0.00061242in}
\begingroup\makeatletter\ifx\SetFigFont\undefined%
\gdef\SetFigFont#1#2#3#4#5{%
  \reset@font\fontsize{#1}{#2pt}%
  \fontfamily{#3}\fontseries{#4}\fontshape{#5}%
  \selectfont}%
\fi\endgroup%
{\renewcommand{\dashlinestretch}{30}
\begin{picture}(8628,4500)(0,-10)
\put(7845,2085){\blacken\ellipse{90}{90}}
\put(7845,2085){\ellipse{90}{90}}
\put(60,2085){\blacken\ellipse{90}{90}}
\put(60,2085){\ellipse{90}{90}}
\put(3165,2085){\blacken\ellipse{90}{90}}
\put(3165,2085){\ellipse{90}{90}}
\put(5235,2085){\blacken\ellipse{90}{90}}
\put(5235,2085){\ellipse{90}{90}}
\put(6270,2085){\blacken\ellipse{90}{90}}
\put(6270,2085){\ellipse{90}{90}}
\put(7305,2085){\blacken\ellipse{90}{90}}
\put(7305,2085){\ellipse{90}{90}}
\put(8520,2085){\blacken\ellipse{90}{90}}
\put(8520,2085){\ellipse{90}{90}}
\put(3165,4200){\blacken\ellipse{90}{90}}
\put(3165,4200){\ellipse{90}{90}}
\put(5235,4200){\blacken\ellipse{90}{90}}
\put(5235,4200){\ellipse{90}{90}}
\put(4200,240){\blacken\ellipse{90}{90}}
\put(4200,240){\ellipse{90}{90}}
\put(1095,2085){\blacken\ellipse{90}{90}}
\put(1095,2085){\ellipse{90}{90}}
\put(555,2085){\blacken\ellipse{90}{90}}
\put(555,2085){\ellipse{90}{90}}
\path(60,2085)(555,2085)
\path(7845,2085)(8520,2085)
\path(1095,2085)(3165,2085)
\blacken\path(3015.000,2047.500)(3165.000,2085.000)(3015.000,2122.500)(3015.000,2047.500)
\path(5235,2085)(6270,2085)
\blacken\path(6120.000,2047.500)(6270.000,2085.000)(6120.000,2122.500)(6120.000,2047.500)
\blacken\path(7455.000,2122.500)(7305.000,2085.000)(7455.000,2047.500)(7455.000,2122.500)
\path(7305,2085)(7845,2085)
\dashline{60.000}(6765,3480)(6765,4155)
\path(6795.000,4035.000)(6765.000,4155.000)(6735.000,4035.000)
\dottedline{45}(6765,3480)(7260,3705)
\path(7163.170,3628.033)(7260.000,3705.000)(7138.342,3682.655)
\dashline{60.000}(5775,2625)(5235,2895)
\path(5355.748,2868.167)(5235.000,2895.000)(5328.915,2814.502)
\dottedline{45}(5775,2625)(5640,2130)
\path(5642.631,2253.665)(5640.000,2130.000)(5700.517,2237.878)
\path(3165,4200)(3165,2085)
\blacken\path(3127.500,2235.000)(3165.000,2085.000)(3202.500,2235.000)(3127.500,2235.000)
\path(5235,4200)(5235,2085)
\blacken\path(5197.500,2235.000)(5235.000,2085.000)(5272.500,2235.000)(5197.500,2235.000)
\dottedline{60}(7350,1455)(6855,2040)
\path(6955.415,1967.772)(6855.000,2040.000)(6909.611,1929.015)
\dottedline{45}(7350,1455)(7665,2040)
\path(7634.522,1920.120)(7665.000,2040.000)(7581.694,1948.566)
\path(3165,2085)(5235,2085)
\blacken\path(5085.000,2047.500)(5235.000,2085.000)(5085.000,2122.500)(5085.000,2047.500)
\thicklines
\path(4200,240)(510,2085)
\blacken\thinlines
\path(660.935,2051.459)(510.000,2085.000)(627.394,1984.377)(660.935,2051.459)
\thicklines
\path(555,2085)(1095,2085)
\blacken\thinlines
\path(945.000,2047.500)(1095.000,2085.000)(945.000,2122.500)(945.000,2047.500)
\thicklines
\path(1095,2085)(1095,4200)
\blacken\thinlines
\path(1132.500,4050.000)(1095.000,4200.000)(1057.500,4050.000)(1132.500,4050.000)
\thicklines
\path(1095,4200)(3165,4200)
\blacken\thinlines
\path(3015.000,4162.500)(3165.000,4200.000)(3015.000,4237.500)(3015.000,4162.500)
\thicklines
\path(3165,4200)(5235,4200)
\blacken\thinlines
\path(5085.000,4162.500)(5235.000,4200.000)(5085.000,4237.500)(5085.000,4162.500)
\thicklines
\path(5235,4200)(7305,4200)
\blacken\thinlines
\path(7155.000,4162.500)(7305.000,4200.000)(7155.000,4237.500)(7155.000,4162.500)
\thicklines
\path(7305,4200)(7305,2085)
\blacken\thinlines
\path(7267.500,2235.000)(7305.000,2085.000)(7342.500,2235.000)(7267.500,2235.000)
\blacken\path(6420.000,2122.500)(6270.000,2085.000)(6420.000,2047.500)(6420.000,2122.500)
\thicklines
\path(6270,2085)(7305,2085)
\path(6270,2085)(4200,240)
\blacken\thinlines
\path(4287.026,367.800)(4200.000,240.000)(4336.928,311.811)(4287.026,367.800)
\put(510,1860){\makebox(0,0)[lb]{\smash{{\SetFigFont{8}{9.6}{\rmdefault}{\mddefault}{\updefault}$x$}}}}
\put(5190,1860){\makebox(0,0)[lb]{\smash{{\SetFigFont{8}{9.6}{\rmdefault}{\mddefault}{\updefault}$v_3$}}}}
\put(6225,1860){\makebox(0,0)[lb]{\smash{{\SetFigFont{8}{9.6}{\rmdefault}{\mddefault}{\updefault}$y$}}}}
\put(7260,1860){\makebox(0,0)[lb]{\smash{{\SetFigFont{8}{9.6}{\rmdefault}{\mddefault}{\updefault}$v_4$}}}}
\put(8475,1860){\makebox(0,0)[lb]{\smash{{\SetFigFont{8}{9.6}{\rmdefault}{\mddefault}{\updefault}$w$}}}}
\put(15,1860){\makebox(0,0)[lb]{\smash{{\SetFigFont{8}{9.6}{\rmdefault}{\mddefault}{\updefault}$v$}}}}
\put(3120,1860){\makebox(0,0)[lb]{\smash{{\SetFigFont{8}{9.6}{\rmdefault}{\mddefault}{\updefault}$v_2$}}}}
\put(3120,4335){\makebox(0,0)[lb]{\smash{{\SetFigFont{8}{9.6}{\rmdefault}{\mddefault}{\updefault}$u_1$}}}}
\put(5190,4335){\makebox(0,0)[lb]{\smash{{\SetFigFont{8}{9.6}{\rmdefault}{\mddefault}{\updefault}$u_2$}}}}
\put(15,2310){\makebox(0,0)[lb]{\smash{{\SetFigFont{8}{9.6}{\rmdefault}{\mddefault}{\updefault}$P_s(S)$}}}}
\put(4200,3615){\makebox(0,0)[lb]{\smash{{\SetFigFont{8}{9.6}{\rmdefault}{\mddefault}{\updefault}$C_2$}}}}
\put(6180,3615){\makebox(0,0)[lb]{\smash{{\SetFigFont{8}{9.6}{\rmdefault}{\mddefault}{\updefault}$C_3$}}}}
\put(2130,3615){\makebox(0,0)[lb]{\smash{{\SetFigFont{8}{9.6}{\rmdefault}{\mddefault}{\updefault}$C_1$}}}}
\put(5325,3165){\makebox(0,0)[lb]{\smash{{\SetFigFont{8}{9.6}{\rmdefault}{\mddefault}{\updefault}$J_t(C_3)$}}}}
\put(6585,3255){\makebox(0,0)[lb]{\smash{{\SetFigFont{8}{9.6}{\rmdefault}{\mddefault}{\updefault}$I(C_3)$}}}}
\put(5820,2535){\makebox(0,0)[lb]{\smash{{\SetFigFont{8}{9.6}{\rmdefault}{\mddefault}{\updefault}$B(C_3)$}}}}
\put(7170,1275){\makebox(0,0)[lb]{\smash{{\SetFigFont{8}{9.6}{\rmdefault}{\mddefault}{\updefault}$J_h(S)$}}}}
\put(1050,1860){\makebox(0,0)[lb]{\smash{{\SetFigFont{8}{9.6}{\rmdefault}{\mddefault}{\updefault}$v_1$}}}}
\put(4155,15){\makebox(0,0)[lb]{\smash{{\SetFigFont{8}{9.6}{\rmdefault}{\mddefault}{\updefault}$z$}}}}
\put(1950,870){\makebox(0,0)[lb]{\smash{{\SetFigFont{8}{9.6}{\rmdefault}{\mddefault}{\updefault}$M_t(S)$}}}}
\put(5235,870){\makebox(0,0)[lb]{\smash{{\SetFigFont{8}{9.6}{\rmdefault}{\mddefault}{\updefault}$M_h(S)$}}}}
\end{picture}
}
\end{center}
\caption{Figure for Lemma \ref{lemthreefs}.} \label{threefsfig}
\end{figure}
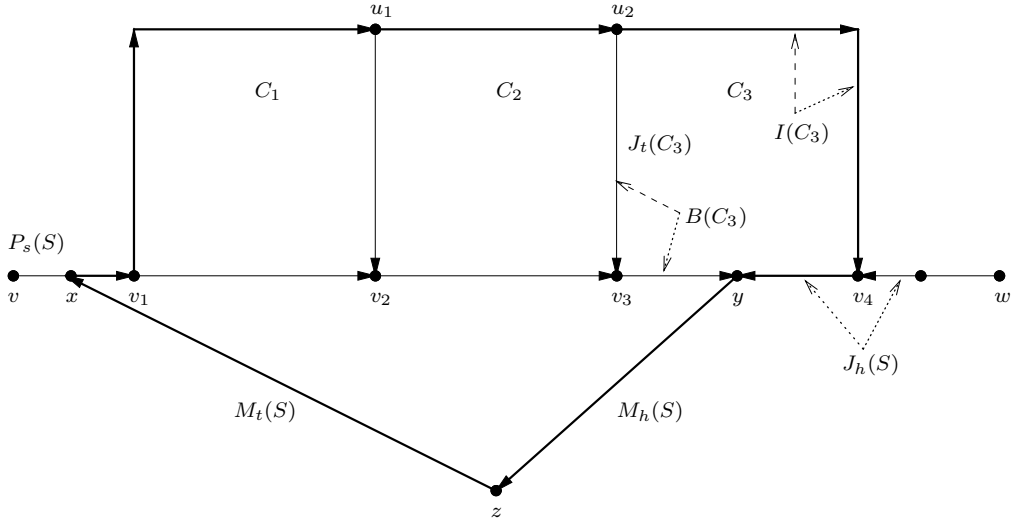

\begin{lem}
\label{lemthreefs} Suppose that $S$ satisfies the induction
hypotheses and $C_1,C_2,C_3$ were oriented $v_1 \rightarrow u_1
\rightarrow u_2 \rightarrow v_4$, $ u_1 \rightarrow v_2$, $u_2
\rightarrow v_3$ (see Figure \ref{threefsfig}). Then $C_3$ satisfies
the induction hypotheses. Moreover $|N(C_3)| = 2$.
\end{lem}

\begin{prf}
By Observation  \ref{winpath} the orientation  $u_1 \rightarrow v_2$
indicates that $l(C_1) < l(C_2)$ and the orientation  $u_2
\rightarrow v_3$ indicates that $l(C_2) < l(C_3)$. It follows that
$U(C_3) \subset [u_2,v_4 ]$. Therefore
\[ I(C_3) = [u_2 \rightarrow v_4], \ t(C_3) = u_2, \ h(C_3)=v_4
.\] Since  $l(C_1) < l(C_2) < l(C_3)$ \footnote{ $[v_1,u_1]$ is the
path in $P_s(C_1)$ and $[u_2,v_4]$ is the path in $P_s(C_3)$}
\begin{equation}
\label{c1sc3} l(v_1,u_1) + l(u_1,v_2) + l(v_3,v_4) < l(v_1,v_2) +
l(v_3,u_2) + l(u_2,v_4),
\end{equation}
\begin{equation}
\label{c2sc3}  l(u_1,v_2)+ l(u_1,u_2) + l(v_3,v_4) < l(v_2,v_3) +
l(u_2,v_4).
\end{equation}

\begin{description}
\item[A1] By A3 on $S$ $x \in [v,v_2]$ and $y \in
[v_3,w]$. In Figure \ref{threefsfig} the case $x \in [v,v_1]$, $y\in
[v_3,v_4]$ is described. By A4 for all these cases we can define:
\[ M_t(C_3) = M_t(S) \cup [x \rightarrow v_1 \rightarrow u_1 \rightarrow u_2],
\ M_h(C_3) =[v_4 \rightarrow y] \cup M_h(S) .\]
 Using A1 on $S$
\begin{eqnarray*}
l(M(C_3)) &=& l(M(S)) + l(x,v_1) + l(v_1 \rightarrow u_1
\rightarrow u_2) +  l(v_4,y) \\
&\leq& 15l(P_c(S)) + 2l(P_s(S)) + l(I(S)) + l(x,v_1) + l(v_1
\rightarrow u_1 \rightarrow u_2 ) +  l(v_4,y)\\
&\leq& 15l(P_c(S)) + 4l(v,v_1) + 3l(v_1 \rightarrow v_2 \rightarrow
v_3 \rightarrow v_4) + 4l(v_4,w) +\\
&&l(v_1 \rightarrow u_1 \rightarrow u_2 ).
\end{eqnarray*}
(The lengths of $[v,v_1]$ and $[v_4,w]$ are taken 4 times to satisfy
all possible  locations of $x,y$.)

>From (\ref{c1sc3}) and (\ref{c2sc3})
\begin{eqnarray*}
l(M(C_3)) &\leq&  15 l(P_c(S)) + 4l(v,v_1) + 4l(v_1,v_2) +
 3l(v_2,v_3)+ 2l(v_3,v_4) + \\
 &&4l(v_4,w) + l(P_s(C_3)) + l(u_1,u_2)\\
&\leq&  15 l(P_c(S)) + 4l(v,v_1) + 4l(v_1,v_2) +
 4l(v_2,v_3)+ l(v_3,v_4) + \\
 &&4l(v_4,w) + l(P_s(C_3)) + l(u_2,v_4).
\end{eqnarray*}
 By definition
\[ l(P_c(C_3)) = l(P_c(S)) + l(v,v_1) + l(v_1,v_2) +
l(v_2,v_3) + l(v_4,w) ,\] so that
\[ l(M(C_3)) \leq 15 l(P_c(C_3)) +l(v_3,v_4) +l(P_s(C_3)) + l(u_2,v_4).\]
Since $C_3$ is a son of $S$, $l(v_3,v_4) \leq l(P_s(C_3))$. $I(C_3)
= [u_2 \rightarrow v_4]$ so
\[ l(M(C_3)) \leq 15 l(P_c(C_3)) +2l(P_s(C_3)) +l(I(C_3)).\]

\item[A2] When $y \in [v_4,w]$ we define
\[ B(C_3) = [u_2 \rightarrow v_3 \rightarrow v_4] .\]
When $y \in [v_3,v_4]$ we define
\[ B(C_3) = [u_2 \rightarrow v_3 \rightarrow y] .\]
In both cases $B(C_3) \subseteq [u_2 \rightarrow v_3 \rightarrow
v_4]$.
 From the definition of $I(C_3)$ and since $C_3$ is a son of
$S$,
\begin{eqnarray*}
l(B(C_3)) &\leq& l(u_2,v_3) + l(v_3,v_4) \\
&=& l(P_s(C_3)) - l(I(C_3)) + l(v_3,v_4) \\
&\leq& l(P_s(C_3)) - l(I(C_3)) +   l(P_s(C_3)) \\
&=& 2l(P_s(C_3)) - l(I(C_3)).
\end{eqnarray*}

  \item [A3] Let $\hat{C}$ be a son of $C_3$. If
$P_f(\hat{C}) \subseteq [u_2,v_3]$ then
 since $[u_2,v_3] \subseteq P_s(C_2)$ and $l(C_2) < l(C_3)$, $\hat{C}$
should have been a son of $C_2$ . Therefore $P_f(\hat{C}) \cap
[u_2,v_4] \neq \phi$.
 \item [A4] In this case
$J_t(C_3) = [u_2 \rightarrow v_3]$. Obviously for every son
$\hat{C}$ of $C_3$ $P_f(\hat{C}) \subset P_s(C_3) =  I(C_3) \cup
J_t(C_3)$.
 \item [A5] $M(C_3)$ is
contained in $P_s(C_3) \cup C_1 \cup C_2 \cup M(S)$. Since $l(C_1) <
l(C_3)$ and $l(C_2) < l(C_3)$ the condition is satisfied.
 \item[A6] $M(C_3)$ is contained in $P_s(C_3)$, $C_1$ ,$C_2$(two
brothers of $C_3$), $S$  and $M(S)$, so $N(C_3) = \{ C_1,C_2 \}$.

\end{description}

\end{prf}

\begin{figure}
\begin{center}
\setlength{\unitlength}{0.00061242in}
\begingroup\makeatletter\ifx\SetFigFont\undefined%
\gdef\SetFigFont#1#2#3#4#5{%
  \reset@font\fontsize{#1}{#2pt}%
  \fontfamily{#3}\fontseries{#4}\fontshape{#5}%
  \selectfont}%
\fi\endgroup%
{\renewcommand{\dashlinestretch}{30}
\begin{picture}(8628,4509)(0,-10)
\put(7845,2085){\blacken\ellipse{90}{90}}
\put(7845,2085){\ellipse{90}{90}}
\put(60,2085){\blacken\ellipse{90}{90}}
\put(60,2085){\ellipse{90}{90}}
\put(555,2085){\blacken\ellipse{90}{90}}
\put(555,2085){\ellipse{90}{90}}
\put(1095,2085){\blacken\ellipse{90}{90}}
\put(1095,2085){\ellipse{90}{90}}
\put(3165,2085){\blacken\ellipse{90}{90}}
\put(3165,2085){\ellipse{90}{90}}
\put(5235,2085){\blacken\ellipse{90}{90}}
\put(5235,2085){\ellipse{90}{90}}
\put(6270,2085){\blacken\ellipse{90}{90}}
\put(6270,2085){\ellipse{90}{90}}
\put(7305,2085){\blacken\ellipse{90}{90}}
\put(7305,2085){\ellipse{90}{90}}
\put(8520,2085){\blacken\ellipse{90}{90}}
\put(8520,2085){\ellipse{90}{90}}
\put(3165,4200){\blacken\ellipse{90}{90}}
\put(3165,4200){\ellipse{90}{90}}
\put(5235,4200){\blacken\ellipse{90}{90}}
\put(5235,4200){\ellipse{90}{90}}
\put(4200,240){\blacken\ellipse{90}{90}}
\put(4200,240){\ellipse{90}{90}}
\path(60,2085)(555,2085)
\path(7845,2085)(8520,2085)
\path(1095,2085)(3165,2085)
\blacken\path(3015.000,2047.500)(3165.000,2085.000)(3015.000,2122.500)(3015.000,2047.500)
\path(3165,2085)(5235,2085)
\blacken\path(5085.000,2047.500)(5235.000,2085.000)(5085.000,2122.500)(5085.000,2047.500)
\path(5235,2085)(6270,2085)
\blacken\path(6120.000,2047.500)(6270.000,2085.000)(6120.000,2122.500)(6120.000,2047.500)
\blacken\path(7455.000,2122.500)(7305.000,2085.000)(7455.000,2047.500)(7455.000,2122.500)
\path(7305,2085)(7845,2085)
\path(3165,4200)(3165,2085)
\blacken\path(3127.500,2235.000)(3165.000,2085.000)(3202.500,2235.000)(3127.500,2235.000)
\dottedline{60}(7350,1455)(6855,2040)
\path(6955.415,1967.772)(6855.000,2040.000)(6909.611,1929.015)
\dottedline{45}(7350,1455)(7665,2040)
\path(7634.522,1920.120)(7665.000,2040.000)(7581.694,1948.566)
\blacken\path(5272.500,4050.000)(5235.000,4200.000)(5197.500,4050.000)(5272.500,4050.000)
\path(5235,4200)(5235,2085)
\dottedline{45}(4290,2445)(5190,2625)
\path(5078.214,2572.049)(5190.000,2625.000)(5066.447,2630.883)
\dottedline{45}(4290,2445)(3210,2625)
\path(3333.299,2634.864)(3210.000,2625.000)(3323.435,2575.680)
\dottedline{45}(4290,2445)(4290,2085)
\path(4260.000,2205.000)(4290.000,2085.000)(4320.000,2205.000)
\thicklines
\path(4200,240)(510,2085)
\blacken\thinlines
\path(660.935,2051.459)(510.000,2085.000)(627.394,1984.377)(660.935,2051.459)
\thicklines
\path(555,2085)(1095,2085)
\blacken\thinlines
\path(945.000,2047.500)(1095.000,2085.000)(945.000,2122.500)(945.000,2047.500)
\thicklines
\path(1095,2085)(1095,4200)
\blacken\thinlines
\path(1132.500,4050.000)(1095.000,4200.000)(1057.500,4050.000)(1132.500,4050.000)
\thicklines
\path(1095,4200)(3165,4200)
\blacken\thinlines
\path(3015.000,4162.500)(3165.000,4200.000)(3015.000,4237.500)(3015.000,4162.500)
\thicklines
\path(3165,4200)(5235,4200)
\blacken\thinlines
\path(5085.000,4162.500)(5235.000,4200.000)(5085.000,4237.500)(5085.000,4162.500)
\thicklines
\path(5235,4200)(7305,4200)
\blacken\thinlines
\path(7155.000,4162.500)(7305.000,4200.000)(7155.000,4237.500)(7155.000,4162.500)
\thicklines
\path(7305,4200)(7305,2085)
\blacken\thinlines
\path(7267.500,2235.000)(7305.000,2085.000)(7342.500,2235.000)(7267.500,2235.000)
\blacken\path(6420.000,2122.500)(6270.000,2085.000)(6420.000,2047.500)(6420.000,2122.500)
\thicklines
\path(6270,2085)(7305,2085)
\path(6270,2085)(4200,240)
\blacken\thinlines
\path(4287.026,367.800)(4200.000,240.000)(4336.928,311.811)(4287.026,367.800)
\put(510,1860){\makebox(0,0)[lb]{\smash{{\SetFigFont{8}{9.6}{\rmdefault}{\mddefault}{\updefault}$x$}}}}
\put(1050,1860){\makebox(0,0)[lb]{\smash{{\SetFigFont{8}{9.6}{\rmdefault}{\mddefault}{\updefault}$v_1$}}}}
\put(5190,1860){\makebox(0,0)[lb]{\smash{{\SetFigFont{8}{9.6}{\rmdefault}{\mddefault}{\updefault}$v_3$}}}}
\put(6225,1860){\makebox(0,0)[lb]{\smash{{\SetFigFont{8}{9.6}{\rmdefault}{\mddefault}{\updefault}$y$}}}}
\put(7260,1860){\makebox(0,0)[lb]{\smash{{\SetFigFont{8}{9.6}{\rmdefault}{\mddefault}{\updefault}$v_4$}}}}
\put(8475,1860){\makebox(0,0)[lb]{\smash{{\SetFigFont{8}{9.6}{\rmdefault}{\mddefault}{\updefault}$w$}}}}
\put(15,1860){\makebox(0,0)[lb]{\smash{{\SetFigFont{8}{9.6}{\rmdefault}{\mddefault}{\updefault}$v$}}}}
\put(3120,1860){\makebox(0,0)[lb]{\smash{{\SetFigFont{8}{9.6}{\rmdefault}{\mddefault}{\updefault}$v_2$}}}}
\put(3120,4335){\makebox(0,0)[lb]{\smash{{\SetFigFont{8}{9.6}{\rmdefault}{\mddefault}{\updefault}$u_1$}}}}
\put(15,2310){\makebox(0,0)[lb]{\smash{{\SetFigFont{8}{9.6}{\rmdefault}{\mddefault}{\updefault}$P_s(S)$}}}}
\put(4200,3615){\makebox(0,0)[lb]{\smash{{\SetFigFont{8}{9.6}{\rmdefault}{\mddefault}{\updefault}$C_2$}}}}
\put(6180,3615){\makebox(0,0)[lb]{\smash{{\SetFigFont{8}{9.6}{\rmdefault}{\mddefault}{\updefault}$C_3$}}}}
\put(2130,3615){\makebox(0,0)[lb]{\smash{{\SetFigFont{8}{9.6}{\rmdefault}{\mddefault}{\updefault}$C_1$}}}}
\put(7170,1275){\makebox(0,0)[lb]{\smash{{\SetFigFont{8}{9.6}{\rmdefault}{\mddefault}{\updefault}$J_h(S)$}}}}
\put(4110,4335){\makebox(0,0)[lb]{\smash{{\SetFigFont{8}{9.6}{\rmdefault}{\mddefault}{\updefault}$I(C_2)$}}}}
\put(5190,4335){\makebox(0,0)[lb]{\smash{{\SetFigFont{8}{9.6}{\rmdefault}{\mddefault}{\updefault}$u_2$}}}}
\put(3255,2985){\makebox(0,0)[lb]{\smash{{\SetFigFont{8}{9.6}{\rmdefault}{\mddefault}{\updefault}$J_t(C_2)$}}}}
\put(4065,2535){\makebox(0,0)[lb]{\smash{{\SetFigFont{8}{9.6}{\rmdefault}{\mddefault}{\updefault}$B(C_2)$}}}}
\put(4650,2985){\makebox(0,0)[lb]{\smash{{\SetFigFont{8}{9.6}{\rmdefault}{\mddefault}{\updefault}$J_h(C_2)$}}}}
\put(4155,15){\makebox(0,0)[lb]{\smash{{\SetFigFont{8}{9.6}{\rmdefault}{\mddefault}{\updefault}$z$}}}}
\put(1950,870){\makebox(0,0)[lb]{\smash{{\SetFigFont{8}{9.6}{\rmdefault}{\mddefault}{\updefault}$M_t(S)$}}}}
\put(5235,870){\makebox(0,0)[lb]{\smash{{\SetFigFont{8}{9.6}{\rmdefault}{\mddefault}{\updefault}$M_h(S)$}}}}
\end{picture}
}
\end{center}
\caption{Figure for Lemma \ref{lemthreefmid}.} \label{threefmidfig}
\end{figure}
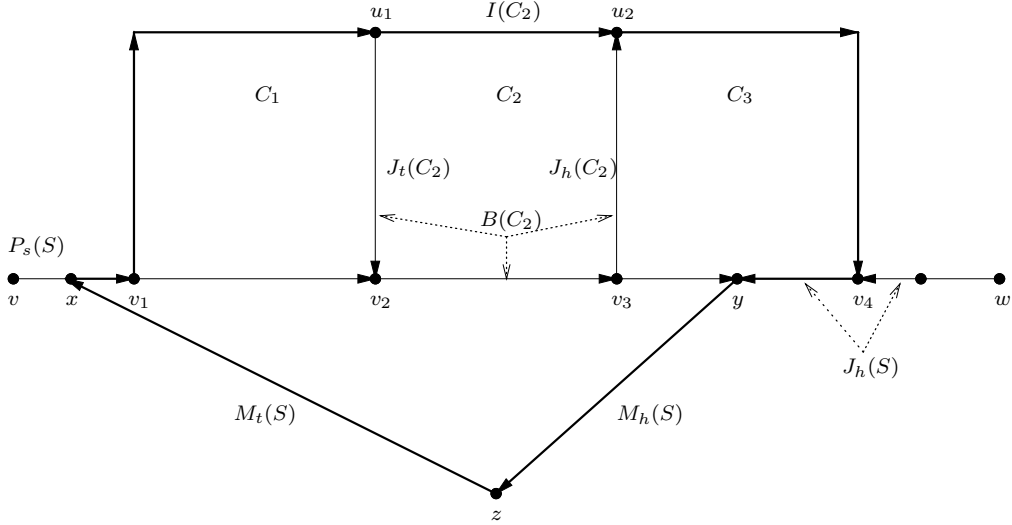

\begin{lem}
\label{lemthreefmid} Suppose that $S$ satisfies the induction
hypotheses, and $C_1,C_2,C_3$ were oriented $v_1 \rightarrow u_1
\rightarrow u_2 \rightarrow v_4$, $ u_1 \rightarrow v_2$, $v_3
\rightarrow u_2$ (see Figure \ref{threefmidfig}). Then $C_2$
satisfies the induction hypotheses. Moreover $|N(C_2)| = 2$.
\end{lem}

\begin{prf}
By Observation  \ref{winpath} the  orientation  $u_1 \rightarrow
v_2$  indicates that $l(C_1) < l(C_2)$ and the orientation $v_3
\rightarrow u_2$ indicates that $l(C_3) < l(C_2)$. It follows that
$U(C_2) \subset [u_1,u_2]$ so
\[ I(C_2) = [u_1 \rightarrow u_2], \ t(C_2) = u_1, \ h(C_2)=u_2 .\]
Since  $l(C_1) < l(C_2)$ and $l(C_3) < l(C_2)$:
\begin{equation}
\label{c1sc2in3} l(v_1,u_1)  + l(v_2,v_3) < l(u_1,u_2) + l(u_2,v_3)
+ l(v_1,v_2),
\end{equation}

\begin{equation}
\label{c3sc2in3}  l(u_2,v_4)+  l(v_2,v_3) < l(u_1,u_2)+l(v_2,u_1) +
l(v_3,v_4).
\end{equation}

\begin{description}
\item[A1]   By A3 on $S$ $x \in [v,v_2]$ and
$y \in [v_3,w]$. In Figure \ref{threefmidfig} the case $x \in
[v,v_1]$, $y\in [v_3,v_4]$ is described. By A4 for all these cases
we can define:
\[ M_t(C_2) = M_t(S) \cup [x \rightarrow v_1 \rightarrow u_1],  \ M_h(C_3) =[u_2 \rightarrow v_4 \rightarrow y] \cup M_h(S).\]
 Using A1 on $S$
\begin{eqnarray*}
l(M(C_2)) &=& l(M(S)) + l(x,v_1) + l(v_1 \rightarrow u_1)
+ l(u_2 \rightarrow v_4) +  l(v_4,y) \\
&\leq& 15l(P_c(S)) + 2l(P_s(S)) + l(I(S)) + l(x,v_1) + l(v_1
\rightarrow u_1) + l(u_2 \rightarrow v_4) +  l(v_4,y)\\
&\leq& 15l(P_c(S)) + 4l(v,v_1) + 3l(v_1 \rightarrow v_2
\rightarrow v_3 \rightarrow v_4) + 4l(v_4,w) +\\
&&l(v_1 \rightarrow u_1) + l(u_2 \rightarrow v_4).
\end{eqnarray*}
(The lengths of $[v,v_1]$ and $[v_4,w]$ are taken 4 times to satisfy
all four cases of locations for $x,y$.)

 From (\ref{c1sc2in3}) and (\ref{c3sc2in3}):
\begin{eqnarray*}
l(M(C_2)) &\leq& 15 l(P_c(S)) + 4l(v,v_1) + 4l(v_1,v_2) +
 2l(v_2,v_3)+ 3l(v_3,v_4) + \\
 && 4l(v_4,w)  + l(u_1,u_2) + l(u_2,v_3) + l(u_2,v_4)\\
 &\leq& 15 l(P_c(S)) + 4l(v,v_1) + 4l(v_1,v_2) +
 l(v_2,v_3)+ 4l(v_3,v_4) + \\
 && 4l(v_4,w)  + 2l(u_1,u_2) + l(u_2,v_3) + l(u_1,v_2).
\end{eqnarray*}
 By definition
\[ l(P_c(C_2)) = l(P_c(S)) + l(v,v_1) + l(v_1,v_2) +
l(v_3,v_4) + l(v_4,w) ,\] so that
\[ l(M(C_2)) \leq 15 l(P_c(C_2)) +l(v_2,v_3)  + 2l(u_1,u_2) + l(u_2,v_3) + l(u_1,v_2).\]
By definition
\[ l(P_s(C_2)) = l(u_1,v_2) + l(u_1,u_2) + l(u_2,v_3) ,\]
giving
\[ l(M(C_2)) \leq 15 l(P_c(C_2)) +l(v_2,v_3)  + l(u_1,u_2) + l(P_s(C_2)) .\]
 Since $C_2$ is a son of $S$, $l(v_2,v_3) \leq
l(P_s(C_2))$. $I(C_2) = [u_1 \rightarrow u_2]$ so
\[ l(M(C_2)) \leq 15 l(P_c(C_2)) +2l(P_s(C_2)) +l(I(C_2)).\]
\item[A2]
For all possible locations of $x$ and $y$
\[ B(C_2) = [u_1 \rightarrow v_2 \rightarrow v_3 \rightarrow u_2] .\]
>From the definition of $I(C_2)$ and since $C_2$ is a son of $S$,
\begin{eqnarray*}
l(B(C_2)) &=& l(u_1,v_2) +  l(v_3,u_2) +l(v_2,v_3)  \\
&=& l(P_s(C_2)) - l(I(C_2)) + l(v_2,v_3) \\
&\leq& l(P_s(C_2)) - l(I(C_2)) +   l(P_s(C_2)) \\
&=& 2l(P_s(C_2)) - l(I(C_2)).
\end{eqnarray*}

  \item [A3] Let $\hat{C}$ be a son of $C_2$. If
$P_f(\hat{C}) \subseteq [u_1,v_2]$
 since $[u_2,v_3] \subseteq P_s(C_1)$ and $l(C_1) < l(C_2)$, $\hat{C}$
should have been a son of $C_1$ . In a similar way if $P_f(\hat{C})
\subseteq [u_2,v_3]$ since $l(C_3) < l(C_2)$ $\hat{C}$ would have
been a son of $C_3$. Therefore $P_f(\hat{C}) \cap [u_1,u_2] \neq
\phi$.
 \item [A4] In this case
$J_t(C_2) = [u_1 \rightarrow v_2]$ and $J_h(C_2) = [v_3 \rightarrow
u_2]$. Obviously for every son $\hat{C}$ of $C_2$ $P_f(\hat{C})
\subset P_s(C_2) = J_t(C_2) \cup I(C_2) \cup J_h(C_2) $.
 \item [A5] $M(C_2)$ is
contained in $P_s(C_2) \cup C_1 \cup C_3 \cup M(S)$. Since $l(C_1) <
l(C_2)$ and $l(C_3) < l(C_2)$ the condition is satisfied.
 \item[A6] $M(C_2)$ is contained in $P_s(C_2)$, $C_1$ ,$C_3$(two
brothers of $C_2$), $S$  and $M(S)$, $N(C_2) = \{ C_1,C_3 \} $.

\end{description}

\end{prf}

\subsection{Four  brothers} \label{secfour} In this
section we consider four neighbor cycles $C_1,\ldots,C_4$
 where the four brothers were all oriented forwards.\footnote{For example,  such
orientation can be the result of DIRECT-MANY when $n=4$, $C_2$ and
$C_3$ are heavy and $l_1=l_2=1$.} In the first lemma we handle the
orientation applied by the algorithm when $l(C_1) < l(C_2) < l(C_3)
< l(C_4)$. In the second lemma we handle the orientation applied by
the algorithm when $l(C_1) < l(C_2) < l(C_3)$ and $l(C_4) < l(C_3)$.
Other equivalent cases can be handled by similar lemmas.

 Mark \bi
\item $u_1:= $ the end node of the path $P_s(C_1) \cap P_s(C_2)$, such that $u_1 \not\in S$.
\item $u_2:= $ the end node of the path $P_s(C_2) \cap P_s(C_3)$, such that $u_2 \not\in S$.
\item $u_3:= $ the end node of the path $P_s(C_3) \cap P_s(C_4)$, such that $u_3 \not\in S$.
\item $v_1:= S \cap (P_s(C_1) \backslash P_s(C_2))$.
\item $v_2:= S \cap (P_s(C_1) \cap P_s(C_2))$.
\item $v_3:= S \cap (P_s(C_2) \cap P_s(C_3))$.
\item $v_4:= S \cap (P_s(C_3) \cap P_s(C_4))$.
\item $v_5:= S \cap (P_s(C_4) \backslash P_s(C_3))$.
\ei

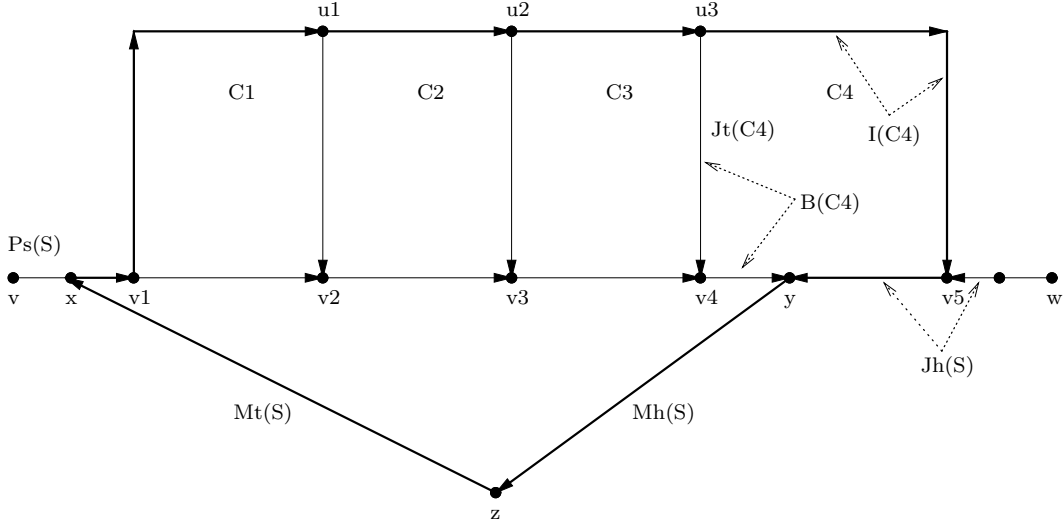
\begin{figure}
\begin{center}
\setlength{\unitlength}{0.00061242in}
\begingroup\makeatletter\ifx\SetFigFont\undefined%
\gdef\SetFigFont#1#2#3#4#5{%
  \reset@font\fontsize{#1}{#2pt}%
  \fontfamily{#3}\fontseries{#4}\fontshape{#5}%
  \selectfont}%
\fi\endgroup%
{\renewcommand{\dashlinestretch}{30}
\begin{picture}(9078,4500)(0,-10)
\put(60,2085){\blacken\ellipse{90}{90}}
\put(60,2085){\ellipse{90}{90}}
\put(555,2085){\blacken\ellipse{90}{90}}
\put(555,2085){\ellipse{90}{90}}
\put(4335,2085){\blacken\ellipse{90}{90}}
\put(4335,2085){\ellipse{90}{90}}
\put(8970,2085){\blacken\ellipse{90}{90}}
\put(8970,2085){\ellipse{90}{90}}
\put(4335,4200){\blacken\ellipse{90}{90}}
\put(4335,4200){\ellipse{90}{90}}
\put(2715,4200){\blacken\ellipse{90}{90}}
\put(2715,4200){\ellipse{90}{90}}
\put(2715,2085){\blacken\ellipse{90}{90}}
\put(2715,2085){\ellipse{90}{90}}
\put(5955,2085){\blacken\ellipse{90}{90}}
\put(5955,2085){\ellipse{90}{90}}
\put(8520,2085){\blacken\ellipse{90}{90}}
\put(8520,2085){\ellipse{90}{90}}
\put(6720,2085){\blacken\ellipse{90}{90}}
\put(6720,2085){\ellipse{90}{90}}
\put(8070,2085){\blacken\ellipse{90}{90}}
\put(8070,2085){\ellipse{90}{90}}
\put(5955,4200){\blacken\ellipse{90}{90}}
\put(5955,4200){\ellipse{90}{90}}
\put(4200,240){\blacken\ellipse{90}{90}}
\put(4200,240){\ellipse{90}{90}}
\put(1095,2085){\blacken\ellipse{90}{90}}
\put(1095,2085){\ellipse{90}{90}}
\path(60,2085)(555,2085)
\path(1095,2085)(2715,2085)
\blacken\path(2565.000,2047.500)(2715.000,2085.000)(2565.000,2122.500)(2565.000,2047.500)
\path(4335,4200)(4335,2085)
\blacken\path(4297.500,2235.000)(4335.000,2085.000)(4372.500,2235.000)(4297.500,2235.000)
\path(2715,4200)(2715,2085)
\blacken\path(2677.500,2235.000)(2715.000,2085.000)(2752.500,2235.000)(2677.500,2235.000)
\path(2715,2085)(4335,2085)
\blacken\path(4185.000,2047.500)(4335.000,2085.000)(4185.000,2122.500)(4185.000,2047.500)
\path(5955,4200)(5955,2085)
\blacken\path(5917.500,2235.000)(5955.000,2085.000)(5992.500,2235.000)(5917.500,2235.000)
\path(4335,2085)(5955,2085)
\blacken\path(5805.000,2047.500)(5955.000,2085.000)(5805.000,2122.500)(5805.000,2047.500)
\dottedline{60}(8025,1455)(7530,2040)
\path(7630.415,1967.772)(7530.000,2040.000)(7584.611,1929.015)
\path(8970,2085)(8520,2085)
\path(8520,2085)(8070,2085)
\blacken\path(8220.000,2122.500)(8070.000,2085.000)(8220.000,2047.500)(8220.000,2122.500)
\dottedline{45}(8025,1455)(8340,2040)
\path(8309.522,1920.120)(8340.000,2040.000)(8256.694,1948.566)
\dottedline{45}(6765,2760)(6000,3075)
\path(6122.384,3057.050)(6000.000,3075.000)(6099.539,3001.570)
\dottedline{45}(6765,2760)(6315,2175)
\path(6364.387,2288.406)(6315.000,2175.000)(6411.944,2251.824)
\dottedline{45}(7575,3480)(7125,4155)
\path(7216.526,4071.795)(7125.000,4155.000)(7166.603,4038.513)
\dottedline{45}(7575,3480)(8025,3795)
\path(7943.896,3701.608)(8025.000,3795.000)(7909.488,3750.761)
\path(5955,2085)(6720,2085)
\blacken\path(6570.000,2047.500)(6720.000,2085.000)(6570.000,2122.500)(6570.000,2047.500)
\thicklines
\path(4200,240)(510,2085)
\blacken\thinlines
\path(660.935,2051.459)(510.000,2085.000)(627.394,1984.377)(660.935,2051.459)
\thicklines
\path(555,2085)(1095,2085)
\blacken\thinlines
\path(945.000,2047.500)(1095.000,2085.000)(945.000,2122.500)(945.000,2047.500)
\thicklines
\path(1095,2085)(1095,4200)
\blacken\thinlines
\path(1132.500,4050.000)(1095.000,4200.000)(1057.500,4050.000)(1132.500,4050.000)
\thicklines
\path(1095,4200)(2715,4200)
\blacken\thinlines
\path(2565.000,4162.500)(2715.000,4200.000)(2565.000,4237.500)(2565.000,4162.500)
\thicklines
\path(2715,4200)(4335,4200)
\blacken\thinlines
\path(4185.000,4162.500)(4335.000,4200.000)(4185.000,4237.500)(4185.000,4162.500)
\thicklines
\path(4335,4200)(5955,4200)
\blacken\thinlines
\path(5805.000,4162.500)(5955.000,4200.000)(5805.000,4237.500)(5805.000,4162.500)
\thicklines
\path(5955,4200)(8070,4200)
\blacken\thinlines
\path(7920.000,4162.500)(8070.000,4200.000)(7920.000,4237.500)(7920.000,4162.500)
\thicklines
\path(8070,4200)(8070,2085)
\blacken\thinlines
\path(8032.500,2235.000)(8070.000,2085.000)(8107.500,2235.000)(8032.500,2235.000)
\thicklines
\path(8070,2085)(6720,2085)
\blacken\thinlines
\path(6870.000,2122.500)(6720.000,2085.000)(6870.000,2047.500)(6870.000,2122.500)
\thicklines
\path(6720,2085)(4200,240)
\blacken\thinlines
\path(4298.877,358.868)(4200.000,240.000)(4343.182,298.353)(4298.877,358.868)
\put(510,1860){\makebox(0,0)[lb]{\smash{{\SetFigFont{8}{9.6}{\rmdefault}{\mddefault}{\updefault}x}}}}
\put(1050,1860){\makebox(0,0)[lb]{\smash{{\SetFigFont{8}{9.6}{\rmdefault}{\mddefault}{\updefault}v1}}}}
\put(6675,1860){\makebox(0,0)[lb]{\smash{{\SetFigFont{8}{9.6}{\rmdefault}{\mddefault}{\updefault}y}}}}
\put(15,1860){\makebox(0,0)[lb]{\smash{{\SetFigFont{8}{9.6}{\rmdefault}{\mddefault}{\updefault}v}}}}
\put(2670,1860){\makebox(0,0)[lb]{\smash{{\SetFigFont{8}{9.6}{\rmdefault}{\mddefault}{\updefault}v2}}}}
\put(2670,4335){\makebox(0,0)[lb]{\smash{{\SetFigFont{8}{9.6}{\rmdefault}{\mddefault}{\updefault}u1}}}}
\put(4290,4335){\makebox(0,0)[lb]{\smash{{\SetFigFont{8}{9.6}{\rmdefault}{\mddefault}{\updefault}u2}}}}
\put(4155,15){\makebox(0,0)[lb]{\smash{{\SetFigFont{8}{9.6}{\rmdefault}{\mddefault}{\updefault}z}}}}
\put(15,2310){\makebox(0,0)[lb]{\smash{{\SetFigFont{8}{9.6}{\rmdefault}{\mddefault}{\updefault}Ps(S)}}}}
\put(3525,3615){\makebox(0,0)[lb]{\smash{{\SetFigFont{8}{9.6}{\rmdefault}{\mddefault}{\updefault}C2}}}}
\put(4290,1860){\makebox(0,0)[lb]{\smash{{\SetFigFont{8}{9.6}{\rmdefault}{\mddefault}{\updefault}v3}}}}
\put(1905,3615){\makebox(0,0)[lb]{\smash{{\SetFigFont{8}{9.6}{\rmdefault}{\mddefault}{\updefault}C1}}}}
\put(8925,1860){\makebox(0,0)[lb]{\smash{{\SetFigFont{8}{9.6}{\rmdefault}{\mddefault}{\updefault}w}}}}
\put(7845,1275){\makebox(0,0)[lb]{\smash{{\SetFigFont{8}{9.6}{\rmdefault}{\mddefault}{\updefault}Jh(S)}}}}
\put(8025,1860){\makebox(0,0)[lb]{\smash{{\SetFigFont{8}{9.6}{\rmdefault}{\mddefault}{\updefault}v5}}}}
\put(5910,1860){\makebox(0,0)[lb]{\smash{{\SetFigFont{8}{9.6}{\rmdefault}{\mddefault}{\updefault}v4}}}}
\put(5910,4335){\makebox(0,0)[lb]{\smash{{\SetFigFont{8}{9.6}{\rmdefault}{\mddefault}{\updefault}u3}}}}
\put(7035,3615){\makebox(0,0)[lb]{\smash{{\SetFigFont{8}{9.6}{\rmdefault}{\mddefault}{\updefault}C4}}}}
\put(5145,3615){\makebox(0,0)[lb]{\smash{{\SetFigFont{8}{9.6}{\rmdefault}{\mddefault}{\updefault}C3}}}}
\put(6045,3300){\makebox(0,0)[lb]{\smash{{\SetFigFont{8}{9.6}{\rmdefault}{\mddefault}{\updefault}Jt(C4)}}}}
\put(6810,2670){\makebox(0,0)[lb]{\smash{{\SetFigFont{8}{9.6}{\rmdefault}{\mddefault}{\updefault}B(C4)}}}}
\put(7395,3255){\makebox(0,0)[lb]{\smash{{\SetFigFont{8}{9.6}{\rmdefault}{\mddefault}{\updefault}I(C4)}}}}
\put(1950,870){\makebox(0,0)[lb]{\smash{{\SetFigFont{8}{9.6}{\rmdefault}{\mddefault}{\updefault}Mt(S)}}}}
\put(5370,870){\makebox(0,0)[lb]{\smash{{\SetFigFont{8}{9.6}{\rmdefault}{\mddefault}{\updefault}Mh(S)}}}}
\end{picture}
}
\end{center}
\caption{Figure for Lemma \ref{lemfourfs}.} \label{fourfsfig}
\end{figure}

\begin{lem}
\label{lemfourfs} Suppose that $S$ satisfies the induction
hypotheses and $C_1,C_2,C_3,C_4$ were oriented $v_1 \rightarrow u_1
\rightarrow u_2 \rightarrow u_3 \rightarrow v_5$, $ u_1 \rightarrow
v_2$, $u_2 \rightarrow v_3$ and $u_3 \rightarrow v_4$ (see Figure
\ref{fourfsfig}). Then $C_4$ satisfies the induction hypotheses.
Moreover $|N(C_4)| = 3$.
\end{lem}

\begin{prf}
By Observation  \ref{winpath} the  orientation   $u_1 \rightarrow
v_2$ indicates that $l(C_1) < l(C_2)$, the orientation $u_2
\rightarrow v_3$ indicates that $l(C_2) < l(C_3)$ and the
orientation $u_3 \rightarrow v_4$ indicates that $l(C_3) < l(C_4)$.
It follows that  $U(C_4) \subseteq [u_3,v_5]$, so
\[ I(C_4) = [u_3 \rightarrow v_5], \ t(C_4) = u_3, \ h(C_4)=v_5 .\]
 Since  $l(C_1) < l(C_4)$ , $l(C_2) < l(C_4)$ and $l(C_3) < l(C_4)$
\begin{equation}
\label{c1sc4} l(v_1,u_1) + l(u_1,v_2) + l(v_4,v_5) < l(P_s(C_4))
+l(v_1,v_2) ,
\end{equation}
\begin{equation}
\label{c2sc4} l(u_1,u_2) + l(u_1,v_2)+ l(u_2,v_3)  + l(v_4,v_5) <
l(P_s(C_4)) +l(v_2,v_3) ,
\end{equation}
\begin{equation}
\label{c3sc4}  l(u_2,u_3)+ l(u_2,v_3) + l(v_4,v_5) < l(u_3,v_5) +
l(v_3,v_4) .
\end{equation}

\begin{description}
\item [A1]  By A3 on $S$ $x \in [v,v_2]$  and
$y \in [v_4,w]$. By A4 for all these cases we can define:
\[ M_t(C_4) = M_t(S) \cup [x \rightarrow v_1 \rightarrow u_1 \rightarrow u_2 \rightarrow u_3], \ M_h(C_3) =[v_5 \rightarrow y] \cup M_h(S) .\]
Using A1 on $S$
\begin{eqnarray*}
l(M(C_4)) &=& l(M(S)) + l(x,v_1) + l(v_1 \rightarrow u_1
\rightarrow u_2 \rightarrow u_3) + l(v_5,y) \\
&\leq& 15l(P_c(S)) + 2l(P_s(S)) + l(I(S)) + l(x,v_1) + l(v_1
\rightarrow u_1 \rightarrow u_2 \rightarrow u_3) + l(v_5,y)\\
&\leq& 15l(P_c(S)) + 4l(v,v_1) + 3l(v_1 \rightarrow v_2 \rightarrow
v_3 \rightarrow v_4 \rightarrow v_5) + 4l(v_5,w) +\\
&&l(v_1 \rightarrow u_1 \rightarrow u_2 \rightarrow u_3).
\end{eqnarray*}
(The lengths of $[v,v_1]$ and $[v_5,w]$ are taken 4 times to satisfy
all four cases of locations for $x,y$.)

>From (\ref{c1sc4}),(\ref{c2sc4}) and (\ref{c3sc4})
\begin{eqnarray*}
 l(M(C_4)) &\leq& 15 l(P_c(S)) + 4l(v,v_1) + 4l(v_1,v_2) +
 3l(v_2,v_3)+ 3l(v_3,v_4) + 2l(v_4,v_5) +\\
 && 4l(v_5,w)  + l(P_s(C_4)) + l(u_1,u_2)+l(u_2,u_3)\\
&\leq& 15 l(P_c(S)) + 4l(v,v_1) + 4l(v_1,v_2) +
 4l(v_2,v_3)+ 3l(v_3,v_4) + l(v_4,v_5) +\\
 && 4l(v_5,w)  + 2l(P_s(C_4)) +l(u_2,u_3)\\
&\leq& 15 l(P_c(S)) + 4l(v,v_1) + 4l(v_1,v_2) +
 4l(v_2,v_3)+ 4l(v_3,v_4) +
 4l(v_5,w)+  \\
 && 2l(P_s(C_4)) +l(u_3,v_5).
\end{eqnarray*}

By definition
\[ l(P_c(C_4)) = l(P_c(S)) + l(v,v_1) + l(v_1,v_2) + l(v_2,v_3) +
l(v_3,v_4) + l(v_5,w) ,\] giving:
\[ l(M(C_4)) \leq 15l(P_c(C_4)) + 2l(P_s(C_4)) +l(u_3,v_5).\]
>From $I(C_4)=[u_3 \rightarrow v_5]$ we get
\[  l(M(C_4)) \leq 15l(P_c(C_4)) + 2l(P_s(C_4)) +l(I(C_4)).\]

\item[A2] When $y \in [v_5,w]$ we define $B(C_4) = [u_3 \rightarrow v_4 \rightarrow v_5]$,
 when $y \in [v_4,v_5]$ we define
$B(C_4) = [u_3 \rightarrow v_4 \rightarrow y]$. In both cases
\[ l(B(C_4)) \leq l(u_3,v_4) + l(v_4,v_5).\]
>From the definition of $I(C_4)$ and since $C_4$ is a son of $S$ we
get that
\begin{eqnarray*}
l(B(C_4)) &\leq& l(u_3,v_4) + l(v_4,v_5)\\
&=& l(P_s(C_4)) - l(I(C_4)) + l(v_4,v_5) \\
&\leq& l(P_s(C_4)) - l(I(C_4)) + l(P_s(C_4))\\
&=& 2l(P_s(C_4)) - l(I(C_4)).
\end{eqnarray*}

  \item [A3] Let $\hat{C}$ be a son of $C_4$.  If
$P_f(\hat{C}) \subseteq [u_3,v_4]$
 then since $[u_3,v_4] \subseteq P_s(C_3)$ and $l(C_3) < l(C_4)$, $\hat{C}$
should have been a son of $C_3$ . Therefore $P_f(\hat{C}) \cap
[u_3,v_5] \neq \phi$.
 \item [A4] In this case
$J_t(C_4) = [u_3 \rightarrow v_4]$. Obviously for every son
$\hat{C}$ of $C_4$ $P_f(\hat{C}) \subset P_s(C_4) =  I(C_4) \cup
J_t(C_4)$.

 \item [A5] $M(C_4)$ is
contained in $P_s(C_4) \cup C_1 \cup C_2 \cup C_3 \cup M(S)$. Since
$l(C_i) < l(C_4), i=1,2,3$ the condition is satisfied.
 \item[A6] $M(C_4)$ is contained in $P_s(C_4)$, $C_1$
 ,$C_2$,$C_3$(three brothers of $C_4$), $S$  and $M(S)$, so
 $N(C_4) = \{ C_1,C_2,C_3 \}$.

\end{description}

\end{prf}

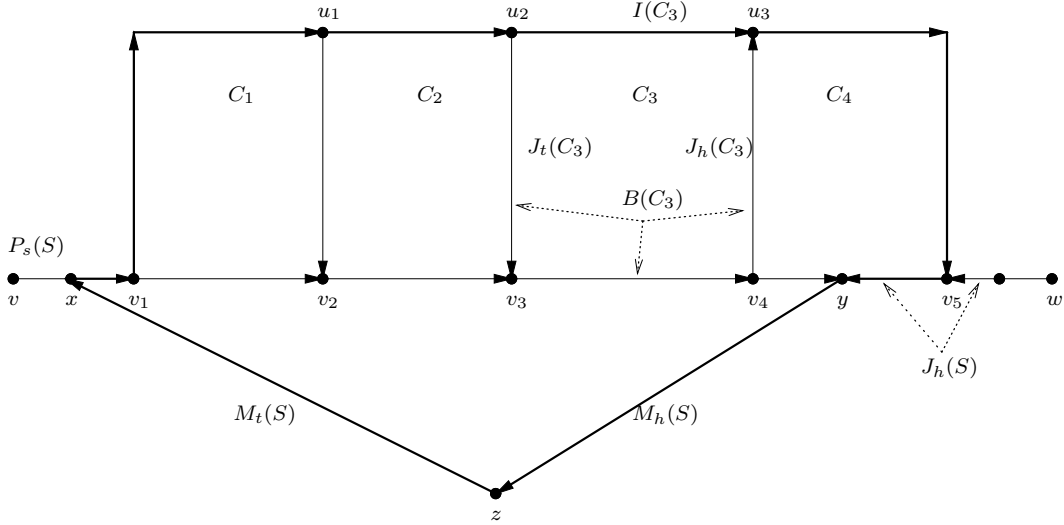
\begin{figure}
\begin{center}
\setlength{\unitlength}{0.00061242in}
\begingroup\makeatletter\ifx\SetFigFont\undefined%
\gdef\SetFigFont#1#2#3#4#5{%
  \reset@font\fontsize{#1}{#2pt}%
  \fontfamily{#3}\fontseries{#4}\fontshape{#5}%
  \selectfont}%
\fi\endgroup%
{\renewcommand{\dashlinestretch}{30}
\begin{picture}(9078,4509)(0,-10)
\put(60,2085){\blacken\ellipse{90}{90}}
\put(60,2085){\ellipse{90}{90}}
\put(555,2085){\blacken\ellipse{90}{90}}
\put(555,2085){\ellipse{90}{90}}
\put(1095,2085){\blacken\ellipse{90}{90}}
\put(1095,2085){\ellipse{90}{90}}
\put(4335,2085){\blacken\ellipse{90}{90}}
\put(4335,2085){\ellipse{90}{90}}
\put(8970,2085){\blacken\ellipse{90}{90}}
\put(8970,2085){\ellipse{90}{90}}
\put(4335,4200){\blacken\ellipse{90}{90}}
\put(4335,4200){\ellipse{90}{90}}
\put(2715,4200){\blacken\ellipse{90}{90}}
\put(2715,4200){\ellipse{90}{90}}
\put(2715,2085){\blacken\ellipse{90}{90}}
\put(2715,2085){\ellipse{90}{90}}
\put(6405,2085){\blacken\ellipse{90}{90}}
\put(6405,2085){\ellipse{90}{90}}
\put(8520,2085){\blacken\ellipse{90}{90}}
\put(8520,2085){\ellipse{90}{90}}
\put(8070,2085){\blacken\ellipse{90}{90}}
\put(8070,2085){\ellipse{90}{90}}
\put(6405,4200){\blacken\ellipse{90}{90}}
\put(6405,4200){\ellipse{90}{90}}
\put(7170,2085){\blacken\ellipse{90}{90}}
\put(7170,2085){\ellipse{90}{90}}
\put(4200,240){\blacken\ellipse{90}{90}}
\put(4200,240){\ellipse{90}{90}}
\path(60,2085)(555,2085)
\path(1095,2085)(2715,2085)
\blacken\path(2565.000,2047.500)(2715.000,2085.000)(2565.000,2122.500)(2565.000,2047.500)
\path(4335,4200)(4335,2085)
\blacken\path(4297.500,2235.000)(4335.000,2085.000)(4372.500,2235.000)(4297.500,2235.000)
\path(2715,4200)(2715,2085)
\blacken\path(2677.500,2235.000)(2715.000,2085.000)(2752.500,2235.000)(2677.500,2235.000)
\path(2715,2085)(4335,2085)
\blacken\path(4185.000,2047.500)(4335.000,2085.000)(4185.000,2122.500)(4185.000,2047.500)
\path(4335,2085)(6405,2085)
\blacken\path(6255.000,2047.500)(6405.000,2085.000)(6255.000,2122.500)(6255.000,2047.500)
\dottedline{60}(8025,1455)(7530,2040)
\path(7630.415,1967.772)(7530.000,2040.000)(7584.611,1929.015)
\path(8970,2085)(8520,2085)
\path(8520,2085)(8070,2085)
\blacken\path(8220.000,2122.500)(8070.000,2085.000)(8220.000,2047.500)(8220.000,2122.500)
\dottedline{45}(8025,1455)(8340,2040)
\path(8309.522,1920.120)(8340.000,2040.000)(8256.694,1948.566)
\path(6405,2085)(7170,2085)
\blacken\path(7020.000,2047.500)(7170.000,2085.000)(7020.000,2122.500)(7020.000,2047.500)
\blacken\path(6442.500,4050.000)(6405.000,4200.000)(6367.500,4050.000)(6442.500,4050.000)
\path(6405,4200)(6405,2085)
\dottedline{45}(5460,2580)(4380,2715)
\path(4502.794,2729.884)(4380.000,2715.000)(4495.352,2670.347)
\dottedline{45}(5460,2580)(5415,2130)
\path(5397.089,2252.390)(5415.000,2130.000)(5456.792,2246.419)
\dottedline{45}(5460,2580)(6315,2670)
\path(6198.800,2627.603)(6315.000,2670.000)(6192.519,2687.273)
\thicklines
\path(4200,240)(510,2085)
\blacken\thinlines
\path(660.935,2051.459)(510.000,2085.000)(627.394,1984.377)(660.935,2051.459)
\thicklines
\path(555,2085)(1095,2085)
\blacken\thinlines
\path(945.000,2047.500)(1095.000,2085.000)(945.000,2122.500)(945.000,2047.500)
\thicklines
\path(1095,2085)(1095,4200)
\blacken\thinlines
\path(1132.500,4050.000)(1095.000,4200.000)(1057.500,4050.000)(1132.500,4050.000)
\thicklines
\path(1095,4200)(2715,4200)
\blacken\thinlines
\path(2565.000,4162.500)(2715.000,4200.000)(2565.000,4237.500)(2565.000,4162.500)
\thicklines
\path(2715,4200)(4335,4200)
\blacken\thinlines
\path(4185.000,4162.500)(4335.000,4200.000)(4185.000,4237.500)(4185.000,4162.500)
\thicklines
\path(4335,4200)(6405,4200)
\blacken\thinlines
\path(6255.000,4162.500)(6405.000,4200.000)(6255.000,4237.500)(6255.000,4162.500)
\thicklines
\path(6405,4200)(8070,4200)
\blacken\thinlines
\path(7920.000,4162.500)(8070.000,4200.000)(7920.000,4237.500)(7920.000,4162.500)
\thicklines
\path(8070,4200)(8070,2085)
\blacken\thinlines
\path(8032.500,2235.000)(8070.000,2085.000)(8107.500,2235.000)(8032.500,2235.000)
\thicklines
\path(8070,2085)(7170,2085)
\blacken\thinlines
\path(7320.000,2122.500)(7170.000,2085.000)(7320.000,2047.500)(7320.000,2122.500)
\thicklines
\path(7170,2085)(4200,240)
\blacken\thinlines
\path(4307.628,351.007)(4200.000,240.000)(4347.204,287.298)(4307.628,351.007)
\put(510,1860){\makebox(0,0)[lb]{\smash{{\SetFigFont{8}{9.6}{\rmdefault}{\mddefault}{\updefault}$x$}}}}
\put(1050,1860){\makebox(0,0)[lb]{\smash{{\SetFigFont{8}{9.6}{\rmdefault}{\mddefault}{\updefault}$v_1$}}}}
\put(15,1860){\makebox(0,0)[lb]{\smash{{\SetFigFont{8}{9.6}{\rmdefault}{\mddefault}{\updefault}$v$}}}}
\put(2670,1860){\makebox(0,0)[lb]{\smash{{\SetFigFont{8}{9.6}{\rmdefault}{\mddefault}{\updefault}$v_2$}}}}
\put(2670,4335){\makebox(0,0)[lb]{\smash{{\SetFigFont{8}{9.6}{\rmdefault}{\mddefault}{\updefault}$u_1$}}}}
\put(4290,4335){\makebox(0,0)[lb]{\smash{{\SetFigFont{8}{9.6}{\rmdefault}{\mddefault}{\updefault}$u_2$}}}}
\put(4155,15){\makebox(0,0)[lb]{\smash{{\SetFigFont{8}{9.6}{\rmdefault}{\mddefault}{\updefault}$z$}}}}
\put(15,2310){\makebox(0,0)[lb]{\smash{{\SetFigFont{8}{9.6}{\rmdefault}{\mddefault}{\updefault}$P_s(S)$}}}}
\put(3525,3615){\makebox(0,0)[lb]{\smash{{\SetFigFont{8}{9.6}{\rmdefault}{\mddefault}{\updefault}$C_2$}}}}
\put(4290,1860){\makebox(0,0)[lb]{\smash{{\SetFigFont{8}{9.6}{\rmdefault}{\mddefault}{\updefault}$v_3$}}}}
\put(1905,3615){\makebox(0,0)[lb]{\smash{{\SetFigFont{8}{9.6}{\rmdefault}{\mddefault}{\updefault}$C_1$}}}}
\put(8925,1860){\makebox(0,0)[lb]{\smash{{\SetFigFont{8}{9.6}{\rmdefault}{\mddefault}{\updefault}$w$}}}}
\put(7845,1275){\makebox(0,0)[lb]{\smash{{\SetFigFont{8}{9.6}{\rmdefault}{\mddefault}{\updefault}$J_h(S)$}}}}
\put(8025,1860){\makebox(0,0)[lb]{\smash{{\SetFigFont{8}{9.6}{\rmdefault}{\mddefault}{\updefault}$v_5$}}}}
\put(6360,1860){\makebox(0,0)[lb]{\smash{{\SetFigFont{8}{9.6}{\rmdefault}{\mddefault}{\updefault}$v_4$}}}}
\put(6360,4335){\makebox(0,0)[lb]{\smash{{\SetFigFont{8}{9.6}{\rmdefault}{\mddefault}{\updefault}$u_3$}}}}
\put(7035,3615){\makebox(0,0)[lb]{\smash{{\SetFigFont{8}{9.6}{\rmdefault}{\mddefault}{\updefault}$C_4$}}}}
\put(5370,3615){\makebox(0,0)[lb]{\smash{{\SetFigFont{8}{9.6}{\rmdefault}{\mddefault}{\updefault}$C_3$}}}}
\put(7125,1860){\makebox(0,0)[lb]{\smash{{\SetFigFont{8}{9.6}{\rmdefault}{\mddefault}{\updefault}$y$}}}}
\put(5370,4335){\makebox(0,0)[lb]{\smash{{\SetFigFont{8}{9.6}{\rmdefault}{\mddefault}{\updefault}$I(C_3)$}}}}
\put(4470,3165){\makebox(0,0)[lb]{\smash{{\SetFigFont{8}{9.6}{\rmdefault}{\mddefault}{\updefault}$J_t(C_3)$}}}}
\put(5820,3165){\makebox(0,0)[lb]{\smash{{\SetFigFont{8}{9.6}{\rmdefault}{\mddefault}{\updefault}$J_h(C_3)$}}}}
\put(5280,2715){\makebox(0,0)[lb]{\smash{{\SetFigFont{8}{9.6}{\rmdefault}{\mddefault}{\updefault}$B(C_3)$}}}}
\put(1950,870){\makebox(0,0)[lb]{\smash{{\SetFigFont{8}{9.6}{\rmdefault}{\mddefault}{\updefault}$M_t(S)$}}}}
\put(5370,870){\makebox(0,0)[lb]{\smash{{\SetFigFont{8}{9.6}{\rmdefault}{\mddefault}{\updefault}$M_h(S)$}}}}
\end{picture}
}
\end{center}
\caption{Figure for Lemma \ref{lemfourfmid}.} \label{fourfmidfig}
\end{figure}

\begin{lem}
\label{lemfourfmid}
 Suppose $S$ satisfies the induction hypotheses, and $C_1,C_2,C_3,C_4$ were oriented $v_1 \rightarrow u_1 \rightarrow
u_2 \rightarrow u_3 \rightarrow v_5$, $ u_1 \rightarrow v_2$, $u_2
\rightarrow v_3$ and $v_4 \rightarrow u_3$ (see Figure
\ref{fourfmidfig}). Then $C_3$ satisfies the induction hypotheses.
Moreover $|N(C_3)| = 3$.
\end{lem}

\begin{prf} By Observation  \ref{winpath}  the orientation
 $u_1 \rightarrow v_2$ indicates that $l(C_1) <
l(C_2)$, the orientation of the path  $u_2 \rightarrow v_3$
indicates that $l(C_2) < l(C_3)$ and  the orientation $v_4
\rightarrow u_3$ indicates that $l(C_4) < l(C_3)$. It follows that
 $U(C_3) \subseteq [u_2,u_3]$
(even if it is only one point) so
\[ I(C_3) = [u_2 \rightarrow u_3], \ t(C_3) = u_2, \ h(C_3)=u_3 .\]
 Since  $l(C_1) < l(C_3)$, $l(C_2) < l(C_3)$ and $l(C_4) < l(C_3)$
\begin{equation}
\label{c1sc3in4} l(v_1,u_1) + l(u_1,v_2) + l(v_3,v_4) < l(P_s(C_3))
+l(v_1,v_2) ,
\end{equation}
\begin{equation}
\label{c2sc3in4} l(u_1,u_2) + l(u_1,v_2)+   l(v_3,v_4) < l(u_2,u_3)
+ l(u_3,v_4) +l(v_2,v_3) ,
\end{equation}
\begin{equation}
\label{c4sc3in4}   l(u_3,v_5) + l(v_3,v_4) < l(u_2,u_3)+ l(u_2,v_3)
+ l(v_4,v_5)  .
\end{equation}

\begin{description}
\item[A1]  By A3 on $S$ $x \in [v,v_2]$ and $y \in
[v_4,w]$. By A4 for all these cases we can define:
\[ M_t(C_3) = M_t(S) \cup [x \rightarrow v_1 \rightarrow u_1 \rightarrow u_2],
\ M_h(C_3) = [u_3 \rightarrow v_5 \rightarrow y] \cup M_h(S) .\]
 Using A1 on $S$
\begin{eqnarray*}
l(M(C_3)) &=& l(M(S)) + l(x,v_1) + l(v_1 \rightarrow u_1
\rightarrow u_2) + l( u_3 \rightarrow v_5) + l(v_5,y) \\
&\leq& 15l(P_c(S)) + 2l(P_s(S)) + l(I(S)) + l(x,v_1) + l(v_1
\rightarrow u_1 \rightarrow u_2 ) + l( u_3 \rightarrow v_5)+ l(v_5,y)\\
&\leq& 15l(P_c(S)) + 4l(v,v_1) + 3l(v_1 \rightarrow v_2 \rightarrow
v_3 \rightarrow v_4 \rightarrow v_5) + 4l(v_5,w) +\\
&&l(v_1 \rightarrow u_1 \rightarrow u_2 ) + l( u_3 \rightarrow v_5).
\end{eqnarray*}
(The lengths of $[v,v_1]$ and $[v_5,w]$ are taken 4 times to satisfy
all four cases of locations for $x,y$.)

>From (\ref{c1sc3in4}),(\ref{c2sc3in4}) and (\ref{c4sc3in4})
\begin{eqnarray*}
 l(M(C_3)) &\leq& 15l(P_c(S)) + 4l(v,v_1) + 4l(v_1,v_2) +
 3l(v_2,v_3)+ 2l(v_3,v_4) + 3l(v_4,v_5) +
 4l(v_5,w) \\
 && + l(P_s(C_3)) + l(u_1,u_2)+l(u_3,v_5)\\
 &\leq& 15l(P_c(S)) + 4l(v,v_1) + 4l(v_1,v_2) +
 4l(v_2,v_3)+ l(v_3,v_4) + 3l(v_4,v_5) +
 4l(v_5,w) \\
 && + l(P_s(C_3))+ l(u_2,u_3) + l(u_3,v_4) +l(u_3,v_5)\\
&\leq& 15l(P_c(S)) + 4l(v,v_1) + 4l(v_1,v_2) +
 4l(v_2,v_3) + 4l(v_4,v_5) +
 4l(v_5,w) \\
 && + l(P_s(C_3))+ 2l(u_2,u_3) + l(u_3,v_4)+ l(u_2,v_3) \\
&=&15l(P_c(S)) + 4l(v,v_1) + 4l(v_1,v_2) +
 4l(v_2,v_3) + 4l(v_4,v_5) +
 4l(v_5,w) \\
 && + 2l(P_s(C_3))+ l(u_2,u_3) .
\end{eqnarray*}

By definition
\[ l(P_c(C_3)) = l(P_c(S)) + l(v,v_1) + l(v_1,v_2) + l(v_2,v_3) +
l(v_4,v_5) + l(v_5,w) ,\] giving:
\[ l(M(C_3)) \leq 15l(P_c(C_3)) + 2l(P_s(C_3)) +l(u_2,u_3).\]
>From $I(C_3)=[u_2 \rightarrow u_3]$ we get
\[  l(M(C_3)) \leq 15l(P_c(C_3)) + 2l(P_s(C_3)) +l(I(C_3)).\]

\item[A2] For all possible locations of $x$ and $y$  we can define
$B(C_3) = [u_2 \rightarrow v_3 \rightarrow v_4 \rightarrow u_3]$
giving
\[ l(B(C_3)) \leq l(u_2,v_3) + l(v_3,v_4) + l(v_4,u_3).\]
>From the definition of $I(C_3)$ and since $C_3$ is a son of $S$ we
get
\begin{eqnarray*}
l(B(C_3)) &=& l(u_2,v_3) + l(v_4,u_3) + l(v_3,v_4)\\
&=& l(P_s(C_3)) - l(I(C_3)) + l(v_3,v_4) \\
&\leq& l(P_s(C_3)) - l(I(C_3)) + l(P_s(C_3))\\
&=& 2l(P_s(C_3)) - l(I(C_3)).
\end{eqnarray*}

  \item [A3] Let $\hat{C}$ be a son of $C_3$. If
$P_f(\hat{C}) \subseteq [u_2,v_3]$ then
 since $[u_2,v_3] \subseteq P_s(C_2)$ and $l(C_2) < l(C_3)$, $\hat{C}$
should have been a son of $C_2$ . Similarly,  if $P_f(\hat{C})
\subseteq [u_3,v_4] \subseteq P_s(C_4)$ then since $l(C_4) < l(C_3)$
$\hat{C}$ would have been a son of $C_4$. Therefore $P_f(\hat{C})
\cap [u_2,u_3] \neq \phi$.
 \item [A4] In this case
$J_t(C_3) = [u_2 \rightarrow v_3]$ and $J_h(C_3) = [v_4 \rightarrow
u_3]$. Obviously for every son $\hat{C}$ of $C_3$ $P_f(\hat{C})
\subset P_s(C_3) = I(C_3) \cup J_t(C_3) \cup J_h(C_3)$.
 \item [A5] $M(C_3)$ is
contained in $P_s(C_3) \cup C_1 \cup C_2 \cup C_4 \cup M(S)$. Since
$l(C_i) < l(C_3), i=1,2,4$ the condition is satisfied.
 \item[A6] $M(C_3)$ is contained in $P_s(C_3)$, $C_1$, $C_2$, $C_4$ (three brothers of $C_3$), $S$  and $M(S)$, so
 $N(C_3) = \{ C_1,C_2,C_4 \}.$

\end{description}

\end{prf}
\newpage
\subsection{DIRECT-ONE}
\label{secdiron} In this section we consider a cycle  oriented by
DIRECT-ONE and prove that is satisfies the induction hypotheses. The
analysis is preformed according to the level of containment of
$C',C''$ defined in DIRECT-ONE. These cycles are defined only when
$C$ should be oriented forwards (heavy or special-contained) in
addition $C'$ is  defined when $l_1=-1$ and $C''$ is  defined when
$l_2 = -1$. When $C'$ and $C''$ are not defined in DIRECT-ONE, Lemma
\ref{lemf} or Lemma \ref{lemb} applies to prove that $C$ satisfies
the induction hypotheses.

\subsubsection{$C'$ and $C$ from the same level of containment
\label{secsamelevel} $(lc(C')=lc(C))$}

\begin{figure}
\begin{center}
\setlength{\unitlength}{0.00061242in}
\begingroup\makeatletter\ifx\SetFigFont\undefined%
\gdef\SetFigFont#1#2#3#4#5{%
  \reset@font\fontsize{#1}{#2pt}%
  \fontfamily{#3}\fontseries{#4}\fontshape{#5}%
  \selectfont}%
\fi\endgroup%
{\renewcommand{\dashlinestretch}{30}
\begin{picture}(6327,1786)(0,-10)
\put(1050,53){\blacken\ellipse{90}{90}}
\put(1050,53){\ellipse{90}{90}}
\put(2535,1718){\blacken\ellipse{90}{90}}
\put(2535,1718){\ellipse{90}{90}}
\put(2535,53){\blacken\ellipse{90}{90}}
\put(2535,53){\ellipse{90}{90}}
\put(4020,53){\blacken\ellipse{90}{90}}
\put(4020,53){\ellipse{90}{90}}
\put(4020,1718){\blacken\ellipse{90}{90}}
\put(4020,1718){\ellipse{90}{90}}
\put(5505,53){\blacken\ellipse{90}{90}}
\put(5505,53){\ellipse{90}{90}}
\blacken\path(2497.500,203.000)(2535.000,53.000)(2572.500,203.000)(2497.500,203.000)
\path(2535,53)(2535,1718)
\path(2535,1718)(4020,1718)
\blacken\path(3870.000,1680.500)(4020.000,1718.000)(3870.000,1755.500)(3870.000,1680.500)
\path(4020,53)(4020,1718)
\blacken\path(4057.500,1568.000)(4020.000,1718.000)(3982.500,1568.000)(4057.500,1568.000)
\path(4020,1718)(5505,1718)
\blacken\path(5355.000,1680.500)(5505.000,1718.000)(5355.000,1755.500)(5355.000,1680.500)
\blacken\path(5467.500,203.000)(5505.000,53.000)(5542.500,203.000)(5467.500,203.000)
\path(5505,53)(5505,1718)
\path(285,53)(1050,53)
\blacken\path(900.000,15.500)(1050.000,53.000)(900.000,90.500)(900.000,15.500)
\path(1050,53)(2535,53)
\blacken\path(2385.000,15.500)(2535.000,53.000)(2385.000,90.500)(2385.000,15.500)
\path(2535,53)(4020,53)
\blacken\path(3870.000,15.500)(4020.000,53.000)(3870.000,90.500)(3870.000,15.500)
\path(4020,53)(5505,53)
\blacken\path(5355.000,15.500)(5505.000,53.000)(5355.000,90.500)(5355.000,15.500)
\path(5505,53)(6315,53)
\blacken\path(6165.000,15.500)(6315.000,53.000)(6165.000,90.500)(6165.000,15.500)
\path(1050,53)(1050,1718)
\blacken\path(1087.500,1568.000)(1050.000,1718.000)(1012.500,1568.000)(1087.500,1568.000)
\path(1050,1718)(2535,1718)
\blacken\path(2385.000,1680.500)(2535.000,1718.000)(2385.000,1755.500)(2385.000,1680.500)
\put(15,188){\makebox(0,0)[lb]{\smash{{\SetFigFont{8}{9.6}{\rmdefault}{\mddefault}{\updefault}$P_s(F(C))$}}}}
\put(1770,818){\makebox(0,0)[lb]{\smash{{\SetFigFont{8}{9.6}{\rmdefault}{\mddefault}{\updefault}$C'$}}}}
\put(3255,818){\makebox(0,0)[lb]{\smash{{\SetFigFont{8}{9.6}{\rmdefault}{\mddefault}{\updefault}$C$}}}}
\put(4740,818){\makebox(0,0)[lb]{\smash{{\SetFigFont{8}{9.6}{\rmdefault}{\mddefault}{\updefault}$C''$}}}}
\end{picture}
}
\end{center}
\caption{Figure for Lemma \ref{longmid}.} \label{s1cs2fig}
\end{figure}
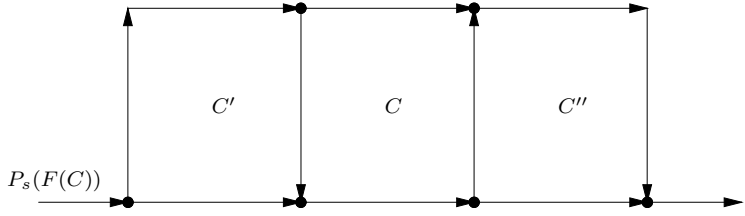

\begin{lem}
\label{longmid} Consider a cycle $C \in \mathcal{C} \backslash \{
C_0 \}$ and suppose that there exist $C'$ and $C''$ in $N(C)$, ($C'$
on the $l_1$ side and $C''$ on the $l_2$ side).(see Figure
\ref{s1cs2fig}). $C'$ and $C''$ are both defined by DIRECT-ONE. Then
$C$ is heavy or outer-crossing with one of these cycles.
\end{lem}
\begin{prf}
We assume that $C$ is not outer-crossing and prove that in this case
$C$ must be heavy. Consider the time when $C$ was in the input of
procedure DIRECT. \be
 \item If the function DIRECT-INNER-CROSSING was used,  $N(C)$ cannot
 contain both $S_1$ and $S_2$  (see Figure \ref{inncrossfig} and \ref{innerlemfig}).
 \item If the function DIRECT-ONE was used and $C$ is not heavy then either this case belongs to
 Lemma \ref{lemf} or by Lemma \ref{lemb} and then
$|N(C)| =0$, or $C$ is a special-contained brother. In the latter
case  $N(C)$ may contain only cycles from one side of $C$.
 \item If the function DIRECT-TWO was used and $C$ is not heavy then  $N(C)$
is constructed by either Lemma \ref{lemtwof} or Lemma \ref{lemtwob},
and then $|N(C)|=1$.
 \item If the function DIRECT-MANY was used:
  \be
 \item \label{scm1} When $C_{m_1}$ is light.
The functions used are DIRECT-BACKWARDS, DIRECT-FORWARDS, and
DIRECT-TWO. The cycles oriented in these functions  satisfy Lemmas
\ref{lemf}, \ref{lemb}, \ref{lemtwof} and \ref{lemtwob} where
$|N(C)| \leq 1$.
 \item When $C_{m_1}$ is heavy ,$C_{m_2}$ is light, and $C_{m_1}$ is not crossing.
 \bi
 \item $m_1 = 1$.  $C_1$ is heavy.
 For the other cycles we perform DIRECT-MANY or DIRECT-TWO on light
cycles, as in (a).
 \item $m_1=n$ similar to previous case.
 \item $1 < m_1 < n$. In this case $C_{m_1}$ is heavy.
 \bi
 \item  If $m_1 > 3$  then we use DIRECT-MANY or
DIRECT-TWO on light cycles  so $|N(C)| \leq 1 $ for all these
cycles.
 \item  If $m_1 = 2$. When $l_1 \neq 0$ $C_1$
is oriented backwards and $N(C_1) = \phi$. When $l_1 = 1$ then
$N(C_1) \subseteq \{ C_2,C_3 \}$.

 \item  If $m_1=n-1$ similar proof holds.
 \ei
  \ei
 \item \label{longc2} When  $C_{m_2}$ is heavy and $C_{m_1}$ is not crossing.
The cycles $C_j$ and $C_k$ are heavy. We consider the other cycles:
  \bi
 \item When $j < k-1$. The cycles $C_{j+1},\ldots,C_{k-1}$ are
oriented  by DIRECT-ONE (the case $v_1 \leftarrow v_2$), DIRECT-TWO
or DIRECT-MANY (on light cycles). In all these cases $|N(C)| \leq 1$
as in (a).
 \item When $j > 2$. The cycles $C_1,\ldots,C_{j-1}$ are oriented  by
DIRECT-TWO or DIRECT-MANY (on light cycles) with $|N(C)| \leq 1$ as
in (a).
 \item When $j=2$ . If $l_1 \neq 1$ then $C_1$ will be oriented
backwards and $|N(C_1)| = 0$. If $l_1 = 1$ then $N(C_1) \subseteq \{
C_2,C_3,C_4 \}$.
 \item When $k < n-2$. The cycles $C_{k+1},\ldots,C_n$ are oriented  by
DIRECT-TWO or DIRECT-MANY (on light cycles) with $|N(C)| \leq 1$.
 \item When $k=n-1$ similar proof like when $j=2$.

 \ei
 \item When $C_{m_1}$ is crossing. In this case $j=k-1$. $C_j$
 and $C_k$ are outer-crossing.  For the
 cycles $C_1,\ldots,C_{j-1}$ and $C_{k+1},\ldots,C_n$ a similar
 proof to case \ref{longc2} holds.
 \ee
 \ee
\end{prf}

\begin{lem}
Consider a non-crossing cycle $C$ oriented by DIRECT-ONE. If the
cycles that set $l_1$ and $l_2$ are in the same level of containment
as $C$ then $C$ satisfies the induction hypotheses.
\end{lem}
\begin{prf}
Since $C$ is non crossing DIRECT-ONE is called by DIRECT-MANY or
DIRECT-TWO. \be
 \item DIRECT-ONE is called by DIRECT-MANY. Note that in this
 case $C_{m_1}$ is heavy. For $C_1$ and $C_n$, $l_1$ and $l_2$ are
set in previous generation or level of containment. Therefore $C
\neq C_1$ and $C \neq C_n$.
  \be
 \item \label{bm1sm2} $C_{m_2}$ is light.
 \bi
 \item  $m_1=1$. $C_2,\ldots,C_{n-1}$ are oriented in DIRECT-K. Since
 $n>2$ DIRECT-K will not call  DIRECT-ONE.
 \item The same proof holds for  $m_1=n$ .
 \item $1 < m_1 < n$. Lemma \ref{lemf}  applies to prove that $C_{m_1}$
 satisfies the induction hypotheses.
 \bi
 \item  $2 < m_1  < n-1$. DIRECT-K will not call DIRECT-ONE
and  the orientation  of $C_{m_1}$ will not be farther changed.
 \item $2=m_1 < n-1$. If $l(C_1) < l(C_2)$ then Lemmas \ref{lemf}
 and \ref{lemtwof} apply to prove that $C_1$ and $C_2$
 satisfy  the induction hypotheses (respectively).
   If $l(C_1) > l(C_2)$  equivalent lemmas  apply.
    DIRECT-K on $C_3,\ldots,C_n$ will  not call DIRECT-ONE.
 \item The same proof holds for  $2 < m_1 =n-1$.
\ei \ei
 \item $C_{m_2}$ is heavy.
 \bi
 \item  $k=j+2$. $C_{j+1}$ is oriented  by DIRECT-ONE : $v_2
 \rightarrow v_1$. Lemma \ref{lemb} applies to prove that $C_{j+1}$
 satisfies  the induction hypotheses. Again it doesn't
 change any previously oriented  cycles, and will not be farther
 changed in this level of containment.
 \item $k=j+1$.
 \bi
 \item When $j=2,k=3,n>4$. Lemmas
 \ref{lemf}, \ref{lemtwof} ,\ref{lemthreefmid} and
 \ref{lemthreefs} (or equivalent lemmas) apply to prove that $C_1,C_2$ and $C_3$ satisfy
 the induction hypotheses.
 \item The same proof holds for $j>2, k=n-1$.
 \item $j=2,k=3,n=4$. In this case  Lemmas \ref{lemf}, \ref{lemtwof} ,\ref{lemthreefmid},
 \ref{lemthreefs}, \ref{lemfourfmid} and \ref{lemfourfs} or
 equivalent lemmas apply to prove that $C_1,C_2,C_3$ and $C_4$
 satisfy  the induction hypotheses.
 \ei
 \ei
 \ee
\item DIRECT-ONE was called by DIRECT-TWO. As in DIRECT-MANY
$C \neq C_1$ and $C \neq C_2$.

\ee
\end{prf}

\subsubsection{$C'$ and $C$ are brothers with $lc(C') < lc(C)$}
\label{secsmalllevl} 
\begin{figure}
\begin{center}
\setlength{\unitlength}{0.00061242in}
\begingroup\makeatletter\ifx\SetFigFont\undefined%
\gdef\SetFigFont#1#2#3#4#5{%
  \reset@font\fontsize{#1}{#2pt}%
  \fontfamily{#3}\fontseries{#4}\fontshape{#5}%
  \selectfont}%
\fi\endgroup%
{\renewcommand{\dashlinestretch}{30}
\begin{picture}(5199,2250)(0,-10)
\put(687,240){\blacken\ellipse{90}{90}}
\put(687,240){\ellipse{90}{90}}
\put(2442,240){\blacken\ellipse{90}{90}}
\put(2442,240){\ellipse{90}{90}}
\put(3387,240){\blacken\ellipse{90}{90}}
\put(3387,240){\ellipse{90}{90}}
\put(4557,240){\blacken\ellipse{90}{90}}
\put(4557,240){\ellipse{90}{90}}
\put(2442,1230){\blacken\ellipse{90}{90}}
\put(2442,1230){\ellipse{90}{90}}
\put(2442,1950){\blacken\ellipse{90}{90}}
\put(2442,1950){\ellipse{90}{90}}
\blacken\path(537.000,202.500)(687.000,240.000)(537.000,277.500)(537.000,202.500)
\path(687,240)(12,240)
\path(687,240)(2442,240)
\blacken\path(2292.000,202.500)(2442.000,240.000)(2292.000,277.500)(2292.000,202.500)
\path(2442,240)(3387,240)
\blacken\path(3237.000,202.500)(3387.000,240.000)(3237.000,277.500)(3237.000,202.500)
\path(3387,240)(4557,240)
\blacken\path(4407.000,202.500)(4557.000,240.000)(4407.000,277.500)(4407.000,202.500)
\path(4557,240)(5187,240)
\blacken\path(5037.000,202.500)(5187.000,240.000)(5037.000,277.500)(5037.000,202.500)
\path(687,240)(687,1950)
\blacken\path(724.500,1800.000)(687.000,1950.000)(649.500,1800.000)(724.500,1800.000)
\path(687,1950)(2442,1950)
\blacken\path(2292.000,1912.500)(2442.000,1950.000)(2292.000,1987.500)(2292.000,1912.500)
\path(2442,1905)(2442,1230)
\blacken\path(2404.500,1380.000)(2442.000,1230.000)(2479.500,1380.000)(2404.500,1380.000)
\path(2442,1230)(2442,240)
\blacken\path(2404.500,390.000)(2442.000,240.000)(2479.500,390.000)(2404.500,390.000)
\path(2487,1230)(3387,1230)
\path(3387,1230)(3387,240)
\path(2442,1950)(4557,1950)
\blacken\path(4407.000,1912.500)(4557.000,1950.000)(4407.000,1987.500)(4407.000,1912.500)
\path(4557,1950)(4557,240)
\blacken\path(4519.500,390.000)(4557.000,240.000)(4594.500,390.000)(4519.500,390.000)
\put(3342,15){\makebox(0,0)[lb]{\smash{{\SetFigFont{8}{9.6}{\rmdefault}{\mddefault}{\updefault}$v_3$}}}}
\put(4512,15){\makebox(0,0)[lb]{\smash{{\SetFigFont{8}{9.6}{\rmdefault}{\mddefault}{\updefault}$v_4$}}}}
\put(642,15){\makebox(0,0)[lb]{\smash{{\SetFigFont{8}{9.6}{\rmdefault}{\mddefault}{\updefault}$v_1$}}}}
\put(2397,15){\makebox(0,0)[lb]{\smash{{\SetFigFont{8}{9.6}{\rmdefault}{\mddefault}{\updefault}$v_2$}}}}
\put(2802,690){\makebox(0,0)[lb]{\smash{{\SetFigFont{8}{9.6}{\rmdefault}{\mddefault}{\updefault}$C$}}}}
\put(1587,2085){\makebox(0,0)[lb]{\smash{{\SetFigFont{8}{9.6}{\rmdefault}{\mddefault}{\updefault}$C'$}}}}
\put(3567,2085){\makebox(0,0)[lb]{\smash{{\SetFigFont{8}{9.6}{\rmdefault}{\mddefault}{\updefault}$D$}}}}
\put(2172,1140){\makebox(0,0)[lb]{\smash{{\SetFigFont{8}{9.6}{\rmdefault}{\mddefault}{\updefault}$u_1$}}}}
\put(2397,2085){\makebox(0,0)[lb]{\smash{{\SetFigFont{8}{9.6}{\rmdefault}{\mddefault}{\updefault}$u_2$}}}}
\end{picture}
}
\end{center}
\caption{$C'$ from previous containment level.} \label{C'prconlfig}
\end{figure}
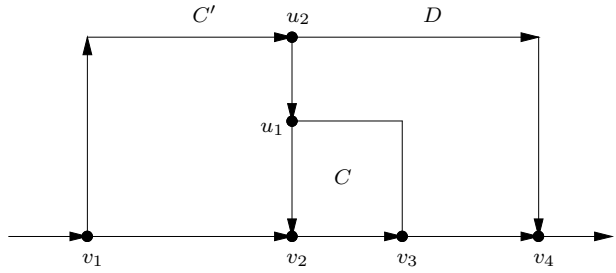

Consider a  cycle $C$ oriented forwards in
DIRECT-ONE even though $l_1=-1$. We observe that $C$ is either heavy
or special-contained brother, otherwise DIRECT-ONE would have
oriented it backwards. Denote by $D$ a containing brother of $C$. In
Figure \ref{C'prconlfig} $C,C'$ and $D$ are described before the
application of DIRECT-ONE.  Since $l_1 = -1$, $C$ is the first
brother in the block of contained brothers of $D$. We can assume
that before $C$ is oriented, $C'$ and $D$ satisfied the induction
hypotheses.
 If $C$ is heavy,
 then $D$ is heavy and it is oriented  forwards. If $C$ is a
 special-contained brother then $D$ is oriented  forwards.
Since $l_1=-1$, $C' \in N(D)$ and $C'$ is oriented  forwards.
 Mark \bi
 \item $v_1 := S \cap (P_s(C') \backslash P_s(D))$.
 \item $v_2 := S \cap (P_s(D) \cap P_s(C) \cap P_s(C')$.
 \item $v_3 := S \cap (P_s(C) \backslash P_s(C'))$.
 \item $v_4 := S \cap (P_s(D) \backslash P_s(C'))$ .
 \item $u_1:=$ the end node of the path $P_s(C) \cap P_s(C')$ such
 that $u_1 \not\in S$.
 \item $u_2:=$ the end node of the path $P_s(D) \cap P_s(C')$ such
 that $u_2 \not\in S$.
\ei .

If $l(C) > l(C')$ then DIRECT-ONE will not change the direction of
$[u_1,v_2]$ and $D$, and we only need to prove that $C$ satisfies
the induction hypotheses. If $l(C) < l(C')$ then orientation  of
$C'$ ,  $J_t(D)$ and $B_t(D)$ are changed. In this case we also need
to prove that after DIRECT-ONE is activated $C'$ and $D$ still
satisfy the induction hypotheses. We prove all these in the
following lemmas.

\begin{figure}
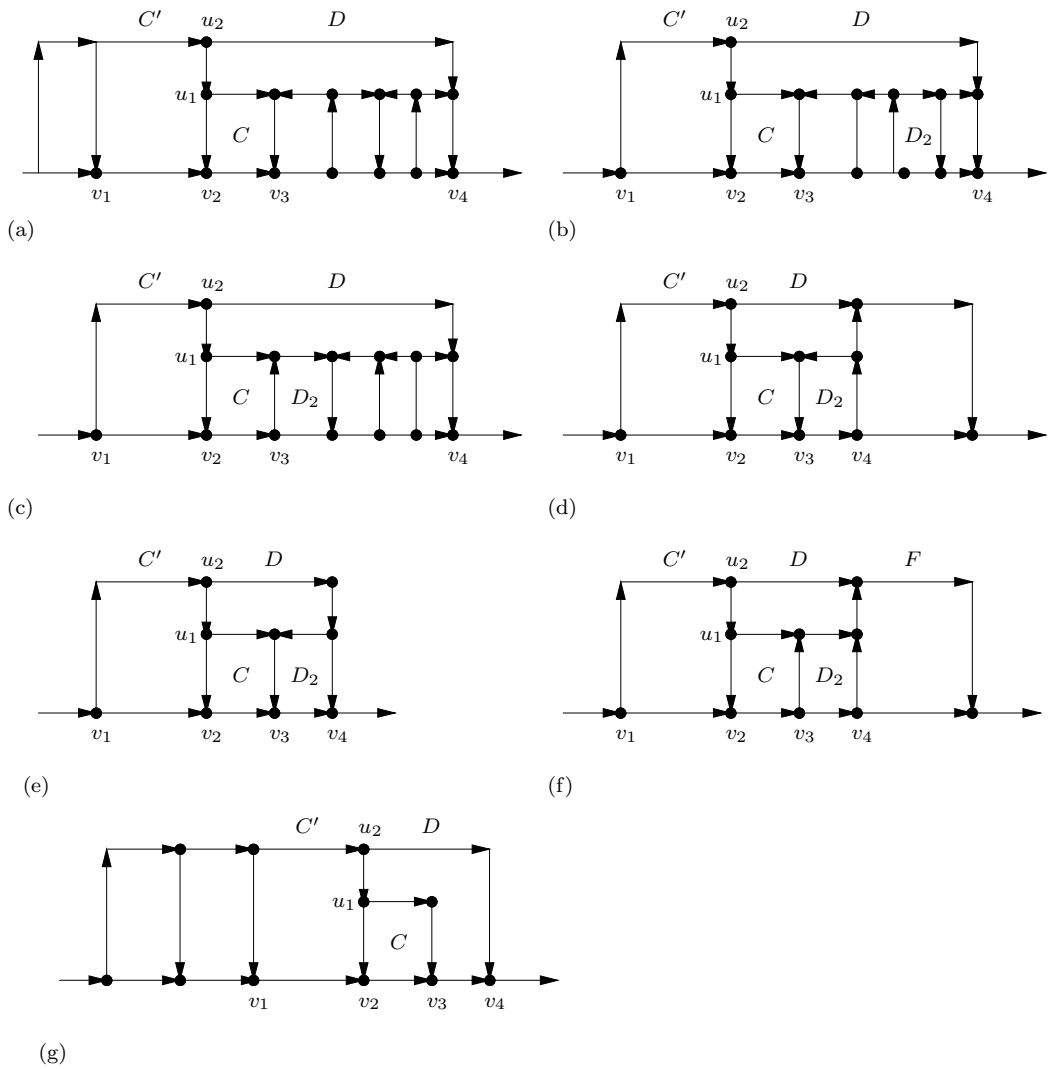

\begin{center}
\setlength{\unitlength}{0.00061242in}
\begingroup\makeatletter\ifx\SetFigFont\undefined%
\gdef\SetFigFont#1#2#3#4#5{%
  \reset@font\fontsize{#1}{#2pt}%
  \fontfamily{#3}\fontseries{#4}\fontshape{#5}%
  \selectfont}%
\fi\endgroup%
{\renewcommand{\dashlinestretch}{30}

}
\end{center}
\caption{Figure for Lemma \ref{CgC'lem}.} \label{SonsofGfig}
\end{figure}

\begin{lem}
\label{CgC'lem} If $l(C)> l(C')$, then after DIRECT-ONE is activated
$C$ satisfies the induction hypotheses.
\end{lem}
\begin{prf}
There are three cases to be considered.
 \be
\item  There are
more then two contained brothers in $D$ (DIRECT-ONE was called by
DIRECT-MANY). See Figure \ref{SonsofGfig} (a) (b) and (c).
 \bi
 \item $C$ is the only heavy contained brother in $D$ (see Figure \ref{SonsofGfig}
(a)). In this case
 (even after all the contained brothers of $D$ are oriented )
 $N(C)$ will not contain any cycle on the $l_2$ side of $C$. By Lemma \ref{longmid} $N(C)
 \subset N(D), |N(C)| \leq 2$, and Lemma \ref{lemtwof} or Lemma \ref{lemthreefs}
  apply to prove that $C$ satisfies the induction hypotheses.
 \item There is another heavy  brother  $D_2$ which is
 contained in $D$,
 such that $P_s(C) \cap P_s(D_2) = \phi$ (see Figure \ref{SonsofGfig}
(b)). In this case $C'$ has to
 be light, and by Lemma \ref{longmid} $N(C') = \phi$. Again $N(C)$ will not contain any
 cycle on the $l_2$ side of $C$ so $N(C) = \{ C'\} $ and Lemma \ref{lemtwof}
 applies to prove that $C$ satisfies the induction hypotheses.
\item There is another heavy  brother  $D_2$ which is
 contained in $D$,
 such that $P_s(C) \cap P_s(D_2) \neq \phi$ (see Figure \ref{SonsofGfig}
(c)).
 In this case $C'$ has to be light, so by Lemma \ref{longmid} $N(C') = \phi$, giving that
 $N(C) = \{ D_2,C'\}$ and Lemma \ref{lemthreefmid} applies to prove
 that $C$ satisfies the induction hypotheses.
\ei
 \item There are exactly two contained brothers in $D$ (DIRECT-ONE
 was called by DIRECT-TWO) (see Figure \ref{SonsofGfig}
(d) (e) and (f)). Mark the other contained
 brother of $D$ as $D_2$.
 \bi
  \item $D_2$ is light.
  \bi
  \item $N(D)$ contains another cycle in
  the $l_2$ side of $D$ (see Figure \ref{SonsofGfig}
(d)). In this case $D_2$ is oriented  backwards
  so $N(C)$ doesn't contain any cycle in the $l_2$ side of $C$,
  giving that $N(C)\subset N(D)$ and $|N(C)| \leq 2$. Lemma \ref{lemtwof} or Lemma \ref{lemthreefs}
  applies to prove that $C$ satisfies the induction hypotheses.
 \item $N(D)$ doesn't contain another cycle in the $l_2$ side
 of $D$ (see Figure \ref{SonsofGfig}
(e)). Since $D$ is heavy, by Lemma \ref{longmid} $|N(D)| \leq 2$.
   In this case $D_2$ will also be oriented  forwards, and $N(C) =
   N(D) \cup \{ D_2 \}$, but still $ |N(C)| \leq 3$. In this case
   Lemma \ref{lemthreefmid} or Lemma \ref{lemfourfmid}  applies to
   prove that $C$ satisfies the induction hypotheses.
 \ei
  \item When $F_2$ is heavy (see Figure \ref{SonsofGfig}(f)). In this case $C'$ has to be light, so
  by Lemma \ref{longmid} $N(D)$ contains $C'$ and possibly one more cycle $F$ at
the $l_2$ side of $D$. In this case $N(C) \subset \{ D_2,C',F
  \}$ and Lemma \ref{lemthreefmid} or Lemma \ref{lemfourfmid} applies to prove that $C$
  satisfies the induction hypotheses.
\ei
 \item There is exactly one contained brother in $D$ ($C$ is a
 special-contained brother of $D$). See Figure \ref{SonsofGfig}(g). Since $D$ satisfies the induction hypotheses $N(D)$ contains
 at most 3 brothers at its $l_1$ side.
  $N(C)$ contains all the cycles from this side (all are shorter
  than $C$) and $C$ will be oriented to satisfy Lemma
  \ref{lemtwof},  Lemma \ref{lemthreefs}, or Lemma \ref{lemfourfs}.
 \ee

\end{prf}

\begin{lem}
Suppose that $l(C)< l(C')$, then after DIRECT-ONE is activated $C$
satisfies the induction hypotheses.
\end{lem}
\begin{prf}
In this case DIRECT-ONE changes a previous direction and $I(C) =
P_s(C)$, so Lemma \ref{lemf} applies to prove that $C$ satisfy the
induction hypotheses.
\end{prf}

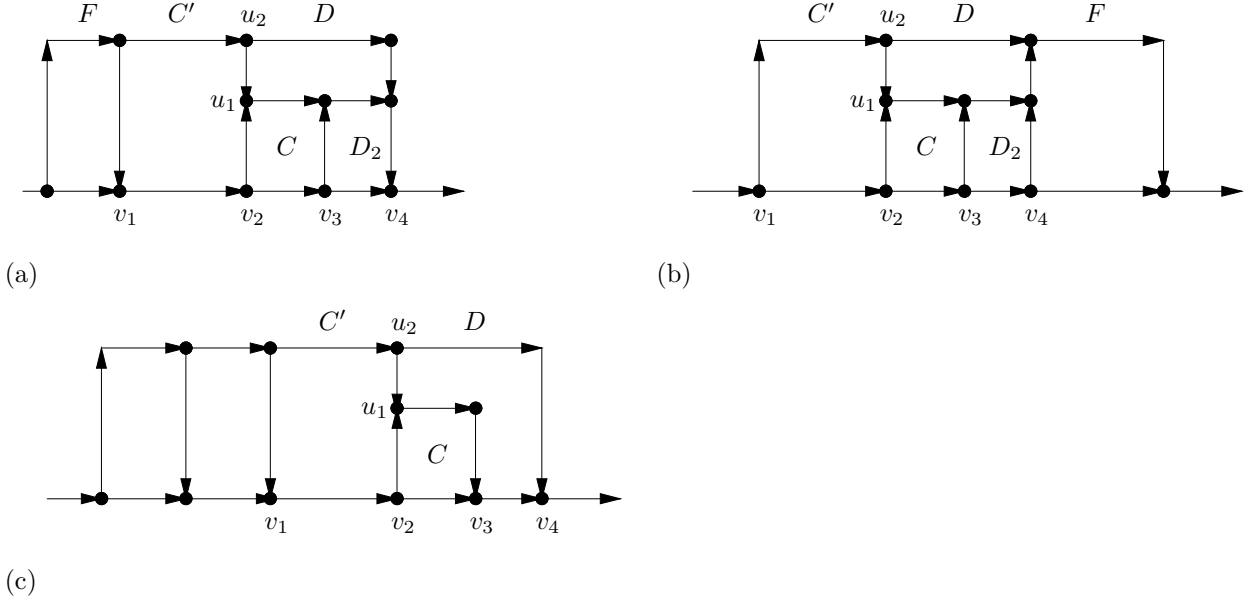
\begin{figure}
\begin{center}
\setlength{\unitlength}{0.00069991in}
\begingroup\makeatletter\ifx\SetFigFont\undefined%
\gdef\SetFigFont#1#2#3#4#5{%
  \reset@font\fontsize{#1}{#2pt}%
  \fontfamily{#3}\fontseries{#4}\fontshape{#5}%
  \selectfont}%
\fi\endgroup%
{\renewcommand{\dashlinestretch}{30}
\begin{picture}(9252,4464)(0,-10)
\put(1995,744){\blacken\ellipse{90}{90}}
\put(1995,744){\ellipse{90}{90}}
\put(2940,744){\blacken\ellipse{90}{90}}
\put(2940,744){\ellipse{90}{90}}
\put(3525,744){\blacken\ellipse{90}{90}}
\put(3525,744){\ellipse{90}{90}}
\put(2940,1419){\blacken\ellipse{90}{90}}
\put(2940,1419){\ellipse{90}{90}}
\put(3525,1419){\blacken\ellipse{90}{90}}
\put(3525,1419){\ellipse{90}{90}}
\put(4020,744){\blacken\ellipse{90}{90}}
\put(4020,744){\ellipse{90}{90}}
\put(2940,1869){\blacken\ellipse{90}{90}}
\put(2940,1869){\ellipse{90}{90}}
\put(1995,1869){\blacken\ellipse{90}{90}}
\put(1995,1869){\ellipse{90}{90}}
\put(1365,1869){\blacken\ellipse{90}{90}}
\put(1365,1869){\ellipse{90}{90}}
\put(1365,744){\blacken\ellipse{90}{90}}
\put(1365,744){\ellipse{90}{90}}
\put(735,744){\blacken\ellipse{90}{90}}
\put(735,744){\ellipse{90}{90}}
\put(870,3039){\blacken\ellipse{90}{90}}
\put(870,3039){\ellipse{90}{90}}
\put(1815,3039){\blacken\ellipse{90}{90}}
\put(1815,3039){\ellipse{90}{90}}
\put(2400,3039){\blacken\ellipse{90}{90}}
\put(2400,3039){\ellipse{90}{90}}
\put(1815,3714){\blacken\ellipse{90}{90}}
\put(1815,3714){\ellipse{90}{90}}
\put(1815,4164){\blacken\ellipse{90}{90}}
\put(1815,4164){\ellipse{90}{90}}
\put(2400,3714){\blacken\ellipse{90}{90}}
\put(2400,3714){\ellipse{90}{90}}
\put(2895,3714){\blacken\ellipse{90}{90}}
\put(2895,3714){\ellipse{90}{90}}
\put(2895,3039){\blacken\ellipse{90}{90}}
\put(2895,3039){\ellipse{90}{90}}
\put(2895,4164){\blacken\ellipse{90}{90}}
\put(2895,4164){\ellipse{90}{90}}
\put(870,4164){\blacken\ellipse{90}{90}}
\put(870,4164){\ellipse{90}{90}}
\put(330,3039){\blacken\ellipse{90}{90}}
\put(330,3039){\ellipse{90}{90}}
\put(5640,3039){\blacken\ellipse{90}{90}}
\put(5640,3039){\ellipse{90}{90}}
\put(6585,3039){\blacken\ellipse{90}{90}}
\put(6585,3039){\ellipse{90}{90}}
\put(7170,3039){\blacken\ellipse{90}{90}}
\put(7170,3039){\ellipse{90}{90}}
\put(6585,3714){\blacken\ellipse{90}{90}}
\put(6585,3714){\ellipse{90}{90}}
\put(6585,4164){\blacken\ellipse{90}{90}}
\put(6585,4164){\ellipse{90}{90}}
\put(7170,3714){\blacken\ellipse{90}{90}}
\put(7170,3714){\ellipse{90}{90}}
\put(7665,3714){\blacken\ellipse{90}{90}}
\put(7665,3714){\ellipse{90}{90}}
\put(7665,3039){\blacken\ellipse{90}{90}}
\put(7665,3039){\ellipse{90}{90}}
\put(8655,3039){\blacken\ellipse{90}{90}}
\put(8655,3039){\ellipse{90}{90}}
\put(7665,4164){\blacken\ellipse{90}{90}}
\put(7665,4164){\ellipse{90}{90}}
\path(1995,744)(2940,744)
\blacken\path(2790.000,706.500)(2940.000,744.000)(2790.000,781.500)(2790.000,706.500)
\path(2940,744)(3525,744)
\blacken\path(3375.000,706.500)(3525.000,744.000)(3375.000,781.500)(3375.000,706.500)
\path(1995,1869)(2940,1869)
\blacken\path(2790.000,1831.500)(2940.000,1869.000)(2790.000,1906.500)(2790.000,1831.500)
\path(2940,1824)(2940,1374)
\blacken\path(2902.500,1524.000)(2940.000,1374.000)(2977.500,1524.000)(2902.500,1524.000)
\path(2940,1419)(3525,1419)
\blacken\path(3375.000,1381.500)(3525.000,1419.000)(3375.000,1456.500)(3375.000,1381.500)
\path(3525,1419)(3525,744)
\blacken\path(3487.500,894.000)(3525.000,744.000)(3562.500,894.000)(3487.500,894.000)
\blacken\path(772.500,1719.000)(735.000,1869.000)(697.500,1719.000)(772.500,1719.000)
\path(735,1869)(735,744)
\path(3525,744)(4020,744)
\blacken\path(3870.000,706.500)(4020.000,744.000)(3870.000,781.500)(3870.000,706.500)
\path(4020,744)(4605,744)
\blacken\path(4455.000,706.500)(4605.000,744.000)(4455.000,781.500)(4455.000,706.500)
\blacken\path(585.000,706.500)(735.000,744.000)(585.000,781.500)(585.000,706.500)
\path(735,744)(330,744)
\blacken\path(1890.000,706.500)(2040.000,744.000)(1890.000,781.500)(1890.000,706.500)
\path(2040,744)(1365,744)
\blacken\path(1215.000,706.500)(1365.000,744.000)(1215.000,781.500)(1215.000,706.500)
\path(1365,744)(735,744)
\blacken\path(1215.000,1831.500)(1365.000,1869.000)(1215.000,1906.500)(1215.000,1831.500)
\path(1365,1869)(735,1869)
\path(1365,1869)(1365,744)
\blacken\path(1327.500,894.000)(1365.000,744.000)(1402.500,894.000)(1327.500,894.000)
\blacken\path(1845.000,1831.500)(1995.000,1869.000)(1845.000,1906.500)(1845.000,1831.500)
\path(1995,1869)(1365,1869)
\blacken\path(1957.500,894.000)(1995.000,744.000)(2032.500,894.000)(1957.500,894.000)
\path(1995,744)(1995,1869)
\path(2940,1869)(4020,1869)
\blacken\path(3870.000,1831.500)(4020.000,1869.000)(3870.000,1906.500)(3870.000,1831.500)
\path(4020,1869)(4020,744)
\blacken\path(3982.500,894.000)(4020.000,744.000)(4057.500,894.000)(3982.500,894.000)
\path(1815,3039)(2400,3039)
\blacken\path(2250.000,3001.500)(2400.000,3039.000)(2250.000,3076.500)(2250.000,3001.500)
\path(870,4164)(1815,4164)
\blacken\path(1665.000,4126.500)(1815.000,4164.000)(1665.000,4201.500)(1665.000,4126.500)
\path(1815,4119)(1815,3669)
\blacken\path(1777.500,3819.000)(1815.000,3669.000)(1852.500,3819.000)(1777.500,3819.000)
\path(1815,3714)(2400,3714)
\blacken\path(2250.000,3676.500)(2400.000,3714.000)(2250.000,3751.500)(2250.000,3676.500)
\path(2400,3039)(2895,3039)
\blacken\path(2745.000,3001.500)(2895.000,3039.000)(2745.000,3076.500)(2745.000,3001.500)
\path(1815,4164)(2895,4164)
\blacken\path(2745.000,4126.500)(2895.000,4164.000)(2745.000,4201.500)(2745.000,4126.500)
\path(2895,3714)(2895,3039)
\blacken\path(2857.500,3189.000)(2895.000,3039.000)(2932.500,3189.000)(2857.500,3189.000)
\path(2895,4164)(2895,3714)
\blacken\path(2857.500,3864.000)(2895.000,3714.000)(2932.500,3864.000)(2857.500,3864.000)
\path(2895,3039)(3435,3039)
\blacken\path(3285.000,3001.500)(3435.000,3039.000)(3285.000,3076.500)(3285.000,3001.500)
\blacken\path(367.500,4014.000)(330.000,4164.000)(292.500,4014.000)(367.500,4014.000)
\path(330,4164)(330,3039)
\blacken\path(720.000,4126.500)(870.000,4164.000)(720.000,4201.500)(720.000,4126.500)
\path(870,4164)(330,4164)
\blacken\path(832.500,3189.000)(870.000,3039.000)(907.500,3189.000)(832.500,3189.000)
\path(870,3039)(870,4164)
\path(2400,3714)(2895,3714)
\blacken\path(2745.000,3676.500)(2895.000,3714.000)(2745.000,3751.500)(2745.000,3676.500)
\blacken\path(2437.500,3564.000)(2400.000,3714.000)(2362.500,3564.000)(2437.500,3564.000)
\path(2400,3714)(2400,3039)
\path(870,3039)(1815,3039)
\blacken\path(1665.000,3001.500)(1815.000,3039.000)(1665.000,3076.500)(1665.000,3001.500)
\path(5640,3039)(6585,3039)
\blacken\path(6435.000,3001.500)(6585.000,3039.000)(6435.000,3076.500)(6435.000,3001.500)
\path(6585,3039)(7170,3039)
\blacken\path(7020.000,3001.500)(7170.000,3039.000)(7020.000,3076.500)(7020.000,3001.500)
\path(5640,4164)(6585,4164)
\blacken\path(6435.000,4126.500)(6585.000,4164.000)(6435.000,4201.500)(6435.000,4126.500)
\path(6585,4119)(6585,3669)
\blacken\path(6547.500,3819.000)(6585.000,3669.000)(6622.500,3819.000)(6547.500,3819.000)
\path(6585,3714)(7170,3714)
\blacken\path(7020.000,3676.500)(7170.000,3714.000)(7020.000,3751.500)(7020.000,3676.500)
\path(7170,3039)(7665,3039)
\blacken\path(7515.000,3001.500)(7665.000,3039.000)(7515.000,3076.500)(7515.000,3001.500)
\path(7665,3039)(8655,3039)
\blacken\path(8505.000,3001.500)(8655.000,3039.000)(8505.000,3076.500)(8505.000,3001.500)
\path(6585,4164)(7665,4164)
\blacken\path(7515.000,4126.500)(7665.000,4164.000)(7515.000,4201.500)(7515.000,4126.500)
\path(7665,4164)(8655,4164)
\blacken\path(8505.000,4126.500)(8655.000,4164.000)(8505.000,4201.500)(8505.000,4126.500)
\path(8655,4164)(8655,3039)
\blacken\path(8617.500,3189.000)(8655.000,3039.000)(8692.500,3189.000)(8617.500,3189.000)
\blacken\path(7702.500,4014.000)(7665.000,4164.000)(7627.500,4014.000)(7702.500,4014.000)
\path(7665,4164)(7665,3714)
\blacken\path(7702.500,3564.000)(7665.000,3714.000)(7627.500,3564.000)(7702.500,3564.000)
\path(7665,3714)(7665,3039)
\path(8700,3039)(9240,3039)
\blacken\path(9090.000,3001.500)(9240.000,3039.000)(9090.000,3076.500)(9090.000,3001.500)
\path(5145,3039)(5640,3039)
\blacken\path(5490.000,3001.500)(5640.000,3039.000)(5490.000,3076.500)(5490.000,3001.500)
\path(5640,3039)(5640,4164)
\blacken\path(5677.500,4014.000)(5640.000,4164.000)(5602.500,4014.000)(5677.500,4014.000)
\blacken\path(7207.500,3564.000)(7170.000,3714.000)(7132.500,3564.000)(7207.500,3564.000)
\path(7170,3714)(7170,3039)
\path(7170,3714)(7665,3714)
\blacken\path(7515.000,3676.500)(7665.000,3714.000)(7515.000,3751.500)(7515.000,3676.500)
\blacken\path(6622.500,3564.000)(6585.000,3714.000)(6547.500,3564.000)(6622.500,3564.000)
\path(6585,3714)(6585,3039)
\blacken\path(1852.500,3564.000)(1815.000,3714.000)(1777.500,3564.000)(1852.500,3564.000)
\path(1815,3714)(1815,3039)
\path(150,3039)(870,3039)
\blacken\path(720.000,3001.500)(870.000,3039.000)(720.000,3076.500)(720.000,3001.500)
\blacken\path(2977.500,1269.000)(2940.000,1419.000)(2902.500,1269.000)(2977.500,1269.000)
\path(2940,1419)(2940,744)
\put(1950,519){\makebox(0,0)[lb]{\smash{{\SetFigFont{10}{12.0}{\rmdefault}{\mddefault}{\updefault}$v_1$}}}}
\put(2895,519){\makebox(0,0)[lb]{\smash{{\SetFigFont{10}{12.0}{\rmdefault}{\mddefault}{\updefault}$v_2$}}}}
\put(3480,519){\makebox(0,0)[lb]{\smash{{\SetFigFont{10}{12.0}{\rmdefault}{\mddefault}{\updefault}$v_3$}}}}
\put(2670,1374){\makebox(0,0)[lb]{\smash{{\SetFigFont{10}{12.0}{\rmdefault}{\mddefault}{\updefault}$u_1$}}}}
\put(3975,519){\makebox(0,0)[lb]{\smash{{\SetFigFont{10}{12.0}{\rmdefault}{\mddefault}{\updefault}$v_4$}}}}
\put(3165,1014){\makebox(0,0)[lb]{\smash{{\SetFigFont{10}{12.0}{\rmdefault}{\mddefault}{\updefault}$C$}}}}
\put(2895,2004){\makebox(0,0)[lb]{\smash{{\SetFigFont{10}{12.0}{\rmdefault}{\mddefault}{\updefault}$u_2$}}}}
\put(3435,2004){\makebox(0,0)[lb]{\smash{{\SetFigFont{10}{12.0}{\rmdefault}{\mddefault}{\updefault}$D$}}}}
\put(1770,4299){\makebox(0,0)[lb]{\smash{{\SetFigFont{10}{12.0}{\rmdefault}{\mddefault}{\updefault}$u_2$}}}}
\put(2310,4299){\makebox(0,0)[lb]{\smash{{\SetFigFont{10}{12.0}{\rmdefault}{\mddefault}{\updefault}$D$}}}}
\put(825,2814){\makebox(0,0)[lb]{\smash{{\SetFigFont{10}{12.0}{\rmdefault}{\mddefault}{\updefault}$v_1$}}}}
\put(1770,2814){\makebox(0,0)[lb]{\smash{{\SetFigFont{10}{12.0}{\rmdefault}{\mddefault}{\updefault}$v_2$}}}}
\put(2355,2814){\makebox(0,0)[lb]{\smash{{\SetFigFont{10}{12.0}{\rmdefault}{\mddefault}{\updefault}$v_3$}}}}
\put(1545,3669){\makebox(0,0)[lb]{\smash{{\SetFigFont{10}{12.0}{\rmdefault}{\mddefault}{\updefault}$u_1$}}}}
\put(2850,2814){\makebox(0,0)[lb]{\smash{{\SetFigFont{10}{12.0}{\rmdefault}{\mddefault}{\updefault}$v_4$}}}}
\put(2040,3309){\makebox(0,0)[lb]{\smash{{\SetFigFont{10}{12.0}{\rmdefault}{\mddefault}{\updefault}$C$}}}}
\put(2580,3309){\makebox(0,0)[lb]{\smash{{\SetFigFont{10}{12.0}{\rmdefault}{\mddefault}{\updefault}$D_2$}}}}
\put(1230,4299){\makebox(0,0)[lb]{\smash{{\SetFigFont{10}{12.0}{\rmdefault}{\mddefault}{\updefault}$C'$}}}}
\put(555,4299){\makebox(0,0)[lb]{\smash{{\SetFigFont{10}{12.0}{\rmdefault}{\mddefault}{\updefault}$F$}}}}
\put(6000,4299){\makebox(0,0)[lb]{\smash{{\SetFigFont{10}{12.0}{\rmdefault}{\mddefault}{\updefault}$C'$}}}}
\put(6540,4299){\makebox(0,0)[lb]{\smash{{\SetFigFont{10}{12.0}{\rmdefault}{\mddefault}{\updefault}$u_2$}}}}
\put(7080,4299){\makebox(0,0)[lb]{\smash{{\SetFigFont{10}{12.0}{\rmdefault}{\mddefault}{\updefault}$D$}}}}
\put(8070,4299){\makebox(0,0)[lb]{\smash{{\SetFigFont{10}{12.0}{\rmdefault}{\mddefault}{\updefault}$F$}}}}
\put(5595,2814){\makebox(0,0)[lb]{\smash{{\SetFigFont{10}{12.0}{\rmdefault}{\mddefault}{\updefault}$v_1$}}}}
\put(6540,2814){\makebox(0,0)[lb]{\smash{{\SetFigFont{10}{12.0}{\rmdefault}{\mddefault}{\updefault}$v_2$}}}}
\put(7125,2814){\makebox(0,0)[lb]{\smash{{\SetFigFont{10}{12.0}{\rmdefault}{\mddefault}{\updefault}$v_3$}}}}
\put(6315,3669){\makebox(0,0)[lb]{\smash{{\SetFigFont{10}{12.0}{\rmdefault}{\mddefault}{\updefault}$u_1$}}}}
\put(7620,2814){\makebox(0,0)[lb]{\smash{{\SetFigFont{10}{12.0}{\rmdefault}{\mddefault}{\updefault}$v_4$}}}}
\put(6810,3309){\makebox(0,0)[lb]{\smash{{\SetFigFont{10}{12.0}{\rmdefault}{\mddefault}{\updefault}$C$}}}}
\put(7350,3309){\makebox(0,0)[lb]{\smash{{\SetFigFont{10}{12.0}{\rmdefault}{\mddefault}{\updefault}$D_2$}}}}
\put(4875,2364){\makebox(0,0)[lb]{\smash{{\SetFigFont{10}{12.0}{\rmdefault}{\mddefault}{\updefault}(b)}}}}
\put(15,69){\makebox(0,0)[lb]{\smash{{\SetFigFont{10}{12.0}{\rmdefault}{\mddefault}{\updefault}(c)}}}}
\put(15,2364){\makebox(0,0)[lb]{\smash{{\SetFigFont{10}{12.0}{\rmdefault}{\mddefault}{\updefault}(a)}}}}
\put(2355,2004){\makebox(0,0)[lb]{\smash{{\SetFigFont{10}{12.0}{\rmdefault}{\mddefault}{\updefault}$C'$}}}}
\end{picture}
}
\end{center}
\caption{Figure for Lemma \ref{C'gClem}.} \label{SonsofGtagfig}
\end{figure}

\begin{lem}
\label{C'gClem} Suppose that $l(C)< l(C')$, then after DIRECT-ONE is
activated $C'$ satisfies the induction hypotheses.
\end{lem}
\begin{prf}
There are two cases to be considered.
 \be
 \item $C$ is heavy. According to Lemma \ref{longmid} after  DIRECT-ONE is activated $N(C')$ contains $C$ and
at most two more cycles.
 \bi
\item If $C'$ is heavy, $N(C')$ contains at most one cycle at the
$l_1$ side of $C'$ and one cycle at the $l_2$ side of $C$. (see
Figure \ref{SonsofGtagfig} (a)). In this case equivalent lemma to
Lemma \ref{lemtwof}, or lemma \ref{lemthreefmid} or Lemma
\ref{lemfourfmid} applies to prove that $C'$ satisfies the induction
hypotheses.
 \item If $C'$ is light, $N(C')$ contains $C$
and at most two more cycles, both at the $l_2$ side of $C$. (see
Figure \ref{SonsofGtagfig} (b)). In this case equivalent lemma to
Lemma \ref{lemtwof}, lemma \ref{lemthreefs} or Lemma \ref{lemfourfs}
applies to prove that $C'$ satisfies the induction hypotheses.
 \ei
 \item $C$ is the only contained brother (in his level of containment) of $D$ (i.e., $C$ is a
 special-contained brother). By Lemma \ref{longmid} $N(D)$
 contains at most 3 cycles in its $l_1$ side, one of them is $C'$
(see Figure \ref{SonsofGtagfig} (c)). Hence, before DIRECT-ONE is
activated $|N(C')| \leq 2$ (all cycles in $N(C')$ are on the $l_1$
side of $C'$ ). After DIRECT-ONE is activated $C$ is added to
$N(C')$ but still $|N(C')| \leq 3$   and Lemma \ref{lemthreefmid},
or Lemma \ref{lemfourfmid} or equivalent lemma to Lemma
\ref{lemtwof} applies to prove that $C'$ satisfies the induction
hypotheses.
 \ee

\end{prf}

\begin{figure}
\begin{center}
\setlength{\unitlength}{0.00061242in}
\begingroup\makeatletter\ifx\SetFigFont\undefined%
\gdef\SetFigFont#1#2#3#4#5{%
  \reset@font\fontsize{#1}{#2pt}%
  \fontfamily{#3}\fontseries{#4}\fontshape{#5}%
  \selectfont}%
\fi\endgroup%
{\renewcommand{\dashlinestretch}{30}
\begin{picture}(7824,3969)(0,-10)
\put(1518,240){\blacken\ellipse{90}{90}}
\put(1518,240){\ellipse{90}{90}}
\put(3093,240){\blacken\ellipse{90}{90}}
\put(3093,240){\ellipse{90}{90}}
\put(5253,240){\blacken\ellipse{90}{90}}
\put(5253,240){\ellipse{90}{90}}
\put(1518,2130){\blacken\ellipse{90}{90}}
\put(1518,2130){\ellipse{90}{90}}
\put(2508,2130){\blacken\ellipse{90}{90}}
\put(2508,2130){\ellipse{90}{90}}
\put(1518,3255){\blacken\ellipse{90}{90}}
\put(1518,3255){\ellipse{90}{90}}
\put(1518,1140){\blacken\ellipse{90}{90}}
\put(1518,1140){\ellipse{90}{90}}
\put(3273,3255){\blacken\ellipse{90}{90}}
\put(3273,3255){\ellipse{90}{90}}
\put(1923,2130){\blacken\ellipse{90}{90}}
\put(1923,2130){\ellipse{90}{90}}
\put(2778,2535){\blacken\ellipse{90}{90}}
\put(2778,2535){\ellipse{90}{90}}
\path(1518,2130)(3093,2130)
\path(3093,2130)(3093,240)
\path(1518,240)(3093,240)
\path(3093,240)(5253,240)
\thicklines
\path(1518,240)(213,240)
\thinlines
\path(1518,1140)(1518,2130)
\path(1518,2130)(1518,3255)
\thicklines
\path(1518,1140)(1518,240)
\path(2508,2130)(1518,1140)
\path(3273,3255)(2508,2130)
\thinlines
\path(1518,3255)(3273,3255)
\thicklines
\path(3273,3255)(5253,3255)
\path(5253,3255)(5253,240)
\path(5253,240)(6513,240)
\put(2103,1005){\makebox(0,0)[lb]{\smash{{\SetFigFont{8}{9.6}{\rmdefault}{\mddefault}{\updefault}$C$}}}}
\put(3903,3390){\makebox(0,0)[lb]{\smash{{\SetFigFont{8}{9.6}{\rmdefault}{\mddefault}{\updefault}$D$}}}}
\put(1473,15){\makebox(0,0)[lb]{\smash{{\SetFigFont{8}{9.6}{\rmdefault}{\mddefault}{\updefault}$v_2$}}}}
\put(3048,15){\makebox(0,0)[lb]{\smash{{\SetFigFont{8}{9.6}{\rmdefault}{\mddefault}{\updefault}$v_3$}}}}
\put(5118,15){\makebox(0,0)[lb]{\smash{{\SetFigFont{8}{9.6}{\rmdefault}{\mddefault}{\updefault}$v_4$}}}}
\put(3048,2670){\makebox(0,0)[lb]{\smash{{\SetFigFont{8}{9.6}{\rmdefault}{\mddefault}{\updefault}$D'$}}}}
\put(2373,2265){\makebox(0,0)[lb]{\smash{{\SetFigFont{8}{9.6}{\rmdefault}{\mddefault}{\updefault}$b$}}}}
\put(1158,3165){\makebox(0,0)[lb]{\smash{{\SetFigFont{8}{9.6}{\rmdefault}{\mddefault}{\updefault}$u_2$}}}}
\put(1158,2040){\makebox(0,0)[lb]{\smash{{\SetFigFont{8}{9.6}{\rmdefault}{\mddefault}{\updefault}$u_1$}}}}
\put(3228,3390){\makebox(0,0)[lb]{\smash{{\SetFigFont{8}{9.6}{\rmdefault}{\mddefault}{\updefault}$b_1$}}}}
\put(1158,1050){\makebox(0,0)[lb]{\smash{{\SetFigFont{8}{9.6}{\rmdefault}{\mddefault}{\updefault}$b_2$}}}}
\put(1878,2265){\makebox(0,0)[lb]{\smash{{\SetFigFont{8}{9.6}{\rmdefault}{\mddefault}{\updefault}$v_c$}}}}
\put(2508,2535){\makebox(0,0)[lb]{\smash{{\SetFigFont{8}{9.6}{\rmdefault}{\mddefault}{\updefault}$v_d$}}}}
\put(6333,3795){\makebox(0,0)[lb]{\smash{{\SetFigFont{8}{9.6}{\rmdefault}{\mddefault}{\updefault}The cycles}}}}
\put(6333,3570){\makebox(0,0)[lb]{\smash{{\SetFigFont{8}{9.6}{\rmdefault}{\mddefault}{\updefault}$S$ $zv_2v_3v_4z$}}}}
\put(6333,3345){\makebox(0,0)[lb]{\smash{{\SetFigFont{8}{9.6}{\rmdefault}{\mddefault}{\updefault}$D$ $zv_2b_2u_1u_2b_1v_4z$}}}}
\put(438,330){\makebox(0,0)[lb]{\smash{{\SetFigFont{8}{9.6}{\rmdefault}{\mddefault}{\updefault}$S$}}}}
\put(6333,2895){\makebox(0,0)[lb]{\smash{{\SetFigFont{8}{9.6}{\rmdefault}{\mddefault}{\updefault}$D'$ $zv_2b_2bb_1v_4z$}}}}
\put(6333,3120){\makebox(0,0)[lb]{\smash{{\SetFigFont{8}{9.6}{\rmdefault}{\mddefault}{\updefault}$C$ $zv_2b_2u_1bv_3v_4z$}}}}
\end{picture}
}
\end{center}
\caption{Figure for Lemma \ref{goodsonb}.} \label{goodsonbfig}
\end{figure}
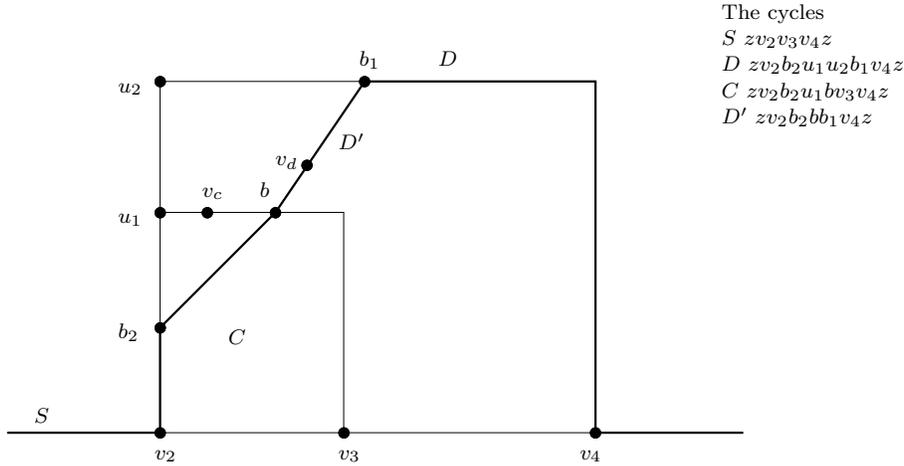

\begin{lem}
\label{goodsonb} Let $D'$ be a son of $D$, then $P_f(D') \cap
[v_2,u_1] = \phi$.
\end{lem}
\begin{prf}
Suppose otherwise, then from  planarity $P_s(D') \cap [u_1,v_3] $
includes at least one node $b$ (see Figure \ref {goodsonbfig}). Let
$v_d$ be a node from $U(D') $ and $v_c$ be a node from $U(C)$. Mark
$b_1,b_2$ the end nodes of $P_s(D')$ such that $b_1 \in [u_2,v_4]$
subpath of $P_s(D)$, $b_2 \in [u_1,v_2]$. If $v_d \in [b_1,b]$ then
the path $[b,b_2]$ is shorter than the path $[b,u_1,b_2]$ so $v_c
\not\in [b,v_3]$ (otherwise it would also use the path $[b,b_2]$
which is not part of $C$). So either $U(D') \subseteq \in [b,b_1]$
and $U(C) \subseteq [b,u_1]$ or $U(D') \subseteq [b,b_2]$ and $U(C)
\subseteq [b,v_3]$.

Suppose that  $v_d \in [b,b_1]$ and $v_c \in [b,u_1]$ (a similar
proof holds for the other case). In this case $l(b,b_2) <
l(b,u_1,b_2)$ and $l(b,v_3,v_4) < l(b,b_1,v_4)$. Mark $NC$ the cycle
$C \backslash [b,u_1,b_2] \cup [b,b_2]$. Then $l(NC) < l(C)$.
Obviously the node $b$ uses this cycle. But then $NC$ is a contained
brother of $D$ and $C$ is the son of $NC$.

\end{prf}

\begin{figure}
\begin{center}
\setlength{\unitlength}{0.00069991in}
\begingroup\makeatletter\ifx\SetFigFont\undefined%
\gdef\SetFigFont#1#2#3#4#5{%
  \reset@font\fontsize{#1}{#2pt}%
  \fontfamily{#3}\fontseries{#4}\fontshape{#5}%
  \selectfont}%
\fi\endgroup%
{\renewcommand{\dashlinestretch}{30}
\begin{picture}(5784,4095)(0,-10)
\put(687,2085){\blacken\ellipse{90}{90}}
\put(687,2085){\ellipse{90}{90}}
\put(2442,2085){\blacken\ellipse{90}{90}}
\put(2442,2085){\ellipse{90}{90}}
\put(3387,2085){\blacken\ellipse{90}{90}}
\put(3387,2085){\ellipse{90}{90}}
\put(4557,2085){\blacken\ellipse{90}{90}}
\put(4557,2085){\ellipse{90}{90}}
\put(2442,3075){\blacken\ellipse{90}{90}}
\put(2442,3075){\ellipse{90}{90}}
\put(2442,3795){\blacken\ellipse{90}{90}}
\put(2442,3795){\ellipse{90}{90}}
\put(1452,2085){\blacken\ellipse{90}{90}}
\put(1452,2085){\ellipse{90}{90}}
\put(4962,2085){\blacken\ellipse{90}{90}}
\put(4962,2085){\ellipse{90}{90}}
\put(2802,240){\blacken\ellipse{90}{90}}
\put(2802,240){\ellipse{90}{90}}
\put(4557,3390){\blacken\ellipse{90}{90}}
\put(4557,3390){\ellipse{90}{90}}
\put(5547,2085){\blacken\ellipse{90}{90}}
\put(5547,2085){\ellipse{90}{90}}
\path(687,2085)(12,2085)
\path(2442,2085)(3387,2085)
\blacken\path(3237.000,2047.500)(3387.000,2085.000)(3237.000,2122.500)(3237.000,2047.500)
\path(3387,2085)(4557,2085)
\blacken\path(4407.000,2047.500)(4557.000,2085.000)(4407.000,2122.500)(4407.000,2047.500)
\path(2442,3750)(2442,3075)
\blacken\path(2404.500,3225.000)(2442.000,3075.000)(2479.500,3225.000)(2404.500,3225.000)
\path(2487,3075)(3387,3075)
\blacken\path(3237.000,3037.500)(3387.000,3075.000)(3237.000,3112.500)(3237.000,3037.500)
\path(3387,3075)(3387,2085)
\blacken\path(3349.500,2235.000)(3387.000,2085.000)(3424.500,2235.000)(3349.500,2235.000)
\path(1452,2085)(2442,2085)
\blacken\path(2292.000,2047.500)(2442.000,2085.000)(2292.000,2122.500)(2292.000,2047.500)
\blacken\path(2479.500,2925.000)(2442.000,3075.000)(2404.500,2925.000)(2479.500,2925.000)
\path(2442,3075)(2442,2085)
\path(4962,2085)(5772,2085)
\dashline{60.000}(4557,3390)(5547,3390)
\blacken\path(5397.000,3352.500)(5547.000,3390.000)(5397.000,3427.500)(5397.000,3352.500)
\dashline{60.000}(5547,3390)(5547,2085)
\blacken\path(5509.500,2235.000)(5547.000,2085.000)(5584.500,2235.000)(5509.500,2235.000)
\thicklines
\path(2802,240)(1452,2085)
\blacken\thinlines
\path(1570.840,1986.090)(1452.000,2085.000)(1510.313,1941.801)(1570.840,1986.090)
\blacken\path(837.000,2122.500)(687.000,2085.000)(837.000,2047.500)(837.000,2122.500)
\thicklines
\path(687,2085)(1452,2085)
\path(687,2085)(687,3795)
\blacken\thinlines
\path(724.500,3645.000)(687.000,3795.000)(649.500,3645.000)(724.500,3645.000)
\thicklines
\path(687,3795)(2442,3795)
\blacken\thinlines
\path(2292.000,3757.500)(2442.000,3795.000)(2292.000,3832.500)(2292.000,3757.500)
\thicklines
\path(2442,3795)(4557,3795)
\blacken\thinlines
\path(4407.000,3757.500)(4557.000,3795.000)(4407.000,3832.500)(4407.000,3757.500)
\thicklines
\path(4557,3795)(4557,2085)
\blacken\thinlines
\path(4519.500,2235.000)(4557.000,2085.000)(4594.500,2235.000)(4519.500,2235.000)
\thicklines
\path(4557,2085)(4962,2085)
\blacken\thinlines
\path(4812.000,2047.500)(4962.000,2085.000)(4812.000,2122.500)(4812.000,2047.500)
\thicklines
\path(4962,2085)(2757,240)
\blacken\thinlines
\path(2847.976,365.019)(2757.000,240.000)(2896.105,307.498)(2847.976,365.019)
\put(3342,1860){\makebox(0,0)[lb]{\smash{{\SetFigFont{10}{12.0}{\rmdefault}{\mddefault}{\updefault}$v_3$}}}}
\put(4512,1860){\makebox(0,0)[lb]{\smash{{\SetFigFont{10}{12.0}{\rmdefault}{\mddefault}{\updefault}$v_4$}}}}
\put(642,1860){\makebox(0,0)[lb]{\smash{{\SetFigFont{10}{12.0}{\rmdefault}{\mddefault}{\updefault}$v_1$}}}}
\put(2397,1860){\makebox(0,0)[lb]{\smash{{\SetFigFont{10}{12.0}{\rmdefault}{\mddefault}{\updefault}$v_2$}}}}
\put(2802,2535){\makebox(0,0)[lb]{\smash{{\SetFigFont{10}{12.0}{\rmdefault}{\mddefault}{\updefault}$C$}}}}
\put(1587,3930){\makebox(0,0)[lb]{\smash{{\SetFigFont{10}{12.0}{\rmdefault}{\mddefault}{\updefault}$C'$}}}}
\put(3567,3930){\makebox(0,0)[lb]{\smash{{\SetFigFont{10}{12.0}{\rmdefault}{\mddefault}{\updefault}$D$}}}}
\put(2172,2985){\makebox(0,0)[lb]{\smash{{\SetFigFont{10}{12.0}{\rmdefault}{\mddefault}{\updefault}$u_1$}}}}
\put(2397,3930){\makebox(0,0)[lb]{\smash{{\SetFigFont{10}{12.0}{\rmdefault}{\mddefault}{\updefault}$u_2$}}}}
\put(2757,15){\makebox(0,0)[lb]{\smash{{\SetFigFont{10}{12.0}{\rmdefault}{\mddefault}{\updefault}$z$}}}}
\put(1452,1140){\makebox(0,0)[lb]{\smash{{\SetFigFont{10}{12.0}{\rmdefault}{\mddefault}{\updefault}$M_t(S)$}}}}
\put(4197,1140){\makebox(0,0)[lb]{\smash{{\SetFigFont{10}{12.0}{\rmdefault}{\mddefault}{\updefault}$M_h(S)$}}}}
\put(1362,2220){\makebox(0,0)[lb]{\smash{{\SetFigFont{10}{12.0}{\rmdefault}{\mddefault}{\updefault}$t(S)$}}}}
\put(4872,2220){\makebox(0,0)[lb]{\smash{{\SetFigFont{10}{12.0}{\rmdefault}{\mddefault}{\updefault}$h(S)$}}}}
\put(5007,3525){\makebox(0,0)[lb]{\smash{{\SetFigFont{10}{12.0}{\rmdefault}{\mddefault}{\updefault}$E$}}}}
\put(4287,3345){\makebox(0,0)[lb]{\smash{{\SetFigFont{10}{12.0}{\rmdefault}{\mddefault}{\updefault}$u_3$}}}}
\end{picture}
}
\end{center}
\caption{Figure for Lemma \ref{lemGisgood}.} \label{Gisgoodfig}
\end{figure}
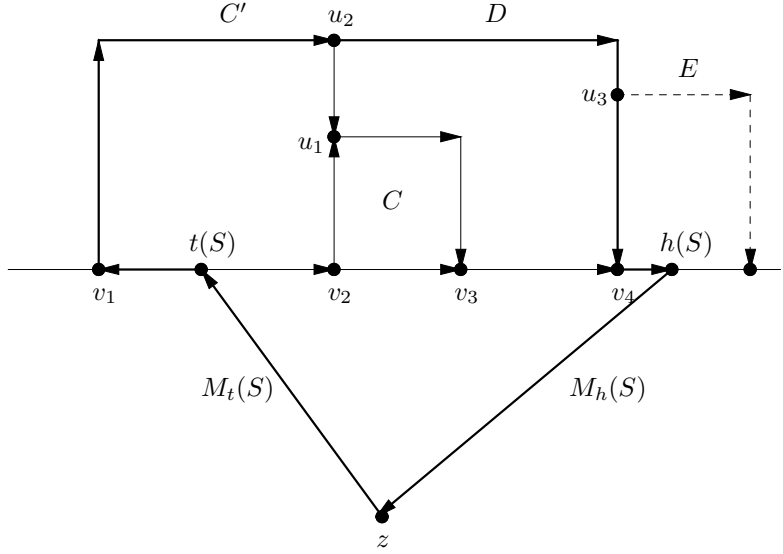

\begin{lem}
\label{lemGisgood} Suppose that  $l(C)< l(C')$, then after
DIRECT-ONE is activated $D$ satisfies the induction hypotheses. See
Figure \ref{Gisgoodfig}.
\end{lem}
\begin{prf}
We go over the induction hypotheses:
\begin{description}
 \item[A1] When DIRECT-ONE is activated on $C$ the direction of the
path $[u_1,v_2]$ is changed.   Since $[u_1,v_2] \cap  (M(D) \cup
I(D))$ doesn't contain an edge, $M(D)$ and  $I(D)$ are not changed.
Since $D$ satisfied A1 before DIRECT-ONE  the bound on $M(D)$ still
holds.
 \item [A2] We
prove this by bounding the  $l(B(D))$ (see Lemma \ref{lbscor}).
Since $C'$ is a son of $S$, from A4 on $S$ $h(S) \not \in P_f(D)$.
W.l.o.g we also assume that $t(S) \not \in P_f(d)$ (if $t(S) \in
P_f(D)$ the length of $B(D)$ is even shorter). There are two cases
to be considered.
 \bi
 \item $N(D)$ doesn't contain another cycle on the $l_2$ side of
 $D$, so $I(D) = [u_2 \rightarrow v_4]$. Define $B(D) = [u_2 \rightarrow u_1 \rightarrow v_3 \rightarrow v_4]$.
 Before DIRECT-ONE was applied $D$ satisfied the induction hypotheses, so by A5 $l(C') < l(D)$. From $l(C) < l(C')$ also
$l(C) < l(D)$ and\footnote{$[u_1,v_3]$ is the subpath of $P_s(C)$.}:
\begin{equation}
\label{CsmC'smD} l(u_1,v_3) < l(u_1,v_2) + l(v_2,v_3).
\end {equation}

Using Equation \ref{CsmC'smD} and since $D$ is a son of $S$:
\begin{eqnarray*}
l(B(D)) &=& l(u_2,u_1) + l(u_1,v_3) + l(v_3,v_4) \\
&<& l(u_2,u_1) + l(u_1,v_2) + l(v_2,v_3) + l(v_3,v_4)  \\
&=& l(P_s(D)) - l(I(D)) + l(v_2,v_3,v_4) \\
&\leq& l(P_s(D)) - l(I(D)) + l(P_s(D)) .
\end{eqnarray*}
 \item $N(D)$ contains another cycle $E$ on the $l_2$ side of $D$.
  Mark $[u_3,v_4] = P_s(D) \cap P_s(E)$. In this case before the
application of  DIRECT-ONE,  $B_{old}(D)$ was $[u_2 \rightarrow u_1
\rightarrow v_2 \rightarrow v_3 \rightarrow v_4 \rightarrow u_3]$.
Define  $B_{new}(D) = [u_2 \rightarrow u_1
 \rightarrow v_3 \rightarrow v_4 \rightarrow
u_3]$.  Using again Equation \ref{CsmC'smD} we conclude that
$B_{new}(D) < B_{old}(D)$.
 \ei

\item [A3] $I(D)$ hasn't changed so A3 still holds.
 \item [A4] $J_t(D)$ has  shortened
to be $[u_2 \rightarrow u_1]$, so $J_t(D) \cup I(D) \cup J_h(D) =
[u_1,u_2,v_4]$. By Lemma \ref{goodsonb} if $D'$ is a son of $D$ then
$P_f(D') \subset [u_1,u_2,v_4]$.
 \item [A5] $M(D)$ hasn't changed so A5 still holds.
 \item [A6] $M(D)$ hasn't changed so A6 still holds.

\end{description}
\end{prf}

\newpage
\subsubsection{$C''$ is an uncle (or from previous generation)}
\label{secuncle}
 In this section we consider the case $C''$ is an uncle (similarly
$C''$ can be a brother of an ancestor). In the following we assume
that $P_s(C)$ is inside the part of the plane surrounded by $P_s(S)
\cup P_f(S)$ $($where $S = F(C))$.  Similar proofs follow for the
case it is outside this part of the plane.

\paragraph{$S=F(C)$ is oriented forwards\\ \\}
\label{secfor}

\begin{figure}
\begin{center}
\setlength{\unitlength}{0.00069991in}
\begingroup\makeatletter\ifx\SetFigFont\undefined%
\gdef\SetFigFont#1#2#3#4#5{%
  \reset@font\fontsize{#1}{#2pt}%
  \fontfamily{#3}\fontseries{#4}\fontshape{#5}%
  \selectfont}%
\fi\endgroup%
{\renewcommand{\dashlinestretch}{30}
\begin{picture}(9742,4015)(0,-10)
\put(3453,250){\blacken\ellipse{90}{90}}
\put(3453,250){\ellipse{90}{90}}
\put(3453,1780){\blacken\ellipse{90}{90}}
\put(3453,1780){\ellipse{90}{90}}
\put(6738,250){\blacken\ellipse{90}{90}}
\put(6738,250){\ellipse{90}{90}}
\put(1833,3715){\blacken\ellipse{90}{90}}
\put(1833,3715){\ellipse{90}{90}}
\put(978,250){\blacken\ellipse{90}{90}}
\put(978,250){\ellipse{90}{90}}
\put(5343,250){\blacken\ellipse{90}{90}}
\put(5343,250){\ellipse{90}{90}}
\put(5343,1780){\blacken\ellipse{90}{90}}
\put(5343,1780){\ellipse{90}{90}}
\put(5343,3715){\blacken\ellipse{90}{90}}
\put(5343,3715){\ellipse{90}{90}}
\path(978,250)(3453,250)
\blacken\path(3303.000,212.500)(3453.000,250.000)(3303.000,287.500)(3303.000,212.500)
\path(3453,250)(5343,250)
\blacken\path(5193.000,212.500)(5343.000,250.000)(5193.000,287.500)(5193.000,212.500)
\thicklines
\path(5343,250)(6738,250)
\blacken\thinlines
\path(6588.000,212.500)(6738.000,250.000)(6588.000,287.500)(6588.000,212.500)
\thicklines
\path(978,250)(978,3715)
\blacken\thinlines
\path(1015.500,3565.000)(978.000,3715.000)(940.500,3565.000)(1015.500,3565.000)
\blacken\path(3415.500,400.000)(3453.000,250.000)(3490.500,400.000)(3415.500,400.000)
\path(3453,250)(3453,1780)
\blacken\path(5380.500,3565.000)(5343.000,3715.000)(5305.500,3565.000)(5380.500,3565.000)
\path(5343,3715)(5343,1780)
\dashline{60.000}(5343,3715)(6738,3715)
\blacken\path(6588.000,3677.500)(6738.000,3715.000)(6588.000,3752.500)(6588.000,3677.500)
\dashline{60.000}(6738,3715)(6738,250)
\blacken\path(6700.500,400.000)(6738.000,250.000)(6775.500,400.000)(6700.500,400.000)
\blacken\path(828.000,212.500)(978.000,250.000)(828.000,287.500)(828.000,212.500)
\thicklines
\path(978,250)(213,250)
\blacken\thinlines
\path(1683.000,3677.500)(1833.000,3715.000)(1683.000,3752.500)(1683.000,3677.500)
\thicklines
\path(1833,3715)(978,3715)
\path(1833,3715)(3453,1780)
\blacken\thinlines
\path(3603.000,1817.500)(3453.000,1780.000)(3603.000,1742.500)(3603.000,1817.500)
\thicklines
\path(3453,1780)(5343,1780)
\blacken\thinlines
\path(5380.500,1630.000)(5343.000,1780.000)(5305.500,1630.000)(5380.500,1630.000)
\thicklines
\path(5343,1780)(5343,250)
\path(6738,250)(7503,250)
\blacken\thinlines
\path(7353.000,212.500)(7503.000,250.000)(7353.000,287.500)(7353.000,212.500)
\path(1833,3715)(5343,3715)
\blacken\path(5193.000,3677.500)(5343.000,3715.000)(5193.000,3752.500)(5193.000,3677.500)
\put(2037,2805){\makebox(0,0)[lb]{\smash{{\SetFigFont{10}{12.0}{\rmdefault}{\mddefault}{\updefault}$C$}}}}
\put(4062,1815){\makebox(0,0)[lb]{\smash{{\SetFigFont{10}{12.0}{\rmdefault}{\mddefault}{\updefault}$C''$}}}}
\put(3207,15){\makebox(0,0)[lb]{\smash{{\SetFigFont{10}{12.0}{\rmdefault}{\mddefault}{\updefault}$v_4$}}}}
\put(1587,3840){\makebox(0,0)[lb]{\smash{{\SetFigFont{10}{12.0}{\rmdefault}{\mddefault}{\updefault}$v_1$}}}}
\put(372,375){\makebox(0,0)[lb]{\smash{{\SetFigFont{10}{12.0}{\rmdefault}{\mddefault}{\updefault}$T$}}}}
\put(732,15){\makebox(0,0)[lb]{\smash{{\SetFigFont{10}{12.0}{\rmdefault}{\mddefault}{\updefault}$v_3$}}}}
\put(5097,15){\makebox(0,0)[lb]{\smash{{\SetFigFont{10}{12.0}{\rmdefault}{\mddefault}{\updefault}$v_5$}}}}
\put(2847,3840){\makebox(0,0)[lb]{\smash{{\SetFigFont{10}{12.0}{\rmdefault}{\mddefault}{\updefault}$S$}}}}
\put(5862,3840){\makebox(0,0)[lb]{\smash{{\SetFigFont{10}{12.0}{\rmdefault}{\mddefault}{\updefault}$S''$}}}}
\put(5097,3840){\makebox(0,0)[lb]{\smash{{\SetFigFont{10}{12.0}{\rmdefault}{\mddefault}{\updefault}$u_2$}}}}
\put(6942,3480){\makebox(0,0)[lb]{\smash{{\SetFigFont{10}{12.0}{\rmdefault}{\mddefault}{\updefault}The cycles}}}}
\put(6942,3255){\makebox(0,0)[lb]{\smash{{\SetFigFont{10}{12.0}{\rmdefault}{\mddefault}{\updefault}$T=F(S)=F(S'')=F(C'')$ $zv_3v_4v_5z$}}}}
\put(6942,3030){\makebox(0,0)[lb]{\smash{{\SetFigFont{10}{12.0}{\rmdefault}{\mddefault}{\updefault}$S=F(C)$ $zv_3v_1u_2v_2v_5z$}}}}
\put(6942,2805){\makebox(0,0)[lb]{\smash{{\SetFigFont{10}{12.0}{\rmdefault}{\mddefault}{\updefault}$C$ $zv_3v_1u_1v_2v_5z$}}}}
\put(6942,2580){\makebox(0,0)[lb]{\smash{{\SetFigFont{10}{12.0}{\rmdefault}{\mddefault}{\updefault}$C''$ $zv_3v_4u_1v_2v_5z$}}}}
\put(5568,1735){\makebox(0,0)[lb]{\smash{{\SetFigFont{10}{12.0}{\rmdefault}{\mddefault}{\updefault}$v_2$}}}}
\put(3183,1735){\makebox(0,0)[lb]{\smash{{\SetFigFont{10}{12.0}{\rmdefault}{\mddefault}{\updefault}$u_1$}}}}
\end{picture}
}
\end{center}
\caption{$C''$ is an uncle, $S$ is oriented forwards.}
\label{fcforfig}
\end{figure}
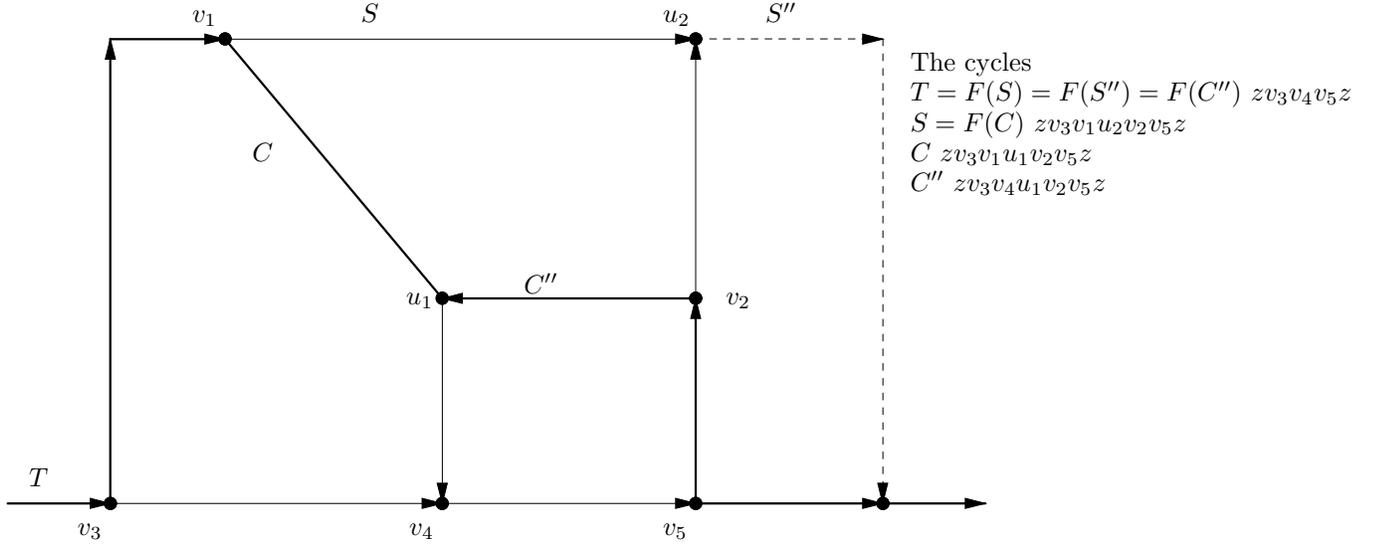

 Here we consider a  cycle $C$ oriented forwards by
DIRECT-ONE even though  $l_2=-1$. In this subsection we assume that
$S=F(C)$ is oriented forwards, $T=F(S)$ and $C''$ a contained
brother of $S$ (see Figure \ref{fcforfig}). $C$ which is a son of
$S$ shares the path $[v_2,u_1]$ at its $l_2$ side with $C''$. Thus
$C$ has to be the last cycle in the block of  sons of $S$, and $C''$
is oriented backwards. In Figure \ref{fcforfig} $C,C'',S$ and $T$
are described before the application of DIRECT-ONE.

 Mark \bi
\item $v_1:= S \cap P_s(C) \backslash  P_s(C'')$.
\item $v_2:= S \cap P_s(C) \cap  P_s(C'')$.
\item $v_3:= T \cap P_s(S) \backslash P_s(C'')$.
\item $v_4:= T \cap P_s(C'') \backslash P_s(S)$.
\item $v_5:= T \cap P_s(C'') \cap P_s(S)$.
\item $u_1:= $ the end node of the path $P_s(C) \cap P_s(C'')$, such that $u_1 \not\in S$.
\ei

Since $C''$ is oriented  backwards $N(S)$ contains at least one
cycle at the $l_2$ side of $P_s(S)$ $(S''$ in Figure
\ref{fcforfig}). We consider the case where  $v_3 = t(S)$ (similar
but longer proof holds otherwise$)$. When DIRECT-ONE is activated on
$C$ it compares $l(C)$ and $l(C'') + 2l(P_c(C'') \backslash C)$ and
decides whether to change the direction of $[u_1,v_2]$.  In the
following two lemmas we assume $[v_1,u_1]$ was directed $v_1
\rightarrow u_1$ ($C$ didn't have heavy brothers at its $l_1$ side
to contradict this direction). In lemmas \ref{lemfourbwmid} and
\ref{lemfourbwsid} we will consider the other case.

\begin{lem}
\label{lemCDforC} Suppose that $l(C'') + 2l(P_c(C'') \backslash C) <
l(C) $, then  $C$ satisfies the induction hypotheses.
\end{lem}
\begin{prf}
In this case when DIRECT-ONE is activated on $C$ it doesn't change
the direction $v_2 \rightarrow u_1$.

 Since $ l(C'') + 2l(P_c(C'') \backslash C)  < l(C) $
\[ l(v_3,v_4) + l(v_4,u_1) + l(u_1,v_2) + l(v_2,v_5) +2l(v_3,v_4)
< l(v_3,v_1) + l(v_1,u_1) + l(u_1,v_2) + l(v_2,v_5) \] giving:
\begin{equation}
\label{csmd2f} 3l(v_3,v_4)+ l(v_4,u_1)  < l(v_3,v_1) + l(v_1,u_1).
\end{equation}

Since $T = F(S)$
\begin{equation}
\label{SsonT} l(v_3,v_4) + l(v_4,v_5) < l(v_3,v_1) + l(v_1,v_2) +
l(v_2,v_5) .
\end{equation}

 Since $C$ is a son of $S$
 \begin{equation}
 \label{CsonB}
 l(v_1,v_2) < l(P_s(C)).
 \end{equation}

\begin{description}
\item [A1] Let $x=t(T),y=h(T)$, and  the end nodes $v,w$ of
$P_s(T)$ be ordered $v,x,y,w$.

Since $C''$ was oriented  backwards by Property \ref{longfwd}:
\begin{equation}
\label{shortD} 2l(v_4,v_5) < l(v,v_3) + l(v_3,v_4) + l(v_5,w).
\end{equation}

Since $C''$ was oriented  backwards and $S$ is oriented  forwards,
by Property \ref{scbprop} $C''$ is not a special-contained brother
of $S$, and there is at least one more brother contained in $S$ in
the same level of containment as $C''$. Hence,  By A3 on $T$ $x \in
[v,v_4]$. Since there is at least one more brother of $S$ on its
$l_2$  side   $y \in [v_5,w]$. $I(C) = [v_1 \rightarrow u_1]$ and by
A4 we can define (see Figure \ref{fcforctagfig})

\begin{figure}
\begin{center}
\setlength{\unitlength}{0.00061242in}
\begingroup\makeatletter\ifx\SetFigFont\undefined%
\gdef\SetFigFont#1#2#3#4#5{%
  \reset@font\fontsize{#1}{#2pt}%
  \fontfamily{#3}\fontseries{#4}\fontshape{#5}%
  \selectfont}%
\fi\endgroup%
{\renewcommand{\dashlinestretch}{30}
\begin{picture}(8673,5940)(0,-10)
\put(3840,2175){\blacken\ellipse{90}{90}}
\put(3840,2175){\ellipse{90}{90}}
\put(3840,3705){\blacken\ellipse{90}{90}}
\put(3840,3705){\ellipse{90}{90}}
\put(2220,5640){\blacken\ellipse{90}{90}}
\put(2220,5640){\ellipse{90}{90}}
\put(1365,2175){\blacken\ellipse{90}{90}}
\put(1365,2175){\ellipse{90}{90}}
\put(5730,2175){\blacken\ellipse{90}{90}}
\put(5730,2175){\ellipse{90}{90}}
\put(5730,3705){\blacken\ellipse{90}{90}}
\put(5730,3705){\ellipse{90}{90}}
\put(5730,5640){\blacken\ellipse{90}{90}}
\put(5730,5640){\ellipse{90}{90}}
\put(7125,2175){\blacken\ellipse{90}{90}}
\put(7125,2175){\ellipse{90}{90}}
\put(7890,2175){\blacken\ellipse{90}{90}}
\put(7890,2175){\ellipse{90}{90}}
\put(60,2175){\blacken\ellipse{90}{90}}
\put(60,2175){\ellipse{90}{90}}
\put(8565,2175){\blacken\ellipse{90}{90}}
\put(8565,2175){\ellipse{90}{90}}
\put(4380,240){\blacken\ellipse{90}{90}}
\put(4380,240){\ellipse{90}{90}}
\put(600,2175){\blacken\ellipse{90}{90}}
\put(600,2175){\ellipse{90}{90}}
\path(1365,2175)(3840,2175)
\blacken\path(3690.000,2137.500)(3840.000,2175.000)(3690.000,2212.500)(3690.000,2137.500)
\blacken\path(3990.000,3742.500)(3840.000,3705.000)(3990.000,3667.500)(3990.000,3742.500)
\path(3840,3705)(5730,3705)
\blacken\path(5767.500,5490.000)(5730.000,5640.000)(5692.500,5490.000)(5767.500,5490.000)
\path(5730,5640)(5730,3705)
\blacken\path(5767.500,3555.000)(5730.000,3705.000)(5692.500,3555.000)(5767.500,3555.000)
\path(5730,3705)(5730,2175)
\path(5730,5640)(7125,5640)
\blacken\path(6975.000,5602.500)(7125.000,5640.000)(6975.000,5677.500)(6975.000,5602.500)
\path(7125,5640)(7125,2175)
\blacken\path(7087.500,2325.000)(7125.000,2175.000)(7162.500,2325.000)(7087.500,2325.000)
\path(600,2175)(60,2175)
\path(7890,2175)(8565,2175)
\dottedline{45}(1950,4560)(1815,5550)
\path(1860.938,5435.154)(1815.000,5550.000)(1801.489,5427.047)
\dottedline{45}(1950,4560)(1455,4650)
\path(1578.431,4658.050)(1455.000,4650.000)(1567.698,4599.018)
\dottedline{45}(2850,4110)(3120,4425)
\path(3064.683,4314.365)(3120.000,4425.000)(3019.127,4353.413)
\dottedline{45}(4920,2850)(3885,3210)
\path(4008.195,3198.912)(3885.000,3210.000)(3988.484,3142.243)
\dottedline{45}(4920,2850)(4920,2220)
\path(4890.000,2340.000)(4920.000,2220.000)(4950.000,2340.000)
\dottedline{45}(4920,2850)(6405,2220)
\path(6282.814,2239.249)(6405.000,2220.000)(6306.247,2294.483)
\thicklines
\path(1365,2175)(1365,5640)
\blacken\thinlines
\path(1402.500,5490.000)(1365.000,5640.000)(1327.500,5490.000)(1402.500,5490.000)
\path(2220,5640)(5730,5640)
\thicklines
\path(1365,5640)(2220,5640)
\path(2220,5640)(3840,3705)
\blacken\thinlines
\path(3714.956,3795.941)(3840.000,3705.000)(3772.463,3844.086)(3714.956,3795.941)
\blacken\path(3802.500,2325.000)(3840.000,2175.000)(3877.500,2325.000)(3802.500,2325.000)
\thicklines
\path(3840,2175)(3840,3705)
\path(3840,2175)(5730,2175)
\blacken\thinlines
\path(5580.000,2137.500)(5730.000,2175.000)(5580.000,2212.500)(5580.000,2137.500)
\thicklines
\path(5730,2175)(7125,2175)
\blacken\thinlines
\path(6975.000,2137.500)(7125.000,2175.000)(6975.000,2212.500)(6975.000,2137.500)
\dottedline{45}(4920,2850)(7575,2220)
\path(7451.316,2218.516)(7575.000,2220.000)(7465.168,2276.895)
\thicklines
\path(7125,2175)(7890,2175)
\blacken\thinlines
\path(7740.000,2137.500)(7890.000,2175.000)(7740.000,2212.500)(7740.000,2137.500)
\blacken\path(4493.257,345.257)(4380.000,240.000)(4529.465,279.577)(4493.257,345.257)
\thicklines
\path(4380,240)(7890,2175)
\thinlines
\dottedline{45}(2535,2805)(2355,1320)
\path(2339.658,1442.738)(2355.000,1320.000)(2399.222,1435.518)
\dottedline{45}(2535,2805)(4470,2355)
\path(4346.324,2352.961)(4470.000,2355.000)(4359.914,2411.402)
\dottedline{45}(2535,2805)(5685,3255)
\path(5570.449,3208.331)(5685.000,3255.000)(5561.963,3267.728)
\dottedline{45}(2535,2805)(4920,3615)
\path(4816.022,3548.004)(4920.000,3615.000)(4796.727,3604.817)
\blacken\path(1215.000,2137.500)(1365.000,2175.000)(1215.000,2212.500)(1215.000,2137.500)
\thicklines
\path(1365,2175)(600,2175)
\path(4380,240)(600,2175)
\blacken\thinlines
\path(750.610,2140.030)(600.000,2175.000)(716.435,2073.269)(750.610,2140.030)
\dottedline{45}(2535,2805)(1005,2220)
\path(1106.372,2290.878)(1005.000,2220.000)(1127.800,2234.835)
\dottedline{45}(1950,4560)(960,2220)
\path(979.128,2342.205)(960.000,2220.000)(1034.386,2318.827)
\dottedline{45}(1950,4560)(2220,1410)
\path(2179.861,1527.000)(2220.000,1410.000)(2239.642,1532.124)
\put(2625,4740){\makebox(0,0)[lb]{\smash{{\SetFigFont{8}{9.6}{\rmdefault}{\mddefault}{\updefault}$C$}}}}
\put(4650,3750){\makebox(0,0)[lb]{\smash{{\SetFigFont{8}{9.6}{\rmdefault}{\mddefault}{\updefault}$C''$}}}}
\put(5910,3660){\makebox(0,0)[lb]{\smash{{\SetFigFont{8}{9.6}{\rmdefault}{\mddefault}{\updefault}$v_2$}}}}
\put(3795,1950){\makebox(0,0)[lb]{\smash{{\SetFigFont{8}{9.6}{\rmdefault}{\mddefault}{\updefault}$v_4$}}}}
\put(3525,3660){\makebox(0,0)[lb]{\smash{{\SetFigFont{8}{9.6}{\rmdefault}{\mddefault}{\updefault}$u_1$}}}}
\put(2175,5775){\makebox(0,0)[lb]{\smash{{\SetFigFont{8}{9.6}{\rmdefault}{\mddefault}{\updefault}$v_1$}}}}
\put(1320,1950){\makebox(0,0)[lb]{\smash{{\SetFigFont{8}{9.6}{\rmdefault}{\mddefault}{\updefault}$v_3$}}}}
\put(5685,1950){\makebox(0,0)[lb]{\smash{{\SetFigFont{8}{9.6}{\rmdefault}{\mddefault}{\updefault}$v_5$}}}}
\put(6450,5775){\makebox(0,0)[lb]{\smash{{\SetFigFont{8}{9.6}{\rmdefault}{\mddefault}{\updefault}$S''$}}}}
\put(5685,5775){\makebox(0,0)[lb]{\smash{{\SetFigFont{8}{9.6}{\rmdefault}{\mddefault}{\updefault}$u_2$}}}}
\put(15,1950){\makebox(0,0)[lb]{\smash{{\SetFigFont{8}{9.6}{\rmdefault}{\mddefault}{\updefault}$v$}}}}
\put(555,1950){\makebox(0,0)[lb]{\smash{{\SetFigFont{8}{9.6}{\rmdefault}{\mddefault}{\updefault}$x$}}}}
\put(7845,1950){\makebox(0,0)[lb]{\smash{{\SetFigFont{8}{9.6}{\rmdefault}{\mddefault}{\updefault}$y$}}}}
\put(8520,1950){\makebox(0,0)[lb]{\smash{{\SetFigFont{8}{9.6}{\rmdefault}{\mddefault}{\updefault}$w$}}}}
\put(1860,1095){\makebox(0,0)[lb]{\smash{{\SetFigFont{8}{9.6}{\rmdefault}{\mddefault}{\updefault}$M_t(T)$}}}}
\put(6315,1095){\makebox(0,0)[lb]{\smash{{\SetFigFont{8}{9.6}{\rmdefault}{\mddefault}{\updefault}$M_h(T)$}}}}
\put(2535,4020){\makebox(0,0)[lb]{\smash{{\SetFigFont{8}{9.6}{\rmdefault}{\mddefault}{\updefault}$I(C)$}}}}
\put(4965,2895){\makebox(0,0)[lb]{\smash{{\SetFigFont{8}{9.6}{\rmdefault}{\mddefault}{\updefault}$M_h(C)$}}}}
\put(3435,5775){\makebox(0,0)[lb]{\smash{{\SetFigFont{8}{9.6}{\rmdefault}{\mddefault}{\updefault}$S$}}}}
\put(2220,2940){\makebox(0,0)[lb]{\smash{{\SetFigFont{8}{9.6}{\rmdefault}{\mddefault}{\updefault}$B_h(C)$}}}}
\put(240,2310){\makebox(0,0)[lb]{\smash{{\SetFigFont{8}{9.6}{\rmdefault}{\mddefault}{\updefault}$T$}}}}
\put(2040,4470){\makebox(0,0)[lb]{\smash{{\SetFigFont{8}{9.6}{\rmdefault}{\mddefault}{\updefault}$M_t(C)$}}}}
\put(4335,15){\makebox(0,0)[lb]{\smash{{\SetFigFont{8}{9.6}{\rmdefault}{\mddefault}{\updefault}$z$}}}}
\end{picture}
}
\end{center}
\caption{Figure for Lemma \ref{lemCDforC}.} \label{fcforctagfig}
\end{figure}

\[ M_t(C) = M_t(T) \cup [x \rightarrow v_3 \rightarrow v_1] \ \
M_h(C) = [u_1 \rightarrow v_4 \rightarrow v_5 \rightarrow y] \cup
M_h(T).\]

Using A1 on $T$:
\begin{eqnarray*}
l(M(C)) &=& l(M(T)) + l(x,v_3) + l(v_3,v_1) + l(u_1,v_4) +
l(v_4,v_5) + l(v_5,y) \\
&\leq& 15l(P_c(T)) +  4l(v,v_3) + 3l(v_3,v_4) + 4l(v_4,v_5) +
4l(v_5,w) + l(v_3,v_1) + l(u_1,v_4).
\end{eqnarray*}

The length of $[v,v_3]$ is  taken 4 times to satisfy both cases of
locations for $x$. Using Equations (\ref{shortD}), (\ref{csmd2f}),
(\ref{SsonT}), the definition of $P_c(C)$, (\ref{CsonB}) and the
definition of $I(C)$

\begin{eqnarray*}
l(M(C)) &\leq& 15l(P_c(T)) +  5l(v,v_3) + 4l(v_3,v_4) + 2l(v_4,v_5)
+ 5l(v_5,w) + l(v_3,v_1)  + l(u_1,v_4) \\
&\leq& 15l(P_c(T)) +  5l(v,v_3) + l(v_3,v_4) + 2l(v_4,v_5)
+ 5l(v_5,w) + 2l(v_3,v_1) +l(v_1,u_1)  \\
&\leq& 15l(P_c(T)) +  5l(v,v_3) +  5l(v_5,w) + 4l(v_3,v_1) +l(v_1,u_1) + 2l(v_1,v_2) + 2l(v_2,v_5) \\
&\leq& 15l(P_c(C)) +    l(v_1,u_1) + 2l(v_1,v_2)  \\
&\leq& 15l(P_c(C)) +    l(v_1,u_1) + 2l(P_s(C))  \\
&=& 15l(P_c(C)) +    l(I(C)) + 2l(P_s(C)).
\end{eqnarray*}

\item [A2] In this case $t(C) = v_1 \in S$ but $h(C) = u_1 \not\in
S$. So we define
\[ B_h(C) = M_t(T) \cup [x \rightarrow v_3 \rightarrow v_4
\rightarrow v_5 \rightarrow v_2 \rightarrow u_1] \] (if $x \in
[v_3,v_4]$ the path is even shorter).

This gives
\begin{eqnarray*}
l(B_h(C)) + l(M_h(C)) &\leq& l(M_t(T)) + l(M_h(T)) + l(x,v_3) +
l(v_3,v_4) + 2l(v_4,v_5) + \\
&& l(v_5,v_2) + l(v_2,u_1)  + l(u_1,v_4) +l(v_5,y) .
\end{eqnarray*}

Using A1 on $T$, the notation we use (again $l(v,v_3)$ is counted 4
times to satisfy all cases of location of $x$), Equations
(\ref{shortD}), (\ref{csmd2f}), (\ref{SsonT}), the definition of
$P_c(C)$, (\ref{CsonB}) and the definition of $P_s(C)$:
\begin{eqnarray*}
l(B_h(C)) + l(M_h(C)) &\leq& 15l(P_c(T)) + 2l(P_s(T)) + l(I(T)) +
l(x,v_3) +l(v_3,v_4) + \\
&&  2l(v_4,v_5) +l(v_5,v_2) + l(v_2,u_1) + l(u_1,v_4) +l(v_5,y) \\
 &\leq& 15l(P_c(T)) +4l(v,v_3) + 4l(v_3,v_4) + 5l(v_4,v_5) + 4l(v_5,w)
  + \\
  &&l(v_5,v_2) + l(v_2,u_1) + l(u_1,v_4)  \\
 &\leq& 15l(P_c(T)) +5l(v,v_3) + 5l(v_3,v_4) + 3l(v_4,v_5) + 5l(v_5,w)
  + \\
  &&l(v_5,v_2) + l(v_2,u_1) + l(u_1,v_4)  \\
 &\leq& 15l(P_c(T)) +5l(v,v_3) + 2l(v_3,v_4) + 3l(v_4,v_5) + 5l(v_5,w)
  + \\
  &&l(v_5,v_2) + l(v_2,u_1)  + l(v_3,v_1) + l(v_1,u_1)  \\
&\leq& 15l(P_c(T)) +5l(v,v_3) +5l(v_5,w)  + 4l(v_5,v_2) +
l(v_2,u_1) + \\
&& 4l(v_3,v_1) + l(v_1,u_1) + 3l(v_1,v_2)  \\
&\leq& 15l(P_c(C)) +l(v_2,u_1) +  l(v_1,u_1) + 3l(v_1,v_2)  \\
&\leq& 15l(P_c(C)) +l(v_2,u_1) +  l(v_1,u_1) + 3l(P_s(C))  \\
&\leq& 15l(P_c(C))  + 4l(P_s(C)) .
\end{eqnarray*}

\item[A3] Since $2l(P_c(C'') \backslash C) + l(C'') < l(C) $ then
$l(C'') < l(C)$. Let $\hat{C}$ be a son of $C$. If $P_f(\hat{C})
\subset [v_2,u_1]$ then since $[v_2,u_1] \subset P_s(C'')$,
$\hat{C}$ should have been a son of $C''$. Therefore $P_f(\hat{C})
\cap [v_1,u_1] \neq \phi$.
 \item [A4] $J_h(C) = [v_2 \rightarrow u_1]$ and for every $\hat{C}$ a son of $C$
 $P_f(\hat{C}) \subset P_s(C) = I(C) \cup J_h(C)$.

\item[A5] $M(C)$ is contained in $P_s(C)$, $S$, $C''$ and $T$ which
are all shorter then $C$.
 \item[A6] $M(C)$ is contained in $P_s(C), P_s(S), M(S)$ and $C''$,
 so $N(C) = \{ C'' \}$.

\end{description}

\end{prf}

\begin{lem}
\label{lemCDforD}  Suppose that  $l(C) < 2l(P_c(C'') \backslash C) +
l(C'') $ then after the application of DIRECT-ONE $C''$ satisfies
the induction hypotheses.
\end{lem}
\begin{prf}
In this case when DIRECT-ONE is activated on $C$ it  changes the
direction of $[u_1,v_2]$ to $v_2 \rightarrow u_1$.

Since $l(C) < 2l(P_c(C'') \backslash P_c(S)) + l(C'')$:
\begin{equation}
\label{d2fsmc}  l(v_3,v_1) + l(v_1,u_1) < 3l(v_3,v_4)+ l(v_4,u_1).
\end{equation}

Since $C''$ was oriented  backwards by Property \ref{longfwd}
Equation (\ref{shortD}) still holds.

\begin{description}
 \item [A1] Let $x=t(T),y=h(T)$, and  the end nodes $v,w$ of
$P_s(T)$ be ordered $v,x,y,w$.

Since $C''$ was oriented  backwards by Property \ref{longfwd}
Equation (\ref{shortD}) still holds.

Since $C''$ was oriented  backwards and $S$ is oriented  forwards,
then by Property \ref{scbprop} $C''$ is not a special-contained
brother of $S$, and there is at least one more brother contained in
$S$ in the same level of containment as $C''$. Hence,  by A3 on $T$
$x \in [v,v_4]$. Since there is at least one more brother of $S$ on
its $l_2$  side  and $y \in [v_5,w]$. See Figure \ref{fcforcfig}.

\begin{figure}
\begin{center}
\setlength{\unitlength}{0.00061242in}
\begingroup\makeatletter\ifx\SetFigFont\undefined%
\gdef\SetFigFont#1#2#3#4#5{%
  \reset@font\fontsize{#1}{#2pt}%
  \fontfamily{#3}\fontseries{#4}\fontshape{#5}%
  \selectfont}%
\fi\endgroup%
{\renewcommand{\dashlinestretch}{30}
\begin{picture}(8673,5940)(0,-10)
\put(3840,2175){\blacken\ellipse{90}{90}}
\put(3840,2175){\ellipse{90}{90}}
\put(3840,3705){\blacken\ellipse{90}{90}}
\put(3840,3705){\ellipse{90}{90}}
\put(2220,5640){\blacken\ellipse{90}{90}}
\put(2220,5640){\ellipse{90}{90}}
\put(1365,2175){\blacken\ellipse{90}{90}}
\put(1365,2175){\ellipse{90}{90}}
\put(5730,2175){\blacken\ellipse{90}{90}}
\put(5730,2175){\ellipse{90}{90}}
\put(5730,3705){\blacken\ellipse{90}{90}}
\put(5730,3705){\ellipse{90}{90}}
\put(5730,5640){\blacken\ellipse{90}{90}}
\put(5730,5640){\ellipse{90}{90}}
\put(600,2175){\blacken\ellipse{90}{90}}
\put(600,2175){\ellipse{90}{90}}
\put(7890,2175){\blacken\ellipse{90}{90}}
\put(7890,2175){\ellipse{90}{90}}
\put(60,2175){\blacken\ellipse{90}{90}}
\put(60,2175){\ellipse{90}{90}}
\put(8565,2175){\blacken\ellipse{90}{90}}
\put(8565,2175){\ellipse{90}{90}}
\put(4380,240){\blacken\ellipse{90}{90}}
\put(4380,240){\ellipse{90}{90}}
\put(7125,2175){\blacken\ellipse{90}{90}}
\put(7125,2175){\ellipse{90}{90}}
\path(1365,2175)(3840,2175)
\blacken\path(3690.000,2137.500)(3840.000,2175.000)(3690.000,2212.500)(3690.000,2137.500)
\blacken\path(5767.500,5490.000)(5730.000,5640.000)(5692.500,5490.000)(5767.500,5490.000)
\path(5730,5640)(5730,3705)
\blacken\path(5767.500,3555.000)(5730.000,3705.000)(5692.500,3555.000)(5767.500,3555.000)
\path(5730,3705)(5730,2175)
\path(5730,5640)(7125,5640)
\blacken\path(6975.000,5602.500)(7125.000,5640.000)(6975.000,5677.500)(6975.000,5602.500)
\path(7125,5640)(7125,2175)
\blacken\path(7087.500,2325.000)(7125.000,2175.000)(7162.500,2325.000)(7087.500,2325.000)
\path(600,2175)(60,2175)
\path(7890,2175)(8565,2175)
\path(3840,3705)(5730,3705)
\blacken\path(5580.000,3667.500)(5730.000,3705.000)(5580.000,3742.500)(5580.000,3667.500)
\dottedline{45}(5190,4740)(6495,5550)
\path(6408.864,5461.227)(6495.000,5550.000)(6377.222,5512.206)
\dottedline{45}(5190,4740)(7035,4695)
\path(6914.304,4667.935)(7035.000,4695.000)(6915.767,4727.917)
\dottedline{45}(5190,4740)(7530,2220)
\path(7426.362,2287.522)(7530.000,2220.000)(7470.330,2328.349)
\dottedline{45}(5190,4740)(5685,3930)
\path(5596.828,4016.750)(5685.000,3930.000)(5648.024,4048.037)
\dottedline{45}(2625,3435)(1770,1680)
\path(1795.587,1801.018)(1770.000,1680.000)(1849.526,1774.740)
\dottedline{45}(2625,3435)(1050,2220)
\path(1126.690,2317.050)(1050.000,2220.000)(1163.338,2269.543)
\dottedline{45}(2625,3435)(1455,4290)
\path(1569.587,4243.420)(1455.000,4290.000)(1534.186,4194.976)
\dottedline{45}(2625,3435)(1860,5595)
\path(1928.340,5491.900)(1860.000,5595.000)(1871.783,5471.869)
\dottedline{45}(2625,3435)(3120,4470)
\path(3095.289,4348.800)(3120.000,4470.000)(3041.161,4374.688)
\dottedline{45}(5415,1680)(5145,2130)
\path(5232.464,2042.536)(5145.000,2130.000)(5181.015,2011.666)
\dottedline{45}(5415,1680)(6270,2085)
\path(6174.394,2006.518)(6270.000,2085.000)(6148.709,2060.742)
\dottedline{45}(5415,1680)(6180,1275)
\path(6059.909,1304.633)(6180.000,1275.000)(6087.982,1357.660)
\thicklines
\path(4380,240)(600,2175)
\blacken\thinlines
\path(750.610,2140.030)(600.000,2175.000)(716.435,2073.269)(750.610,2140.030)
\blacken\path(1215.000,2137.500)(1365.000,2175.000)(1215.000,2212.500)(1215.000,2137.500)
\thicklines
\path(1365,2175)(600,2175)
\thinlines
\path(2220,5640)(5730,5640)
\thicklines
\path(1365,5640)(2220,5640)
\blacken\thinlines
\path(2070.000,5602.500)(2220.000,5640.000)(2070.000,5677.500)(2070.000,5602.500)
\thicklines
\path(1365,2175)(1365,5640)
\blacken\thinlines
\path(1402.500,5490.000)(1365.000,5640.000)(1327.500,5490.000)(1402.500,5490.000)
\thicklines
\path(2220,5640)(3840,3705)
\blacken\thinlines
\path(3714.956,3795.941)(3840.000,3705.000)(3772.463,3844.086)(3714.956,3795.941)
\blacken\path(3802.500,2325.000)(3840.000,2175.000)(3877.500,2325.000)(3802.500,2325.000)
\thicklines
\path(3840,2175)(3840,3705)
\path(3840,2175)(5730,2175)
\blacken\thinlines
\path(5580.000,2137.500)(5730.000,2175.000)(5580.000,2212.500)(5580.000,2137.500)
\thicklines
\path(5730,2175)(7125,2175)
\blacken\thinlines
\path(6975.000,2137.500)(7125.000,2175.000)(6975.000,2212.500)(6975.000,2137.500)
\dottedline{45}(5415,1680)(7575,2130)
\path(7463.641,2076.156)(7575.000,2130.000)(7451.404,2134.895)
\thicklines
\path(7125,2175)(7890,2175)
\blacken\thinlines
\path(7740.000,2137.500)(7890.000,2175.000)(7740.000,2212.500)(7740.000,2137.500)
\blacken\path(4493.257,345.257)(4380.000,240.000)(4529.465,279.577)(4493.257,345.257)
\thicklines
\path(4380,240)(7890,2175)
\thinlines
\dottedline{45}(5190,4740)(5100,3750)
\path(5080.987,3872.223)(5100.000,3750.000)(5140.741,3866.791)
\put(2625,4740){\makebox(0,0)[lb]{\smash{{\SetFigFont{8}{9.6}{\rmdefault}{\mddefault}{\updefault}$C$}}}}
\put(4650,3750){\makebox(0,0)[lb]{\smash{{\SetFigFont{8}{9.6}{\rmdefault}{\mddefault}{\updefault}$C''$}}}}
\put(5910,3660){\makebox(0,0)[lb]{\smash{{\SetFigFont{8}{9.6}{\rmdefault}{\mddefault}{\updefault}$v_2$}}}}
\put(3795,1950){\makebox(0,0)[lb]{\smash{{\SetFigFont{8}{9.6}{\rmdefault}{\mddefault}{\updefault}$v_4$}}}}
\put(3525,3660){\makebox(0,0)[lb]{\smash{{\SetFigFont{8}{9.6}{\rmdefault}{\mddefault}{\updefault}$u_1$}}}}
\put(2175,5775){\makebox(0,0)[lb]{\smash{{\SetFigFont{8}{9.6}{\rmdefault}{\mddefault}{\updefault}$v_1$}}}}
\put(960,2310){\makebox(0,0)[lb]{\smash{{\SetFigFont{8}{9.6}{\rmdefault}{\mddefault}{\updefault}$T$}}}}
\put(1320,1950){\makebox(0,0)[lb]{\smash{{\SetFigFont{8}{9.6}{\rmdefault}{\mddefault}{\updefault}$v_3$}}}}
\put(5685,1950){\makebox(0,0)[lb]{\smash{{\SetFigFont{8}{9.6}{\rmdefault}{\mddefault}{\updefault}$v_5$}}}}
\put(3435,5775){\makebox(0,0)[lb]{\smash{{\SetFigFont{8}{9.6}{\rmdefault}{\mddefault}{\updefault}$S$}}}}
\put(6450,5775){\makebox(0,0)[lb]{\smash{{\SetFigFont{8}{9.6}{\rmdefault}{\mddefault}{\updefault}$S''$}}}}
\put(5685,5775){\makebox(0,0)[lb]{\smash{{\SetFigFont{8}{9.6}{\rmdefault}{\mddefault}{\updefault}$u_2$}}}}
\put(15,1950){\makebox(0,0)[lb]{\smash{{\SetFigFont{8}{9.6}{\rmdefault}{\mddefault}{\updefault}$v$}}}}
\put(555,1950){\makebox(0,0)[lb]{\smash{{\SetFigFont{8}{9.6}{\rmdefault}{\mddefault}{\updefault}$x$}}}}
\put(7845,1950){\makebox(0,0)[lb]{\smash{{\SetFigFont{8}{9.6}{\rmdefault}{\mddefault}{\updefault}$y$}}}}
\put(8520,1950){\makebox(0,0)[lb]{\smash{{\SetFigFont{8}{9.6}{\rmdefault}{\mddefault}{\updefault}$w$}}}}
\put(1860,1095){\makebox(0,0)[lb]{\smash{{\SetFigFont{8}{9.6}{\rmdefault}{\mddefault}{\updefault}$M_t(T)$}}}}
\put(6315,1095){\makebox(0,0)[lb]{\smash{{\SetFigFont{8}{9.6}{\rmdefault}{\mddefault}{\updefault}$M_h(T)$}}}}
\put(3300,2895){\makebox(0,0)[lb]{\smash{{\SetFigFont{8}{9.6}{\rmdefault}{\mddefault}{\updefault}$I(C'')$}}}}
\put(4920,4830){\makebox(0,0)[lb]{\smash{{\SetFigFont{8}{9.6}{\rmdefault}{\mddefault}{\updefault}$B_t(C'')$}}}}
\put(2760,3255){\makebox(0,0)[lb]{\smash{{\SetFigFont{8}{9.6}{\rmdefault}{\mddefault}{\updefault}$M_t(C'')$}}}}
\put(4695,1545){\makebox(0,0)[lb]{\smash{{\SetFigFont{8}{9.6}{\rmdefault}{\mddefault}{\updefault}$M_h(C'')$}}}}
\put(7215,2265){\makebox(0,0)[lb]{\smash{{\SetFigFont{8}{9.6}{\rmdefault}{\mddefault}{\updefault}$v_6$}}}}
\put(4335,15){\makebox(0,0)[lb]{\smash{{\SetFigFont{8}{9.6}{\rmdefault}{\mddefault}{\updefault}$z$}}}}
\end{picture}
}
\end{center}
\caption{Figure for Lemma \ref{lemCDforD}.} \label{fcforcfig}
\end{figure}

$I(C'') = [u_1 \rightarrow v_4]$,  by A4 we can  define
\[ M_t(C'') = M_t(T) \cup [x \rightarrow v_3 \rightarrow v_1 \rightarrow u_1]
\ \ M_h(C'') = [v_4 \rightarrow v_5 \rightarrow y] \cup M_h(T).\]

Using A1 on $T$:
\begin{eqnarray*}
l(M(C'')) &=& l(M(T)) + l(x,v_3) + l(v_3,v_1) + l(v_1,u_1) +
l(v_4,v_5) + l(v_5,y) \\
&\leq& 15l(P_c(T)) +  4l(v,v_3) + 3l(v_3,v_4) + 4l(v_4,v_5) +
3l(v_5,w) + l(v_3,v_1) +l(v_1,u_1).
\end{eqnarray*}

The length of $[v,v_3]$ is  taken 4 times to satisfy both cases of
possible locations for $x$.

 Using Equations (\ref{d2fsmc}) and (\ref{shortD}):
\begin{eqnarray*}
l(M(C'') &\leq& 15l(P_c(T)) +  4l(v,v_3) + 6l(v_3,v_4) + 4l(v_4,v_5)
+ 3l(v_5,w) + 3l(v_1,u_1) \\
&\leq& 15l(P_c(T)) +  6l(v,v_3) + 8l(v_3,v_4) + 5l(v_5,w) +
3l(v_1,u_1) .
\end{eqnarray*}

By definition
\[ l(P_c(C'')) = l(P_c(T)) + l(v,v_3) + l(v_3,v_4)  + l(v_5,w) .\]

Using the definition of $I(C'')$ we get
 \begin{eqnarray*}
l(M(C'') &\leq& 15l(P_c(C'')) +   l(u_1,v_4) \\
 &=& 15l(P_c(C)) + l(I(C)) + 2l(P_s(C)) .
\end{eqnarray*}

\item [A2] $N(S)$ contains at least one more brother at the $l_2$
side of $C''$. W.l.o.g we assume that that $N(S) = 1$ (again, longer
but similar proofs can be given for the other cases). Mark this
brother as $S''$. Mark
 \bi
 \item $u_2 :=$ the end node of the path $P_s(S) \cap P_s(S'')$ such
 that $u_2\not \in T$.
  \item $v_6 = T \cap (P_s(S'') \backslash P_s(S))$ .
  \ei
Since $E \in N(S)$ by A5 on $S$ $l(S'') < l(S)$. Since $C$ is a son
of $S$, $l(S) < l(C)$ and by the lemma's assumption
\[ l(S'') < l(S) < l(C) < l(C'') + 2l(P_c(C'') \backslash C) , \] giving:
\begin{equation}
\label{EsmD2pf}  l(v_4,v_5) + l(v_2,u_2) + l(u_2,v_6) < l(v_4,u_1) +
l(u_1,v_2) + l(v_5,v_6) + 2l(v_3,v_4).
\end{equation}

In this case $h(C'') = v_4 \in T$. Define
\[ B_t(C'') = [u_1 \rightarrow v_2  \rightarrow u_2 \rightarrow v_6 \rightarrow y] \cup M_h(T) .\]

Using A1 on $T$ (again $l(v,v_3)$ and $l(v_6,w)$ are taken 4 times
to satisfy all possible locations of $x$ and $y$), (\ref{d2fsmc})
(\ref{EsmD2pf}), (\ref{shortD}), definition of $P_c(C'')$,the fact
that $C''$ is a son of $T$ and definition of $P_s(C'')$:
\begin{eqnarray*}
l(M_t(C'')) + l(B_t(C'')) &=& l(M(T)) + l(x,v_3) + l(v_3,v_1) +
l(v_1,u_1) + l(u_1,v_2)+ \\
&&  l(v_2,u_2) +l(u_2,v_6) + l(v_6,y)\\
 &\leq&15 l(P_c(T)) + 4l(v,v_3) + 3l(v_3,v_4)  + 3l(v_4,v_5) + 3l(v_5,v_6) + \\
 && 4l(v_6,w) +l(v_3,v_1) +l(v_1,u_1) + l(u_1,v_2) + l(v_2,u_2) + l(u_2,v_6) \\
&\leq&15 l(P_c(T)) + 4l(v,v_3) + 6l(v_3,v_4)  + 3l(v_4,v_5) + 3l(v_5,v_6) + \\
 && 4l(v_6,w) + + l(u_1,v_2) + l(v_2,u_2) + l(u_2,v_6)+ l(v_4,u_1) \\
&\leq&15 l(P_c(T)) + 4l(v,v_3) + 8l(v_3,v_4)  + 2l(v_4,v_5) + 4l(v_5,v_6) + \\
 && 4l(v_6,w) + + 2l(u_1,v_2) +  2l(v_4,u_1) \\
&\leq&15 l(P_c(C'')) + 2l(v_4,v_5)   + 2l(u_1,v_2) +  2l(v_4,u_1) \\
&\leq&15 l(P_c(C'')) + 2l(v_4,v_5)   + 2l(P_s(C'')) \\
&\leq&15 l(P_c(C''))   + 4l(P_s(C'')).
\end{eqnarray*}

 \item[A3] There are two cases to be considered:
 \bi
 \item $l(C) < l(C'')$. In this case $l(S'') < l(C'')$. Let $\hat{C}$ be a son of
 $C''$. If $P_f(\hat{C}) \subset ([u_1,v_2] \cup [v_2,v_5])$ then $\hat{C}$
 should have been a son of $C$ or $S''$ but not of $C''$.
 \item $l(C) > l(C'')$ . Let $\hat{C}$ be a son of $C''$. If $(P_f(\hat{C})
 \subset [u_1,v_2,v_5]$ DIRECT-ONE changes the undirected cycle for $U(\hat{C})$
 to be a son of $C$ (we replace the  path $[v_3,v_4,u_1]$ by the
 path $[v_3,v_1,u_1]$).

\ei
 \item [A4] $J_t(C'') = [u_1 \rightarrow v_2]$ and for every $\hat{C}$ a son of $C''$
 (which remains a son after the change we described in A3)
 $P_f(\hat{C}) \subset [v_4,u_1] \cup [u_1,v_2] = I(C'') \cup J_t(C'')$.
\item[A5] $M(C'')$ is contained in $P_s(C'')$, $C$, and $T$ which are
all shorter then $C$.
 \item[A6] $M(C)$ is contained in $P_s(C''), P_s(T), M(T)$ and $C$,
 so $N(C'') = \{ C \}$.

\end{description}
\end{prf}

\begin{remark} \label{undrem}
We note that after DIRECT-ONE $U(C'') \subseteq [u_1,v_4]$ and for
every $\hat{C}$ son of $C''$ $P_f(\hat{C})$ shares an edge with
$[u_1,v_4]$. For the other nodes (and sons) we changed their
undirected cycle. The length of the new undirected cycle is at most
three times the length of the original undirected cycle.
\end{remark}
\begin{prf}  The change in the undirected cycle was to replace the path
$[v_3,v_4,u_1]$ by the
 path $[v_3,v_1,u_1]$. According to Equation (\ref{d2fsmc}) length of
the new path is at most three times the length of the old path. So
 this changes  the length of the undirected
cycle by at most three times. We also note that since $C$ is
oriented forwards and we perform these changes on sons of a cycle
oriented backwards $(C'')$ the next time we might perform such as
change can only happen in the grandsons. So no overlap occurs.
\end{prf}

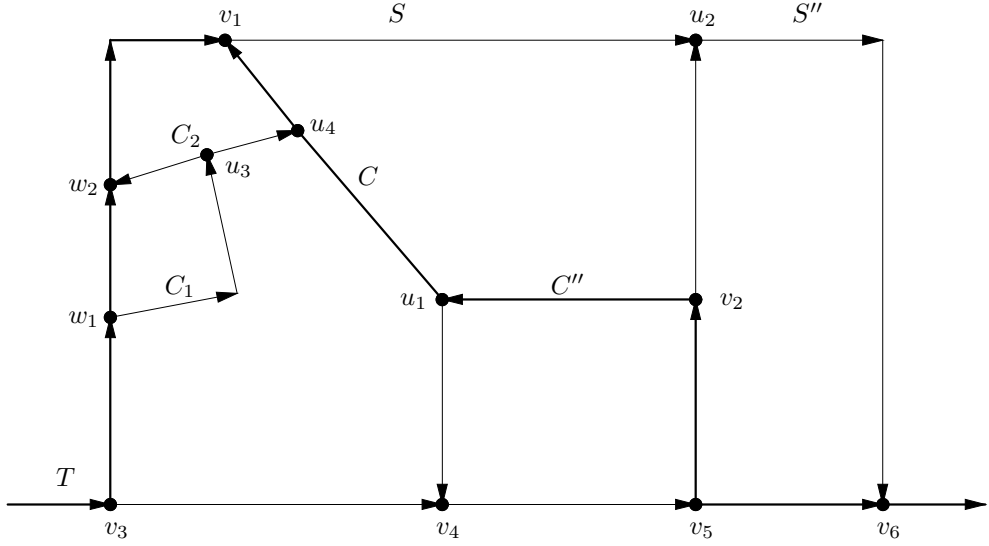
\begin{figure}
\begin{center}
\setlength{\unitlength}{0.00069991in}
\begingroup\makeatletter\ifx\SetFigFont\undefined%
\gdef\SetFigFont#1#2#3#4#5{%
  \reset@font\fontsize{#1}{#2pt}%
  \fontfamily{#3}\fontseries{#4}\fontshape{#5}%
  \selectfont}%
\fi\endgroup%
{\renewcommand{\dashlinestretch}{30}
\begin{picture}(10307,5419)(0,-10)
\put(3453,250){\blacken\ellipse{90}{90}}
\put(3453,250){\ellipse{90}{90}}
\put(3453,1780){\blacken\ellipse{90}{90}}
\put(3453,1780){\ellipse{90}{90}}
\put(6738,250){\blacken\ellipse{90}{90}}
\put(6738,250){\ellipse{90}{90}}
\put(1833,3715){\blacken\ellipse{90}{90}}
\put(1833,3715){\ellipse{90}{90}}
\put(978,250){\blacken\ellipse{90}{90}}
\put(978,250){\ellipse{90}{90}}
\put(5343,250){\blacken\ellipse{90}{90}}
\put(5343,250){\ellipse{90}{90}}
\put(5343,1780){\blacken\ellipse{90}{90}}
\put(5343,1780){\ellipse{90}{90}}
\put(5343,3715){\blacken\ellipse{90}{90}}
\put(5343,3715){\ellipse{90}{90}}
\put(2373,3040){\blacken\ellipse{90}{90}}
\put(2373,3040){\ellipse{90}{90}}
\put(978,1645){\blacken\ellipse{90}{90}}
\put(978,1645){\ellipse{90}{90}}
\put(978,2635){\blacken\ellipse{90}{90}}
\put(978,2635){\ellipse{90}{90}}
\put(1698,2860){\blacken\ellipse{90}{90}}
\put(1698,2860){\ellipse{90}{90}}
\path(978,250)(3453,250)
\blacken\path(3303.000,212.500)(3453.000,250.000)(3303.000,287.500)(3303.000,212.500)
\path(3453,250)(5343,250)
\blacken\path(5193.000,212.500)(5343.000,250.000)(5193.000,287.500)(5193.000,212.500)
\blacken\path(3415.500,400.000)(3453.000,250.000)(3490.500,400.000)(3415.500,400.000)
\path(3453,250)(3453,1780)
\blacken\path(5380.500,3565.000)(5343.000,3715.000)(5305.500,3565.000)(5380.500,3565.000)
\path(5343,3715)(5343,1780)
\path(5343,3715)(6738,3715)
\blacken\path(6588.000,3677.500)(6738.000,3715.000)(6588.000,3752.500)(6588.000,3677.500)
\path(6738,3715)(6738,250)
\blacken\path(6700.500,400.000)(6738.000,250.000)(6775.500,400.000)(6700.500,400.000)
\blacken\path(1782.666,1760.095)(1923.000,1825.000)(1768.633,1833.771)(1782.666,1760.095)
\path(1923,1825)(978,1645)
\blacken\path(1766.509,2721.390)(1698.000,2860.000)(1693.220,2705.457)(1766.509,2721.390)
\path(1698,2860)(1923,1825)
\blacken\path(1109.987,2715.534)(978.000,2635.000)(1132.357,2643.948)(1109.987,2715.534)
\path(978,2635)(1698,2860)
\path(1698,2860)(2373,3040)
\blacken\path(2237.727,2965.117)(2373.000,3040.000)(2218.402,3037.584)(2237.727,2965.117)
\blacken\path(828.000,212.500)(978.000,250.000)(828.000,287.500)(828.000,212.500)
\thicklines
\path(978,250)(213,250)
\path(978,250)(978,1645)
\blacken\thinlines
\path(1015.500,1495.000)(978.000,1645.000)(940.500,1495.000)(1015.500,1495.000)
\thicklines
\path(978,1645)(978,2635)
\blacken\thinlines
\path(1015.500,2485.000)(978.000,2635.000)(940.500,2485.000)(1015.500,2485.000)
\thicklines
\path(978,2635)(978,3715)
\blacken\thinlines
\path(1015.500,3565.000)(978.000,3715.000)(940.500,3565.000)(1015.500,3565.000)
\blacken\path(1683.000,3677.500)(1833.000,3715.000)(1683.000,3752.500)(1683.000,3677.500)
\thicklines
\path(1833,3715)(978,3715)
\thinlines
\path(1833,3715)(5343,3715)
\blacken\path(5193.000,3677.500)(5343.000,3715.000)(5193.000,3752.500)(5193.000,3677.500)
\blacken\path(1955.987,3621.296)(1833.000,3715.000)(1897.422,3574.444)(1955.987,3621.296)
\thicklines
\path(1833,3715)(2373,3040)
\path(2373,3040)(3453,1780)
\blacken\thinlines
\path(3603.000,1817.500)(3453.000,1780.000)(3603.000,1742.500)(3603.000,1817.500)
\thicklines
\path(3453,1780)(5343,1780)
\blacken\thinlines
\path(5380.500,1630.000)(5343.000,1780.000)(5305.500,1630.000)(5380.500,1630.000)
\thicklines
\path(5343,1780)(5343,250)
\path(5343,250)(6738,250)
\blacken\thinlines
\path(6588.000,212.500)(6738.000,250.000)(6588.000,287.500)(6588.000,212.500)
\thicklines
\path(6738,250)(7503,250)
\blacken\thinlines
\path(7353.000,212.500)(7503.000,250.000)(7353.000,287.500)(7353.000,212.500)
\put(4263,1825){\makebox(0,0)[lb]{\smash{{\SetFigFont{10}{12.0}{\rmdefault}{\mddefault}{\updefault}$C''$}}}}
\put(5523,1735){\makebox(0,0)[lb]{\smash{{\SetFigFont{10}{12.0}{\rmdefault}{\mddefault}{\updefault}$v_2$}}}}
\put(3408,25){\makebox(0,0)[lb]{\smash{{\SetFigFont{10}{12.0}{\rmdefault}{\mddefault}{\updefault}$v_4$}}}}
\put(3138,1735){\makebox(0,0)[lb]{\smash{{\SetFigFont{10}{12.0}{\rmdefault}{\mddefault}{\updefault}$u_1$}}}}
\put(1788,3850){\makebox(0,0)[lb]{\smash{{\SetFigFont{10}{12.0}{\rmdefault}{\mddefault}{\updefault}$v_1$}}}}
\put(573,385){\makebox(0,0)[lb]{\smash{{\SetFigFont{10}{12.0}{\rmdefault}{\mddefault}{\updefault}$T$}}}}
\put(933,25){\makebox(0,0)[lb]{\smash{{\SetFigFont{10}{12.0}{\rmdefault}{\mddefault}{\updefault}$v_3$}}}}
\put(5298,25){\makebox(0,0)[lb]{\smash{{\SetFigFont{10}{12.0}{\rmdefault}{\mddefault}{\updefault}$v_5$}}}}
\put(3048,3850){\makebox(0,0)[lb]{\smash{{\SetFigFont{10}{12.0}{\rmdefault}{\mddefault}{\updefault}$S$}}}}
\put(6063,3850){\makebox(0,0)[lb]{\smash{{\SetFigFont{10}{12.0}{\rmdefault}{\mddefault}{\updefault}$S''$}}}}
\put(5298,3850){\makebox(0,0)[lb]{\smash{{\SetFigFont{10}{12.0}{\rmdefault}{\mddefault}{\updefault}$u_2$}}}}
\put(6693,25){\makebox(0,0)[lb]{\smash{{\SetFigFont{10}{12.0}{\rmdefault}{\mddefault}{\updefault}$v_6$}}}}
\put(1428,2950){\makebox(0,0)[lb]{\smash{{\SetFigFont{10}{12.0}{\rmdefault}{\mddefault}{\updefault}$C_2$}}}}
\put(1383,1825){\makebox(0,0)[lb]{\smash{{\SetFigFont{10}{12.0}{\rmdefault}{\mddefault}{\updefault}$C_1$}}}}
\put(2823,2635){\makebox(0,0)[lb]{\smash{{\SetFigFont{10}{12.0}{\rmdefault}{\mddefault}{\updefault}$C$}}}}
\put(663,1600){\makebox(0,0)[lb]{\smash{{\SetFigFont{10}{12.0}{\rmdefault}{\mddefault}{\updefault}$w_1$}}}}
\put(663,2590){\makebox(0,0)[lb]{\smash{{\SetFigFont{10}{12.0}{\rmdefault}{\mddefault}{\updefault}$w_2$}}}}
\put(1833,2725){\makebox(0,0)[lb]{\smash{{\SetFigFont{10}{12.0}{\rmdefault}{\mddefault}{\updefault}$u_3$}}}}
\put(2463,3040){\makebox(0,0)[lb]{\smash{{\SetFigFont{10}{12.0}{\rmdefault}{\mddefault}{\updefault}$u_4$}}}}
\put(7808,5245){\makebox(0,0)[lb]{\smash{{\SetFigFont{10}{12.0}{\rmdefault}{\mddefault}{\updefault}The cycles}}}}
\put(7808,5020){\makebox(0,0)[lb]{\smash{{\SetFigFont{10}{12.0}{\rmdefault}{\mddefault}{\updefault}$T$ $zv_3v_4v_5v_6z$}}}}
\put(7808,4795){\makebox(0,0)[lb]{\smash{{\SetFigFont{10}{12.0}{\rmdefault}{\mddefault}{\updefault}$S$ $zv_3w_1w_2v_1u_2v_2v_5v_6z$}}}}
\put(7808,4570){\makebox(0,0)[lb]{\smash{{\SetFigFont{10}{12.0}{\rmdefault}{\mddefault}{\updefault}$S''$ $zv_3v_4v_5v_2u_2v_6z$}}}}
\put(7808,4120){\makebox(0,0)[lb]{\smash{{\SetFigFont{10}{12.0}{\rmdefault}{\mddefault}{\updefault}$C$ $zv_3w_1w_2v_1u_4u_1v_2v_5v_6z$}}}}
\put(7808,3895){\makebox(0,0)[lb]{\smash{{\SetFigFont{10}{12.0}{\rmdefault}{\mddefault}{\updefault}$C_1$ $zv_3w_1u_3w_2v_1u_2v_2v_5v_6z$}}}}
\put(7808,3670){\makebox(0,0)[lb]{\smash{{\SetFigFont{10}{12.0}{\rmdefault}{\mddefault}{\updefault}$C_2$ $zv_3w_1w_2u_3u_4v_1u_2v_2v_5v_6z$}}}}
\put(7808,4345){\makebox(0,0)[lb]{\smash{{\SetFigFont{10}{12.0}{\rmdefault}{\mddefault}{\updefault}$C''$ $zv_3v_4u_1v_2v_5v_6z$}}}}
\end{picture}
}
\end{center}
\caption{$N(C)$ contains two brothers at its $l_1$ side.}
\label{fcforthreefig}
\end{figure}

Now assume that $C$ has two brothers $C_1,C_2$ at its $l_1$ side,
ordered $C_1,C_2,C$,  such that $ N(C) = \{ C_1,C_2,C'' \}$ (see
Figure \ref{fcforthreefig}). This happens when $C$ and $C_2$ are
heavy, $C,C_1,C_2$ are oriented forwards, and $l(C_1) < l(C_2) <
l(C)$. If $C_1$ doesn't exist then similar but simpler proofs apply.

Since $C''$ was oriented  backwards $N(S)$ contains at least one
more brother at the $l_2$ side of $C''$. W.l.o.g we assume that that
$N(S) = 1$ (again longer but similar proofs follows for the other
cases). Mark this brother as $S''$.

Mark
 \bi
 \item $u_2 :=$ the end node of the path $P_s(S) \cap P_s(S'')$ such
 that $u_2\not \in T$.
 \item $v_6 = T \cap (P_s(S'') \backslash P_s(S))$ .
 \item $w_1 := S \cap (P_s(C_1) \backslash P_s(C_2))$.
 \item $w_2 := S \cap P_s(C_1) \cap P_s(C_2)$.
 \item $u_3 :=$ the end node of the path $P_s(C_1) \cap P_s(C_2)$,
 such that $u_3 \not\in C$.
 \item $u_4 :=$ the end node of the path $P_s(C_2) \cap P_s(C)$,
 such that $u_4 \not\in C$.
 \ei

Since $C_1 \in N(C), C_2 \in N(C_1)$,  $l(C_1) < l(C_2) < l(C)$, and
since $S$ was oriented  forwards, according to Observation
\ref{winpath} the algorithm in previous stages (before activating
DIRECT-ONE on $C$) oriented $[w_1 \rightarrow u_3 \rightarrow u_4],
[u_3 \rightarrow w_2], [u_4 \rightarrow v_1]$. Since $l(C_2) < l(C)$
DIRECT-ONE will not change the direction of $[u_4 \rightarrow
v_1]$. So we only need to consider what happens to the path
$[u_1,v_2]$ (whether or not its current direction $[v_2 \rightarrow
u_1]$  changes).

\begin{lem}
\label{lemfourbwmid} Suppose that $l(C_1) < l(C_2) < l(C)$ and $
l(C'') + 2l(P_c(C'') \backslash C) < l(C)$, then $C$ satisfies the
induction hypotheses.
\end{lem}

\begin{prf}
Since $  l(C'') + 2l(P_c(C'') \backslash C) < l(C)$, when DIRECT-ONE
is activated on $C$ it will not change the direction $[v_2
\rightarrow u_1]$, Moreover Equation \ref{csmd2f} holds. Since
$l(C_1) < l(C_2) < l(C)$ by Observation \ref{winpath} the
orientation of the paths will be: $[w_1 \rightarrow u_3 \rightarrow
w_2]$ and $[u_3 \rightarrow u_4 \rightarrow v_1]$.

 Since $T=F(S)$
Equation (\ref{SsonT}) holds. Since $l(C_1)  < l(C)$
\begin{equation}
\label{C1sC} l(w_1,u_3) + l(u_3,w_2) + l(v_1,v_2) < l(w_1,w_2) +
l(P_s(C)).
\end{equation}
>From $l(C_2) < l(C)$
\begin{equation}
\label{C2sC} l(w_2,u_3) + l(u_3,u_4) + l(v_1,v_2) < l(u_4,u_1) +
l(u_1,v_2) + l(w_2,v_1).
\end{equation}
Since $S'' \in N(S)$ by A5 on $S$, $l(S'') < l(S)$, and since $C$ is
a son of $S$ $l(S) < l(C)$. Therefore $l(S'')  < l(C)$, giving:
\begin{equation}
\label{EsC} l(v_3,v_4) + l(v_4,v_5) + l(v_2,u_2) + l(u_2,v_6) <
l(v_3,w_1) + l(w_1,w_2) + l(w_2,v_1) + l(v_1,u_1) + l(u_1,v_2) +
l(v_5,v_6).
\end{equation}

\begin{description}
\item [A1] Let $x=t(T),y=h(T)$, and  the end nodes $v,w$ of
$P_s(T)$ be ordered $v,x,y,w$. Since $C''$ is oriented  backwards by
Property \ref{longfwd} Equation (\ref{shortD})  holds.

 Again from A3 on $T$ $x \in
[v,v_4]$ and Since there is at least one more brother of $S$ on its
$l_2$ side $y \in [v_5,w]$. $I(C) = [u_4 \rightarrow u_1]$, by A4 we
can  define
\[ M_t(C) = M_t(T) \cup [x \rightarrow v_3 \rightarrow w_1 \rightarrow u_3 \rightarrow u_3,u_4], \
  M_h(C) = [u_1 \rightarrow v_4 \rightarrow v_5 \rightarrow y] \cup
M_h(T).\]
 Using A1 on $T$:
\begin{eqnarray*}
l(M(C) &=& l(M(T)) + l(x,v_3) + l(v_3,w_1) + l(w_1,u_3) +
l(u_3,u_4) + l(u_1,v_4) + l(v_4,v_5) + l(v_5,y) \\
&\leq& 15l(P_c(T)) +  4l(v,v_3) + 3l(v_3,v_4) + 4l(v_4,v_5) +
4l(v_5,w) + l(v_3,w_1) + l(w_1,u_3) + \\
&& l(u_3,u_4) + l(u_1,v_4)
\end{eqnarray*}

The length of $[v,v_3]$ is  taken 4 times to satisfy both cases of
possible locations for $x$.
 Using Equations (\ref{shortD}),(\ref{csmd2f}), (\ref{SsonT}),
 (\ref{C1sC}),(\ref{C2sC}), the definition of $P_c(C)$, the definition of
 $P_c(C)$ and the definition of $I(C)$
\begin{eqnarray*}
l(M(C)) &\leq& 15l(P_c(T)) +  5l(v,v_3) + 4l(v_3,v_4) + 2l(v_4,v_5)
+5l(v_5,w) + l(v_3,w_1) +   \\
&& l(w_1,u_3) +l(u_3,u_4) + l(u_1,v_4)\\
&\leq& 15l(P_c(T)) +  5l(v,v_3) + l(v_3,v_4) + 2l(v_4,v_5)
+5l(v_5,w) + 2l(v_3,w_1) +  \\
&&  l(w_1,u_3) +l(u_3,u_4) + l(w_1,w_2) + l(w_2,v_1) + l(v_1,u_1) \\
&\leq& 15l(P_c(T)) +  5l(v,v_3) +5l(v_5,w) + 4l(v_3,w_1) +
l(w_1,u_3) +  l(u_3,u_4) + \\
&&3l(w_1,w_2) +3l(w_2,v_1) + l(v_1,u_1) + 2l(v_1,v_2) + 2l(v_2,v_5) \\
&\leq& 15l(P_c(T)) +  5l(v,v_3) +5l(v_5,w) + 4l(v_3,w_1) +
4l(w_1,w_2) +4l(w_2,v_1) +  \\
&& l(v_1,u_1)  +2l(v_2,v_5) +l(P_s(C))+l(u_4,u_1) + l(u_1,v_2)\\
&\leq& 15l(P_c(C)) + l(v_1,u_1)   +l(P_s(C))+l(u_4,u_1) + l(u_1,v_2)\\
&\leq& 15l(P_c(C)) + 2l(P_s(C))+l(u_4,u_1) \\
&\leq& 15l(P_c(C)) + 2l(P_s(C))+l(I(C)).
\end{eqnarray*}

\item [A2] In this case $t(C) = u_4 \not\in S$ and $h(C) = u_1 \not\in
S$. So we define
\[ B_h(C) = M_t(T) \cup [x \rightarrow v_3 \rightarrow v_4
\rightarrow v_5 \rightarrow v_2 \rightarrow u_1] \] (if $x \in
[v_3,v_4]$ the path is even shorter), and
\[ B_t(C) = [u_4 \rightarrow v_1 \rightarrow u_2 \rightarrow v_6
\rightarrow y] \cup M_h(T) .\]

This gives
\begin{eqnarray*}
l(B_h(C)) + l(B_t(C)) &\leq& l(M_t(T)) + l(M_h(T)) + l(x,v_3) +
l(v_3,v_4) + l(v_4,v_5) + l(v_5,v_2) + \\
&& l(v_2,u_1) + l(u_4,v_1) + l(v_1,u_2) + l(u_2,v_6) + l(v_6,y).
\end{eqnarray*}

Using A1 on $T$, the notation we use (again $l(v,v_3)$ and
$l(v_6,w)$ are counted 4 times to satisfy all cases of locations of
$x$ and $y$), Equations (\ref{shortD}), (\ref{csmd2f}), (\ref{EsC}),
(\ref{SsonT}), the definition of $P_c(C)$, $[v_1,u_2] \subset
[v_1,v_2]$, (\ref{CsonB}), the definition of $P_s(C)$ and the
definition of $I(C)$
\begin{eqnarray*}
l(B_h(C)) + l(B_t(C)) &\leq& 15(l(P_c(T)) + 2l(P_s(T)) + l(I(T))+
l(x,v_3) + l(v_3,v_4) + l(v_4,v_5) +\\
&& l(v_5,v_2) +  l(v_2,u_1) +l(u_4,v_1) + l(v_1,u_2) + l(u_2,v_6) + l(v_6,y)\\
 &\leq& 15(l(P_c(T)) + 4l(v,v_3) + 4l(v_3,v_4) +  4l(v_4,v_5) + 3l(v_5,v_6) + \\
&& 4l(v_6,w)+l(v_5,v_2) +  l(v_2,u_1) +l(u_4,v_1) + l(v_1,u_2) + l(u_2,v_6)\\
 &\leq& 15(l(P_c(T)) + 5l(v,v_3) + 5l(v_3,v_4) + 2l(v_4,v_5) + 4(l(v_5,v_6) + \\
&& 5l(v_6,w)+l(v_5,v_2) +  l(v_2,u_1) +l(u_4,v_1) + l(v_1,u_2) + l(u_2,v_6)\\
 &\leq& 15(l(P_c(T)) + 5l(v,v_3) + 2l(v_3,v_4) + 2l(v_4,v_5) + 4(l(v_5,v_6) + \\
&& 5l(v_6,w)+ l(v_5,v_2) +  l(v_2,u_1) +l(u_4,v_1) + l(v_1,u_2) + l(u_2,v_6) + \\
&& l(v_3,v_1) + l(v_1,u_1) \\
 &\leq& 15(l(P_c(T)) + 5l(v,v_3) + l(v_3,v_4) + l(v_4,v_5) + 5(l(v_5,v_6) + 5l(v_6,w)+\\
&& l(v_5,v_2) + 2l(v_2,u_1) +l(u_4,v_1) + l(v_1,u_2) + 2l(v_3,v_1)  + 2l(v_1,u_1)\\
 &\leq& 15(l(P_c(T)) + 5l(v,v_3) +  5(l(v_5,v_6) + 5l(v_6,w)+
2l(v_5,v_2) +  \\
&& 2l(v_2,u_1) + l(u_4,v_1) + l(v_1,u_2) + 3l(v_3,v_1)  + 2l(v_1,u_1)+ l(v_1,v_2) \\
 &\leq& 15(l(P_c(C)) + 2l(v_2,u_1) +l(u_4,v_1) + l(v_1,u_2) + 2l(v_1,u_1)+ l(v_1,v_2) \\
 &\leq& 15(l(P_c(C)) + 2l(v_2,u_1) +l(u_4,v_1) +  2l(v_1,u_1)+ 2l(v_1,v_2) \\
&=& 15(l(P_c(C)) + 2l(v_2,u_1) +l(u_4,v_1) +  2l(v_1,u_1)+ 2l(P_s(C)) \\
&\leq& 15(l(P_c(C)) + l(P_s(C) \backslash I(C)) + 4l(P_s(C)) \\
&=& 15(l(P_c(C)) - l(I(C)) + 5l(P_s(C)).
\end{eqnarray*}

Also
\begin{eqnarray*}
l(B_h(C)) + l(M_h(C)) &\leq&  l(M_t(T)) + l(M_h(s)) +l(x,v_3) +
l(v_3,v_4) + l(v_4,v_5)  + l(v_5,v_2) +\\
&& l(v_2,u_1) + l(u_1,v_4) + l(v_4,v_5) + l(v_5,y).
\end{eqnarray*}

Using A1 on $T$, the notation we use (again $l(v,v_3)$ and
$l(v_5,w)$ are counted 4 times to satisfy all cases of locations of
$x$ and $y$), Equations (\ref{shortD}), (\ref{csmd2f}),
(\ref{SsonT}), the definition of $P_c(C)$, (\ref{CsonB}) and the
definition of $P_s(C)$:
\begin{eqnarray*}
l(B_h(C)) + l(M_h(C)) &\leq&  15l(P_c(T)) + 2l(P_s(T)) + l(I(T))
+l(x,v_3) + l(v_3,v_4) +  \\
&& l(v_4,v_5) +l(v_5,v_2) + l(v_2,u_1) +l(u_1,v_4) + l(v_4,v_5) + l(v_5,y) \\
 &\leq&  15l(P_c(T)) + 5l(v,v_3) + 5l(v_3,v_4) + 3l(v_4,v_5) + 5l(v_5,w)
  +  \\
  && l(v_5,v_2) +l(v_2,u_1) +l(u_1,v_4)  \\
&\leq&  15l(P_c(T)) + 5l(v,v_3) + 2l(v_3,v_4) + 3l(v_4,v_5) +
5l(v_5,w)  + \\
&& l(v_5,v_2) +l(v_2,u_1) + l(v_3,v_1) + l(v_1,u_1) \\
&\leq&  15l(P_c(T)) + 5l(v,v_3) +
5l(v_5,w)  + 3l(v_5,v_2) + l(v_2,u_1) +  \\
&& 3l(v_3,v_1) +l(v_1,u_1)+ 3l(v_1,v_2) \\
&\leq&  15l(P_c(C))  + l(v_2,u_1) +  l(v_1,u_1)+ 3l(v_1,v_2) \\
&\leq&  15l(P_c(C))  + l(v_2,u_1) +  l(v_1,u_1)+ 3l(P_s(C)) \\
&=& 15l(P_c(C)) + 4l(P_s(C)).
\end{eqnarray*}

and:

\begin{eqnarray*}
l(B_t(C)) + l(M_t(C)) &=& l(M_t(T)) + l(M_h(T)) + l(x,v_3) +
l(v_3,w_1) + l(w_1,u_3) + l(u_3,u_4) +\\
&& l(u_4,v_1) + l(v_1,u_2) + l(u_2,v_4) + l(v_6,y).
\end{eqnarray*}

Using A1 on $T$, the notation we use (again $l(v,v_3)$ and
$l(v_6,w)$ are counted 4 times to satisfy all cases of locations of
$x$ and $y$), Equations (\ref{EsC}),(\ref{SsonT}),(\ref{C1sC}),
(\ref{C2sC}), the definition of $P_c(C)$, the definition of
$P_s(C)$, $[v_1,u_2] \subset [v_1,v_2]$ and Equation (\ref{CsonB})

\begin{eqnarray*}
l(B_t(C)) + l(M_t(C)) &\leq& 15l(P_c(T))+ 2l(P_s(T)) + l(I(T))   +
l(x,v_3) + l(v_3,w_1) + \\
&& l(w_1,u_3) + l(u_3,u_4) + l(u_4,v_1) +l(v_1,u_2) + l(u_2,v_6) + l(v_6,y)\\
 &\leq& 15l(P_c(T))+ 4l(v,v_3) + 3l(v_3,v_4) + 3l(v_4,v_5) + 3l(v_5,v_6) + \\
 && 4l(v_6,w) +l(v_3,w_1) + l(w_1,u_3) + l(u_3,u_4) + l(u_4,v_1) +l(v_1,u_2) +  \\
 && l(u_2,v_6) \\
 &\leq& 15l(P_c(T))+ 4l(v,v_3) + 2l(v_3,v_4) + 2l(v_4,v_5) + 4l(v_5,v_6) +  \\
 && 4l(v_6,w) + 2l(v_3,w_1) + l(w_1,u_3) + l(u_3,u_4) + l(u_4,v_1) + \\
 && l(v_1,u_2) + l(w_1,w_2) +l(w_2,v_1) + l(P_s(C))   \\
 &\leq& 15l(P_c(T))+ 4l(v,v_3)  + 4l(v_5,v_6) + 4l(v_6,w)
 + 4l(v_3,w_1) + \\
 && l(w_1,u_3) +l(u_3,u_4) + l(u_4,v_1) +l(v_1,u_2) + 3l(w_1,w_2) + \\
 && 3l(w_2,v_1) + l(P_s(C)) + 2l(v_1,v_2) + 2l(v_2,v_5)   \\
 &\leq& 15l(P_c(T))+ 4l(v,v_3)  + 4l(v_5,v_6) + 4l(v_6,w)
 + 4l(v_3,w_1) +   \\
 && l(u_3,u_4) + l(u_4,v_1) +l(v_1,u_2) + 4l(w_1,w_2) + 3l(w_2,v_1) +  \\
&& 2l(P_s(C)) + l(v_1,v_2) +2l(v_2,v_5)   \\
 &\leq& 15l(P_c(T))+ 4l(v,v_3)  + 4l(v_5,v_6) + 4l(v_6,w)
 + 4l(v_3,w_1) +   \\
&& l(u_4,v_1) +l(v_1,u_2) + 4l(w_1,w_2) + 4l(w_2,v_1) + 2l(P_s(C))  + \\
&& 2l(v_2,v_5)+ l(u_4,u_1)+ l(u_1,v_2)   \\
 &\leq& 15l(P_c(C))+    l(u_4,v_1) +l(v_1,u_2) + 2l(P_s(C))  + l(u_4,u_1) + l(u_1,v_2)   \\
&=& 15l(P_c(C))+    l(v_1,u_2) + 3l(P_s(C))     \\
&\leq& 15l(P_c(C))+     4l(P_s(C)).
\end{eqnarray*}

\item[A3] From   $ l(C'') +2l(P_c(C'') \backslash C ) < l(C)$   we get that $l(C'') < l(C)$.
 Let $\hat{C}$ be a son of $C$. If $P_f(\hat{C}) \subset [v_2,u_1]$ then since $[v_2,u_1]
\subset P_s(C'')$ $\hat{C}$ should have been a son of $C''$. If
$P_f(\hat{C}) \subset [v_1,u_4]$ then $\hat{C}$ should have been a
son of $C_2$ (since $[v_1,u_4] \subset P_s(C_2)$ and $l(C_2) <
l(C)$). Therefore $P_f(\hat{C}) \cap [u_4,u_1] \neq \phi$.
 \item [A4] $J_h(C) = [v_2 \rightarrow u_1]$ and $J_t(C) = [u_4 \rightarrow v_1]$ so for every  son $\hat{C}$ of
 $C$,  $P_f(\hat{C}) \subset P_s(C) = J_t(C) \cup I(C) \cup J_h(C)$.

\item[A5] $M(C)$ is contained in $P_s(C)$, $S$,$C_1$,$C_2$, $C''$ and $T$ which
are all shorter then $C$.
 \item[A6] $M(C)$ is contained in $P_s(C), P_s(S), M(S)$ and $C_1,C_2$ and $C''$,
 so $N(C) = \{ C_1,C_2,C'' \}$.

\end{description} \end{prf}

\begin{remark}
Please note that in the following lemma the 15 bound is required.
\end{remark}

\begin{lem}
\label{lemfourbwsid} Suppose that $l(C_1) < l(C_2) < l(C)$, $l(C) <
l(C'') + 2l(P_c(C'') \backslash C)$, and the direction is $u_1
\rightarrow v_2$ then $C''$  satisfies the induction hypotheses.
\end{lem}

\begin{prf}
Since $l(C) < 2l(P_c(C'') \backslash C) + l(C'') $, Equation
(\ref{d2fsmc}) still holds.

Since $l(C_1) < l(C) < 2l(P_c(C'') \backslash C) + l(C''),$
\begin{equation}
\label{C1sD} l(v_3,w_1) + l(P_s(C_1)) + l(w_2,v_1,v_2) < 3l(v_3,v_4)
+ l(u_1,v_4) + l(u_1,v_2).
\end{equation}

Since $l(C_2) < l(C) < 2l(P_c(C'') \backslash C) + l(C'')$,
\begin{equation}
\label{C2sD} l(v_3,w_1,w_2) + l(P_s(C_2)) + l(v_1,v_2) < 3l(v_3,v_4)
+ l(u_1,v_4) + l(u_1,v_2).
\end{equation}

\begin{description}
 \item [A1] Let $x=t(T),y=h(T)$, and  the end nodes $v,w$ of
$P_s(T)$ be ordered $v,x,y,w$.

 Since $C''$ was oriented  backwards by Property \ref{longfwd} Equation (\ref{shortD}) still holds.

According to Property \ref{scbprop} $C''$ is not a special contained
brother, by A3 on $T$ $x \in [v,v_4]$, and $y \in [v_5,w]$ (since
$S''$ exists).

 By A4 we can always define
\[ M_t(C'') = M_t(T) \cup [x \rightarrow v_3 \rightarrow w_1 \rightarrow u_3 \rightarrow u_4 \rightarrow u_1]
\ \ M_h(C'') = [v_4 \rightarrow v_5 \rightarrow y] \cup M_h(T).\]
 We get that $I(C'') = [u_1 \rightarrow v_4]$.
Using A1 on $T$:

The bound of $l(M(C))$ can be achieved as in lemma \ref{lemCDforD}.



\item [A2]

Since $S'' \in N(S)$ by A5 on $S$ then $l(S'') < l(S)$. Since $C$ is
a son of $S$, $l(S) < l(C)$, so Equation (\ref{EsmD2pf}) still
holds.

 In this case $h(C'') = v_4 \in T$. Define
\[ B_t(C'') = [u_1 \rightarrow v_2 \rightarrow u_2 \rightarrow v_6 \rightarrow y] \cup M_h(T) .\]

Using A1 on $T$ (again $l(v,v_3)$ and $l(v_6,w)$ are taken 4 times
to satisfy all possible locations of $x$ and $y$), Equations
(\ref{C1sD}), (\ref{C2sD}), (\ref{d2fsmc}),(\ref{EsmD2pf}),
(\ref{shortD}), definition of $P_c(C'')$ and the definition of
$P_s(C'')$ :
\begin{eqnarray*}
l(M_t(C'')) + l(B_t(C'')) &=& l(M_t(T))  + l(x,v_3) + l(v_3,w_1) +
l(w_1,u_3) + l(u_3,u_4) +  \\
&& l(u_4,u_1) +l(u_1,v_2) + l(v_2,u_2) +l(u_2,v_6) + l(v_6,y)\\
&\leq& 15l(P_c(T)) + 4l(v,v_3) + 3l(v_3,v_4) + 3l(v_4,v_5) +3l(v_5,v_6) +  \\
&& 4l(v_6,w)  + l(v_3,w_1) + l(w_1,u_3) + l(u_3,u_4) +l(u_4,u_1) + l(u_1,v_2) +  \\
 && l(v_2,u_2) + l(u_2,v_6) \\
&\leq& 15l(P_c(T)) + 4l(v,v_3) + 6l(v_3,v_4) + 3l(v_4,v_5) +3l(v_5,v_6) + 4l(v_6,w)   + \\
&& l(u_3,u_4) +l(u_4,u_1) + l(u_1,v_2) + l(v_2,u_2) + l(u_2,v_6) +  l(u_1,v_4)\\
&\leq& 15l(P_c(T)) + 4l(v,v_3) + 6l(v_3,v_4) + 3l(v_4,v_5) +3l(v_5,v_6) + 4l(v_6,w)   + \\
&& l(u_3,u_4) +l(u_4,u_1) + l(u_1,v_2) + l(v_2,u_2) + l(u_2,v_6) +  l(u_1,v_4)\\
&\leq& 15l(P_c(T)) + 4l(v,v_3) + 9l(v_3,v_4) + 3l(v_4,v_5) +3l(v_5,v_6) + \\
&& 4l(v_6,w)  +l(u_4,u_1) + l(u_1,v_2) + l(v_2,u_2) + l(u_2,v_6) +  2l(u_1,v_4)\\
&\leq& 15l(P_c(T)) + 4l(v,v_3) + 12l(v_3,v_4) + 3l(v_4,v_5) +3l(v_5,v_6) +\\
&&  4l(v_6,w)  +l(u_1,v_2) + l(v_2,u_2) + l(u_2,v_6) +  3l(u_1,v_4)\\
&\leq& 15l(P_c(T)) + 4l(v,v_3) + 14l(v_3,v_4) + 2l(v_4,v_5) +4l(v_5,v_6) + \\
&& 4l(v_6,w)  + 2l(u_1,v_2) +   4l(u_1,v_4) \\
&\leq& 15l(P_c(T)) + 5l(v,v_3) + 15l(v_3,v_4) + 4l(v_5,v_6) +5l(v_6,w)  + \\
&& l(u_1,v_2) +  4l(u_1,v_4) + l(u_1,v_2)\\
&\leq& 15l(P_c(C''))  + 2l(u_1,v_2) +   4l(u_1,v_4) \\
&\leq& 15l(P_c(C''))  + 2l(P_s(C'')).
\end{eqnarray*}

 \item[A3] Same proof like in Lemma \ref{lemCDforD}.
 \item [A4] Same proof like in Lemma \ref{lemCDforD}.
\item[A5] $M(C'')$ is contained in $P_s(C'')$, $C$,$C_1$,$C_2$ and $T$ which are
all shorter then $C$.
 \item[A6] $M(C)$ is contained in $P_s(C''), P_s(T), M(T)$ and $C$,$C_1,C_2$,
 so $N(C'') = \{ C,C_1,C_2 \}$.

\end{description}

\end{prf}

\begin{cor}
\label{uncle'fatf} After activating DIRECT-ONE on a cycle $C$ whose
father is oriented  forwards and when $C''$ is an uncle, $C$ and
$C''$ will satisfy the induction hypotheses.
\end{cor}

\paragraph{$S=F(C)$ is oriented backwards\\ \\}
\label{secback}

\begin{figure}
\begin{center}
\setlength{\unitlength}{0.00069991in}
\begingroup\makeatletter\ifx\SetFigFont\undefined%
\gdef\SetFigFont#1#2#3#4#5{%
  \reset@font\fontsize{#1}{#2pt}%
  \fontfamily{#3}\fontseries{#4}\fontshape{#5}%
  \selectfont}%
\fi\endgroup%
{\renewcommand{\dashlinestretch}{30}
\begin{picture}(8347,3439)(0,-10)
\put(1068,250){\blacken\ellipse{90}{90}}
\put(1068,250){\ellipse{90}{90}}
\put(2373,250){\blacken\ellipse{90}{90}}
\put(2373,250){\ellipse{90}{90}}
\put(2373,1690){\blacken\ellipse{90}{90}}
\put(2373,1690){\ellipse{90}{90}}
\put(2373,2815){\blacken\ellipse{90}{90}}
\put(2373,2815){\ellipse{90}{90}}
\put(3633,1690){\blacken\ellipse{90}{90}}
\put(3633,1690){\ellipse{90}{90}}
\put(3633,250){\blacken\ellipse{90}{90}}
\put(3633,250){\ellipse{90}{90}}
\put(5388,250){\blacken\ellipse{90}{90}}
\put(5388,250){\ellipse{90}{90}}
\put(4668,2815){\blacken\ellipse{90}{90}}
\put(4668,2815){\ellipse{90}{90}}
\path(2373,250)(3633,250)
\blacken\path(3483.000,212.500)(3633.000,250.000)(3483.000,287.500)(3483.000,212.500)
\path(3633,250)(5388,250)
\blacken\path(5238.000,212.500)(5388.000,250.000)(5238.000,287.500)(5238.000,212.500)
\blacken\path(2523.000,2852.500)(2373.000,2815.000)(2523.000,2777.500)(2523.000,2852.500)
\path(2373,2815)(4668,2815)
\path(2373,1690)(2373,2815)
\blacken\path(2410.500,2665.000)(2373.000,2815.000)(2335.500,2665.000)(2410.500,2665.000)
\blacken\path(1218.000,2852.500)(1068.000,2815.000)(1218.000,2777.500)(1218.000,2852.500)
\dashline{60.000}(1068,2815)(2373,2815)
\blacken\path(1030.500,400.000)(1068.000,250.000)(1105.500,400.000)(1030.500,400.000)
\dashline{60.000}(1068,250)(1068,2815)
\path(3633,1690)(3633,250)
\blacken\path(3595.500,400.000)(3633.000,250.000)(3670.500,400.000)(3595.500,400.000)
\blacken\path(918.000,212.500)(1068.000,250.000)(918.000,287.500)(918.000,212.500)
\thicklines
\path(1068,250)(213,250)
\path(1068,250)(2373,250)
\blacken\thinlines
\path(2223.000,212.500)(2373.000,250.000)(2223.000,287.500)(2223.000,212.500)
\thicklines
\path(2373,250)(2373,1690)
\blacken\thinlines
\path(2410.500,1540.000)(2373.000,1690.000)(2335.500,1540.000)(2410.500,1540.000)
\thicklines
\path(2373,1690)(3633,1690)
\blacken\thinlines
\path(3483.000,1652.500)(3633.000,1690.000)(3483.000,1727.500)(3483.000,1652.500)
\thicklines
\path(3633,1690)(4668,2815)
\blacken\thinlines
\path(4818.000,2852.500)(4668.000,2815.000)(4818.000,2777.500)(4818.000,2852.500)
\thicklines
\path(4668,2815)(5388,2815)
\blacken\thinlines
\path(5425.500,2665.000)(5388.000,2815.000)(5350.500,2665.000)(5425.500,2665.000)
\thicklines
\path(5388,2815)(5388,250)
\path(5388,250)(6513,250)
\blacken\thinlines
\path(6363.000,212.500)(6513.000,250.000)(6363.000,287.500)(6363.000,212.500)
\put(2868,1105){\makebox(0,0)[lb]{\smash{{\SetFigFont{10}{12.0}{\rmdefault}{\mddefault}{\updefault}$C''$}}}}
\put(4398,2185){\makebox(0,0)[lb]{\smash{{\SetFigFont{10}{12.0}{\rmdefault}{\mddefault}{\updefault}$C$}}}}
\put(1518,2950){\makebox(0,0)[lb]{\smash{{\SetFigFont{10}{12.0}{\rmdefault}{\mddefault}{\updefault}$S''$}}}}
\put(3633,2950){\makebox(0,0)[lb]{\smash{{\SetFigFont{10}{12.0}{\rmdefault}{\mddefault}{\updefault}$S$}}}}
\put(4623,2950){\makebox(0,0)[lb]{\smash{{\SetFigFont{10}{12.0}{\rmdefault}{\mddefault}{\updefault}$v_1$}}}}
\put(3588,25){\makebox(0,0)[lb]{\smash{{\SetFigFont{10}{12.0}{\rmdefault}{\mddefault}{\updefault}$v_4$}}}}
\put(3858,1645){\makebox(0,0)[lb]{\smash{{\SetFigFont{10}{12.0}{\rmdefault}{\mddefault}{\updefault}$u_1$}}}}
\put(2058,1645){\makebox(0,0)[lb]{\smash{{\SetFigFont{10}{12.0}{\rmdefault}{\mddefault}{\updefault}$v_2$}}}}
\put(2328,25){\makebox(0,0)[lb]{\smash{{\SetFigFont{10}{12.0}{\rmdefault}{\mddefault}{\updefault}$v_3$}}}}
\put(5343,25){\makebox(0,0)[lb]{\smash{{\SetFigFont{10}{12.0}{\rmdefault}{\mddefault}{\updefault}$v_5$}}}}
\put(2328,2950){\makebox(0,0)[lb]{\smash{{\SetFigFont{10}{12.0}{\rmdefault}{\mddefault}{\updefault}$u_2$}}}}
\put(213,385){\makebox(0,0)[lb]{\smash{{\SetFigFont{10}{12.0}{\rmdefault}{\mddefault}{\updefault}$T$}}}}
\put(5748,3265){\makebox(0,0)[lb]{\smash{{\SetFigFont{10}{12.0}{\rmdefault}{\mddefault}{\updefault}The cycles}}}}
\put(5748,3040){\makebox(0,0)[lb]{\smash{{\SetFigFont{10}{12.0}{\rmdefault}{\mddefault}{\updefault}$T=F(S)=F(S'')=F(C'')$ $zv_3v_4v_5z$}}}}
\put(5748,2815){\makebox(0,0)[lb]{\smash{{\SetFigFont{10}{12.0}{\rmdefault}{\mddefault}{\updefault}$S=F(C)$ $zv_3v_2u_2v_1v_5z$}}}}
\put(5748,2590){\makebox(0,0)[lb]{\smash{{\SetFigFont{10}{12.0}{\rmdefault}{\mddefault}{\updefault}$C$ $zv_3v_2u_1v_1v_5z$}}}}
\put(5748,2365){\makebox(0,0)[lb]{\smash{{\SetFigFont{10}{12.0}{\rmdefault}{\mddefault}{\updefault}$C''$ $zv_3v_2u_1v_4v_5z$}}}}
\end{picture}
}
\end{center}
\caption{$C''$ is an uncle, $S$ is oriented backwards.}
\label{fcbackfig}
\end{figure}
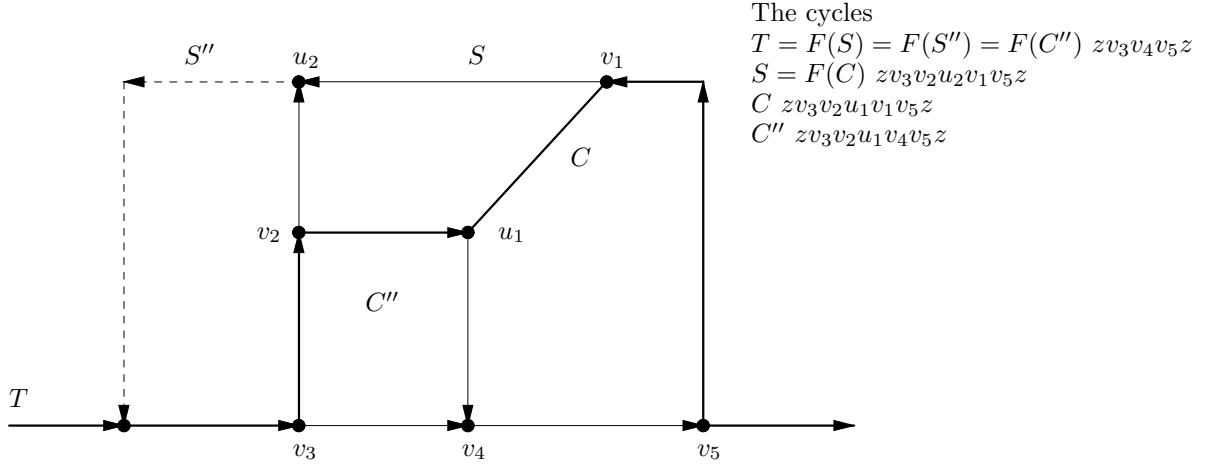

 Here we consider a  cycle $C$ oriented forwards in
DIRECT-ONE even though $l_2=-1$. In this subsection we assume that
$S=F(C)$ is oriented backwards, $T=F(S)$ and $C''$ a contained
brother of $S$.  $C$ which is a son of $S$ shares the path
$[v_2,u_1]$ at its $l_2$ side with $C''$. Thus $C$ has to be the
last cycle  in the block of  sons of $S$, and $C''$ is oriented
forwards. In Figure \ref{fcbackfig} $C,C'',S$ and $T$ are described
before the application of DIRECT-ONE.

 Mark \bi
\item $v_1:= S \cap P_s(C) \backslash  P_s(C'')$.
 \item $v_2:= S \cap P_s(C) \cap  P_s(C'')$.
 \item $v_3:= T \cap P_s(C'') \cap P_s(S)$.
 \item $v_4:= T \cap P_s(C'') \backslash P_s(S)$.
 \item $v_5:= T \cap P_s(S) \backslash P_s(C'')$.
 \item $u_1:= $ the end node of the path $P_s(C) \cap P_s(C'')$, such that $u_1 \not\in S$.
\ei

Since $C''$ is oriented  forwards $N(S)$ contains at least one cycle
at the $l_2$ side of $P_s(S)$, $S''$ in Figure \ref{fcbackfig} (such
orientation is discussed in Lemma
(\ref{lemtwob}) where $N(S)=\{ S'' \}$). When DIRECT-ONE is
activated on $C$ it compares $l(C)$ and $l(C'')$ (both are oriented
forwards) and decides whether to change the direction of
$[u_1,v_2]$. In the following two lemmas we assume  $[v_1,u_1]$ was
directed $v_1 \rightarrow u_1$ ($C$ didn't have heavy brothers at
its $l_1$ side to contradict this direction). In lemmas
\ref{lemCC1C2Dbacc} and \ref{lemCC1C2Dbacc} we will consider the
other case.

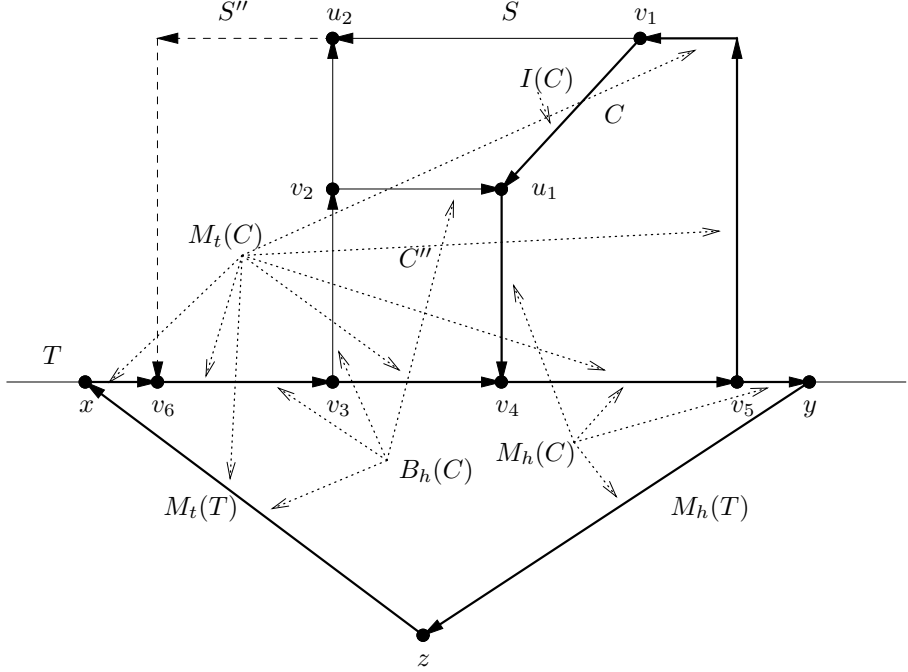
\begin{figure}
\begin{center}
\setlength{\unitlength}{0.00069991in}
\begingroup\makeatletter\ifx\SetFigFont\undefined%
\gdef\SetFigFont#1#2#3#4#5{%
  \reset@font\fontsize{#1}{#2pt}%
  \fontfamily{#3}\fontseries{#4}\fontshape{#5}%
  \selectfont}%
\fi\endgroup%
{\renewcommand{\dashlinestretch}{30}
\begin{picture}(6729,4995)(0,-10)
\put(2442,2130){\blacken\ellipse{90}{90}}
\put(2442,2130){\ellipse{90}{90}}
\put(2442,3570){\blacken\ellipse{90}{90}}
\put(2442,3570){\ellipse{90}{90}}
\put(2442,4695){\blacken\ellipse{90}{90}}
\put(2442,4695){\ellipse{90}{90}}
\put(3702,3570){\blacken\ellipse{90}{90}}
\put(3702,3570){\ellipse{90}{90}}
\put(3702,2130){\blacken\ellipse{90}{90}}
\put(3702,2130){\ellipse{90}{90}}
\put(5457,2130){\blacken\ellipse{90}{90}}
\put(5457,2130){\ellipse{90}{90}}
\put(4737,4695){\blacken\ellipse{90}{90}}
\put(4737,4695){\ellipse{90}{90}}
\put(597,2130){\blacken\ellipse{90}{90}}
\put(597,2130){\ellipse{90}{90}}
\put(5997,2130){\blacken\ellipse{90}{90}}
\put(5997,2130){\ellipse{90}{90}}
\put(3117,240){\blacken\ellipse{90}{90}}
\put(3117,240){\ellipse{90}{90}}
\put(1137,2130){\blacken\ellipse{90}{90}}
\put(1137,2130){\ellipse{90}{90}}
\blacken\path(2592.000,4732.500)(2442.000,4695.000)(2592.000,4657.500)(2592.000,4732.500)
\path(2442,4695)(4737,4695)
\path(2442,2130)(2442,3570)
\blacken\path(2479.500,3420.000)(2442.000,3570.000)(2404.500,3420.000)(2479.500,3420.000)
\path(2442,3570)(2442,4695)
\blacken\path(2479.500,4545.000)(2442.000,4695.000)(2404.500,4545.000)(2479.500,4545.000)
\path(2442,3570)(3702,3570)
\blacken\path(3552.000,3532.500)(3702.000,3570.000)(3552.000,3607.500)(3552.000,3532.500)
\dottedline{45}(3972,4290)(4062,4065)
\path(3989.579,4165.275)(4062.000,4065.000)(4045.287,4187.559)
\dottedline{45}(1767,3075)(777,2130)
\path(843.088,2234.558)(777.000,2130.000)(884.517,2191.156)
\dottedline{45}(1767,3075)(1497,2175)
\path(1502.747,2298.560)(1497.000,2175.000)(1560.217,2281.319)
\dottedline{45}(1767,3075)(1677,1410)
\path(1653.521,1531.444)(1677.000,1410.000)(1713.433,1528.206)
\dottedline{45}(1767,3075)(2937,2220)
\path(2822.413,2266.580)(2937.000,2220.000)(2857.814,2315.024)
\dottedline{45}(1767,3075)(4467,2220)
\path(4343.542,2227.627)(4467.000,2220.000)(4361.656,2284.827)
\dottedline{45}(1767,3075)(5322,3255)
\path(5203.671,3218.970)(5322.000,3255.000)(5200.636,3278.893)
\dottedline{45}(1767,3075)(5142,4605)
\path(5045.093,4528.130)(5142.000,4605.000)(5020.320,4582.777)
\path(5997,2130)(6717,2130)
\dottedline{45}(4242,1680)(3792,2850)
\path(3863.078,2748.768)(3792.000,2850.000)(3807.077,2727.229)
\dottedline{45}(4242,1680)(4602,2085)
\path(4544.699,1975.380)(4602.000,2085.000)(4499.854,2015.242)
\dottedline{45}(4242,1680)(5682,2085)
\path(5574.604,2023.631)(5682.000,2085.000)(5558.360,2081.390)
\dottedline{45}(4242,1680)(4557,1275)
\path(4459.647,1351.304)(4557.000,1275.000)(4507.008,1388.140)
\dottedline{45}(2847,1545)(1992,1185)
\path(2090.955,1259.216)(1992.000,1185.000)(2114.238,1203.918)
\dottedline{45}(2847,1545)(2037,2085)
\path(2153.487,2043.397)(2037.000,2085.000)(2120.205,1993.474)
\dottedline{45}(2847,1545)(2487,2355)
\path(2563.151,2257.527)(2487.000,2355.000)(2508.322,2233.158)
\dottedline{45}(2847,1545)(3342,3480)
\path(3341.324,3356.309)(3342.000,3480.000)(3283.196,3371.179)
\thicklines
\path(3117,240)(597,2130)
\blacken\thinlines
\path(739.500,2070.000)(597.000,2130.000)(694.500,2010.000)(739.500,2070.000)
\path(597,2130)(12,2130)
\thicklines
\path(597,2130)(1137,2130)
\blacken\thinlines
\path(987.000,2092.500)(1137.000,2130.000)(987.000,2167.500)(987.000,2092.500)
\thicklines
\path(1137,2130)(2442,2130)
\blacken\thinlines
\path(2292.000,2092.500)(2442.000,2130.000)(2292.000,2167.500)(2292.000,2092.500)
\thicklines
\path(2442,2130)(3702,2130)
\blacken\thinlines
\path(3552.000,2092.500)(3702.000,2130.000)(3552.000,2167.500)(3552.000,2092.500)
\thicklines
\path(3702,2130)(5457,2130)
\blacken\thinlines
\path(5307.000,2092.500)(5457.000,2130.000)(5307.000,2167.500)(5307.000,2092.500)
\blacken\path(5494.500,4545.000)(5457.000,4695.000)(5419.500,4545.000)(5494.500,4545.000)
\thicklines
\path(5457,4695)(5457,2130)
\blacken\thinlines
\path(4887.000,4732.500)(4737.000,4695.000)(4887.000,4657.500)(4887.000,4732.500)
\thicklines
\path(4737,4695)(5457,4695)
\blacken\thinlines
\path(3775.961,3705.779)(3702.000,3570.000)(3831.156,3655.000)(3775.961,3705.779)
\thicklines
\path(3702,3570)(4737,4695)
\path(3702,3570)(3702,2130)
\blacken\thinlines
\path(3664.500,2280.000)(3702.000,2130.000)(3739.500,2280.000)(3664.500,2280.000)
\thicklines
\path(5502,2130)(5997,2130)
\blacken\thinlines
\path(5847.000,2092.500)(5997.000,2130.000)(5847.000,2167.500)(5847.000,2092.500)
\thicklines
\path(5997,2130)(3117,240)
\blacken\thinlines
\path(3221.833,353.650)(3117.000,240.000)(3262.982,290.947)(3221.833,353.650)
\blacken\path(1287.000,4732.500)(1137.000,4695.000)(1287.000,4657.500)(1287.000,4732.500)
\dashline{60.000}(1137,4695)(2442,4695)
\blacken\path(1099.500,2280.000)(1137.000,2130.000)(1174.500,2280.000)(1099.500,2280.000)
\dashline{60.000}(1137,2130)(1137,4695)
\put(2937,2985){\makebox(0,0)[lb]{\smash{{\SetFigFont{10}{12.0}{\rmdefault}{\mddefault}{\updefault}$C''$}}}}
\put(4467,4065){\makebox(0,0)[lb]{\smash{{\SetFigFont{10}{12.0}{\rmdefault}{\mddefault}{\updefault}$C$}}}}
\put(1587,4830){\makebox(0,0)[lb]{\smash{{\SetFigFont{10}{12.0}{\rmdefault}{\mddefault}{\updefault}$S''$}}}}
\put(3702,4830){\makebox(0,0)[lb]{\smash{{\SetFigFont{10}{12.0}{\rmdefault}{\mddefault}{\updefault}$S$}}}}
\put(4692,4830){\makebox(0,0)[lb]{\smash{{\SetFigFont{10}{12.0}{\rmdefault}{\mddefault}{\updefault}$v_1$}}}}
\put(3657,1905){\makebox(0,0)[lb]{\smash{{\SetFigFont{10}{12.0}{\rmdefault}{\mddefault}{\updefault}$v_4$}}}}
\put(3927,3525){\makebox(0,0)[lb]{\smash{{\SetFigFont{10}{12.0}{\rmdefault}{\mddefault}{\updefault}$u_1$}}}}
\put(2127,3525){\makebox(0,0)[lb]{\smash{{\SetFigFont{10}{12.0}{\rmdefault}{\mddefault}{\updefault}$v_2$}}}}
\put(2397,1905){\makebox(0,0)[lb]{\smash{{\SetFigFont{10}{12.0}{\rmdefault}{\mddefault}{\updefault}$v_3$}}}}
\put(5412,1905){\makebox(0,0)[lb]{\smash{{\SetFigFont{10}{12.0}{\rmdefault}{\mddefault}{\updefault}$v_5$}}}}
\put(2397,4830){\makebox(0,0)[lb]{\smash{{\SetFigFont{10}{12.0}{\rmdefault}{\mddefault}{\updefault}$u_2$}}}}
\put(282,2265){\makebox(0,0)[lb]{\smash{{\SetFigFont{10}{12.0}{\rmdefault}{\mddefault}{\updefault}$T$}}}}
\put(3837,4335){\makebox(0,0)[lb]{\smash{{\SetFigFont{10}{12.0}{\rmdefault}{\mddefault}{\updefault}$I(C)$}}}}
\put(5952,1905){\makebox(0,0)[lb]{\smash{{\SetFigFont{10}{12.0}{\rmdefault}{\mddefault}{\updefault}$y$}}}}
\put(552,1905){\makebox(0,0)[lb]{\smash{{\SetFigFont{10}{12.0}{\rmdefault}{\mddefault}{\updefault}$x$}}}}
\put(1092,1905){\makebox(0,0)[lb]{\smash{{\SetFigFont{10}{12.0}{\rmdefault}{\mddefault}{\updefault}$v_6$}}}}
\put(3072,15){\makebox(0,0)[lb]{\smash{{\SetFigFont{10}{12.0}{\rmdefault}{\mddefault}{\updefault}$z$}}}}
\put(1182,1140){\makebox(0,0)[lb]{\smash{{\SetFigFont{10}{12.0}{\rmdefault}{\mddefault}{\updefault}$M_t(T)$}}}}
\put(4962,1140){\makebox(0,0)[lb]{\smash{{\SetFigFont{10}{12.0}{\rmdefault}{\mddefault}{\updefault}$M_h(T)$}}}}
\put(1362,3165){\makebox(0,0)[lb]{\smash{{\SetFigFont{10}{12.0}{\rmdefault}{\mddefault}{\updefault}$M_t(C)$}}}}
\put(3657,1545){\makebox(0,0)[lb]{\smash{{\SetFigFont{10}{12.0}{\rmdefault}{\mddefault}{\updefault}$M_h(C)$}}}}
\put(2937,1410){\makebox(0,0)[lb]{\smash{{\SetFigFont{10}{12.0}{\rmdefault}{\mddefault}{\updefault}$B_h(C)$}}}}
\end{picture}
}
\end{center}
\caption{Figure for Lemma \ref{lemCDbacc}.} \label{fcbackcfig}
\end{figure}

\begin{lem}
\label{lemCDbacc} Suppose that  $ l(C'') < l(C) $, then $C$
satisfies the induction hypotheses.
\end{lem}
\begin{prf}
In this case when DIRECT-ONE is activated on $C$ it doesn't change
the direction $v_2 \rightarrow u_1$.

 Since $l(C'') < l(C) $
\begin{equation}
\label{DsmalC} l(u_1,v_4) + l(v_4,v_5) < l(u_1,v_1) + l(v_1,v_5).
\end{equation}

\begin{description}
\item [A1] Let $x=t(T),y=h(T)$, and  the end nodes $v,w$ of
$P_s(T)$ be ordered $v,x,y,w$ (See Figure \ref{fcbackcfig}).

Since $S$ was oriented backwards by Property \ref{longfwd}:
\begin{equation}
\label{shortB} 2l(v_3,v_4) + 2l(v_4,v_5) < l(v,v_3) + l(v_5,w).
\end{equation}

Since there is at least one more brother of $S$ on its $l_2$ side $x
\in [v,v_3]$. By A4 on $T$  $y \in [v_3,v_5]$ or $y \in \cup
[v_5,w]$.

We consider these two possible locations:
 \bi
 \item $y \in [v_5,w]$, $I(C) = [v_1 \rightarrow u_1]$ so we can
define:
\[ M_t(C) = M_t(T) \cup [x  \rightarrow v_3 \rightarrow v_4
\rightarrow v_5 \rightarrow v_1 ] \ M_h(C) = [u_1 \rightarrow v_4
\rightarrow v_5 \rightarrow y ] \cup M_h(T) \]

Using A1 on $T$:
\begin{eqnarray*}
l(M(C) &=& l(M(T)) + l(x,v_3) + l(v_3,v_4) + 2l(v_4,v_5) +
l(v_5,v_1)  + l(u_1,v_4) + l(v_5,y) \\
&\leq& 15 l(P_c(T)) + 4 l(v,v_3) + 4l(v_3,v_4) + 5l(v_4,v_5) +
4l(v_5,w) +  l(v_5,v_1) + l(u_1,v_4)
\end{eqnarray*}

 \item $y \in [v_3,v_5]$ again $I(C) = [v_1 \rightarrow u_1]$ and we can
define (Using A4 on $T$):
\[ M_t(C) = (B_h(T) \backslash [y,v_5]) \cup [v_5 \rightarrow v_1]
\ M_h(C) = [u_1 \rightarrow v_4 \rightarrow y] \cup M_h(T)\]

Using A2 on $T$ ($y = h(T) \not\in F(T)$)
\begin{eqnarray*}
l(M(C)) &\leq& l(B_h(T)) + l(M_h(T)) + l(v_5,v_1) + l(u_1,v_4)  + l(v_4,v_5) \\
&\leq& 15 l(P_c(T)) + 4l(v,v_3) + 4l(v_3,v_4) + 5l(v_4,v_5) +
4l(v_5,w)  + l(v_5,v_1) + l(u_1,v_4) .
\end{eqnarray*}
(same bound as before)

 \ei

Using Equations (\ref{shortB}) (\ref{DsmalC}), the definition of
$P_c(C)$ and the definition of $I(C)$ :
\begin{eqnarray*}
l(M(C)) &\leq& 15 l(P_c(T)) + 6 l(v,v_3) + l(v_4,v_5) + 6l(v_5,w)
+ l(v_5,v_1) + l(u_1,v_4) \\
&\leq& 15 l(P_c(T)) + 6 l(v,v_3)  + 6l(v_5,w)+ 2l(v_5,v_1) + l(u_1,v_1)  \\
&\leq& 15 l(P_c(C)) + l(u_1,v_1)  \\
&\leq& 15 l(P_c(C)) + l(I(C))  \\
\end{eqnarray*}

 \item [A2] In this case $t(C) = v_1 \in S$ but $h(C) = u_1
\not\in S$. So we define:
\[ B_h(C) = M_t(T) \cup [x \rightarrow v_3 \rightarrow v_2
\rightarrow u_1] \]

 Giving (for all possible locations of $y$)
\begin{eqnarray*}
l(B_h(C)) + l(M_h(C)) &\leq& l(M_t(T)) + l(M_h(T)) + l(x,v_3) +
l(v_3,v_2) + l(v_2,u_1) + l(u_1,v_4) +\\
&&  l(v_3,v_4) + l(v_4,v_5) +l(v_5,w) \\
\end{eqnarray*}

Using A1 on $T$, the notation we use, Equations (\ref{shortB})
(\ref{DsmalC}), the definition of $P_c(C)$ and the definition of
$P_s(C)$ :
\begin{eqnarray*}
l(B_h(C)) + l(M_h(C)) &\leq& 15l(P_c(T)) + 2l(P_s(T)) + l(I(T)) +
 + l(x,v_3) +l(v_3,v_2) + \\
 && l(v_2,u_1) + l(u_1,v_4) + l(v_3,v_4) + l(v_4,v_5) +l(v_5,w) \\
 &\leq& 15l(P_c(T)) + 4l(v,v_3) + 4l(v_3,v_4)
+ 4l(v_4,v_5) + 4l(v_5,w) + \\
&& l(v_3,v_2) + l(v_2,u_1) + l(u_1,v_4) +\\
 &\leq& 15l(P_c(T)) + 6l(v,v_3) +  6l(v_5,w) + l(v_3,v_2) + l(v_2,u_1) + l(u_1,v_4) \\
&\leq& 15l(P_c(T)) + 6l(v,v_3) +  6l(v_5,w) + l(v_3,v_2) + l(v_2,u_1)  + \\
&&  l(u_1,v_1) +l(v_1,v_5) \\
&\leq& 15l(P_c(C))  + l(v_2,u_1)  + l(u_1,v_1) \\
&=& 15l(P_c(C)) + l(I(C)).
\end{eqnarray*}

\item[A3] In the Lemma assumption $l(C'') < l(C)$. Let $\hat{C}$ be a son
of $C$. If $P_f(\hat{C}) \subset [v_2,u_1]$ since $[v_2,u_1] \subset
P_s(C'')$ $\hat{C}$ should have been a son of $C''$. Therefore
$P_f(\hat{C}) \cap [v_1,u_1] \neq \phi$.
 \item [A4] $J_h(C) = [v_2 \rightarrow u_1]$ and for every $\hat{C}$ a son of $C$
 $P_f(\hat{C}) \subset P_s(C) = I(C) \cup J_h(C)$.

\item[A5] $M(C)$ is contained in $P_s(C)$, $S$, $C''$ and $T$ which
are all shorter then $C$.
 \item[A6] $M(C)$ is contained in $P_s(C), P_s(S), M(S)$ and $C''$,
 so $N(C) = \{ C'' \}$.

\end{description}

\end{prf}

\begin{figure}
\begin{center}
\setlength{\unitlength}{0.00069991in}
\begingroup\makeatletter\ifx\SetFigFont\undefined%
\gdef\SetFigFont#1#2#3#4#5{%
  \reset@font\fontsize{#1}{#2pt}%
  \fontfamily{#3}\fontseries{#4}\fontshape{#5}%
  \selectfont}%
\fi\endgroup%
{\renewcommand{\dashlinestretch}{30}
\begin{picture}(6594,4995)(0,-10)
\put(1002,2130){\blacken\ellipse{90}{90}}
\put(1002,2130){\ellipse{90}{90}}
\put(2307,2130){\blacken\ellipse{90}{90}}
\put(2307,2130){\ellipse{90}{90}}
\put(2307,3570){\blacken\ellipse{90}{90}}
\put(2307,3570){\ellipse{90}{90}}
\put(2307,4695){\blacken\ellipse{90}{90}}
\put(2307,4695){\ellipse{90}{90}}
\put(3567,3570){\blacken\ellipse{90}{90}}
\put(3567,3570){\ellipse{90}{90}}
\put(3567,2130){\blacken\ellipse{90}{90}}
\put(3567,2130){\ellipse{90}{90}}
\put(5322,2130){\blacken\ellipse{90}{90}}
\put(5322,2130){\ellipse{90}{90}}
\put(4602,4695){\blacken\ellipse{90}{90}}
\put(4602,4695){\ellipse{90}{90}}
\put(462,2130){\blacken\ellipse{90}{90}}
\put(462,2130){\ellipse{90}{90}}
\put(5862,2130){\blacken\ellipse{90}{90}}
\put(5862,2130){\ellipse{90}{90}}
\put(2982,240){\blacken\ellipse{90}{90}}
\put(2982,240){\ellipse{90}{90}}
\blacken\path(2457.000,4732.500)(2307.000,4695.000)(2457.000,4657.500)(2457.000,4732.500)
\path(2307,4695)(4602,4695)
\path(2307,2130)(2307,3570)
\blacken\path(2344.500,3420.000)(2307.000,3570.000)(2269.500,3420.000)(2344.500,3420.000)
\path(2307,3570)(2307,4695)
\blacken\path(2344.500,4545.000)(2307.000,4695.000)(2269.500,4545.000)(2344.500,4545.000)
\dottedline{45}(1632,3075)(642,2130)
\path(708.088,2234.558)(642.000,2130.000)(749.517,2191.156)
\dottedline{45}(1632,3075)(1362,2175)
\path(1367.747,2298.560)(1362.000,2175.000)(1425.217,2281.319)
\dottedline{45}(1632,3075)(1542,1410)
\path(1518.521,1531.444)(1542.000,1410.000)(1578.433,1528.206)
\dottedline{45}(1632,3075)(2802,2220)
\path(2687.413,2266.580)(2802.000,2220.000)(2722.814,2315.024)
\dottedline{45}(1632,3075)(4332,2220)
\path(4208.542,2227.627)(4332.000,2220.000)(4226.656,2284.827)
\dottedline{45}(1632,3075)(5187,3255)
\path(5068.671,3218.970)(5187.000,3255.000)(5065.636,3278.893)
\dottedline{45}(1632,3075)(5007,4605)
\path(4910.093,4528.130)(5007.000,4605.000)(4885.320,4582.777)
\path(5862,2130)(6582,2130)
\dottedline{45}(4107,1680)(4467,2085)
\path(4409.699,1975.380)(4467.000,2085.000)(4364.854,2015.242)
\dottedline{45}(4107,1680)(5547,2085)
\path(5439.604,2023.631)(5547.000,2085.000)(5423.360,2081.390)
\dottedline{45}(4107,1680)(4422,1275)
\path(4324.647,1351.304)(4422.000,1275.000)(4372.008,1388.140)
\blacken\path(2457.000,3607.500)(2307.000,3570.000)(2457.000,3532.500)(2457.000,3607.500)
\path(2307,3570)(3567,3570)
\dottedline{45}(1632,3075)(3837,3930)
\path(3735.962,3858.646)(3837.000,3930.000)(3714.271,3914.587)
\thicklines
\path(2982,240)(462,2130)
\blacken\thinlines
\path(604.500,2070.000)(462.000,2130.000)(559.500,2010.000)(604.500,2070.000)
\thicklines
\path(1002,2130)(2307,2130)
\blacken\thinlines
\path(2157.000,2092.500)(2307.000,2130.000)(2157.000,2167.500)(2157.000,2092.500)
\path(462,2130)(12,2130)
\thicklines
\path(462,2130)(1002,2130)
\blacken\thinlines
\path(852.000,2092.500)(1002.000,2130.000)(852.000,2167.500)(852.000,2092.500)
\thicklines
\path(2307,2130)(3567,2130)
\blacken\thinlines
\path(3417.000,2092.500)(3567.000,2130.000)(3417.000,2167.500)(3417.000,2092.500)
\thicklines
\path(3567,2130)(5322,2130)
\blacken\thinlines
\path(5172.000,2092.500)(5322.000,2130.000)(5172.000,2167.500)(5172.000,2092.500)
\blacken\path(5359.500,4545.000)(5322.000,4695.000)(5284.500,4545.000)(5359.500,4545.000)
\thicklines
\path(5322,4695)(5322,2130)
\blacken\thinlines
\path(4752.000,4732.500)(4602.000,4695.000)(4752.000,4657.500)(4752.000,4732.500)
\thicklines
\path(4602,4695)(5322,4695)
\blacken\thinlines
\path(3640.961,3705.779)(3567.000,3570.000)(3696.156,3655.000)(3640.961,3705.779)
\thicklines
\path(3567,3570)(4602,4695)
\path(3567,3570)(3567,2130)
\blacken\thinlines
\path(3529.500,2280.000)(3567.000,2130.000)(3604.500,2280.000)(3529.500,2280.000)
\thicklines
\path(5367,2130)(5862,2130)
\blacken\thinlines
\path(5712.000,2092.500)(5862.000,2130.000)(5712.000,2167.500)(5712.000,2092.500)
\thicklines
\path(5862,2130)(2982,240)
\blacken\thinlines
\path(3086.833,353.650)(2982.000,240.000)(3127.982,290.947)(3086.833,353.650)
\blacken\path(1152.000,4732.500)(1002.000,4695.000)(1152.000,4657.500)(1152.000,4732.500)
\dashline{60.000}(1002,4695)(2307,4695)
\blacken\path(964.500,2280.000)(1002.000,2130.000)(1039.500,2280.000)(964.500,2280.000)
\dashline{60.000}(1002,2130)(1002,4695)
\put(2802,2985){\makebox(0,0)[lb]{\smash{{\SetFigFont{10}{12.0}{\rmdefault}{\mddefault}{\updefault}$C''$}}}}
\put(4332,4065){\makebox(0,0)[lb]{\smash{{\SetFigFont{10}{12.0}{\rmdefault}{\mddefault}{\updefault}$C$}}}}
\put(1452,4830){\makebox(0,0)[lb]{\smash{{\SetFigFont{10}{12.0}{\rmdefault}{\mddefault}{\updefault}$S''$}}}}
\put(3567,4830){\makebox(0,0)[lb]{\smash{{\SetFigFont{10}{12.0}{\rmdefault}{\mddefault}{\updefault}$S$}}}}
\put(4557,4830){\makebox(0,0)[lb]{\smash{{\SetFigFont{10}{12.0}{\rmdefault}{\mddefault}{\updefault}$v_1$}}}}
\put(3522,1905){\makebox(0,0)[lb]{\smash{{\SetFigFont{10}{12.0}{\rmdefault}{\mddefault}{\updefault}$v_4$}}}}
\put(3792,3525){\makebox(0,0)[lb]{\smash{{\SetFigFont{10}{12.0}{\rmdefault}{\mddefault}{\updefault}$u_1$}}}}
\put(1992,3525){\makebox(0,0)[lb]{\smash{{\SetFigFont{10}{12.0}{\rmdefault}{\mddefault}{\updefault}$v_2$}}}}
\put(2262,1905){\makebox(0,0)[lb]{\smash{{\SetFigFont{10}{12.0}{\rmdefault}{\mddefault}{\updefault}$v_3$}}}}
\put(5277,1905){\makebox(0,0)[lb]{\smash{{\SetFigFont{10}{12.0}{\rmdefault}{\mddefault}{\updefault}$v_5$}}}}
\put(2262,4830){\makebox(0,0)[lb]{\smash{{\SetFigFont{10}{12.0}{\rmdefault}{\mddefault}{\updefault}$u_2$}}}}
\put(147,2265){\makebox(0,0)[lb]{\smash{{\SetFigFont{10}{12.0}{\rmdefault}{\mddefault}{\updefault}$T$}}}}
\put(5817,1905){\makebox(0,0)[lb]{\smash{{\SetFigFont{10}{12.0}{\rmdefault}{\mddefault}{\updefault}$y$}}}}
\put(417,1905){\makebox(0,0)[lb]{\smash{{\SetFigFont{10}{12.0}{\rmdefault}{\mddefault}{\updefault}$x$}}}}
\put(957,1905){\makebox(0,0)[lb]{\smash{{\SetFigFont{10}{12.0}{\rmdefault}{\mddefault}{\updefault}$v_6$}}}}
\put(2937,15){\makebox(0,0)[lb]{\smash{{\SetFigFont{10}{12.0}{\rmdefault}{\mddefault}{\updefault}$z$}}}}
\put(1047,1140){\makebox(0,0)[lb]{\smash{{\SetFigFont{10}{12.0}{\rmdefault}{\mddefault}{\updefault}$M_t(T)$}}}}
\put(4827,1140){\makebox(0,0)[lb]{\smash{{\SetFigFont{10}{12.0}{\rmdefault}{\mddefault}{\updefault}$M_h(T)$}}}}
\put(1227,3165){\makebox(0,0)[lb]{\smash{{\SetFigFont{10}{12.0}{\rmdefault}{\mddefault}{\updefault}$M_t(C'')$}}}}
\put(3522,1545){\makebox(0,0)[lb]{\smash{{\SetFigFont{10}{12.0}{\rmdefault}{\mddefault}{\updefault}$M_h(C'')$}}}}
\put(3702,2850){\makebox(0,0)[lb]{\smash{{\SetFigFont{10}{12.0}{\rmdefault}{\mddefault}{\updefault}$I(C'')$}}}}
\end{picture}
}
\end{center}
\caption{Figure for Lemma \ref{lemCDbacD}.} \label{fcbackctagfig}
\end{figure}
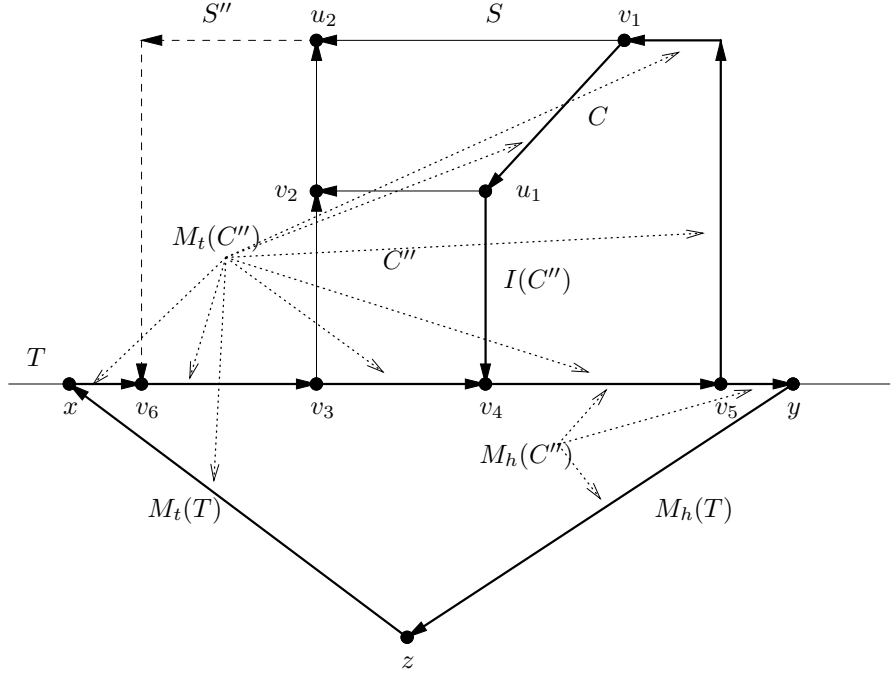

\begin{lem}
\label{lemCDbacD}  Suppose that $l(C) < l(C'') $ then after the
application of DIRECT-ONE  $C''$ satisfies the induction hypotheses.
\end{lem}
\begin{prf}
In this case when DIRECT-ONE is activated on $C$ it changes the
direction of $[u_1,v_2]$ to $v_2 \rightarrow u_1$.

Since $l(C) <  l(C'')$:
\begin{equation}
\label{CsmalD} l(u_1,v_1) + l(v_1,v_5) < l(u_1,v_4) + l(v_4,v_5).
\end{equation}

\begin{description}
 \item [A1]
Let $x=t(T),y=h(T)$, and  the end nodes $v,w$ of $P_s(T)$ be ordered
$v,x,y,w$ (see Figure \ref{fcbackctagfig}).  Since $S$ was oriented
backwards Equation (\ref{shortB}) still holds.

Since there is at least one more brother of $S$ and its $l_2$ side
$x \in [v,v_3]$.  By A4 on $T$  $y \in [v_3,v_5]$ or $y \in
[v_5,w]$.

We consider the two possible locations of $y$:\bi
\item $y \in [v_5,w]$ and $I(C'') = [u_1 \rightarrow v_4]$ so we can
define:
\[ M_t(C'') = M_t(T) \cup [x \rightarrow v_3 \rightarrow v_4 \rightarrow v_5 \rightarrow v_1 \rightarrow u_1] \
M_h(C'') = [v_4 \rightarrow v_5 \rightarrow y] \cup M_h(T)\]
 Using A1 on $T$:
\begin{eqnarray*}
l(M(C'') &=& l(M(T)) + l(x,v_3) + l(v_3,v_4) + 2l(v_4,v_5) +
l(v_5,v_1) + l(v_1,u_1) + l(v_5,y) \\
&\leq& 15 l(P_c(T)) + 4l(v,v_3) + 4l(v_3,v_4) + 5l(v_4,v_5) +
4l(v_5,w) + l(v_5,v_1) + l(v_1,u_1) .
\end{eqnarray*}

 \item $y \in [v_3,v_4] \cup [v_4,v_5]$ and $I(C'') = [u_1 \rightarrow
v_4]$. So we we can define:
\[ M_t(C'') = (B_h(T) \backslash [y,v_5]) \cup [v_5 \rightarrow v_1 \rightarrow u_1]
\ M_h(C'') = [ v_4 \rightarrow y] \cup M_h(T)\]

Using A2 on $T$ ($y = h(T) \not\in F(T)$)

\begin{eqnarray*}
l(M(C'')) &\leq& l(B_h(T)) + l(M_h(T)) + l(v_5,v_1) + l(v_1,u_1) +
l(v_4,v_5) \\
&\leq& 15 l(P_c(T)) + 4l(v,v_3) + 4l(v_3,v_4) + 5l(v_4,v_5) +
4l(v_5,w) + l(v_5,v_1) + l(v_1,u_1).
\end{eqnarray*}
(same bound as before) \ei

 Using Equations (\ref{shortB}), (\ref{CsmalD}), the definition of $P_c(C'')$  and
 the definition of $I(C'')$:
\begin{eqnarray*}
l(M(C'') &\leq& 15 l(P_c(T)) + 6l(v,v_3) + l(v_4,v_5) + 6l(v_5,w) +
l(v_5,v_1) + l(v_1,u_1) \\
&\leq& 15 l(P_c(T)) + 6l(v,v_3) + 2l(v_4,v_5) + 6l(v_5,w) +
l(u_1,v_4) \\
&\leq& 15 l(P_c(C''))  +l(u_1,v_4) \\
&=& 15 l(P_c(C''))  +l(I(C''))\\
\end{eqnarray*}

\item [A2] Mark:
 \bi
 \item $u_2 :=$ the end node of the path $P_s(S) \cap P_s(S'')$ such
 that $u_2\not \in T$.
  \item $v_6 = T \cap (P_s(S'') \backslash P_s(S))$ .
  \ei
Since $S'' \in N(S)$ by A5 on $S$ $l(S'') < l(S)$, since $C$ is a
son of $S$ $l(S) < l(C)$ and by the Lemma assumption
\[ l(S'') < l(S) < l(C) < l(C'')   \] giving:
\begin{equation}
\label{EsmDbac} l(v_6,u_2) + l(u_2,v_2) + l(v_3,v_4) < l(v_2,u_1) +
l(u_1,v_4) + l(v_6,v_3) .
\end{equation}

In this case $h(C'') = v_4 \in T$, but $t(C'') = u_1 \not\in T$.

Two possible locations of $x$:
 \bi
  \item $x \in [v,v_6]$. In this case we can define:
\[ B_t(C'') = [u_1 \rightarrow v_2  \rightarrow u_2 \rightarrow v_6 \rightarrow v_3 \rightarrow y] \cup M_h(T) .\]

 Two possible locations of $y$ to be considered:
 \bi
 \item $y \in [v_5,w]$.

 Using A1 on $T$
 \begin{eqnarray*}
 l(M_t(C'')) + l(B_t(C'')) &\leq& l(M_t(T)) + l(M_h(T)) + l(x,v_3) +
 2l(v_3,v_4) + 2l(v_4,v_5) + \\
 &&l(v_5,v_1) + l(v_1,u_1) + l(u_1,v_2) +
 l(v_2,u_2) + l(u_2,v_6) + l(v_5,y) \\
 &\leq& 15 l(P_c(T)) + 4l(v,v_6) + 4l(v_6,v_3) + 5l(v_3,v_4) +
 5l(v_4,v_5) + \\
 && 4l(v_5,w) + l(v_5,v_1) + l(v_1,u_1) + l(u_1,v_2) +
 l(v_2,u_2) + l(u_2,v_6)
 \end{eqnarray*}

 \item $y \in [v_3,v_5]$.

Using A2 on $T$
\begin{eqnarray*}
 l(M_t(C'')) + l(B_t(C'')) &\leq& l(B_h(T)) + l(M_h(T)) + l(u_1,v_2) +
 l(v_2,u_2) + l(u_2,v_6) + \\
 && l(v_6,v_3) + l(v_3,y) + l(v_5,v_1) + l(v_1,u_1)  \\
 &\leq& 15 l(P_c(T)) + 4l(v,v_6) + 4l(v_6,v_3) + 5l(v_3,v_4) +
 5l(v_4,v_5) + \\
 && 4l(v_5,w) + l(v_5,v_1) + l(v_1,u_1)  + l(u_1,v_2) +
 l(v_2,u_2) + l(u_2,v_6)
\end{eqnarray*}

 \ei

\item $x \in [v_6,v_3]$. In this case we define:
\[ B_t(C'') = [u_1 \rightarrow v_2  \rightarrow u_2 \rightarrow v_6 ]\cup B_t(T) \backslash [v_6,x] .\]

 Two possible locations of $y$:
 \bi
 \item $y \in [v_5,w]$. Using A2
on $T$:
\begin{eqnarray*}
l(B_t(C'')) + l(M_t(C'')) &\leq& l(M_t(T)) + l(B_t(T)) + l(x,v_3) +
l(v_3,v_4) + l(v_4,v_5) +  \\
&& l(v_5,v_1) +l(v_1,u_1) + l(u_1,v_2) +l(v_2,u_2) + l(u_2,v_6) \\
&\leq& 15l(P_c(T)) + 4l(v,v_6) + 4l(v_6,v_3) + 5l(v_3,v_4) +
5l(v_4,v_5) + \\
&& 4l(v_5,w) +  l(v_5,v_1) + l(v_1,u_1) + l(u_1,v_2) + l(v_2,u_2) +
l(u_2,v_6)
\end{eqnarray*}

\item $y \in [v_3,v_5]$. Using A2 on $T$:
\begin{eqnarray*}
l(B_t(C'')) + l(M_t(C'')) &=& l(B_h(T)) + l(B_t(T))  - l(v_6,x) -
l(y,v_5) + l(u_1,v_2) + \\
&&  l(v_2,u_2) +l(u_2,v_6) + l(v_5,v_1) +l(v_1,u_1) \\
&\leq& 15 l(P_c(T)) + 6l(v,v_6) + 5 l(v_6,v_3) + 5l(v_3,v_4) +
5l(v_4,v_5) + \\
&& 6l(v_5,w) + l(u_1,v_2) + l(v_2,u_2) + l(u_2,v_6) + l(v_5,v_1)
+l(v_1,u_1)
\end{eqnarray*}

\ei
 So in all  possible locations of $x$ and $y$:
\begin{eqnarray*}
l(B_t(C'')) + l(M_t(C''))  &\leq& 15 l(P_c(T)) + 6l(v,v_6) + 5
l(v_6,v_3) + 5l(v_3,v_4) + 5l(v_4,v_5) + \\
&& 6l(v_5,w) +l(u_1,v_2) + l(v_2,u_2) + l(u_2,v_6) + l(v_5,v_1)
+l(v_1,u_1)
\end{eqnarray*}

Using Equations (\ref{shortB}), (\ref{EsmDbac}), (\ref{CsmalD}), the
definition of $P_c(C'')$ and the definition of $P_s(C'')$
\begin{eqnarray*}
l(B_t(C'')) + l(M_t(C''))  &\leq& 15 l(P_c(T)) + 8l(v,v_3) +
7l(v_6,v_3) + l(v_3,v_4) + l(v_4,v_5) + \\
&& 8l(v_5,w) + l(u_1,v_2) +l(v_2,u_2) + l(u_2,v_6) + l(v_5,v_1) +l(v_1,u_1)\\
&\leq& 15 l(P_c(T)) + 8l(v,v_3) + 7l(v_6,v_3) +l(v_4,v_5) + 8l(v_5,w) + \\
&& 2l(u_1,v_2)  + l(v_5,v_1) +l(v_1,u_1) + l(u_1,v_4)\\
&\leq& 15 l(P_c(T)) + 8l(v,v_3) + 7l(v_6,v_3) +2l(v_4,v_5) + 8l(v_5,w) +\\
&& 2l(u_1,v_2)   + 2l(u_1,v_4)\\
&\leq& 15 l(P_c(C'')) + 2l(u_1,v_2)   + 2l(u_1,v_4)\\
&\leq& 15 l(P_c(C'')) + 2l(P_s(C'')).
\end{eqnarray*}

 \ei

 \item[A3] From the Lemma assumption $l(S'') < l(C) <
 l(C'')$. Let $\hat{C}$ be a son of
 $C''$ if $P_f(\hat{C}) \subset ([u_1,v_2] \cup [v_2,v_3])$ then $\hat{C}$
 should have been a son of $C$ or $S''$ but not $C''$.
 \item [A4] Same proof like in Lemma \ref{lemCDforD}.
\item[A5] $M(C'')$ is contained in $P_s(C'')$, $C$, and $T$ which are
all shorter then $C$.
 \item[A6] $M(C)$ is contained in $P_s(C''), P_s(T), M(T)$ and $C$,
 so $N(C'') = \{ C \}$.
\end{description}

\end{prf}

Now assume that $C$ had two brothers $C_1,C_2$ at its $l_1$ side,
ordered $C_1,C_2,C$,  such that $N(C)= \{ C_1,C_2,C'' \}$ (see
Figure \ref{fcbackthreefig}). This happens when $C$ and $C_2$ are
heavy, $C,C_1,C_2$ are oriented
 forwards and $l(C_1) < l(C_2) < l(C)$. If $C_1$ doesn't exists then
similar but simpler proofs apply.

\begin{figure}
\begin{center}
\setlength{\unitlength}{0.00069991in}
\begingroup\makeatletter\ifx\SetFigFont\undefined%
\gdef\SetFigFont#1#2#3#4#5{%
  \reset@font\fontsize{#1}{#2pt}%
  \fontfamily{#3}\fontseries{#4}\fontshape{#5}%
  \selectfont}%
\fi\endgroup%
{\renewcommand{\dashlinestretch}{30}
\begin{picture}(8662,3439)(0,-10)
\put(1068,250){\blacken\ellipse{90}{90}}
\put(1068,250){\ellipse{90}{90}}
\put(2373,250){\blacken\ellipse{90}{90}}
\put(2373,250){\ellipse{90}{90}}
\put(2373,1690){\blacken\ellipse{90}{90}}
\put(2373,1690){\ellipse{90}{90}}
\put(2373,2815){\blacken\ellipse{90}{90}}
\put(2373,2815){\ellipse{90}{90}}
\put(3633,1690){\blacken\ellipse{90}{90}}
\put(3633,1690){\ellipse{90}{90}}
\put(3633,250){\blacken\ellipse{90}{90}}
\put(3633,250){\ellipse{90}{90}}
\put(5388,250){\blacken\ellipse{90}{90}}
\put(5388,250){\ellipse{90}{90}}
\put(4668,2815){\blacken\ellipse{90}{90}}
\put(4668,2815){\ellipse{90}{90}}
\put(4353,2455){\blacken\ellipse{90}{90}}
\put(4353,2455){\ellipse{90}{90}}
\put(4848,2230){\blacken\ellipse{90}{90}}
\put(4848,2230){\ellipse{90}{90}}
\put(5388,2365){\blacken\ellipse{90}{90}}
\put(5388,2365){\ellipse{90}{90}}
\put(5388,1645){\blacken\ellipse{90}{90}}
\put(5388,1645){\ellipse{90}{90}}
\path(2373,250)(3633,250)
\blacken\path(3483.000,212.500)(3633.000,250.000)(3483.000,287.500)(3483.000,212.500)
\path(3633,250)(5388,250)
\blacken\path(5238.000,212.500)(5388.000,250.000)(5238.000,287.500)(5238.000,212.500)
\blacken\path(2523.000,2852.500)(2373.000,2815.000)(2523.000,2777.500)(2523.000,2852.500)
\path(2373,2815)(4668,2815)
\path(2373,1690)(2373,2815)
\blacken\path(2410.500,2665.000)(2373.000,2815.000)(2335.500,2665.000)(2410.500,2665.000)
\path(3633,1690)(3633,250)
\blacken\path(3595.500,400.000)(3633.000,250.000)(3670.500,400.000)(3595.500,400.000)
\blacken\path(4818.000,1682.500)(4668.000,1645.000)(4818.000,1607.500)(4818.000,1682.500)
\path(4668,1645)(5388,1645)
\blacken\path(4839.729,2075.605)(4848.000,2230.000)(4768.045,2097.661)(4839.729,2075.605)
\path(4848,2230)(4668,1645)
\path(4848,2230)(5388,2365)
\blacken\path(5251.574,2292.239)(5388.000,2365.000)(5233.384,2365.000)(5251.574,2292.239)
\blacken\path(4505.073,2427.068)(4353.000,2455.000)(4474.037,2358.791)(4505.073,2427.068)
\path(4353,2455)(4848,2230)
\blacken\path(918.000,212.500)(1068.000,250.000)(918.000,287.500)(918.000,212.500)
\thicklines
\path(1068,250)(213,250)
\path(1068,250)(2373,250)
\blacken\thinlines
\path(2223.000,212.500)(2373.000,250.000)(2223.000,287.500)(2223.000,212.500)
\thicklines
\path(2373,250)(2373,1690)
\blacken\thinlines
\path(2410.500,1540.000)(2373.000,1690.000)(2335.500,1540.000)(2410.500,1540.000)
\thicklines
\path(2373,1690)(3633,1690)
\blacken\thinlines
\path(3483.000,1652.500)(3633.000,1690.000)(3483.000,1727.500)(3483.000,1652.500)
\thicklines
\path(3633,1690)(4353,2455)
\path(4353,2455)(4668,2815)
\blacken\thinlines
\path(4597.446,2677.420)(4668.000,2815.000)(4541.003,2726.807)(4597.446,2677.420)
\blacken\path(4818.000,2852.500)(4668.000,2815.000)(4818.000,2777.500)(4818.000,2852.500)
\thicklines
\path(4668,2815)(5388,2815)
\path(5388,2365)(5388,2815)
\blacken\thinlines
\path(5425.500,2665.000)(5388.000,2815.000)(5350.500,2665.000)(5425.500,2665.000)
\thicklines
\path(5388,1645)(5388,2365)
\blacken\thinlines
\path(5425.500,2215.000)(5388.000,2365.000)(5350.500,2215.000)(5425.500,2215.000)
\thicklines
\path(5388,250)(5388,1645)
\blacken\thinlines
\path(5425.500,1495.000)(5388.000,1645.000)(5350.500,1495.000)(5425.500,1495.000)
\thicklines
\path(5388,250)(6513,250)
\blacken\thinlines
\path(6363.000,212.500)(6513.000,250.000)(6363.000,287.500)(6363.000,212.500)
\blacken\path(1218.000,2852.500)(1068.000,2815.000)(1218.000,2777.500)(1218.000,2852.500)
\dashline{60.000}(1068,2815)(2373,2815)
\blacken\path(1030.500,400.000)(1068.000,250.000)(1105.500,400.000)(1030.500,400.000)
\dashline{60.000}(1068,250)(1068,2815)
\put(2868,1105){\makebox(0,0)[lb]{\smash{{\SetFigFont{10}{12.0}{\rmdefault}{\mddefault}{\updefault}$C''$}}}}
\put(1518,2950){\makebox(0,0)[lb]{\smash{{\SetFigFont{10}{12.0}{\rmdefault}{\mddefault}{\updefault}$S''$}}}}
\put(3633,2950){\makebox(0,0)[lb]{\smash{{\SetFigFont{10}{12.0}{\rmdefault}{\mddefault}{\updefault}$S$}}}}
\put(4623,2950){\makebox(0,0)[lb]{\smash{{\SetFigFont{10}{12.0}{\rmdefault}{\mddefault}{\updefault}$v_1$}}}}
\put(3588,25){\makebox(0,0)[lb]{\smash{{\SetFigFont{10}{12.0}{\rmdefault}{\mddefault}{\updefault}$v_4$}}}}
\put(3858,1645){\makebox(0,0)[lb]{\smash{{\SetFigFont{10}{12.0}{\rmdefault}{\mddefault}{\updefault}$u_1$}}}}
\put(2058,1645){\makebox(0,0)[lb]{\smash{{\SetFigFont{10}{12.0}{\rmdefault}{\mddefault}{\updefault}$v_2$}}}}
\put(2328,25){\makebox(0,0)[lb]{\smash{{\SetFigFont{10}{12.0}{\rmdefault}{\mddefault}{\updefault}$v_3$}}}}
\put(5343,25){\makebox(0,0)[lb]{\smash{{\SetFigFont{10}{12.0}{\rmdefault}{\mddefault}{\updefault}$v_5$}}}}
\put(2328,2950){\makebox(0,0)[lb]{\smash{{\SetFigFont{10}{12.0}{\rmdefault}{\mddefault}{\updefault}$u_2$}}}}
\put(213,385){\makebox(0,0)[lb]{\smash{{\SetFigFont{10}{12.0}{\rmdefault}{\mddefault}{\updefault}$T$}}}}
\put(1023,25){\makebox(0,0)[lb]{\smash{{\SetFigFont{10}{12.0}{\rmdefault}{\mddefault}{\updefault}$v_6$}}}}
\put(4758,2410){\makebox(0,0)[lb]{\smash{{\SetFigFont{10}{12.0}{\rmdefault}{\mddefault}{\updefault}$C_2$}}}}
\put(4983,1825){\makebox(0,0)[lb]{\smash{{\SetFigFont{10}{12.0}{\rmdefault}{\mddefault}{\updefault}$C_1$}}}}
\put(3813,2185){\makebox(0,0)[lb]{\smash{{\SetFigFont{10}{12.0}{\rmdefault}{\mddefault}{\updefault}$C$}}}}
\put(6063,3265){\makebox(0,0)[lb]{\smash{{\SetFigFont{10}{12.0}{\rmdefault}{\mddefault}{\updefault}The cycles}}}}
\put(6063,3040){\makebox(0,0)[lb]{\smash{{\SetFigFont{10}{12.0}{\rmdefault}{\mddefault}{\updefault}$T=F(S)=F(S'')=F(C'')$ $zv_3v_4v_5z$}}}}
\put(6063,2815){\makebox(0,0)[lb]{\smash{{\SetFigFont{10}{12.0}{\rmdefault}{\mddefault}{\updefault}$S=F(C)$ $zv_3v_2u_2v_1v_5z$}}}}
\put(6063,2590){\makebox(0,0)[lb]{\smash{{\SetFigFont{10}{12.0}{\rmdefault}{\mddefault}{\updefault}$C$ $zv_6v_3v_2u_1v_1w_2w_1v_5z$}}}}
\put(6063,2365){\makebox(0,0)[lb]{\smash{{\SetFigFont{10}{12.0}{\rmdefault}{\mddefault}{\updefault}$C''$ $zv_6v_3v_2u_1v_4v_5z$}}}}
\put(4083,2500){\makebox(0,0)[lb]{\smash{{\SetFigFont{10}{12.0}{\rmdefault}{\mddefault}{\updefault}$u_4$}}}}
\put(4533,2095){\makebox(0,0)[lb]{\smash{{\SetFigFont{10}{12.0}{\rmdefault}{\mddefault}{\updefault}$u_3$}}}}
\put(5523,1600){\makebox(0,0)[lb]{\smash{{\SetFigFont{10}{12.0}{\rmdefault}{\mddefault}{\updefault}$w_1$}}}}
\put(5523,2320){\makebox(0,0)[lb]{\smash{{\SetFigFont{10}{12.0}{\rmdefault}{\mddefault}{\updefault}$w_2$}}}}
\put(6063,2140){\makebox(0,0)[lb]{\smash{{\SetFigFont{10}{12.0}{\rmdefault}{\mddefault}{\updefault}$C_1$ $zv_6v_3v_2u_2v_1w_2u_3w_1v_5z$}}}}
\put(6063,1915){\makebox(0,0)[lb]{\smash{{\SetFigFont{10}{12.0}{\rmdefault}{\mddefault}{\updefault}$C_2$ $zv_6v_3v_2u_2v_1u_4u_3w_2w_1v_5z$}}}}
\end{picture}
}
\end{center}
\caption{$F(C)$ is oriented backwards and $N(C)$ contains two
brothers at its $l_1$ side.} \label{fcbackthreefig}
\end{figure}

 Since $S$ was oriented
backwards $N(S)$ contains  one more brother $S''$ at its $l_2$ side.
Mark:
 \bi
 \item $u_2 :=$ the end node of the path $P_s(S) \cap P_s(S'')$ such
 that $u_2\not \in T$.
 \item $v_6 = T \cap (P_s(S'') \backslash P_s(S))$ .
 \item $w_1 := S \cap (P_s(C_1) \backslash P_s(C_2))$.
 \item $w_2 := S \cap P_s(C_1) \cap P_s(C_2)$.
 \item $u_3 :=$ the end node of the path $P_s(C_1) \cap P_s(C_2)$,
 such that $u_3 \not\in C$.
 \item $u_4 :=$ the end node of the path $P_s(C_2) \cap P_s(C)$,
 such that $u_4 \not\in C$.
 \ei

Since $C_1 \in N(C_2)$ and $C_2 \in N(C)$,  $l(C_1) < l(C_2) <
l(C)$. According to Observation \ref{winpath} the algorithm in
previous stage $($before activating DIRECT-ONE on $C)$ oriented
$[w_1 \rightarrow u_3 \rightarrow u_4], [u_3 \rightarrow w_2], [u_4
\rightarrow v_1]$. Since $l(C_2) < l(C)$ DIRECT-ONE will not change
the direction of $[u_4 \rightarrow v_1]$. So we only need to
consider what happens to the path $[u_1,v_2]$ (whether or not its
current direction $[v_2 \rightarrow u_1]$ changes).

\begin{lem}
\label{lemCC1C2Dbacc} Suppose that  $ l(C'') < l(C) $  and $l(C_1) <
l(C_2) < l(C)$,  then $C$ satisfies the induction hypotheses.
\end{lem}
\begin{prf}
Since $l(C'') < l(C)$  when DIRECT-ONE is activated on $C$ it will
not change the direction $v_2 \rightarrow u_1$. Moreover  Equation
(\ref{DsmalC}) holds. Since $l(C_1) < l(C_2) < l(C)$ by Observation
\ref{winpath} the orientation of the paths will be: $[w_1
\rightarrow u_3 \rightarrow w_2]$ and $[u_3 \rightarrow u_4
\rightarrow v_1]$.

Since $l(C_1) < l(C)$
\begin{equation}
\label{C1sCb} l(P_s(C_1)) + l(v_1,v_2) < l(P_s(C)) + l(w_1,w_2) .
\end{equation}

Since $l(C_2) < l(C)$
\begin{equation}
\label{C2sCb} l(u_3,w_2) + l(u_3,u_4) + l(v_1,v_2) < l(u_4,u_1) +
l(u_1,v_2)  + l(w_2,v_1) .
\end{equation}

>From A5 on $S$ $l(S'') < l(S) < l(C)$ giving
\begin{equation}
\label{EsmalC} l(v_2,u_2) + l(u_2,v_6) + l(v_3,v_4) + l(v_4,v_5) <
l(v_6,v_3) + l(v_5,w_1) + l(w_1,w_2) + l(w_2,v_1) + l(P_s(C))
\end{equation}

\begin{description}
\item [A1] Let $x=t(T),y=h(T)$, and  the end nodes $v,w$ of
$P_s(T)$ be ordered $v,x,y,w$. From Property \ref{longfwd} on $S$
Equation (\ref{shortB}) still holds.

Since $N(S)$ contains one more brother $S''$ at the $l_2$ side of
$S$, $x \in [v,v_3]$. By A4 on $T$,  $y \in [v_3,v_5]$ or $y \in
[v_5,w]$.

We consider the two possible locations of $y$:
 \bi
 \item $y \in [v_5,w]$ and $I(C) = [u_4 \rightarrow u_1]$, so we
define:
\[ M_t(C) = M_t(T) \cup [x  \rightarrow v_3 \rightarrow v_4
\rightarrow v_5 \rightarrow w_1 \rightarrow u_3 \rightarrow u_4 ] \
M_h(C) = [u_1 \rightarrow v_4 \rightarrow v_5 \rightarrow y ] \cup
M_h(T) \]

Using A1 on $T$:
\begin{eqnarray*}
l(M(C) &=& l(M(T)) + l(x,v_3) + l(v_3,v_4) + 2l(v_4,v_5)   +
 l(v_1,w_1) + l(w_1,u_3) + \\
 && l(u_3,u_4) +l(u_1,v_4) + l(v_5,y) \\
&\leq& 15 l(P_c(T)) + 4 l(v,v_3) + 4l(v_3,v_4) + 5l(v_4,v_5) +
4l(v_5,w) +  l(v_1,w_1) +  \\
&& l(w_1,u_3) +l(u_3,u_4) + l(u_1,v_4)
\end{eqnarray*}

 \item $y \in [v_3,v_5]$ and $I(C) = [u_4 \rightarrow u_1]$.
 Using A4 on $T$ we can define:
\[ M_t(C) = (B_h(T) \backslash [y,v_5]) \cup [v_5 \rightarrow w_1 \rightarrow u_3 \rightarrow u_4]
\ M_h(C) = [u_1 \rightarrow v_4 \rightarrow y] \cup M_h(T)\]

Using A2 on $T$ ($y = h(T) \not\in F(T)$)
\begin{eqnarray*}
l(M(C)) &\leq& l(B_h(T)) + l(M_h(T)) + l(v_5,w_1) + l(w_1,u_3)
+ l(u_3,u_4) + l(u_1,v_4)  + l(v_4,v_5) \\
&\leq& 15 l(P_c(T)) + 4l(v,v_3) + 4l(v_3,v_4) + 5l(v_4,v_5) +
4l(v_5,w)  + l(v_5,w_1) +  \\
&& l(w_1,u_3) +l(u_3,u_4) + l(u_1,v_4) .
\end{eqnarray*}
(same bound as before)

 \ei

Using Equations (\ref{shortB}), (\ref{C1sCb}), (\ref{C2sCb}),
(\ref{DsmalC}), the definition of $P_c(C)$, the definition of
$P_c(C)$ and the definition of $I(C)$ :
\begin{eqnarray*}
l(M(C)) &\leq& 15 l(P_c(T)) + 4l(v,v_3) + 4l(v_3,v_4) +
5l(v_4,v_5) + 4l(v_5,w)  + l(v_5,w_1) +  \\
&& l(w_1,u_3) +l(u_3,u_4) +l(u_1,v_4) \\
&\leq& 15 l(P_c(T)) + 6l(v,v_3) +  l(v_4,v_5) + 6l(v_5,w) +
l(v_5,w_1)  + l(u_3,u_4) +  \\
&& l(u_1,v_4) +l(P_s(C)) + l(w_1,w_2) \\
&\leq& 15 l(P_c(T)) + 6l(v,v_3) +  l(v_4,v_5) + 6l(v_5,w) +
l(v_5,w_1)  +  l(u_1,v_4) +  \\
&& l(P_s(C)) +l(w_1,w_2) +l(w_2,v_1) + l(u_4,u_1) + l(u_1,v_2) \\
&\leq& 15 l(P_c(T)) + 6l(v,v_3) +  6l(v_5,w) +
2l(v_5,w_1)   + l(P_s(C)) + 2l(w_1,w_2) + \\
&& 2l(w_2,v_1) +2l(u_4,u_1) + l(u_1,v_2) + l(v_1,u_1) \\
&\leq& 15 l(P_c(C))  + l(P_s(C)) +2l(u_4,u_1) + l(u_1,v_2) + l(v_1,u_1) \\
&=& 15 l(P_c(C))  + 2l(P_s(C)) +l(u_4,u_1) \\
&=& 15 l(P_c(C))  + 2l(P_s(C)) +l(I(C)) \\
\end{eqnarray*}

 \item [A2] In this case $t(C) = u_4 \not\in S$ and $h(C) = u_1
\not\in S$.

There are three lengths to consider:

\be
\item $l(B_h(C)) + l(M_h(C))$.

Define
\[ B_h(C) = M_t(T) \cup [x \rightarrow v_3 \rightarrow v_2
\rightarrow u_1] \]

 Giving (for all possible locations of $y$)
\begin{eqnarray*}
l(B_h(C)) + l(M_h(C)) &\leq& l(M_t(T)) + l(M_h(T)) + l(x,v_3) +
l(v_3,v_2) + l(v_2,u_1) +  \\
&&  l(u_1,v_4) +l(v_3,v_4) + l(v_4,v_5) +l(v_5,w) \\
\end{eqnarray*}

Using A1 on $T$, the notation we use, Equations (\ref{shortB})
(\ref{DsmalC}), the definition of $P_c(C)$ and the definition of
$P_s(C)$ :
\begin{eqnarray*}
l(B_h(C)) + l(M_h(C)) &\leq& 15l(P_c(T)) + 2l(P_s(T)) + l(I(T)) +
 + l(x,v_3) +l(v_3,v_2) +  \\
 && l(v_2,u_1) +l(u_1,v_4) + l(v_3,v_4) + l(v_4,v_5) +l(v_5,w) \\
 &\leq& 15l(P_c(T)) + 4l(v,v_3) + 4l(v_3,v_4)
+ 4l(v_4,v_5) + 4l(v_5,w) +  \\
&& l(v_3,v_2) +l(v_2,u_1) + l(u_1,v_4) \\
 &\leq& 15l(P_c(T)) + 6l(v,v_3) +  6l(v_5,w) + l(v_3,v_2) + l(v_2,u_1) + l(u_1,v_4) \\
&\leq& 15l(P_c(T)) + 6l(v,v_3) +  6l(v_5,w) + l(v_3,v_2) + l(v_2,u_1)  + \\
&& l(u_1,v_1) + l(v_1,v_5)\\
&\leq& 15l(P_c(C))  + l(v_2,u_1)  + l(u_1,v_1) \\
&=& 15l(P_c(C)) + l(I(C)).
\end{eqnarray*}

\item  $l(B_t(C)) + l(B_h(C))$.

Two possible locations for $x$: \bi
\item $x \in [v,v_6]$.
In this case we can define (for all locations of $y$)
\[ B_t(C) = [u_4 \rightarrow v_1 \rightarrow u_2 \rightarrow v_6
\rightarrow v_3 \rightarrow y] \cup M_h(T)\] Giving:
\begin{eqnarray*}
l(B_t(C)) + l(B_h(C)) &\leq& l(M_t(T)) + l(M_h(T)) + l(x,v_3) +
l(v_3,v_2) + l(v_2,u_1) +  \\
&& l(u_4,v_1) +l(v_1,u_2) + l(u_2,v_6) +l(v_6,v_3) + l(v_3,y)\\
 &\leq& 15l(P_c(T)) +4l(v,v_6) +
5l(v_6,v_3) + 4l(v_3,v_4) + 4l(v_4,v_5) + \\
&& 4l(v_5,w) +l(v_3,v_2) + l(v_2,u_1) + l(u_4,v_1) + l(v_1,u_2) +
l(u_2,v_6)
\end{eqnarray*}
Where the last inequality follows from A1 on $T$.

\item $x \in [v_6,v_3]$.
In this case we can define (for all locations of $y$)
\[ B_t(C) = [u_4 \rightarrow v_1 \rightarrow u_2 \rightarrow v_6] \cup (B_t(T) \backslash [v_6,x])\] Giving:
Giving (for all possible locations of $y$)
\begin{eqnarray*}
l(B_t(C)) + l(B_h(C)) &\leq& l(B_t(T)) + l(M_t(T)) + l(x,v_3) +
l(v_3,v_2) + l(v_2,u_1) + \\
&&  l(u_4,v_1) +l(v_1,u_2) + l(u_2,v_6)\\
 &\leq& 15 l(P_c(T)) + 4l(v,v_6) +
5l(v_6,v_3) + 4l(v_3,v_4) + 4l(v_4,v_5) +\\
&& 4l(v_5,w)  + l(v_3,v_2) + l(v_2,u_1) + l(u_4,v_1) + l(v_1,u_2) +
l(u_2,v_6)
\end{eqnarray*}
Where the last inequality follows from A2 on $T$.

\ei

So for both locations of $x$:
\begin{eqnarray*}
l(B_t(C)) + l(B_h(C)) &\leq& 15 l(P_c(T)) + 4l(v,v_6) +
5l(v_6,v_3) + 4l(v_3,v_4) + 4l(v_4,v_5) +\\
&& 4l(v_5,w)  + l(v_3,v_2) + l(v_2,u_1) + l(u_4,v_1) + l(v_1,u_2) +
l(u_2,v_6).
\end{eqnarray*}

Using Equations (\ref{EsmalC}), (\ref{shortB}), the definition of
$P_c(C)$,(\ref{CsonB}) and the definition of $P_s(C)$:
\begin{eqnarray*}
l(B_t(C)) + l(B_h(C)) &\leq& 15 l(P_c(T)) + 4l(v,v_6) +
5l(v_6,v_3) + 4l(v_3,v_4) + 4l(v_4,v_5) + \\
&& 4l(v_5,w)  +l(v_3,v_2) +l(v_2,u_1) + l(u_4,v_1) + l(v_1,u_2) + l(u_2,v_6) \\
 &\leq& 15l(P_c(T)) + 4l(v,v_6) + 6l(v_6,v_3) + 3l(v_3,v_4) +3l(v_4,v_5)
+\\
&& 4l(v_5,w)  + l(v_3,v_2) + l(v_2,u_1) + l(u_4,v_1) + l(v_1,u_2) +  \\
&& l(v_5,w_1) + l(w_1,w_2) + l(w_2,v_1) + l(P_s(C)) \\
&\leq& 15 l(P_c(T)) + 5.5l(v,v_6) + 7.5l(v_6,v_3) +5.5l(v_5,w)  +
l(v_3,v_2) +\\
&&  l(v_2,u_1) + l(u_4,v_1) +  l(v_1,u_2) + l(v_5,w_1) + l(w_1,w_2) + \\
&&  l(w_2,v_1) +l(P_s(C)) \\
&\leq& 15 l(P_c(C))  + l(v_2,u_1) + l(u_4,v_1) + l(v_1,u_2)  + l(P_s(C)) \\
&\leq& 15 l(P_c(C))  + l(v_2,u_1) + l(u_4,v_1) + 2 l(P_s(C)) \\
&\leq& 15 l(P_c(C))   + 3 l(P_s(C)) \\
\end{eqnarray*}

\item $l(B_t(C)) + l(M_t(C))$.

There are four possible locations for $x$ and $y$ \bi
 \item $x \in [v,v_6]$ and $y \in [v_5,w]$.
 In this case
\begin{eqnarray*}
l(B_t(C)) + l(M_t(C)) &=& l(M_t(T)) + l(M_h(T))+ l(x,v_3) +
2l(v_3,v_4) + 2l(v_4,v_5) + \\
&&  l(v_5,w_1) +l(w_1,u_3) + l(u_3,u_4) +l(u_4,v_1) +l(v_1,u_2) +  \\
&& l(u_2,v_6) + l(v_6,v_3) +l(v_5,w) \\
 &\leq& 15l(P_c(T)) + 4l(v,v_6) + 5l(v_6,v_3) + 5l(v_3,v_4) + 5l(v_4,v_5) +
 \\
&& 4l(v_5,w) + l(v_5,w_1) + l(w_1,u_3) + l(u_3,u_4) + l(u_4,v_1)+\\
&& l(v_1,u_2) +l(u_2,v_6).
\end{eqnarray*}
Where the last inequality follows from A1 on $T$.

\item $x \in [v,v_6]$ and $y \in [v_3,v_5]$.
In this case
\begin{eqnarray*}
l(B_t(C)) + l(M_t(C)) &\leq& l(B_h(T)) + l(M_h(T)) + l(v_5,w_1) +
l(w_1,u_3) + l(u_3,u_4) + \\
&& l(u_4,v_1) + l(v_1,u_2) + l(u_2,v_6)+
l(v_6,v_3) + l(v_3,v_4) + l(v_4,v_5) \\
 &\leq& 15l(P_c(T)) + 4l(v,v_6) +5l(v_6,v_3) + 5l(v_3,v_4) + 5l(v_4,v_5) + \\
 &&4l(v_5,w)+ l(v_5,w_1) + l(w_1,u_3) + l(u_3,u_4) + l(u_4,v_1) +\\
&& l(v_1,u_2) + l(u_2,v_6).
\end{eqnarray*}
Where the last inequality follows from A2 on $T$.

 \item $x \in [v_6,v_3]$ and $y \in [v_5,w]$.
In this case
\begin{eqnarray*}
l(B_t(C)) + l(M_t(C)) &\leq& l(M_t(T)) + l(B_t(T)) + l(x,v_3) +
l(v_3,v_4) + l(v_4,v_5) +  \\
&& l(v_5,w_1) +l(w_1,u_3) + l(u_3,u_4) +
l(u_4,v_1) +  \\
&& l(v_1,u_2) + l(u_2,v_6) \\
 &\leq& 15 l(P_c(T)) + 5l(v,v_6) +4l(v_6,v_3) + 5l(v_3,v_4) + 5l(v_4,v_5) + \\
&& 4l(v_5,y) + l(v_5,w_1) + l(w_1,u_3) + l(u_3,u_4) +l(u_4,v_1) +\\
&& l(v_1,u_2) + l(u_2,v_6).
\end{eqnarray*}
Where the last inequality follows from A2 on $T$.

\item $x \in [v_6,v_3]$  $y \in [v_3,v_5]$.
In this case
\begin{eqnarray*}
l(B_t(C)) + l(M_t(C)) &=& l(B_h(T)) + l(B_t(T))  - l(y,v_5) -
l(v_6,x) + l(v_5,w_1) +  \\
&& l(w_1,u_3) +l(u_3,u_4) + l(u_4,v_1) +l(v_1,u_2) + l(u_2,v_6)\\
&\leq&  15l(P_c(T)) + 6l(v,v_6) + 5l(v_6,v_3) + 5l(v_3,v_4) +
5l(v_4,v_5) +  \\
&& 6l(v_5,w) +l(v_5,w_1) + l(w_1,u_3) + l(u_3,u_4) + l(u_4,v_1)+\\
&& l(v_1,u_2) + l(u_2,v_6). \\
\end{eqnarray*}
Where the last inequality follows from A2 on $T$.
 \ei

 So for all locations of $x$ and $y$
\begin{eqnarray*}
l(B_t(C)) + l(M_t(C)) &\leq& 15l(P_c(T)) + 6l(v,v_6) + 5l(v_6,v_3)
+ 5l(v_3,v_4) + 5l(v_4,v_5) +  \\
&& 6l(v_5,w) +l(v_5,w_1) + l(w_1,u_3) + l(u_3,u_4) + l(u_4,v_1)+\\
&&l(v_1,u_2) + l(u_2,v_6).
\end{eqnarray*}

 Using Equations (\ref{C1sCb}), (\ref{C2sCb}),
(\ref{EsmalC}), (\ref{shortB}), the definition of $P_c(C)$ and the
definition of $P_s(C)$
\begin{eqnarray*}
l(B_t(C)) + l(M_t(C)) &\leq& 15l(P_c(T)) + 6l(v,v_6) + 5l(v_6,v_3)
+ 5l(v_3,v_4) + 5l(v_4,v_5) + \\
&& 6l(v_5,w) + l(v_5,w_1) + l(w_1,u_3)
+ l(u_3,u_4) + l(u_4,v_1) + \\
&&l(v_1,u_2) + l(u_2,v_6) \\
&\leq& 15l(P_c(T)) + 6l(v,v_6) + 5l(v_6,v_3) + 5l(v_3,v_4) +5l(v_4,v_5) + \\
&& 6l(v_5,w) + l(v_5,w_1)+ l(u_3,u_4) + l(u_4,v_1) +
l(u_2,v_6) +\\
&&  l(P_s(C)) + l(w_1,w_2) \\
&\leq& 15l(P_c(T)) + 6l(v,v_6) + 5l(v_6,v_3) + 5l(v_3,v_4) + 5l(v_4,v_5) + \\
&& 6l(v_5,w) + l(v_5,w_1) + l(u_4,v_1) + l(u_2,v_6) +l(P_s(C)) +  \\
&& l(w_1,w_2)  +l(w_2,v_1)+ l(u_4,u_1) +l(u_1,v_2)   \\
&\leq& 15l(P_c(T)) + 6l(v,v_6) + 6l(v_6,v_3) + 4(v_3,v_4) +4l(v_4,v_5) +\\
&&  6l(v_5,w) + 2l(v_5,w_1) + l(u_4,v_1)  +2l(P_s(C)) + 2l(w_1,w_2)  + \\
&& 2l(w_2,v_1)+l(u_4,u_1) +l(u_1,v_2)   \\
&\leq& 15l(P_c(T)) + 8l(v,v_6) + 8l(v_6,v_3)  + 8l(v_5,w) +2l(v_5,w_1) + \\
&& l(u_4,v_1)  +2l(P_s(C)) + 2l(w_1,w_2)  + 2l(w_2,v_1)+ l(u_4,u_1) +\\
&& l(u_1,v_2)   \\
&\leq& 15l(P_c(C))+ l(u_4,v_1)  + 2l(P_s(C)) + l(u_4,u_1) +l(u_1,v_2)   \\
&\leq& 15l(P_c(C)) + 3l(P_s(C)).   \\
\end{eqnarray*}

\ee

\item[A3] In the Lemma assumption $l(C'') < l(C)$ and $l(C_2) < l(C)$. Let $\hat{C}$ be a son
of $C$. If $P_f(\hat{C}) \subset [v_2,u_1]$ since $[v_2,u_1] \subset
P_s(C'')$, $\hat{C}$ should have been a son of $C''$. Similarly if
$P_f(\hat{C}) \subset [v_1,u_4]$ then $\hat{C}$ should have been a
son of $C_2$. Therefore $P_f(C') \cap [u_4,u_1] \neq \phi$.
\item [A4] $J_h(C) = [v_2 \rightarrow u_1]$, $J_t(C) = [u_4 \rightarrow v_1]$ and for every $\hat{C}$ a son of $C$
 $P_f(\hat{C}) \subset P_s(C) = I(C) \cup J_h(C) \cup J_t(C)$.

\item[A5] $M(C)$ is contained in $P_s(C)$,$P_s(C_1)$,$P_s(C_2)$, $S$, $C''$ and $T$ which
are all shorter then $C$.
 \item[A6] $M(C)$ is contained in $P_s(C)$,$P_s(C_1)$,$P_s(C_2)$ $P_s(S), M(S)$ and $C''$,
 so $N(C) = \{ C'',C_1,C_2 \}$.

\end{description}

\end{prf}

\begin{lem}
\label{lemCC1C2DbacD}  Suppose that $l(C) < l(C'')$ then after the
application of DIRECT-ONE  $C''$ satisfies the induction hypotheses.
\end{lem}
\begin{prf}
In this case when DIRECT-ONE is activated on $C$ it changes the
direction of $[u_1,v_2]$ to $v_2 \rightarrow u_1$. Since $l(C)
<l(C'') $ we get that Equation (\ref{CsmalD}) still holds.

Since $l(C_1) < l(C) < l(C'')$:
\begin{equation}
\label{C1sDf} l(P_s(C_1)) + l(w_2,v_1) + l(v_1,v_2) + l(v_5,w_1) <
l(v_2,u_1) + l(u_1,v_4) + l(v_4,v_5)
\end{equation}

Since $l(C_2) < l(C) < l(C'')$:

\begin{equation}
\label{C2sDf} l(P_s(C_2)) +  l(v_1,v_2) + l(v_5,w_1) + l(w_1,w_2) <
l(v_2,u_1) + l(u_1,v_4) + l(v_4,v_5)
\end{equation}

\begin{description}
 \item [A1]
Let $x=t(T),y=h(T)$, and  the end nodes $v,w$ of $P_s(T)$ be ordered
$v,x,y,w$.
 Since $S$ was oriented  backwards Equation (\ref{shortB}) still holds.
Since $N(S)$ contains one more brother $S''$ at the $l_2$ side of
$S$  $x \in [v,v_3]$, and  $y \in [v_3,v_5]$ or $y \in [v_5,w]$.

There are two cases to be considered:
 \bi

\item $y \in [v_5,w]$. Since $I(C'') = [u_1 \rightarrow v_4]$ we can
define:
\[ M_t(C'') = M_t(T) \cup [x \rightarrow v_3 \rightarrow v_4 \rightarrow v_5 \rightarrow w_1 \rightarrow u_3 \rightarrow
u_4 \rightarrow u_1] \ M_h(C'') = [v_4 \rightarrow v_5 \rightarrow
y] \cup M_h(T)\] Using A1 on $T$:
\begin{eqnarray*}
l(M(C'') &=& l(M(T)) + l(x,v_3) + l(v_3,v_4) + 2l(v_4,v_5) +
l(v_5,w_1) + l(w_1,u_3) +  \\
&& l(u_3,u_4) +l(u_4,u_1)  + l(v_5,y) \\
&\leq& 15 l(P_c(T)) + 4l(v,v_3) + 4l(v_3,v_4) + 5l(v_4,v_5) +
4l(v_5,w) +   \\
&& l(v_5,w_1) + l(w_1,u_3) +l(u_3,u_4) + l(u_4,u_1).
\end{eqnarray*}

 \item $y \in [v_3,v_5]$.

 Using A4 on $T$ we can define
\[ M_t(C'') = (B_h(T) \backslash [y,v_5]) \cup [v_5 \rightarrow w_1 \rightarrow u_3 \rightarrow
u_4 \rightarrow u_1] \ M_h(C'') = [ v_4 \rightarrow y] \cup M_h(T)\]

Using A2 on $T$ ($y = h(T) \not\in F(T)$)

\begin{eqnarray*}
l(M(C'')) &\leq& l(B_h(T)) + l(M_h(T)) + l(v_5,w_1) + l(w_1,u_3) +
l(u_3,u_4) + l(u_4,u_1) +l(v_4,v_5) \\
&\leq& 15 l(P_c(T)) + 4l(v,v_3) + 4l(v_3,v_4) + 5l(v_4,v_5) +
4l(v_5,w) + l(v_5,w_1) +  \\
&& l(w_1,u_3) + l(u_3,u_4) + l(u_4,u_1).
\end{eqnarray*}
(same bound as before) \ei

 Using Equations (\ref{shortB}),(\ref{C1sDf}),(\ref{C2sDf}), (\ref{CsmalD}), the definition of $P_c(C'')$,
 the definition of $P_s(C'')$  and the definition of $I(C'')$:
\begin{eqnarray*}
l(M(C'') &\leq& 15 l(P_c(T)) + 6l(v,v_3) + l(v_4,v_5) + 6l(v_5,w) +
l(v_5,w_1) + l(w_1,u_3) + \\
&& l(u_3,u_4) + l(u_4,u_1) \\
 &\leq& 15 l(P_c(T)) + 6l(v,v_3) + 2l(v_4,v_5) +
6l(v_5,w) +  l(u_3,u_4) + l(u_4,u_1)+ \\
&& l(v_2,u_1) + l(u_1,v_4)\\
 &\leq& 15 l(P_c(T)) + 6l(v,v_3) +2l(v_4,v_5) + 6l(v_5,w) +
 l(u_4,u_1)+2l(v_2,u_1) +  \\
&& 2l(u_1,v_4)\\
 &\leq& 15 l(P_c(T)) + 6l(v,v_3) + 3(v_4,v_5) +
6l(v_5,w)  +2l(v_2,u_1) + 3l(u_1,v_4) \\
&\leq& 15 l(P_c(C''))  +2l(v_2,u_1) + 3l(u_1,v_4) \\
&\leq& 15 l(P_c(C''))  +2l(P_s(C'')) + l(u_1,v_4) \\
&=& 15 l(P_c(C''))  +2l(P_s(C'')) + l(I(C'')). \\
\end{eqnarray*}

\item [A2] $N(S)$ contains a brother at the $l_2$ side of $S$. We
marked this brother as $S''$.

Since $S'' \in N(S)$ by A5 on $S$ $l(S'') < l(S)$, since $C$ is a
son of $S$ $l(S) < l(C)$ and by the Lemma assumption Equation
(\ref{EsmDbac}) still holds.

In this case $h(C'') = v_4 \in T$, but $t(C'') = u_1 \not\in T$.

Two possible locations of $x$ to be considered:
 \bi
  \item $x \in [v,v_6]$
\[ B_t(C'') = [u_1 \rightarrow v_2  \rightarrow u_2 \rightarrow v_6 \rightarrow v_3 \rightarrow y] \cup M_h(T) .\]

 Two possible locations of $y$:

 \bi
 \item $y \in [v_5,w]$

 Using A1 on $T$
 \begin{eqnarray*}
 l(M_t(C'')) + l(B_t(C'')) &\leq& l(M_t(T)) + l(M_h(T)) + l(x,v_3) + 2l(v_3,v_4) + 2l(v_4,v_5) + \\
 &&l(v_5,w_1) + l(w_1,u_3) +l(u_3,u_4) + l(u_4,u_1) + l(u_1,v_2) + \\
 && l(v_2,u_2) + l(u_2,v_6) + l(v_5,y) \\
 &\leq& 15 l(P_c(T)) + 4l(v,v_6) + 4l(v_6,v_3) + 5l(v_3,v_4) + 5l(v_4,v_5) + \\
 && 4l(v_5,w) + l(v_5,w_1) + l(w_1,u_3) +l(u_3,u_4) + l(u_4,u_1) + \\
 && l(u_1,v_2) +l(v_2,u_2) + l(u_2,v_6).
 \end{eqnarray*}

 \item $y \in [v_3,v_5]$:

 Using A2 on $T$
\begin{eqnarray*}
 l(M_t(C'')) + l(B_t(C'')) &\leq& l(B_h(T)) + l(M_h(T)) + l(u_1,v_2) +
 l(v_2,u_2) + l(u_2,v_6) + \\
 && l(v_6,v_3) + l(v_3,y) + l(v_5,w_1) + l(w_1,u_3) + l(u_3,u_4) +  \\
&& l(u_4,u_1)\\
 &\leq& 15 l(P_c(T)) + 4l(v,v_6) + 5l(v_6,v_3) + 5l(v_3,v_4) + 5l(v_4,v_5) + \\
 && 4l(v_5,w) +  l(u_1,v_2) + l(v_2,u_2) + l(u_2,v_6) + l(v_5,w_1)+\\
 && l(w_1,u_3) + l(u_3,u_4) + l(u_4,u_1).
\end{eqnarray*}

 \ei

\item $x \in [v_6,v_3]$. In this case we define:
\[ B_t(C'') = [u_1 \rightarrow v_2  \rightarrow u_2 \rightarrow v_6 ]\cup B_t(T) \backslash [v_6,x] .\]

Two possible locations of $y$:
 \bi
 \item $y \in [v_5,w]$. Using A2 on $T$:
\begin{eqnarray*}
l(B_t(C'')) + l(M_t(C'')) &\leq& l(M_t(T)) + l(B_t(T)) + l(x,v_3) +
l(v_3,v_4) + l(v_4,v_5) +  \\
&& l(v_5,w_1)+l(w_1,u_3) +l(u_3,u_4) + l(u_4,u_1) + l(u_1,v_2) + \\
&& l(v_2,u_2) + l(u_2,v_6)\\
 &\leq& 15l(P_c(T)) + 4l(v,v_6) +4l(v_6,v_3) + 5l(v_3,v_4) +
5l(v_4,v_5) + \\
&& 4l(v_5,w) +  l(v_5,w_1)+l(w_1,u_3) + l(u_3,u_4) + l(u_4,u_1) +
\\
&& l(u_1,v_2) + l(v_2,u_2) + l(u_2,v_6).
\end{eqnarray*}

\item $y \in [v_3,v_5]$. Using A2 on $T$
\begin{eqnarray*}
l(B_t(C'')) + l(M_t(C'')) &=& l(B_h(T)) + l(B_t(T))  - l(v_6,x) -
l(y,v_5) + l(u_1,v_2) + \\
&&  l(v_2,u_2) +l(u_2,v_6) + l(v_5,w_1)+l(w_1,u_3) + l(u_3,u_4) +  \\
&& l(u_4,u_1)\\
 &\leq& 15 l(P_c(T)) + 6l(v,v_6) + 5 l(v_6,v_3) +
5l(v_3,v_4) +
5l(v_4,v_5) + \\
&& 6l(v_5,w) + l(u_1,v_2) + l(v_2,u_2) + l(u_2,v_6) +
l(v_5,w_1)+\\
&&  l(w_1,u_3) +l(u_3,u_4) + l(u_4,u_1).
\end{eqnarray*}

\ei \ei

So for all possible locations of $x$ and $y$:
\begin{eqnarray*}
l(B_t(C'')) + l(M_t(C''))  &\leq& 15 l(P_c(T)) + 6l(v,v_6) + 5
l(v_6,v_3) + 5l(v_3,v_4) + 5l(v_4,v_5) + \\
&& 6l(v_5,w) +l(u_1,v_2) + l(v_2,u_2) + l(u_2,v_6) +
l(v_5,w_1)+l(w_1,u_3) + \\
&& l(u_3,u_4) + l(u_4,u_1).
\end{eqnarray*}

Using Equations (\ref{shortB}),(\ref{C1sDf}),(\ref{C2sDf})
(\ref{EsmDbac}), (\ref{CsmalD}), the definition of $P_c(C'')$ and
the definition of $P_s(C'')$
\begin{eqnarray*}
l(B_t(C'')) + l(M_t(C''))  &\leq& 15 l(P_c(T)) + 8l(v,v_6) +
7l(v_6,v_3) + l(v_3,v_4) + l(v_4,v_5) +   \\
&& 8l(v_5,w) +l(u_1,v_2)+l(v_2,u_2) + l(u_2,v_6) +l(v_5,w_1)+l(w_1,u_3) +\\
&& l(u_3,u_4) + l(u_4,u_1)\\
 &\leq& 15 l(P_c(T)) + 8l(v,v_6)+7l(v_6,v_3) + l(v_3,v_4) +
2l(v_4,v_5) +   \\
&& 8l(v_5,w) +2l(u_1,v_2) +l(v_2,u_2) + l(u_2,v_6)+l(u_3,u_4) + \\
&& l(u_4,u_1)+ l(u_1,v_4)\\
 &\leq& 15 l(P_c(T)) + 8l(v,v_6)+7l(v_6,v_3) + l(v_3,v_4) +
3l(v_4,v_5) +  \\
&& 8l(v_5,w) + 3l(u_1,v_2) +l(v_2,u_2) + l(u_2,v_6) + \\
&& l(u_4,u_1)+ 2l(u_1,v_4)\\
&\leq& 15 l(P_c(T)) + 8l(v,v_6) +8l(v_6,v_3) +  3l(v_4,v_5) + 8l(v_5,w) + \\
&& 4l(u_1,v_2) +l(u_4,u_1)+ 3l(u_1,v_4) \\
&\leq& 15 l(P_c(T)) + 8l(v,v_6) +8l(v_6,v_3) +  4l(v_4,v_5) + 8l(v_5,w) +  \\
&& 4l(u_1,v_2) +4l(u_1,v_4) \\
&\leq& 15 l(P_c(C''))  + 4l(u_1,v_2) + 4l(u_1,v_4) \\
&\leq& 15 l(P_c(C''))  + 4l(P_s(C''))
\end{eqnarray*}

 \item[A3] From the Lemma assumption we get that $l(S'') < l(C) <
 l(C'')$. Let $\hat{C}$ be a son of
 $C''$. if $P_f(\hat{C}) \subset ([u_1,v_2] \cup [v_2,v_3])$ then $\hat{C}$
 should have been a son of $C$ or $S''$ but not of $C''$.
 \item [A4] Same proof like in Lemma \ref{lemCDforD}.

\item[A5] $M(C'')$ is contained in $P_s(C'')$, $C$,$C_1$,$C_2$ and $T$
which are all shorter then $C$.
 \item[A6] $M(C)$ is contained in $P_s(C''), P_s(T), M(T)$ and $C$,$C_1$,$C_2$,
 so $N(C'') = \{ C,C_1,C_2 \}$.
\end{description}

\end{prf}

\begin{cor}
\label{uncle'fatb} After activating DIRECT-ONE on a cycle $C$ whose
father is oriented  backwards  and when $C''$ is an uncle, $C$ and
$C''$ will satisfy the induction hypotheses.
\end{cor}

\subsection{Inner-Crossing brothers}
\label{seccross} Suppose that $C_1$ and $C_2$ are inner-crossing
brothers. According to Observation (\ref{croosobs}) they are have a
containing brother $C_3$ such that $C_1,C_2$ are the only maximal
uncontained brothers of $C_3$ and $l(C_3) < min \{ l(C_1),l(C_2)
\}$. Since $C_3$ contains $C_1$ and $C_2$ it will be oriented before
them. Since they are the only maximal uncontained brothers of $C_3$
they will be oriented together in the next level of containment.
$C_1$ and $C_2$ will be oriented in the same direction as $C_3$. In
the next proof we assume that $C_3$ was oriented forwards (and thus
so would $C_1$ and $C_2$). Similar proof applies when $C_3$ is
oriented backwards.

\begin{lem}
Suppose that $C_3$ is oriented forwards. When DIRECT-ONE is applied
on $C_1$, it doesn't change  previous orientation and $C_1$ will be
oriented to satisfy the induction hypotheses.
\end{lem}

\begin{figure}
\begin{center}
\setlength{\unitlength}{0.00069991in}
\begingroup\makeatletter\ifx\SetFigFont\undefined%
\gdef\SetFigFont#1#2#3#4#5{%
  \reset@font\fontsize{#1}{#2pt}%
  \fontfamily{#3}\fontseries{#4}\fontshape{#5}%
  \selectfont}%
\fi\endgroup%
{\renewcommand{\dashlinestretch}{30}
\begin{picture}(6474,2799)(0,-10)
\put(2149.500,-348.553){\arc{2904.239}{3.5589}{5.8659}}
\put(4039.500,-348.553){\arc{2904.239}{3.5589}{5.8659}}
\put(822,240){\blacken\ellipse{90}{90}}
\put(822,240){\ellipse{90}{90}}
\put(2712,240){\blacken\ellipse{90}{90}}
\put(2712,240){\ellipse{90}{90}}
\put(3477,240){\blacken\ellipse{90}{90}}
\put(3477,240){\ellipse{90}{90}}
\put(5367,240){\blacken\ellipse{90}{90}}
\put(5367,240){\ellipse{90}{90}}
\put(3117,735){\blacken\ellipse{90}{90}}
\put(3117,735){\ellipse{90}{90}}
\put(1047,600){\blacken\ellipse{90}{90}}
\put(1047,600){\ellipse{90}{90}}
\path(1182,735)(1272,825)
\blacken\path(1192.450,692.417)(1272.000,825.000)(1139.417,745.450)(1192.450,692.417)
\path(2172,1095)(2307,1095)
\blacken\path(2157.000,1057.500)(2307.000,1095.000)(2157.000,1132.500)(2157.000,1057.500)
\path(4062,1095)(4197,1095)
\blacken\path(4047.000,1057.500)(4197.000,1095.000)(4047.000,1132.500)(4047.000,1057.500)
\path(2937,600)(3027,690)
\blacken\path(2947.450,557.417)(3027.000,690.000)(2894.417,610.450)(2947.450,557.417)
\path(822,240)(2712,240)
\blacken\path(2562.000,202.500)(2712.000,240.000)(2562.000,277.500)(2562.000,202.500)
\path(2712,240)(3477,240)
\blacken\path(3327.000,202.500)(3477.000,240.000)(3327.000,277.500)(3327.000,202.500)
\path(3477,240)(5367,240)
\blacken\path(5217.000,202.500)(5367.000,240.000)(5217.000,277.500)(5217.000,202.500)
\path(5367,240)(5772,240)
\blacken\path(5622.000,202.500)(5772.000,240.000)(5622.000,277.500)(5622.000,202.500)
\blacken\path(4919.417,769.550)(5052.000,690.000)(4972.450,822.583)(4919.417,769.550)
\path(5052,690)(4962,780)
\blacken\path(3190.934,618.640)(3297.000,555.000)(3233.360,661.066)(3190.934,618.640)
\path(3297,555)(3207,645)
\path(12,240)(822,240)
\blacken\path(672.000,202.500)(822.000,240.000)(672.000,277.500)(672.000,202.500)
\path(912,420)(867,375)
\blacken\path(946.550,507.583)(867.000,375.000)(999.583,454.550)(946.550,507.583)
\blacken\path(191.592,1036.953)(147.000,1185.000)(116.677,1033.386)(191.592,1036.953)
\dashline{60.000}(147,1185)(192,240)
\blacken\path(672.000,1147.500)(822.000,1185.000)(672.000,1222.500)(672.000,1147.500)
\dashline{60.000}(822,1185)(147,1185)
\blacken\path(958.153,726.540)(1047.000,600.000)(1028.154,753.464)(958.153,726.540)
\dashline{60.000}(1047,600)(822,1185)
\put(2082,1185){\makebox(0,0)[lb]{\smash{{\SetFigFont{10}{12.0}{\rmdefault}{\mddefault}{\updefault}$C_1$}}}}
\put(3972,1185){\makebox(0,0)[lb]{\smash{{\SetFigFont{10}{12.0}{\rmdefault}{\mddefault}{\updefault}$C_2$}}}}
\put(777,15){\makebox(0,0)[lb]{\smash{{\SetFigFont{10}{12.0}{\rmdefault}{\mddefault}{\updefault}$v_1$}}}}
\put(5322,15){\makebox(0,0)[lb]{\smash{{\SetFigFont{10}{12.0}{\rmdefault}{\mddefault}{\updefault}$v_4$}}}}
\put(2667,15){\makebox(0,0)[lb]{\smash{{\SetFigFont{10}{12.0}{\rmdefault}{\mddefault}{\updefault}$v_3$}}}}
\put(3477,15){\makebox(0,0)[lb]{\smash{{\SetFigFont{10}{12.0}{\rmdefault}{\mddefault}{\updefault}$v_2$}}}}
\put(3072,870){\makebox(0,0)[lb]{\smash{{\SetFigFont{10}{12.0}{\rmdefault}{\mddefault}{\updefault}$u$}}}}
\put(5367,2625){\makebox(0,0)[lb]{\smash{{\SetFigFont{10}{12.0}{\rmdefault}{\mddefault}{\updefault}The cycles}}}}
\put(5367,2400){\makebox(0,0)[lb]{\smash{{\SetFigFont{10}{12.0}{\rmdefault}{\mddefault}{\updefault}$C_1$ $zv_1uv_2v_4z$}}}}
\put(5367,2175){\makebox(0,0)[lb]{\smash{{\SetFigFont{10}{12.0}{\rmdefault}{\mddefault}{\updefault}$C_2$ $zv_1v_3uv_4z$}}}}
\put(5367,1950){\makebox(0,0)[lb]{\smash{{\SetFigFont{10}{12.0}{\rmdefault}{\mddefault}{\updefault}$C_3$ $zv_1uv_4z$}}}}
\end{picture}
}
\end{center}
\caption{Figure for Lemma \ref{leminner}.} \label{innerlemfig}
\end{figure}
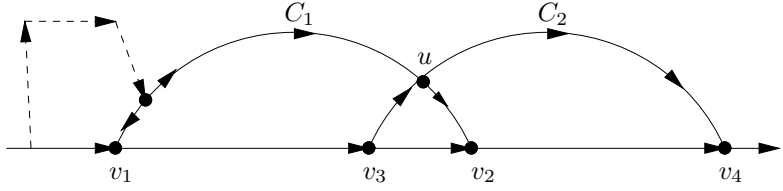

\begin{prf}
\label{leminner} Before DIRECT-ONE was applied on $C_1$, $C_3$
satisfied the induction hypotheses. According to the lemma
assumption  it is oriented forwards. Mark (see Figure
\ref{innerlemfig}):\bi
 \item  $S := F(C_1,C_2)$.
 \item $v_1,v_2:=$ the end nodes of $P_s(C_1)\cap P_s(S)$ (ordered according to direction of
$I(S))$.
 \item $v_3,v_4:=$ the end nodes of $P_s(C_2)\cap P_s(S)$ (ordered according to direction of
$I(S))$.
 \item $u:=(P_s(C_1) \cap P_s(C_2)) \backslash (P_s(S))$.
 \ei
Since $C_1$ and $C_2$ are inner-crossing  $P_s(C_3) =[v_1,u,v_4]$.
Hence $u \in U(C_3)$ and $U(C_1) \subset [u,v_2]$ ($U(C_2) \subset
[v_3,u]$). When DIRECT-ONE is activated on $C_1$ it compares $C_1$
and $C_3$,  since $l(C_3) < l(C_1)$ it doesn't change previous
orientation. DIRECT-ONE  orients $[u \rightarrow v_2]$.  $C_3$ and
any cycle $C$ such that $C_3 \in N(C)$ will not be changed. Since
$C_3$ satisfied the induction hypotheses before DIRECT-ONE was
applied on $C_1$, $|N(C_3)| \leq 3$. For every cycle $C^* \in
N(C_3)$,  $l(C^*) < l(C_3)$. After DIRECT-ONE $N(C_1) = N(C)$ so
$|N(C)| \leq 3$ and for every cycle $C^* \in N(C)$, $l(C^*) < l(C_3)
< l(C_1)$. Hence $C_1$ will be oriented to satisfy the conditions of
lemmas (\ref{lemf}), (\ref{lemtwof}), (\ref{lemthreefs}) or
(\ref{lemfourfs}) and will satisfy the induction hypotheses.
\end{prf}

\end{document}